\documentclass[twocolumn,twocolappendix]{aastex701}

\received{2025 Sep 25}
\revised{2026 Feb 13}
\accepted{2026 Feb 15}
\published{2026 Mar 12: AJ, vol 171, 223}
\reportnum{DOI: 10.3847/1538-3881/ae46a2}
\shorttitle{Cold Debris Disks, Host Star Properties}
\usepackage{epsf,amssymb,amsmath}
\usepackage{wrapfig}
\def\deg{\ifmmode {^\circ}\else {$^\circ$}\fi}
\def\degree{\ifmmode {^\circ}\else {$^\circ$}\fi}
\def\mum{\ifmmode {\rm \,\mu {\rm m}}\else $\rm \,\mu {\rm m}$\fi}

\def\inch{\ifmmode ^{\prime \prime}\else $^{\prime \prime}$\fi}
\def\gs{\ifmmode {{\rm g~s^{-1}}}\else ${\rm g~s^{-1}}$\fi}
\def\msunyr{\ifmmode {M_{\odot}~{\rm yr^{-1}}}\else $M_{\odot}~{\rm yr^{-1}}$\fi}
\def\msun{\ifmmode {M_{\odot}}\else $M_{\odot}$\fi}
\def\rsun{\ifmmode {R_{\odot}}\else $R_{\odot}$\fi}
\def\lsun{\ifmmode {L_{\odot}}\else $L_{\odot}$\fi}
\def\mstar{\ifmmode {M_{\star}}\else $M_{\star}$\fi}
\def\rstar{\ifmmode {R_{\star}}\else $R_{\star}$\fi}
\def\taustar{\ifmmode {\tau_{\star}}\else $\tau_{\star}$\fi}
\def\tstar{\ifmmode {T_{\star}}\else $T_{\star}$\fi}
\def\lstar{\ifmmode {L_{\star}}\else $L_{\star}$\fi}
\def\mwd{\ifmmode {M_{wd}}\else $M_{wd}$\fi}
\def\rwd{\ifmmode {R_{wd}}\else $R_{wd}$\fi}
\def\twd{\ifmmode {T_{wd}}\else $T_{wd}$\fi}
\def\lwd{\ifmmode {L_{wd}}\else $L_{wd}$\fi}
\def\md{\ifmmode {M_d}\else $M_d$\fi}
\def\ld{\ifmmode {L_d}\else $L_d$\fi}
\def\ad{\ifmmode A_d\else $A_d$\fi}
\def\ldlwd{\ifmmode L_d / L_{wd}\else $L_d / L_{wd}$\fi}
\def\ldlstar{\ifmmode L_d / L_\star\else $L_d / L_{\star}$\fi}
\def\lxlstar{\ifmmode L_X / L_\star\else $L_X / L_{\star}$\fi}
\def\qabs{\ifmmode Q_a\else $Q_a$\fi}
\def\qsca{\ifmmode Q_s\else $Q_s$\fi}
\def\qext{\ifmmode Q_e\else $Q_e$\fi}
\def\rearth{\ifmmode {\rm R_{\oplus}}\else $\rm R_{\oplus}$\fi}
\def\mearth{\ifmmode {\rm M_{\oplus}}\else $\rm M_{\oplus}$\fi}
\def\qc{\ifmmode Q_c\else $Q_c$\fi}
\def\qdstar{\ifmmode Q_D^\star\else $Q_D^\star$\fi}
\def\rt{\ifmmode r_t\else $r_t$\fi}
\def\vc{\ifmmode v_c\else $v_c$\fi}
\def\vsqd{\ifmmode v^2 / Q_D^\star\else $v^2 / Q_D^\star$\fi}
\def\kms{\ifmmode {\rm km~s^{-1}}\else $\rm km~s^{-1}$\fi}
\def\masyr{\ifmmode {\rm mas~yr^{-1}}\else $\rm mas~yr^{-1}$\fi}
\def\ms{\ifmmode {\rm m~s^{-1}}\else $\rm m~s^{-1}$\fi}
\def\vrel{\ifmmode v_{rel}\else $v_{rel}$\fi}
\def\mdot{\ifmmode \dot{M}\else $\dot{M}$\fi}
\def\mdotz{\ifmmode \dot{M}_0\else $\dot{M}_0$\fi}
\def\mesc{\ifmmode m_{esc}\else $m_{esc}$\fi}
\def\rmin{\ifmmode r_{min}\else $r_{min}$\fi}
\def\rmax{\ifmmode r_{max}\else $r_{max}$\fi}
\def\xmax{\ifmmode x_{max}\else $x_{max}$\fi}
\def\mmin{\ifmmode m_{min}\else $m_{min}$\fi}
\def\mmax{\ifmmode m_{max}\else $m_{max}$\fi}
\def\rmind{\ifmmode r_{min,d}\else $r_{min,d}$\fi}
\def\rmaxd{\ifmmode r_{max,d}\else $r_{max,d}$\fi}
\def\mmaxd{\ifmmode m_{max,d}\else $m_{max,d}$\fi}
\def\mura{\ifmmode \mu_{RA}\else $\mu_{RA}$\fi}
\def\mudec{\ifmmode \mu_{Dec}\else $\mu_{Dec}$\fi}
\def\vrad{\ifmmode v_{rad}\else $v_{rad}$\fi}
\def\qz{\ifmmode q_{0}\else $q_{0}$\fi}
\def\qi{\ifmmode q_{i}\else $q_{i}$\fi}
\def\ql{\ifmmode q_{l}\else $q_{l}$\fi}
\def\qs{\ifmmode q_{s}\else $q_{s}$\fi}
\def\vhill{\ifmmode v_H\else $r_H$\fi}
\def\rhill{\ifmmode r_H\else $r_H$\fi}
\def\Rhill{\ifmmode R_H\else $R_H$\fi}
\def\rbrk{\ifmmode r_{brk}\else $r_{brk}$\fi}
\def\rdamp{\ifmmode r_{damp}\else $r_{damp}$\fi}
\def\rin{\ifmmode r_{in}\else $r_{in}$\fi}
\def\rout{\ifmmode r_{out}\else $r_{out}$\fi}
\def\tin{\ifmmode t_{in}\else $t_{in}$\fi}
\def\tout{\ifmmode t_{out}\else $t_{out}$\fi}
\def\ain{\ifmmode a_{in}\else $a_{in}$\fi}
\def\aout{\ifmmode a_{out}\else $a_{out}$\fi}
\def\r0{\ifmmode r_{0}\else $r_{0}$\fi}
\def\R0{\ifmmode R_{0}\else $R_{0}$\fi}
\def\m0{\ifmmode m_{0}\else $m_{0}$\fi}
\def\mone{\ifmmode m_{1}\else $m_{1}$\fi}
\def\mtwo{\ifmmode m_{2}\else $m_{2}$\fi}
\def\atwo{\ifmmode a_{2}\else $a_{2}$\fi}
\def\etwo{\ifmmode e_{2}\else $e_{2}$\fi}
\def\mf{\ifmmode m_{f}\else $m_{f}$\fi}
\def\af{\ifmmode a_{f}\else $a_{f}$\fi}
\def\ef{\ifmmode e_{f}\else $e_{f}$\fi}
\def\M0{\ifmmode M_{0}\else $M_{0}$\fi}
\def\amax{\ifmmode a_{max}\else $a_{max}$\fi}
\def\a0{\ifmmode a_{0}\else $a_{0}$\fi}
\def\e0{\ifmmode e_{0}\else $e_{0}$\fi}
\def\v0{\ifmmode v_{0}\else $v_{0}$\fi}
\def\xm{\ifmmode x_{m}\else $x_{m}$\fi}
\def\ag{\ifmmode A_{G}\else $A_{G}$\fi}
\def\ak{\ifmmode A_{K}\else $A_{K}$\fi}
\def\av{\ifmmode A_{V}\else $A_{V}$\fi}
\def\ebmv{\ifmmode E_{B-V}\else $E_{B-V}$\fi}
\def\sigz{\ifmmode \Sigma_0\else $\Sigma_0$\fi}
\def\ergg{\ifmmode {\rm erg~g^{-1}}\else ${\rm erg~g^{-1}}$\fi}
\def\ergs{\ifmmode {\rm erg~s^{-1}}\else ${\rm erg~s^{-1}}$\fi}
\def\gyr{\ifmmode {\rm g~yr^{-1}}\else ${\rm g~yr^{-1}}$\fi}
\def\cms{\ifmmode {\rm cm~s^{-1}}\else ${\rm cm~s^{-1}}$\fi}
\def\gcms{\ifmmode {\rm g~cm^{-2}}\else $\rm g~cm^{-2}$\fi}
\def\gcmc{\ifmmode {\rm g~cm^{-3}}\else $\rm g~cm^{-3}$\fi}
\def\cmsg{\ifmmode {\rm cm^{2}~g^{-1}}\else $\rm cm^{2}~g^{-1}$\fi}
\def\atil{\ifmmode {\tilde{a}}\else $\tilde{a}$\fi}
\def\ttil{\ifmmode {\tilde{t}}\else $\tilde{t}$\fi}
\def\sqrttt{\ifmmode {\tilde{t}^{1/2}}\else $\tilde{t}^{1/2}$\fi}

\def\iras{{\it IRAS}}

\def\bt{$\rm B_T$}
\def\vt{$\rm V_T$}
\def\hip{{\it Hipparcos}}
\def\gaia{{\it Gaia}}
\def\ty2{{\it Tycho-2}}
\def\herschel{{\it Herschel}}
\def\hipparcos{{\it Hipparcos}}
\def\spitz{{\it Spitzer}}

\def\rosat{{\it ROSAT}}
\def\chandra{{\it Chandra}}
\def\xmm{{\it XMM--Newton}}
\def\erosita{{\it SRG/eROSITA}}
\def\2mass{{\it 2MASS}}
\def\wise{{\it WISE}}

\def\vizier{{\it Vizier}}

\def\teff{$\rm T_{eff}$}
\def\logg{log~$g$}
\def\lih{[Li/H]}
\def\feh{[Fe/H]}
\def\vsini{$v$~sin~$i$}
\def\prot{$P_{rot}$}
\def\rhk{$R_{HK}$}
\def\rhkp{$R_{HK}^{\prime}$}
\def\lrhkp{log~$R_{HK}^{\prime}$}
\def\rirt{$R_{IRT}$}

\def\rirtp{$R_{IRT}^{\prime}$}
\def\lrirtp{log~$R_{IRT}^{\prime}$}
\def\fx{$F_X$}
\def\prot{$P_{rot}$}

\def\gaia{{\it Gaia}}
\def\fs{\ifmmode f_S\else $f_S$\fi}
\def\ms{\ifmmode m_S\else $m_S$\fi}
\def\mp{\ifmmode m_P\else $m_P$\fi}
\def\mc{\ifmmode m_C\else $m_C$\fi}
\def\mh{\ifmmode m_H\else $m_H$\fi}
\def\mk{\ifmmode m_K\else $m_K$\fi}
\def\mn{\ifmmode m_N\else $m_N$\fi}
\def\rp{\ifmmode r_P\else $r_P$\fi}
\def\rc{\ifmmode r_C\else $r_C$\fi}
\def\apc{\ifmmode a_{PC}\else $a_{PC}$\fi}
\def\mpc{\ifmmode m_{PC}\else $m_{PC}$\fi}
\def\epc{\ifmmode e_{PC}\else $e_{PC}$\fi}
\def\rgc{\ifmmode r_{GC}\else $r_{GC}$\fi}
\def\qgc{\ifmmode q_{GC}\else $q_{GC}$\fi}
\def\Qgc{\ifmmode Q_{GC}\else $Q_{GC}$\fi}
\def\ag{\ifmmode a_{g}\else $a_{g}$\fi}
\def\eg{\ifmmode e_{g}\else $e_{g}$\fi}
\def\ageo{\ifmmode a_{geo}\else $a_{geo}$\fi}
\def\egeo{\ifmmode e_{geo}\else $e_{geo}$\fi}
\def\efree{\ifmmode e_{free}\else $e_{free}$\fi}
\def\ebin{\ifmmode e_{bin}\else $e_{bin}$\fi}
\def\abin{\ifmmode a_{bin}\else $a_{bin}$\fi}

\def\bvic{BVI$_{\rm C}$}
\def\jhks{JHK$_{\rm s}$}
\def\ncdds{3675}
\def\npi{3652}

\def\nbailer{3349}
\def\npig{3478}
\def\npih{2219}
\def\npigb{3461}
\def\npihb{169}
\def\npmg{3456}
\def\npmh{187}
\def\nvrad{3568}
\def\nspectype{3334}
\def\nb{3496}
\def\nbg{3043}
\def\nv{3551}
\def\nvg{3355}
\def\nic{2686}

\def\nteffc{3675}
\def\nteffg{2664}
\def\nteffl{2890} 
\def\nteffgl{3391}
\def\nteffs{3443}
\def\ngravl{2955}
\def\ngravg{2664}
\def\ngravt{3401}
\def\nfeh{2359}
\def\nfehg{1400}
\def\nfehl{1792}
\def\nfehc{833}
\def\nskym{764}
\def\npstar{182}
\def\nty{2292}
\def\ntyb{1682}
\def\ngaia{3658}
\def\ngfour{149}
\def\ngthree{68}
\def\ngeight{1660}
\def\ngthirteen{486}
\def\nbprp{3642}

\def\ngspec{2754}
\def\ngspecg{2527}
\def\ngspecgn{1195}
\def\nabspec{1235}
\def\nfgkspec{1703}
\def\nmspec{737}
\def\nxray{1996}
\def\nxrayul{130}
\def\nerosita{1157}
\def\nerositap{2562}
\def\nprot{1383}
\def\nvrot{2275}
\def\nvrotl{271}
\def\nlith{1104}
\def\nlithl{197}
\def\nlitht{1301}
\def\nrhk{1132}
\def\nirtp{1187}
\def\nirt{1501}
\def\nirtul{314}
\def\nabrhk{74}
\def\nmrhk{222}
\def\nfgkrhk{1132}
\def\nrosat{1226}
\def\nchandra{225}
\def\nxmm{716}
\def\nerosita{1153}

\def\nhrd{3675}
\def\nfhrd{2042}
\def\nchrd{1633}
\def\napo{791}
\def\ngaiaeso{185}
\def\nrave{262}
\def\ngalah{345}
\def\nsubmm{302}
\def\agri{$g^\prime r^\prime i^\prime$}
\def\sgriz{$g^\prime r^\prime i^\prime z^\prime$}
\def\griz{$g r i z$}

\def\fcl{44\%}
\def\ffld{56\%}

\def\nxa{\textit{Exoplanet Archive}} % wow, Where is \gaia\ defined?

\begin{document}

\title{The Cold Debris Disk Surveys I. Host Star Properties}

%\correspondingauthor{Benjamin C. Bromley}
%\email{bromley@physics.utah.edu}

\author[0000-0003-0214-609X]{Scott J. Kenyon}
\affil{Smithsonian Astrophysical Observatory,
60 Garden Street,
Cambridge, MA 02138, USA}
\email{kenyon@cfa.harvard.edu}

\author[0000-0001-7558-343X]{Benjamin C. Bromley}
\affil{Department of Physics \& Astronomy,
University of Utah, 201 JFB,
Salt Lake City, DC 20006, USA}
\email{bromley@physics.utah.edu}

\author[0000-0002-5758-150X]{Joan R. Najita}
\affiliation{NSF's NOIRLab, 950 N. Cherry Avenue, Tucson, AZ 85719, USA}
\email{joan.najita@noirlab.edu}

\begin{abstract}
We describe the dynamical, photometric, and spectroscopic data available for stars targeted by \spitz\ and \herschel\ to search for cold circumstellar dust emission from debris disks, a collection that we name the Cold Debris Disk Surveys (CDDS). These data include \hipparcos\ and \gaia\ parallaxes, 0.4~\mum\ to 1250~\mum\ photometry, spectral types, effective temperatures, gravities, bolometric luminosities, visual extinctions, metallicities, lithium abundances, rotational periods, projected rotational velocities, the Ca~II HK and IR triplet activity indicators, and X-ray luminosities for \ncdds\ stars. Within this sample, we investigate the frequency of stellar and planetary companions (including potential new proper motion companions); use the data to assign CDDS stars to the field or one of many moving groups, open clusters, or stellar associations; and investigate correlations between stellar activity indicators. In future papers, we plan to explore the magnitude and frequency of infrared excess emission as a function of host star properties; to search for new companions with \gaia; and to examine the evolution of infrared excesses with the ages of stars in clusters and the field.
\end{abstract}

%\begin{multicols}{2}

\section{Introduction}

%an intro!!

%{\it What are debris disks and why do we care?} 

In the core accretion picture of planet formation, planets grow from the accumulation of many planetesimals, small solid bodies a few to hundreds of kilometers in size. Planetesimals that survive the first few Myr of planet formation, avoiding incorporation into large bodies, should later collide and generate 
%debris disks, i.e., 
disks of collisional debris.
%i.e., disks of second-generation dust. 
First detected 40 years ago around the nearby A star Vega \citep{aumann84}, these debris disks have since been detected and studied around thousands of stars spanning a wide range of stellar ages and spectral types \citep{hughes18}.  

%First detected in 1984 around the nearby A star Vega (REF), in the intervening 40 years since that discovery, debris disks have been detected and studied around many (thousands?) stars, with detections spanning a wide range of stellar properties (REFs).  

%Studies of debris disks 
Debris disks lend unique insights into the architectures of planetary systems 
%as a signpost of ongoing planet formation, 
and the planet formation process. 
Substructure in the debris (gaps, rings, and spiral arms) may reveal the presence of planets not seen directly; the detailed properties of the substructure can be used to infer planetary masses and orbits \citep[e.g.,][]{kalas05, golimowski06,follette2017,olofsson2018,pearce2022}.
Debris production has been proposed as a signpost of late-stage icy or rocky planet formation \citep[e.g.,][]{kb2002signpost,kb2004}.
%e.g., terrestrial planet formation is hypothesized to generate detectable debris as the most visible signpost of this process.
More generally, as the product of collisions within the surviving solid mass reservoir of early planet formation, debris disks give us a handle on the efficiency of the planet formation process, i.e., what the first few Myr of planet formation leaves behind.

%\textcolor{red}{Connect to WD pollution?}
Planetesimals (and their collisional debris) that survive beyond a star's main sequence lifetime may eventually fall onto the central white dwarf. The resulting chemical (metal) ``pollution'' of the white dwarf's atmosphere provides clues to the late-time evolution of planetary systems \citep[e.g.,][]{farihi16,veras2021}. The properties of younger debris disks, the presumed evolutionary progenitors of such systems, provide context for understanding where the metals originate and how they are delivered to the star. 

%{\it Previous surveys.}

The debris disk population known today results from numerous studies designed to search for and characterize the properties of debris disks and their occurrence rates. Studies have targeted stars
spanning wide ranges in stellar spectral type (B through M) and age ($\sim 10$ Myr to $\sim 10$ Gyr), including stars 
located in both clusters and the field.  Observations have been carried out 
at a wide range of wavelengths (optical through millimeter), notably in work with the {\it Hubble Space Telescope}, {\it Spitzer Space Telescope}, the {\it Herschel Space Telescope} and the {\it Atacama Large Millimeter Array} \citep[e.g.,][]{matthews2014,chen2020,wyatt2021,manara2023}.

%{\it Examples of what syntheses of data can tell us.}

While many of these studies focus on the interpretation of individual samples, new insights can come from integrating the results of the ensemble.
As one example from our own work, we recently illustrated how the %ensemble of 
debris disk properties studied by {\it Spitzer}/FEPS \citep[e.g.,][]{meyer2006} and {\it Herschel}/DEBRIS/DUNE \citep{matthews2010,eiroa2013,montesinos2016,sibthorpe2018}, when combined with evolutionary models of rings of solids, are consistent with a simple evolutionary picture in which the known population of cold debris disks surrounding FGK stars is a single evolutionary population, the descendants of protoplanetary disks with large, massive rings \citep{nkb2022}.
That is, the $\sim 25$\% of protoplanetary disks with large rings evolve to produce detectable debris throughout their lives, from $\sim 10$ Myr to $\sim 10$ Gyr, producing the known debris disk population. 
%The remaining $\sim 75$\% of FGK stars are born with smaller residual planetesimal reservoirs and live their lives producing much less cold collisional debris (NKB2022).   
%are consistent with about 20\% of FGK stars being born with distant planetesimal belts that generate bright debris; the debris grinds down over time, producing all the known DDs; the other 80\% of stars are born with much lower mass residual mass in planetesimal belts.  

%indicate that the detected DD systems are consistent with about 20\% of FGK stars being born with distant planetesimal belts that generate bright debris; the debris grinds down over time, producing all the known DDs; the other 80\% of stars are born with much lower mass residual mass in planetesimal belts.  

%{\it Good time for a synthesis of samples.}

With new datasets now in hand, the time appears ripe for a new synthesis of the available data on debris disk samples. 
%The time appears ripe for a new synthesis of the available data on debris disk samples acquired to date. 
In particular, {\it Gaia} data make it much easier to obtain stellar distances and ages and to identify cluster members \citep[e.g.,][and references therein]{gaiadr3,hunt2024}.
%With the availability of data from {\it Gaia}, it is now much easier to obtain stellar distances and ages and to identify cluster members. 
%making it feasible to combine  samples from different surveys. 
Additional valuable photometric datasets include Pan-STARRS \citep{chambers2016}, the Skymapper Southern Sky Survey \citep{onken2024}, and WISE \citep{wright2010,cutri2013}.
%Additional datasets are also now available (PANSTARRS, skymapper, WISE).
%\textcolor{cyan}{The availability of {\it Gaia} distances further suggests the possibility of revising the way we estimate ages, through the use isochronal ages rather than the assorted techniques of the past.}
%\textcolor{cyan}{There might even be a revision in the way we estimate ages, if we can use isochronal ages rather than the assorted techniques of the past!}
Along with many individual studies, the large optical spectroscopic surveys APOGEE \citep{abdurrouf2022}, Gaia--ESO \citep{randich2022}, GALAH \citep{buder2021}, LAMOST \citep{xiang2019,wang2020}, and RAVE \citep{steinmetz2020} yield complementary physical properties for debris disk host stars.

To build on these developments, here we combine the available data on existing debris disk samples to enable studies of integrated debris disk populations.
In subsequent papers in this series, we use these data to derive the extent of IR excess among CDDS stars and examine how debris properties depend on host star properties (Paper II), 
%examine possibilities for the detecting new companions with Gaia
search for dynamical companions to CDDS stars using {\it Gaia} data and a relation (if any) of companions to debris (Paper III), and estimate stellar ages and assess the evolution of debris with age (Paper IV). The present paper lays the foundation for these investigations by collecting the needed data. 

To support the study of IR excesses in Papers II and III, here we compile optical and IR photometry for the sample and their parallaxes (\S\ref{sec: sample}). 
We also collect information on host star properties, including companions, stellar and planetary (\S\ref{sec: companions}), and stellar metallicity (\S\ref{sec: host-stars}).

To support the assessment of stellar ages needed for Paper IV, we collect information needed for estimates of isochronal ages (\teff, \logg, \mstar, \rstar), as well as ages from gyrochronology (\vsini, \prot),
activity (\lxlstar, $\log$\,\lrhkp, 
\lrirtp), and Li (\S\ref{sec: host-stars})
We also update the membership of CDDS stars in stellar associations, clusters, and moving groups, relying on recent Gaia-based studies and supplementing these with isochronal constraints (section 6) and stellar activity measures (section 5).

%a good time to examine debris disk observations from space-based missions and to use this analysis to plan for the future.

%{\it What is in this paper?} Discuss the sample and additional photometric data; stellar photospheric properties; chromospheric properties; companions; and membership in associations, clusters, and moving groups. finally, we estimate ages for all of the stars based on the chromospheric properties and the membership data.

%In this paper, we discuss the physical properties of the host stars, any planetary or stellar companions, and membership in associations, clusters, or moving groups. In paper II, we derive the extent of infrared excess from a debris disk and and consider the properties of this debris as a function of host star properties. Paper III discusses estimates of stellar age and the evolution of debris with age. Paper IV examines possibilities for detection of new companions with \gaia. 

We begin with an introduction to the sample of stars, a summary of available photometric data, and a description of parallaxes and proper motions (\S\ref{sec: sample}). We continue with an accounting of stellar and planetary companions (\S\ref{sec: companions}); physical properties of the host stars (\S\ref{sec: host-stars}); membership in stellar associations, clusters, and moving groups (\S\ref{sec: clusters}); and a short discussion of HR diagrams for cluster and field stars (\S\ref{sec: hrdiagram}). We conclude with a discussion (\S\ref{sec: discussion}) and summary (\S\ref{sec: summary}).

\section{The Cold Debris Disk Surveys Sample}
\label{sec: sample}

\subsection{Sample Selection}
To compile a set of stars with strong constraints on cold dust emission, we select published data from \spitz\ and \herschel\ photometric surveys of relatively nearby stars. For \spitz\ programs, we require observations with the MIPS instrument \citep{rieke2004} at 24~$\mum$ (and preferably also at 70~\mum) to isolate cold dust emission from warm dust emission \citep[e.g.,][]{knb2016}. We do not require robust detections at 24--70~\mum. Often, \spitz\ studies report observations of some stars with the IRAC instrument at 3.6--8~\mum\ \citep{fazio2004} along with upper limits at 24~\mum. We include these stars in the compilation. With observations at wavelengths $\gtrsim$ 70~\mum, \herschel\ programs with the PACS \citep{poglitsch2010} and SPIRE \citep{griffin2010} instruments automatically provide constraints on cold dust emission. With {\spitz} and {\herschel} observations, we make no requirement on the level of IR excess emission above the stellar photosphere. 

In collecting targets from \spitz\ and \herschel, we include programs focused on stars with ages $\gtrsim$ 10~Myr. This criterion eliminates \spitz\ surveys of (i) very young, nearby star-forming regions such as Ophiuchus \citep[e.g.,][]{padgett2008}, Perseus \citep[e.g.,][]{jorgensen2006}, and Taurus-Auriga \citep[e.g.,][]{furlan2006}, where the frequency of pre-main sequence stars with optically thick protoplanetary disks is much larger than the frequency of stars with debris disks; and (ii) more distant regions of ongoing star formation such as Orion~B1 and $\sigma$ Ori, which also have a much larger frequency of protoplanetary disks than debris disks \citep[e.g.,][]{hernandez2006,hernandez2007}. This approach allows us to focus on one of the main goals of this study: relating the stellar properties of \spitz\ and \herschel\ targets at shorter wavelengths, $\lesssim$ 2--5~\mum, to the properties of debris disk emission at longer wavelengths, $\gtrsim$ 10--20~\mum. Among stars younger than 5--10~Myr, emission from optically thick protoplanetary disks veils optical emission from the central star and complicates estimates of stellar effective temperature \teff, luminosity \lstar, and surface gravity \logg\ \citep[e.g.,][]{hartigan1989,hartmann2016,herczeg2023,sousa2023}. In these systems, it is also challenging to separate stellar chromospheric and coronal emission from accretion energy \citep[e.g.,][]{calvet1998,manara2013,pittman2025}. We prefer to avoid these extra issues.

Excluding targets younger than $\sim$ 10~Myr also allows us to focus on relating the properties of stellar and planetary companions to the properties of debris disk emission. As discussed below, stellar and planetary companions are common among the \spitz\ and \herschel\ targets, where the lack of protoplanetary disk emission eases detection of radial velocity variability and transits. Adding a set of much younger stars with less robust constraints on planetary and stellar companions would complicate the analysis planned in Paper III. 
 
 To avoid confusion, we include all of the stars in the \spitz\ compilations listed below even when a few stars have ages $\lesssim$ 10~Myr. These additional stars have negligible impact on the analysis described below. The youngest stars among the \spitz\ and \herschel\ targets are within or in the vicinity of Taurus-Auriga \citep[1--3~Myr;][]{luhman2023a}, CrA \citep[4.5~Myr;][]{gennaro2012}, $\epsilon$ Cha \citep[$\sim$5~Myr;][]{dickson2021}, Orion \citep[3--11~Myr;][]{hernandez2023}, Cha-Near \citep[10~Myr;][]{rhee2007}, TW Hya \citep[$\sim$10~Myr;][]{bell2015}, Sco--Cen \citep[10--16~Myr;][]{pecaut2016}, and $\eta$ Cha \citep[11~Myr;][]{bell2015}. Table~\ref{tab: clusters} in \S\ref{sec: clusters} summarizes aspects of the moving groups, open clusters, and stellar associations included in the CDDS. In all, nineteen stars (0.5\% of the CDDS sample) have nominal ages less than 10~Myr and would not be in the compilation without inclusion in one of the publications listed below. 

Published \spitz\ studies have various targeting strategies. Several programs focused on older, apparently single \citep{beichman2005,chen2005b,beichman2006,bryden2006,riaz2006,gautier2007,trill2008,bryden2009,plavchan2009,koerner2010,moor2011a} or binary \citep{trill2007} main sequence stars. Others concentrated on main sequence stars in nearby stellar associations, open clusters, and moving groups \citep[e.g.,][]{gorlova2004,young2004,chen2005a,low2005,rieke2005,stauffer2005,gorlova2006,su2006,gorlova2007,stauffer2007,siegler2007,cieza2008,currie2008,gautier2008,rebull2008,balog2009,carpenter2009b,gaspar2009,sierchio2010,stauffer2010,chen2011,zuckerman2011,chen2012,urban2012}. The FEPS survey \citep{kim2005,meyer2006,silverstone2006,carpenter2008,hillen2008} observed nearby stars from the field and somewhat more distant stars in clusters (e.g., $\alpha$ Per and the Pleiades). 

Among these \spitz\ programs, most centered on FGK stars \citep[or a subset thereof;][]{beichman2005,chen2005a,stauffer2005,beichman2006,bryden2006,meyer2006,trill2007,meyer2008,koerner2010,sierchio2010,chen2011,moor2011a}. Bucking this trend, \citet{rieke2005} concentrated on B-type and A-type stars \citep[see also][]{su2006,chen2012}; \citet{riaz2006} and \citet{gautier2007} selected M-type stars. Others cast a wider net for a broader range of spectral types that could be detected at 24--70~\mum\ with reasonable signal-to-noise \citep{chen2005a,low2005,gorlova2006,siegler2007,rebull2008,trill2008,plavchan2009,stauffer2010,urban2012}. Surveys of associations and clusters probed as far down the main sequence as possible within a predefined area. For regions within 100--250~pc (250--500~pc), these observations yielded good samples of BAFG (BA) stars. 

\citet{chen2014} compiled spectroscopic observations of 571 debris disk candidates acquired with the \spitz\ IRS instrument \citep{houck2004}. Aside from a detailed analysis of spectral energy distributions, this study reports photometric measurements at 13, 24, 31, and 70~\mum\ \citep[for other IRS studies see][]{jura2004,chen2006,chen2009,morales2009}. The host stars of these systems have spectral types of B9--K5 and ages of $\sim$ 10~Myr to $\sim$ 10 Gyr \citep[for an analysis of IRS spectra of stars in the Sco-Cen association, see also][]{jangcondell2015}. Some are field stars; others are members of nearby associations, clusters, and moving groups. 

Compared to \spitz\ programs, \herschel\ surveys of main sequence stars are less extensive. The DEBRIS \citep{matthews2010,thureau2014,kennedy2018,sibthorpe2018,lestrade2025} and DUNES \citep{eiroa2013,montesinos2016} surveys selected targets from the \citet{phillips2010} list of $\sim$ 250 main sequence stars roughly equally divided among AFGKM spectral types. The DUNES program included additional nearby FGK main sequence stars selected from {\it Hipparcos}. Several programs focused on smaller samples of nearby AFGK \citep[e.g.,][]{dodrob2016,morales2016,vican2016,hengst2017} and M-type \citep{tanner2020} main sequence stars. Others focused on the presence of debris disks in binary \citep{yelverton2019} and planetary \citep{yelverton2020} systems. Nearly all \herschel\ programs targeted field stars. \citet[][Tuc--Hor]{donaldson2012}, \citet[][Upper Sco;]{mathews2013}, \citet[][TW Hya]{cieza2013,riviere2013}, \citet[][$\beta$ Pic]{riviere2014}, and \citet[][$\eta$ Cha]{riviere2015} surveyed stars in specific nearby clusters or moving groups; \citet{bonsor2014} focused on planet-hosting subgiants with GK spectral types and masses $\sim$ 1.5~\msun.

In the sections that follow, we discuss the observations included in the compilation and the conclusions we draw about the sample from these observations. In this section, we consider photometric data (infrared, submillimeter, and optical), the \gaia\ catalog, and astrometric data (distances, parallaxes, and proper motions). We then explore the frequency of stellar and binary companions. Stellar companions are common (43\%); planetary companions are much less common (9\%). In \S\ref{sec: host-stars}, we summarize the derived extinction \av, physical properties of stellar photospheres (\teff, \lstar, \logg, metallicity \feh, lithium abundance A(Li), rotational period \prot, and projected rotational velocity \vsini), and observations of stellar activity (X-ray emission and Ca~II H \& K and IR triplet indices). Finally, we conduct a census of CDDS stars that are members of moving groups, open clusters, and stellar associations (\S\ref{sec: clusters}).

Throughout this discussion, we demonstrate that CDDS stars span a broad range of spectral types (B5--M9), luminosity classes (main sequence stars to bright giants) and are typically younger and more active than stars in the vicinity of the Sun. This result is a feature of the \spitz\ and \herschel\ targeting strategies for debris disk stars. Approximately 44\% of CDDS stars are members of a stellar association, open cluster, or moving group with an age $\lesssim$ 1~Gyr (see below: Table~\ref{tab: clusters}, \S\ref{sec: hrdiagram}, and \S\ref{sec: discussion}). The fraction of CDDS stars in a young stellar group is roughly 30 times larger than a random set of stars within 100--200~pc \citep[\S\ref{sec: discussion}; see also][]{gaianearby2021,hunt2024}. With a focus on groups with ages less than $\sim$ 1~Gyr to maximize the probability of detecting cold dust emission, the \spitz\ and \herschel\ surveys guaranteed that targeted stars would have ages considerably less than the age of a typical field star \citep[e.g., $\sim$ 4--10~Gyr;][and references therein]{andrae2023a,fouesneau2023,rathsam2023}. 

\subsection{Infrared and Submillimeter Data}\label{sec: IRdata}

To analyze \spitz\ and \herschel\ data, we compile near-infrared (NIR) JHK data from the {\it Two Micron All Sky Survey} \citep[2MASS;][]{cutri2003,skrutskie2006}, the COBE DIRBE Point Source Catalog \citep{smith2004}, the DENIS catalogs \citep{kimeswenger2004,denis2005,borsenberger2006}, and the Catalog of Infrared Observations \citep[CIO;][]{gezari1993}. We supplement and verify these data with results from other compilations \citep{engels1981,koorneef1983,koen2010}. Although the DIRBE and CIO catalogs are not as deep as 2MASS or DENIS, they provide more accurate photometry for JHK $\lesssim$ 4 where 2MASS observations are saturated. The DENIS catalog of bright stars serves a similar purpose. In the mid-infrared (MIR), the AllWISE catalog \citep{cutri2013} from the Wide-field Infrared Survey Explorer \citep[WISE;][]{wright2010} supplies all-sky photometry at 3.4~\mum\ (W1), 4.6~\mum\ (W2), 12~\mum\ (W3), and 22~\mum\ (W4). Saturation also plagues bright WISE sources with W1--W2 $\lesssim$ 6--7. IRAC measurements and photometry for bright sources at 3.5~\mum\ and at 4.8~\mum\ in the CIO provide good alternatives to W1 and W2 for bright stars. 

At longer wavelengths, data from the {\it Infrared Astronomical Satellite} \citep[IRAS;][]{neugebauer1984} supplement WISE, MIPS, and PACS photometry. We collect measurements from the Point Source \citep{beichman1985} and Faint Source \citep{moshir1992} catalogs. As part of the MSX Astrometric Catalog, \citet{egan1996} match IRAS sources from both catalogs to high quality optical positions. For many of the brightest CDDS stars, good \iras\ measurements at 12--100~\mum\ constrain the spectral energy distributions (SED) for systems without complete \spitz\ or \herschel\ data. Similarly, data from the {\it Infrared Astronomical Mission} \citep[AKARI;][]{murakami2007} and the {\it Infrared Space Observatory} \citep[ISO;][]{kessler1996} complement \spitz\ and \herschel\ data. For CDDS stars, AKARI photometry in the 18~\mum\ and 90~\mum\ bands have much smaller uncertainties than in other AKARI bands. 

The \spitz\ IRS Enhanced Products provides calibrated synthetic photometry at 8, 12, 16, 22, 24, and 25~\mum. The synthetic filters were designed to mimic the IRAC 8~\mum, IRAS 12 and 25~\mum, and the MIPS 24~\mum\ filters. The 22~\mum\ filter is similar to the WISE W4 filter.

Observations at 450--1250~\mum\ generally trace the Rayleigh--Jeans tail of the spectral energy distribution of circumstellar dust \citep[e.g.,][and references therein]{macgregor2016,macgregor2017,marshall2017,holland2017,wilner2018,white2018,sepulveda2019,booth2021,faramaz2021,sullivan2022,matra2025}. Care is required to distinguish dust emission from a background galaxy. Measurements at multiple wavelengths may constrain the slopes of the opacity law and the grain size distribution. To complement the shorter wavelength \spitz\ and \herschel\ observations, we collect data for \nsubmm\ stars at wavelengths 450--9000~\mum\ \citep{greaves1998,greaves2004,liu2004,sheret2004,williams2004,najita2005,williams2006,liseau2008,roccatagliata2009,nilsson2010,mathews2012,carpenter2014,barenfield2016,lieman2016,steele2016,holland2017,marino2017,marino2018,faramaz2019,marino2019,marino2020,matra2020,nederlander2021,pawellek2021,norfolk2021,lovell2021,hales2022,macgregor2022,sullivan2022,marshall2023,roccatagliata2024,carpenter2025,matra2025,marino2026}.

\subsection{Optical Data}
To enable construction of broadband spectral energy distributions (SEDs), we supplement IR data with ground-based optical photometry. For the brightest stars, we rely on \bvic\ data in the Johnson--Cousins system. Starting with observations in the Gliese \citep{gliese1969,bessell1990,gliese1991} and \hip\ \citep{turon1993,perryman1997,vanleeuwen1997} catalogs, we compile data available in original articles, in {\it Vizier} data sets, and in the {\it WEBDA} \citep{mermilliod1995} database \citep{mermilliod1987a,hamdy1993,weis1993,ducati2001,naylor2002,reid2002,sung2002,prisinzano2003,costa2006,lyra2006,stauffer2007,carpenter2008,balog2009,lepine2009,kiraga2012,kiraga2013,kamai2014,kovacs2014,lurie2014,oelkers2016,hendon2018,fritzewski2020}. As in \citet{mermilliod1987b}, we carefully curate these data to avoid associating incorrect measurements with a specific star and to minimize measurement errors. When multiple accurate measurements for a star are available, we adopt the median. 
Of the \nb, \nv, and \nic\ observations respectively available in the \bvic\ filters, \nbg, \nvg, and \nic\ have quoted photometric uncertainties $\lesssim$ 0.05~mag in the original literature. The remainder have uncertainties as large as 0.25~mag.

For fainter CDDS stars without \bvic\ data, we rely on Pan-STARRS \citep{kaiser2010,chambers2016} and the SkyMapper Southern Sky Survey \citep{keller2007,onken2024}. We retrieve {\it grizY} (\sgriz) photometry for \npstar\ (\nskym) CDDS stars with $\delta \ge -30$\deg\ ($\delta \le +30$\deg) in Pan-STARRS DR2 (SkyMapper DR4). For SkyMapper data, we verify targets with associated \gaia, \2mass, and \wise\ identifications. Stars with \sgriz\ $\lesssim$ 9.5 are often saturated \citep{onken2024}. To allow for changing sensitivity across the survey, we check the individual measurements, errors, and colors of each star; observations that fall outside the well-defined main sequence locus of fainter stars are rejected. Pan-STARRS has a much fainter saturation limit, {\it grizY} $\lesssim$ 14 \citep{chambers2016} and therefore yields fewer useful measurements of CDDS stars. As with SkyMapper, we verify saturation for each star using colors and measurement errors. 
To supplement Pan-STARRS and SkyMapper, we include \agri\ data from APASS for stars with V $\approx$ 10.5--15 \citep{hendon2014,hendon2018}. Stars with V $<$ 10 are often saturated; we use colors and measurement errors to eliminate saturated observations.

For bright CDDS stars, the first and second Tycho catalogs from the \hip\ mission provide \bt, \vt, and their associated errors \citep{hog1997,hog2000}. Stars with \vt\ $\lesssim$ 9 have typical uncertainties $\lesssim$ 0.02~mag. Errors grow for fainter stars and are $\gtrsim$ 0.10~mag for \vt\ $\gtrsim$ 11. Nearly 68\% (\nty) of CDDS stars have a Tycho-2 measurement; 73\% (\ntyb) have a high quality measurement with \vt\ $\lesssim$ 9.

\subsection{The \gaia\ DR3 Catalog: Basic Data}

\gaia\ DR3 is a treasure trove \citep{gaiadr3}. We compile \gaia\ identifications along with G, BP, and RP magnitudes for \ngaia\ CDDS stars with G $\gtrsim$ 2. The G-band data have exquisite precision, with uncertainties $\sigma_G \lesssim$ 0.01 mag \citep{riello2021}. Stars with G $<$ 8 (G $\gtrsim$ 13) require a saturation (color) correction that depends on G (BP--RP) as summarized in Appendix C (Fig. 28 and Table 5) of \citet{riello2021}. Both corrections are small, $\Delta G \approx$ $-0.02$ to $+0.01$ for \ngeight\ bright stars and $\Delta G \lesssim$ 0.025~mag for \ngthirteen\ faint stars. Although BP and RP also have issues, most do not apply to CDDS stars. BP and to a lesser extent RP are too bright relative to expectations for G $\gtrsim$ 20 \citet{riello2021}. All CDDS stars have G $\lesssim$ 18.4. For G $\lesssim$ 4, BP and RP must be adjusted for saturation: $\Delta BP \approx -0.6$~mag to zero for \ngfour\ stars with G = 2--4 and $\Delta RP \approx -0.8$~mag to zero for \ngthree\ stars with G = 2--3.45 \citep{riello2021}. We apply these corrections as needed in the following analysis.

To measure the relative quality of the BP and RP fluxes for \gaia\ DR2, \citet{evans2018} defined a dimensionless ratio $C(G) = (I_{BP} + I_{RP}) / I_G$, where $I_i$ is the flux in band $i$ = G, BP, or RP. High quality sources have $C(G) \simeq$ 1; those with much larger $C(G)$ are often extended, variable, or near a much brighter star. For \gaia\ DR3, \citet{riello2021} introduce a correction to $C(G)$ that allows a better measure of the relative quality of BP and RP. We compile $C(G)$ = \verb|phot_bp_rp_excess_factor| from \gaia\ DR3 for each source and calculate $C^*(G) = C(G) - f(x)$, where $x$ = BP--RP and $f(x)$ is a polynomial with coefficients listed in Table 2 of \citet{riello2021}. With correction, high quality sources have $C^*(G) \simeq$ 0; stars with $C^*(G) > 0$ ($C^*(G) < 0$) have more flux in BP and RP (G) than in G (BP and RP). From a set of `gold' stars, \citet{riello2021} also derive the dimensionless scatter in $C^*(G)$ about zero 
\begin{equation}
\sigma(G) = 5.9898 \times 10^{-3} + 8.817481 \times 10^{-12} G^{7.618399}
\end{equation}
High quality measurements typically have a dimensionless noise measure $S_C(G) = C^*(G) / \sigma(G)$ $\lesssim$ 3--5 \citep{riello2021}.

Among the 3500 stars with G $\gtrsim$ 4, 87\% (91\%) have $S_C(G) \lesssim$ 3 (5). Another 2\% (1.7\%) have $S_C(G) \approx$ 5--10 ($S_C(G) \gtrsim$ 100). All stars with $S_C(G) \gtrsim$ 10 are either variables (GJ~866, V1176~Tau, and V1299~Tau) or very close to a much brighter nearby star (e.g., $\alpha$ Cae B, $\psi$ Vel B, and GJ~542.1B). Many of these do not have parallax, proper motion, or stellar photospheric properties within \gaia\ DR3. 

As another check on G, BP, and RP photometry, we compare Johnson--Cousins and 2MASS photometry derived from G, BP, and RP \citep[Tables C.1 and C.2 in][]{riello2021} with the data discussed above. When the difference between \gaia-synthesized data and published optical or near-IR data is smaller than or comparable to the $\sigma$ in the \citet{riello2021} relations, we accept \gaia\ photometry. Otherwise, we flag the observation as contaminated and do not use it in the analysis that follows. This test demonstrates that the conversion from BP and RP to V and I is (reasonably) robust when $S_C(G)$ $\lesssim$ 3 ($S_C(G)$ = 3--5), with a typical uncertainty of 0.03 (0.06) in V and 0.05 (0.10) in I. However, for J and K, the conversion is less accurate. The typical uncertainty is 0.10 (0.14) for J (K) when $S_C(G) \lesssim 3$.  
 
We also collect XP spectra for \ngspec\ stars. Instead of converting these to standard spectra with flux as a function of wavelength, we compile the spectra in the \gaia\ natural system as coefficients of a set of basis vectors \citep[e.g.][]{weiler2023,deangeli2023}. As discussed in \citet{gaiaphot2023}, the `generate' routine enables calculation of broadband and narrowband photometry from XP spectra using project-supplied or user-supplied bandpasses. With fairly complete photometry from some combination of ground-based \bvic, Pan-STARRS, Skymapper, and Tycho, we do not consider `generated' photometry. \citet{bhuang2024} discusses this aspect of \gaia\ data in detail.

\subsection{Photometry Statistics}

\begin{figure}[t]
\begin{center}
\hspace*{0.25cm}
\includegraphics[width=3.25in]{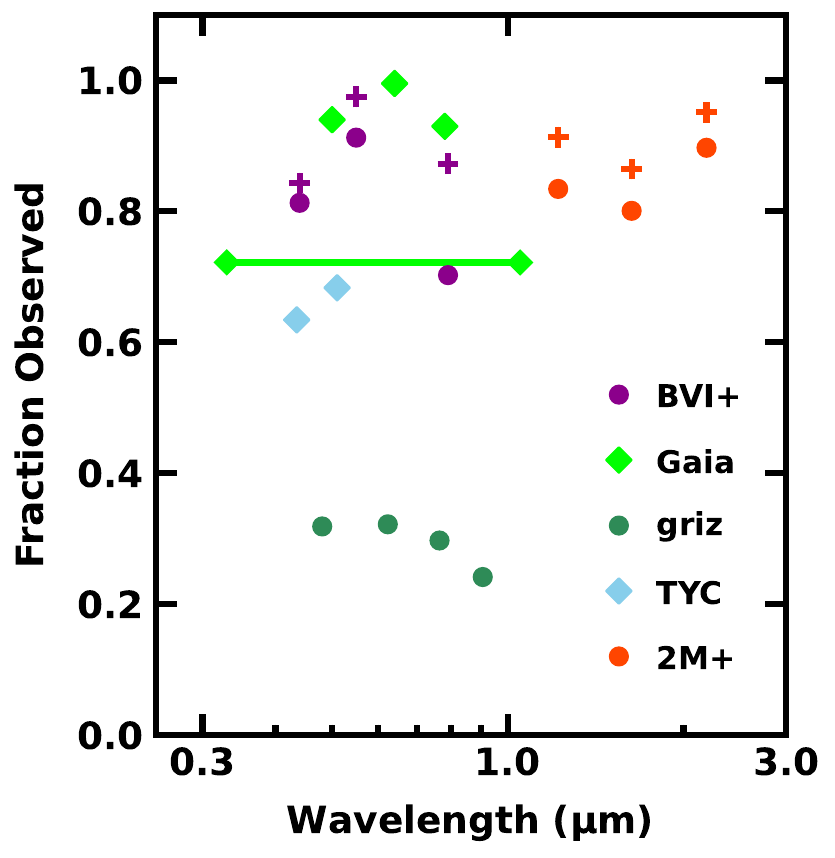}
%\vskip -1ex
\caption{
\label{fig: stats1}
Fraction of CDDS targets with photometric uncertainty $\delta m \lesssim$ 0.05 mag as a function of wavelength for \bvic\ (purple), \gaia\ (light green), Pan-Starrs and SkyMapper {\it griz} (dark green), Tycho (light blue), and 2MASS (orange-red). The purple plus signs illustrate the fraction of stars with (i) \bt\ and B, (ii) $g$, \vt, and V, (iii) $i$\ and I, and (iv) 2MASS JHK plus other JHK data. The light green solid line indicates the fraction of CDDS stars with Gaia low resolution XP spectra. The fraction of targets with high quality optical and near-IR data ranges from $\sim$ 22\% ({\it griz}) to $\sim$ 70\% (Tycho) to $\sim$ 80--90\% (\bvic\ and JHK) to close to 100\% (Gaia).
}
\end{center}
\end{figure}

Fig.~\ref{fig: stats1} and the upper half of Table~\ref{tab: phot-stats} summarize the fraction of CDDS stars with reliable optical and near-IR photometry as a function of wavelength. Table~\ref{tab: phot-stats} lists the fraction of all CDDS stars compiled for each entry. In parentheses, we include the fraction of stars where the photometric uncertainty is $\delta m \lesssim$ 0.05 mag. For the `All Optical' entry, we calculate the fraction of stars with at least one of (i) B or \bt, (ii) V, \vt, or {\it g}, and (iii) I$_{\rm C}$ or {\it i}. For `All JHK', we select stars with the smallest errors among the CIO, COBE, and 2MASS measurements. The `All Optical' and `All JHK' entries illustrate how adding measurements with multiple origins enhances the compilation. 
\begin{deluxetable}{lc}
\tablecolumns{2}
\tablewidth{15cm}
\tabletypesize{\scriptsize}
\tablecaption{Photometry Statistics}
\tablehead{
  \colhead{Passbands} &
  \colhead{$f_{obs}$ ($f_{obs}, err < 0.05$ mag\tablenotemark{\scriptsize \rm 1})}
}
\label{tab: phot-stats}
\startdata
\bt, \vt\ & 0.768,0.781 (0.634,0.683) \\
\bvic\ & 0.951,0.966,0.731 (0.828,0.913,0.703) \\
\griz\ & 0.319,0.322,0.297,0.241 (0.23,0.23,0.23,0.23) \\
All Optical & 0.958,0.995,0.917 (0.828,0.975,0.873) \\
BP, G, RP & 0.995,0.995,0.995 (0.94,0.99,0.93) \\
2MASS JHK & 0.998,0.998,0.998 (0.834,0.801,0.897) \\
All JHK & 1.000,1.000,1.000 (0.913,0.865,0.951) \\
WISE W1--W4 & 0.978,0.980,0.979,0.980 (0.475,0.673,0.823,0.466) \\
IRAC B1--B4 & 0.245,0.275,0.162,0.245 (0.245,0.275,0.155,0.235) \\
All 3.4--4.6 & 0.985,0.981 (0.645,0.706) \\
IRS 8,12,16 & 0.185,0.316,0.316 (0.171,0.267,0.189) \\
IRS 22,24,25 & 0.320,0.320,0.320 (0.155,0.163,0.159) \\
IRAS 12,25,60,100 & 0.379,0.202,0.060,0.009 (0.356,0.123,0.032,0.003) \\
All 12,22,24 & 0.985,0.979,0.803 (0.764,0.461,0.751) \\
MIPS 24,70 & 0.772,0.335 (0.742,0.171)\\
PACS 70,100,160 & 0.077,0.191,0.098 (0.071,0.176,0.069) \\
All 24, 70 & 0.842,0.369 (0.745,0.203) \\
SPIRE 250,350,500 & 0.0207,0.0180,0.0103 (0.0074,0.0011,0.0000) \\
submm 450-1200 & 0.0035,0.0302,0.0199 (0.0035,0.0229,0.0169) \\
\enddata
\tablenotetext{1}{At wavelengths $\gtrsim$ 20~\mum, numbers in parentheses correspond to the fraction of stars with $\delta m \le$ 0.10 mag (MIPS 24~\mum), $\delta m \le$ 0.15 mag (MIPS 70\mum), SNR = 5 (PACS), and SNR = 3 (SPIRE and submm).}
\end{deluxetable}

\begin{figure}[t]
\begin{center}
\includegraphics[width=3.25in]{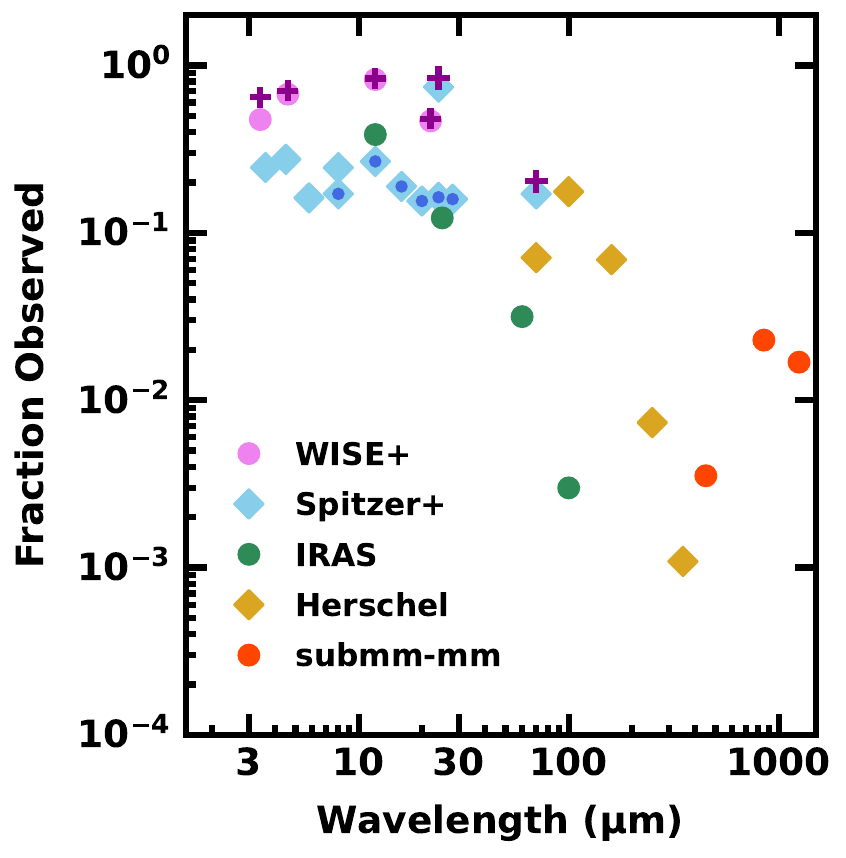}
%\vskip -2ex
\caption{
\label{fig: stats2}
As in Fig.~\ref{fig: stats1} for data from WISE (purple), \spitz\ IRAC  and MIPS (light blue), IRAS (dark green), \herschel\ PACS and SPIRE (gold), and submm--mm (red). The light blue symbols with a dark blue center indicate the fraction of CDDS stars with high quality synthetic photometry at 8--25~\mum\ from \spitz\ IRS. The fraction of observed sources ranges from $\sim$ 1--2\% at the longest wavelengths to $\sim$ 10--20\% for \spitz\ IRAC, \spitz\ IRS, and IRAS 25~\mum\ to $\gtrsim$ 50\% for W3, W4, and MIPS 24~\mum\ measurements. With a fraction observed of zero for 500~\mum\ SPIRE data with SNR $\ge$ 3, this point is not shown.
}
\end{center}
\end{figure}

In Fig.~\ref{fig: stats1}, filled symbols illustrate the fraction of stars with high quality observations at a particular wavelength. Typically, the fraction of stars with high quality photometry ($\delta m \lesssim$ 0.05 mag) is 5\% to 20\% smaller than the complete set of photometry at that wavelength (Table~\ref{tab: phot-stats}). For the `All Optical' set, we plot symbols as plus signs. With this convention, \gaia\ G-band, all K-band data, and the `All Optical' V-band data are nearly complete for the full sample of CDDS stars. The good set of BP and RP data are almost as complete; more than 80\% of CDDS stars have data at B, I$_C$ or {\it i}, J, and H. Although we compile high quality \griz\ data for less than 25\% of the CDDS sample, these data help to make the V-band and I-band data more complete. 
The lower half of Table~\ref{tab: phot-stats} and Fig.~\ref{fig: stats2} summarize the fraction of CDDS stars with measurements at longer wavelengths. With observations for $\sim$ 98\% of the CDDS sample, WISE provides nearly a full set of mid-IR data at 3.4--22~\mum. Most of the 12~\mum\ data have uncertainties $\lesssim$ 0.05~mag; saturation of bright stars at W1 and W2 and low signal-to-noise for faint stars at W4 limits the fraction of high quality measurements. Ground-based L-band and \spitz\ IRAC data nicely supplement the WISE data, increasing the fraction of sources with high quality measurements from 45\% to 65\%. At the other WISE bands, ground-based M-band data along with IRAC, IRS, and IRAS data add a small amount of high quality observations ($\lesssim$ 1\%) to WISE data at 4.6, 12, and 22~\mum.
Moving to the far-IR, sub-mm, and mm, we collect photometry from the \spitz\ MIPS instrument for 78\% (34\%) of the sample at 24~\mum\ (70~\mum). Most of the 24~\mum\ data have errors $\lesssim$ 0.10 mag; roughly half of the 70~\mum\ data have errors $\lesssim$ 0.15 mag. With much lower sensitivity, \iras\ detected $\sim$ 6\% (1\%) of the sample at 60~\mum\ (100~\mum). \herschel\ added observations at 70~\mum\ (8\% of the sample), 100~\mum\ (19\%), and 160~\mum\ (10\%) with PACS. Although \herschel's SPIRE instrument added some high quality observations at longer wavelengths, only $\sim$ 2\% (1\%) of the sample has measurements at 250--350~\mum\ (500~\mum). Combining the set of 22--25~\mum\ data yields data for more than 82\% of the sample. Similarly, roughly 36\% of the sources have at least one 60--70~\mum\ measurement with IRAS, MIPS or PACS. Finally, $\sim$ 2\% to 3\% of CDDS stars have at least one measurement at 450--9000~\mum.        

\subsection{Parallaxes and Proper Motions} \label{sec: plx}

We compile parallaxes $\pi$ and proper motions $\mu$ from \gaia, \hip, and \ty2\ \citep{hog2000,vanleeuwen2007,gaiadr3}. Of the \ncdds\ CDDS stars, \npig\ (\npih) have a \gaia\ (\hip) parallax with $\pi > 0$. Defining the error in the parallax $\delta \pi$, we select observations with the largest signal to noise, $S_{\pi} = \pi / \delta \pi$. This choice yields \npigb\ (\npihb) stars with a \gaia\ (\hip) parallax. For the remaining stars, 25 have no parallax in \gaia, \hip, or SIMBAD; others have measurements from the literature \citep{vanaltena1995,dittman2014,riedel2014,henry2018}. Stars with $\pi \gtrsim$ 3~mas ($\pi \gtrsim$ 10~mas) have a typical $S_{\pi} \gtrsim 50$ ($\gtrsim$ 200). For the few CDDS stars with \gaia\ $\pi \lesssim$ 0.2~mas, the typical $S_{\pi} \lesssim$ 5--10. 

Prioritizing \gaia\ parallaxes over \hip\ parallaxes for any $S_{\pi}$ yields 17 additional (fewer) parallaxes from \gaia\ (\hip). For these 17 stars, the \gaia\ and \hip\ $\pi$ values are similar; however, the \gaia\ errors are larger. Thus, we favor parallax choices with $S_{\pi}$.

Parallaxes from \gaia\ have a mean systematic offset of $-0.021$~mas \citep[e.g.,][]{stassun2016,stassun2018,lindegren2021,groenewegen2021}. Among others, \citet{lindegren2021} and \citet{groenewegen2021} offer methods to correct for the zero-point offset, which depends on brightness, color, sky position, and perhaps other variables. \citet{bailerjones2021} discuss two Bayesian algorithms to treat the zero-point offset and the error distribution for observations with $S_{\pi} \sim 1$. For the ensemble of \nbailer\ CDDS stars with geometric distances $d_g$ and photogeometric distances $d_p$ derived in \citet{bailerjones2021}, $d_g$ and $d_p$ are typically $\sim$ 0.2\% smaller at 100~pc than the simple estimate $d_s = 1000 / \pi$ where $d_s$ is in pc and $\pi$ is in mas. For 99\% of CDDS stars, the ratio of the geometric distance to the simple distance is 0.996--1.007 (0.989--1.013, 0.979--1.024) with median ratios of 0.999 (0.996, 0.992) for distances $d \le$ 100~pc ($d$ = 100-200~pc, 200--400~pc). Beyond 400~pc, the ratio has a much larger range, 0.58--1.27, and a somewhat smaller median of 0.981. Approximately 90\% of CDDS stars without $d_g$ or $d_p$ lie within 100~pc; none have $d_s \gtrsim$ 500~pc.

For the applications discussed here, we generally adopt $d_g$ to place stars on an HR diagram and as a check for astrometric companions and membership in moving groups, open clusters, and stellar associations. In the few cases of a distant star without $d_g$, we apply a correction factor derived from the median ratio $d_g / d_s$ for stars with a range of parallaxes that bracket the parallax of the star without $d_g$. For nearly all stars, the difference in stellar luminosity between estimates based on $d_g$ or $d_s$ is less than 1\%, which is smaller than the overall uncertainty in luminosity. Similarly, the impact of distance uncertainties on astrometric companions and cluster membership is negligible.

Due to uncertain parallaxes, several stars in the CDDS have $d_g$ significantly {\it larger} than $d_s$. In five cases, $\delta$ Cru, $\upsilon$ Phe, $\lambda$ UMa, Cl* NGC 2632 S 169, and HD 110698, we adopt $d_s$. The smaller distances result in more appropriate kinematics for stars clearly in associations or clusters based on their lithium abundances or stellar activity measures and luminosities for pre-main sequence or main sequence stars. Other stars with $d_g \gg d_s$ are giant stars with $d \gtrsim$ 500~pc and have no spectroscopic features that allow favoring one distance over another. Thus, we adopt $d_g$ for these stars. Hereafter, we use $d$ for the adopted distance.

\begin{figure}[t]
\begin{center}
\includegraphics[width=3.25in]{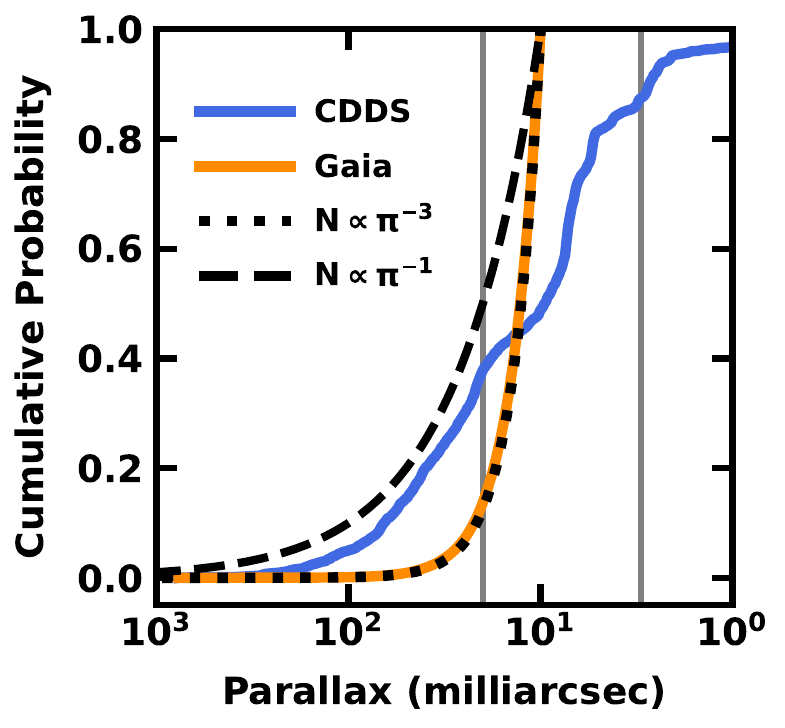}
%\vskip -2ex
\caption{
\label{fig: plx}
Cumulative probability distribution of parallaxes for CDDS stars (blue curve) and stars in the \gaia\ 100 pc sample \citep{gaianearby2021} that lie close to the \citet{pecaut2013} main sequence (orange curve). The set of CDDS (Gaia) stars follows an $n \propto \pi^0$ ($n \propto \pi^{-2}$) volume density distribution, as indicated by the dashed (dotted) black curves.
}
\end{center}
\end{figure}

Fig.~\ref{fig: plx} illustrates the cumulative probability distribution of the \npi\ CDDS stars with measured parallaxes. The distribution follows a trend $N \propto \pi^{-1}$ for $\pi \gtrsim$ 20~mas ($d \lesssim$ 50~pc), rises more slowly to $\pi \sim$ 10 mas ($d \lesssim$ 100~pc), follows the linear trend again to 5~mas, and then levels off. Approximately 40\% of the sample has $\pi \gtrsim$ 20~mas ($d \lesssim$ 50~pc); another 45\% has $\pi \approx$ 5--20~mas. Fewer than 3\% of the sample has $\pi \lesssim$ 1~mas ($d \gtrsim$ 1~kpc).  

To compare CDDS stars with a complete sample, we select stars from the \gaia\ catalog of nearby stars \citep[GCNS;][]{gaianearby2021} that lie within a band 0.25 mag below and 0.75 mag above the \citet{pecaut2013} main sequence locus. The 187749 stars in this sample closely follow the $N \propto \pi^{-3}$ curve expected for an ensemble of stars with $\pi \gtrsim$ 10~mas ($d \lesssim$ 100 pc) (Fig.~\ref{fig: plx}, orange and dot-dashed curves). Relative to the GCNS, the shape of the distribution for CDDS stars is a result of the \spitz\ and \herschel\ selection strategies, which favor nearby, apparently older field stars and stars in younger and somewhat more distant open clusters.

To select proper motions for analysis, we follow procedures outlined above for parallaxes. Choosing by signal-to-noise yields \npmg\ \gaia\ (\npmh\ \hip\ and \ty2, 29 SIMBAD) proper motions. Prioritizing \gaia\ data over \hip\ and \ty2\ data yields twenty-six additional stars with \gaia\ proper motion. As with parallax, \gaia\ data for the extra twenty-six \gaia\ stars in the second example have similar proper motions and lower signal-to-noise than \ty2\ data. Thus, the \ty2\ data provide a better measure of the proper motions for these stars. Although the proper motions of bright stars with $G \lesssim$ 13 show a residual spin relative to fainter stars by $\lesssim$ 0.080~\masyr\ \citep{cantat2021}, this difference has a negligible impact on our analysis. In \S\ref{sec: clusters}, we discuss proper motion data in the context of membership in stellar associations, clusters, and moving groups. 

\subsection{Source Identifications}\label{sec: ids}

To facilitate cross-matching of CDDS sources with other catalogs, we compile a limited set of source identifications. The main source identifier always succeeds in returning information on a particular star from a SIMBAD query. We select other SIMBAD-compatible source identifiers from the BD, CD, CPD, \gaia, Gliese, HD, \hip, 2MASS, \rosat, TESS Input Catalog, Tycho, UCAC4, and WISE catalogs \citep[e.g.,][]{gliese1991,perryman1997,voges1999,voges2000,hog2000,cutri2003,cutri2013,zacharias2013,boller2016,stassun2019,gaiadr3}. We prioritize HD, \hip, and 2MASS source identifiers when available. Aside from Bayer/Flamsteed designations and variable star names from the {\it General Catalog of Variable Stars} \citep[e.g.,][]{samus2017}, we include specific cluster names such as `Cl* Blanco 1 ZS 165' which serve to distinguish stars in various compilations for individual star clusters.

\begin{figure}[t]
\begin{center}
\includegraphics[width=3.25in]{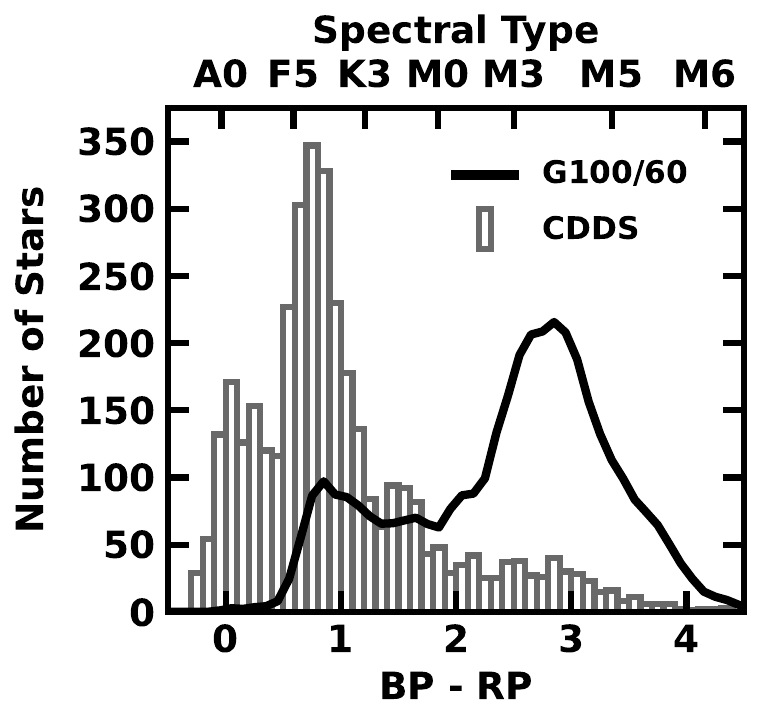}
%\vskip -2ex
\caption{
\label{fig: gaiabprphist}
Frequency of BP--RP colors (lower x-axis labels) and main sequence spectral types \citep[upper x-axis labels;][]{pecaut2013} of stars in the CDDS sample (histogram) and main sequence stars in the Gaia 100 pc sample \citep[solid line;][]{gaianearby2021}. The frequency of Gaia stars has been normalized by a factor of 1/60.  Relative to the Gaia sample of MS stars within 100 pc, the CDDS sample 
has many more AFGK stars and many fewer M-type and later stars.
}
\end{center}
\end{figure}

\subsection{First Look}

To characterize the host stars in the CDDS sample, we compile data on the photospheric and chromospheric properties, stellar and planetary companions, and membership in associations, clusters, and moving groups. We target the spectral type SpT, effective temperature \teff, luminosity \lstar, visual extinction \av, surface gravity \logg, metallicity \feh, projected rotational velocity \vsini, rotational period \prot, the Ca~II H and K emission index \rhk, the Ca~II infrared triplet emission index \rirt, and the X-ray flux \fx. Although we do not consider stellar ages, we illustrate the breadth of the compilation with HR diagrams of cluster and field stars.

To set the stage for the rest of the discussion, Fig.~\ref{fig: gaiabprphist} summarizes the frequency of the observed \gaia\ BP--RP color for the \nbprp\ CDDS stars with BP and RP magnitudes. There is a modest peak at BP--RP $\sim$ -0.1 to 0.25 (A-type stars), a large peak at BP--RP $\sim$ 0.8 (late F-type and G-type stars), and a long tail with larger BP--RP. Middle K-type stars produce the small rise at BP--RP $\sim$ 1.3--1.6. The roughly linear decline in the number of stars for BP--RP $\sim$ 2--4.5 results from cool stars with spectral types M1--M9. The solid line in Fig.~\ref{fig: gaiabprphist} represents the scaled frequency of main sequence stars within 100~pc \citep{gaianearby2021}. This ensemble has a small G star peak at BP--RP $\sim$ 0.8 and a large peak at BP--RP $\approx$ 2.75 (M3--M6 stars). 

Stars in the CDDS are not a fair sample of stars within 100~pc. Within the CDDS sample, 21\% are B-type and A-type stars, 67\% are FGK stars, and 12\% are M-type stars. Among the \gaia\ nearby stars, only 0.3\% have BP--RP colors earlier than an F0 star. There are more FGK stars in the 100~pc sample, 27\%, but this fraction is $\sim$ 2.5 times smaller than the frequency of FGK stars in the CDDS. Finally, slightly more than 72\% of all stars within 100~pc have \gaia\ colors that correspond to M0 stars and later. Only 12\% of CDDS stars have \gaia\ colors as cool as M-type stars; thus the CDDS has a factor of six fewer M-type stars than the nearby star sample.

With the basic character of the CDDS set, we now delve deeper into the properties of the ensemble. We begin with the frequency of companion stars and planets.

\section{Companions}
\label{sec: companions}

% \subsection{Astrometric Binaries}
\subsection{Non-Single Stars in the CDDS Sample}

The interpretation of observations of a star often depends on whether it is a single star or a member of a binary or multiple system. Depending on the separation of companion stars relative to disk dimensions, debris disk emission may also be sensitive to stellar multiplicity. Although it is not possible to have complete knowledge of binaries and multiples within the CDDS, we can compile a set of known binary and multiple systems from published catalogs to quantify stellar properties and the connection between debris disks and their stellar hosts in more detail.  As a starting point, we query the SIMBAD database, which lists object types for each star (the \texttt{otype} field). Object types  \texttt{**}, \texttt{EB*}, and \texttt{SB*} denote double or multiple stars, eclipsing binaries, and spectroscopic binaries. Of our \ncdds\ samples, SIMBAD flags 1334 sources --- close to a third of the total sample --- with one of these object types.

Individual star catalogs referenced by SIMBAD provide more detail.  The Washington Double Star Catalog (WDS), maintained at the U.S.~Naval Observatory \citep{WDS1997}, offers data from visual binaries with separations of a fraction of an arcsecond and larger. 
% bcb added
We accessed the WDS 2020 version through the \textit{Vizier} service. The Sixth Catalog of Orbits of Visual Binary Stars \citep[ORB6;][]{ORB62001a, ORB62001b} offers orbital elements of a subset of WDS stars. The \gaia\ DR3 \texttt{non\_single\_star} (NSS) field for each source in the main catalog references the \texttt{nss\_two\_body\_orbit} table with stellar binary orbit solutions from astrometry, spectroscopy and photometry \citep{gaiamulti2023, gaiabin2023}. We leave consideration of other \gaia\ non-single-star tables, which do not provide complete orbit solutions and which may include stars with substellar companions, for a separate work.  We also consider the Spectroscopic Binary Orbits Ninth Catalog \citep[SB9, 2013 version from \textit{Vizier};][]{SBC2004} that includes orbital periods, as well as the General Catalog of Variable Stars \citep[GCVS, 2020 version (5.1);][]{samus2017} with the periods of eclipsing binaries. 

Table~\ref{tab: non-single-stars} gives a summary of CDDS sources that are cross-listed in these catalogs, a total of 1398 objects. We also provide additional, candidate binary and triple stars that we identified in searches of stars with common proper motion (Appendix, Table~\ref{tab: binary stars}). In both tables, we consider whether a star is in the field or identified as a member of a star cluster where confusion between bound stellar pairs and plentiful interlopers is high. If a CDDS star is in a cluster, then we ignore its membership in the WDS catalog, and do not consider it when searching for common-proper-motion companions.

% \subsection{Spectroscopic Binaries}

\begin{deluxetable}{lrrrrrrr}
%\tablecolumns{2}
\tablewidth{15cm}
%\tabletypesize{\normalsize}
\tablecaption{Stellar Companions\tablenotemark{\scriptsize \rm 1}}
\tablehead{\colhead{Catalog} &
\colhead{**} & \colhead{WDS} &  \colhead{ORB6} &  \colhead{SB9} &  
\colhead{Gaia} &  \colhead{GCVS} & \colhead{not **}
}
\label{tab: non-single-stars}
\startdata % updated for 3675 stars...bcb
 SIMBAD ** &   1385  &    713  &    338  &    270  &    121  &     75  &      0  \\  
 WDS (field) &    --   &    726  &    264  &    104  &     29  &     18  &     13  \\  
 ORB6      &    --   &    --   &    342  &     96  &     20  &     17  &      4  \\  
 SB9       &    --   &    --   &    --   &    270  &     46  &     52  &      0  \\  
 Gaia NSS  &    --   &    --   &    --   &    --   &    144  &     15  &     23  \\  
 GCVS      &    --   &    --   &    --   &    --   &    --   &     90  &     15  \\  
 not **    &    --   &    --   &    --   &    --   &    --   &    --   &   2290  \\  
\enddata
\tablenotetext{1}{SIMBAD ** sources refer to objects listed as belonging to a binary or multiple star systems. The WDS stars include only those that are not part of a star cluster (see \S\ref{sec: clusters} below). The Gaia sources are those with the \texttt{non\_single\_star} flag set to a non-zero value.}
\end{deluxetable}

\begin{figure}[t]
\begin{center}
\includegraphics[width=3.25in]{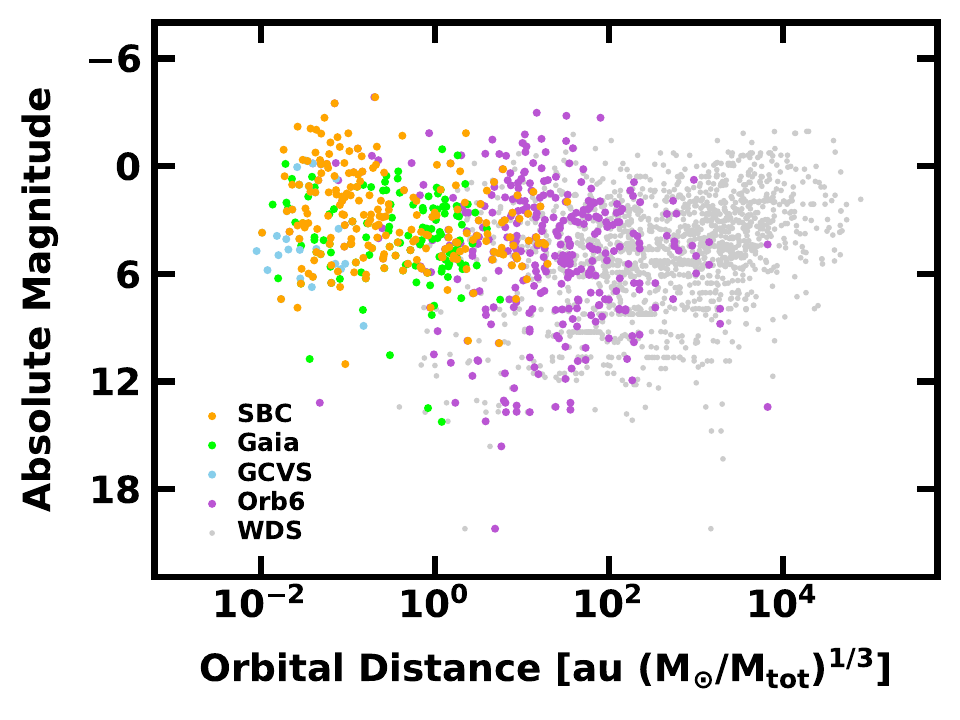}
%\vskip -2ex
\caption{
\label{fig: non-single-stars}
Summary of CDDS stars with companions. Each data point has some uncertainty or ambiguity, as described in the text, but is anticipated to be within a factor of a few of a true or well-determined value. Since each coordinate axis spans orders of magnitude, the broad characteristics of known non-single stars in the CDDS are well represented here.
}
\end{center}
\end{figure}

Figure~\ref{fig: non-single-stars} illustrates the orbital separation between companions of non-single CDDS sources. For sources with reported orbital periods (ORB6, SBC, the \gaia~\texttt{two\_body\_orbit} table, and eclipsing binaries in the GCVS), we estimate orbital distance with Kepler's Third Law assuming that the total stellar mass is that of the Sun. With the range of masses in our catalog spanning $\sim$0.3~\msun\ to $\sim$5~\msun, the true semimajor axis of these sources is within a factor of three of the values in the plot, a small shift compared with the seven orders of magnitude covered by the plot's horizontal axis. The WDS sources, the gray points in the figure, have orbital distance estimated from observed angular separation and parallax. When WDS data provide multiple separations, we use the largest value, ignoring complications from detailed orbital configurations and projection effects. For stars on circular orbits, the points shown are lower limits to the orbital semimajor axis.

As a proxy for more meaningful stellar properties, Figure~\ref{fig: non-single-stars} plots absolute magnitude on the vertical axis. For most sources (1380 of 1398), the \gaia~G~band magnitude is shown, while for the few remaining objects we use the V~band magnitude as listed in SIMBAD. Absolute magnitudes here may be either that of individual stars, or, if unresolved, are from the combined starlight of system components.

Overall Figure~\ref{fig: non-single-stars} shows no particular trends other than the sensitivity of various observational techniques to orbital distance. For example, many visual binaries in WDS data have orbital separations well above 1~au, while eclipsing binaries from the GCVS have orbital separations below 1~au. We have not eliminated or otherwise connected sources that are repeated in individual surveys, leading to some horizontally aligned points in the plot that correspond to the same source. In some cases, individual WDS sources have a half dozen or more ``components,'' leading to the horizontal groups in Figure~\ref{fig: non-single-stars}. These groups emerge from crowded regions of the sky, and do not signify multiple stars that are in isolated, bound systems.

\subsection{Planets}

A principal goal of this work is to understand the connection between debris disks --- signposts of planet formation \citep{kb2002signpost} --- 
% per referee
% and planetary systems. 
and planets.
To identify known planet hosts in the CDDS sample, we use the NASA \nxa\ (NXA). From 
a query to this database (15-Dec 2025), we find that 302 CDDS sources host a total of 471 planets. The planetary system around Trappist~1, with seven confirmed planets, is the most populated \citep{gillon2016}. Four binary stars -- HD~202206 \citep{correia2005}, SCR J0103-5515 \citep[2MASS~J01033563-5515561;][]{delorme2013}, HIP 71865 \citep[b Cen;][]{janson2021}, and HD 143811 \citep{jones2025} -- host a large circumbinary planet.

\begin{figure}[t]
\begin{center}
\includegraphics[width=3.25in]{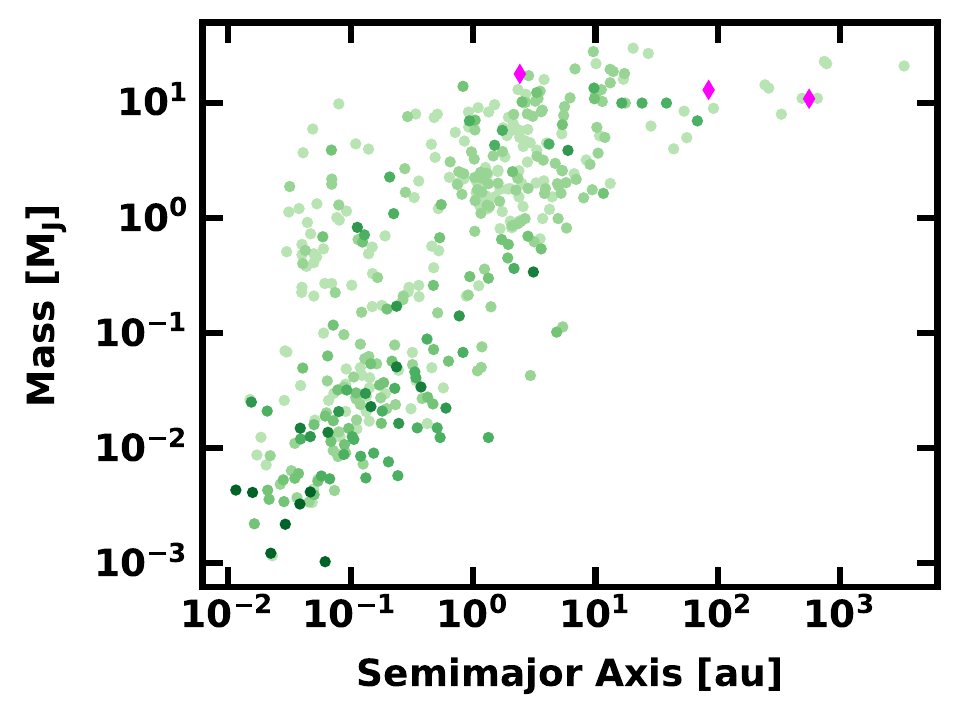}
%\vskip -2ex
\caption{
\label{fig: planets}
The semimajor axis and mass of exoplanets around stars in the CDDS from the NASA \nxa\ database. Here we adopt the database's best estimate for planetary mass (\texttt{bmassj}). The shade of the green circles in the plot correlates with the number of planets around a stellar host. The lightest shade is a single planet around a single host and the shade darkens with increasing number of planets around a single host. The darkest green points (lower left) correspond to the seven planets around Trappist~1. The three magenta diamonds (upper center-right) represent single planets orbiting binary stars.}
\end{center}
\end{figure}

Figure~\ref{fig: planets} reveals broad trends for planet mass and semimajor axis in the CDDS. A wide range of planetary masses, from less than one Earth mass to greater than a Jupiter mass, is found at semimajor axes well inside of 1~au. As the semimajor axis increases, lower mass planets become increasingly scarce. By 10~au, all known CDDS planets are more massive than Jupiter. This trend, seen also in the census of all known exoplanets \citep[figure~1 therein]{zhu2021}, is consistent with the limitations of exoplanet discovery methods. Transits are less sensitive to smaller planets, radial velocity surveys are less sensitive to lower-mass planets, and the two survey types lose sensitivity with increasing orbital distance from the host star. Direct imaging favors larger planets at greater distances. The empty region at small planetary mass and large semimajor axis in the figure (lower right region of the plot) is expected to fill in with space-based gravitational lensing measurements \citep[and references therein]{zhu2021}.

% bcb should revisit this
% sjk revised
Table~\ref{tab: companion stats} summarizes the statistics of stellar and planetary companions as a function of spectral type. Anticipating the results of \S\ref{sec: host-stars} where we derive effective temperatures \teff\ for all CDDS stars, we use the \citet{pecaut2013} relation between \teff\ and spectral type to bin CDDS stars in spectral type. In this tabulation, B-type stars have \teff\ = 9700--31400. The \teff\ ranges for other spectral types are 7220--9700~K (A stars), 5930--7220~K (F stars), 5270--5930~K (G stars), 3850--5270 (K stars), and 2270--3850~K (M stars). 

Among B and A stars, the binary frequency for CDDS stars, 53\% and 41\%, is somewhat smaller than observed in the LAMOST survey \citep[60\% and 48\%;][]{guo2022}. The fraction of binaries for CDDS FGK stars, 41\% to 44\%, is similar to the 43\% fraction in SDSS SEGUE and somewhat larger than the $\sim$ 30\% from LAMOST \citep{gao2014}. A later analysis derives a 42\% binary fraction for GK dwarfs in LAMOST and \gaia\ data \citep{niu2021}. Approximately 46\% of all M dwarfs are in binaries \citep{susemiehl2022}, which is only slightly larger than the fraction among CDDS M dwarfs. Overall, the binary fraction of CDDS stars is similar to the fractions derived as a function of spectral type among larger samples of stars, with perhaps a deficit of binaries among the B and A stars. 

At present, the complete census of exoplanets in the \nxa\ consists of approximately 10\%\ hot Jupiters (mass $m_p$ greater than 100~\mearth\ and semimajor axis $a_p$ less than 0.1~au), 17\%\ cool Jupiters ($m_p > 100$, $a_p \geq 0.1$~au), and 73\%\ Earths and super-Earths at any orbital distance. In contrast, the CDDS planets consist of 6\%\ hot Jupiters, 52\%\ cool Jupiters, and 42\%\ Earths and super-Earths. The differences may reflect multiple factors associated with the selection of stellar hosts.  For example, the CDDS sample contains a significant number of young stars whose atmospheric activity may make Earths and super-Earths harder to detect than Jupiters. These demographics are reflected in Table~\ref{tab: companion stats}, which highlights that the frequency of exoplanets among CDDS stars depends on spectral type as observed in larger samples of main sequence stars \citep[e.g.,][]{ghezzi2018}. We will consider these differences in more detail in Papers II and III.

\begin{deluxetable}{lrrrrrr}
 \tablewidth{15cm}
%\tabletypesize{\normalsize}
% bcb added "Planetary"
\tablecaption{Companion Statistics\tablenotemark{{\scriptsize \rm 1}}}
\tablehead{\colhead{type} &
\colhead{$N_\text{total}$} & \colhead{$N_\text{bin}$} &  \colhead{$f_\text{bin}$} &  \colhead{$N_\text{hosts}$} &  
\colhead{$f_\text{hosts}$} &  \colhead{$f_\text{J}$}
}
\label{tab: companion stats}
\startdata % updated but missing "Adopted  Spect Type" info for the newest stars.
%B &  382 & 201 & 0.526 &   7 & 0.018 & 1.000 \\
%A &  512 & 210 & 0.410 &   4 & 0.008 & 1.000 \\
%F &  868 & 385 & 0.444 &  38 & 0.044 & 0.774 \\
%G &  749 & 322 & 0.430 & 132 & 0.176 & 0.669 \\
%K &  725 & 304 & 0.419 &  86 & 0.119 & 0.554 \\
%M &  451 & 186 & 0.413 &  58 & 0.129 & 0.160 \\
B &  359 & 171 & 0.476 &   5 & 0.014 & 1.000 \\
A &  516 & 207 & 0.401 &   6 & 0.012 & 1.000 \\
F &  860 & 375 & 0.436 &  52 & 0.060 & 0.750 \\
G &  675 & 297 & 0.440 & 110 & 0.163 & 0.618 \\
K &  771 & 262 & 0.340 &  75 & 0.097 & 0.613 \\
M &  492 & 163 & 0.331 &  54 & 0.110 & 0.148 
% two sources have T_eff below the M-star threshold
\enddata
\tablenotetext{1}{The first column lists the spectral type derived from \teff. The next two columns are the total number of stars in each temperature range and the number with binary partners (Table~\ref{tab: non-single-stars}). The planet host information follows, indicating the number of hosts ($N_\text{hosts}$), the fraction of the total of each type known to host planets ($f_\text{hosts}$), and the fraction of hosts with at least one planet that is Jupiter-mass or greater ($f_\text{J}$). A value of $f_\text{J} = 1$ indicates that all stars with planets host at least one giant planet.}
\end{deluxetable}

\section{Host Stars: Data and Analysis}
\label{sec: host-stars}

To characterize the host stars, we collect data from large wide-field surveys and more focused studies of individual stars or clusters. The large surveys usually provide \teff, \logg, \feh, and radial velocity \vrad. Some approaches include \av\ along with \mstar, \rstar, and ages derived from isochrone fitting. More focused efforts add some combination of SpT, A(Li), \prot, \vsini, \lxlstar, \lrhkp\ and \lrirtp. We start with pertinent details for the sources of \teff, \logg, and \feh\ and then discuss observations of these parameters, extinction, and spectral type for CDDS stars. Later, we describe observations of \prot, \vsini, and measures of chromospheric and coronal activity. 

The \gaia\ astrophysical parameters inference system (Apsis) derives stellar parameters from the \gaia\ low resolution XP spectra and the high resolution radial velocity spectrometer (RVS) data \citep{bailerjones2013,creevey2023,andrae2023a,fouesneau2023}. Within the Apsis framework, the General Stellar Parameterizer from Photometry (GSP-Phot) uses an MCMC approach to derive atmospheric parameters -- \teff, \logg, \feh, \rstar, $M_G$, distance, and extinction in the \gaia\ bands ($A_G$, $A_{BP}$, and $A_{RP}$) -- from XP spectra. For an adopted extinction curve, the module constructs a large set of reddened model atmospheres with known \teff, \logg, and \ag, and finds the set of parameters (and their uncertainties) that yields the best match to an individual XP spectrum. The Final Luminosity and Age Estimator (FLAME) derives \lstar\ from \teff\ and distance and then infers the age from fitting model isochrones to \teff, \lstar, and \feh. This analysis yields results for \ngspec\ stars. 

The APOGEE \citep{abdurrouf2022}, \gaia--ESO \citep{randich2022}, GALAH \citep{buder2021}, and RAVE \citep{steinmetz2020} programs are high resolution spectroscopic surveys with a variety of goals. Although APOGEE mainly acquired H-band spectra of red giant stars in the northern hemisphere, many programs surveyed main sequence stars of all spectral types. \gaia--ESO, GALAH, and RAVE collected optical spectra of main sequence and giant stars in the southern hemisphere. \gaia--ESO made a special effort to target stars in young clusters. All of these programs employed a variety of analysis tools to derive stellar parameters from high signal-to-noise spectra. Our reconnaissance yields \teff, \logg, \feh, \vsini, and \vrad\ for \napo\ stars in APOGEE, \ngaiaeso\ in \gaia--ESO, \ngalah\ in GALAH, and \nrave\ in RAVE. 

The optical surveys with the Large sky Area Multi-Object Spectroscopy Telescope (LAMOST) consist of low resolution \citep[e.g.,][]{xiang2019} and medium resolution \citep[e.g.,][]{wang2020} modes in the northern hemisphere. As with the various high resolution surveys, several approaches to analyzing LAMOST data provide estimates of stellar parameters with uncertainties that are a function of the signal-to-noise of the spectra. The low and medium resolution catalogs from LAMOST DR7 yields 391 measurements from 878 spectra.

\citet{queiroz2023} derive a set of stellar parameters -- \mstar, \teff, \logg, distance, \av, and age -- with StarHorse, a Bayesian isochrone fitting tool that analyzes results from stellar spectroscopic surveys. With astrometric data from \gaia\ eDR3 and photometric data from 2MASS, WISE, Pan-STARRS, and SkyMapper, they analyze curated data from APOGEE, \gaia\ RVS, \gaia--ESO, GALAH, LAMOST DR7 low and medium resolution data, RAVE, and SDSS DR12/SEGUE \citep{yanny2009}. Among more than $10^7$ stars, this study includes 1025 estimates of \mstar, \teff, \logg, \av, and distance for 630 CDDS stars. Results for \teff\ and \logg\ agree with those from the original surveys to better than $\sim$ 3\%. The CDDS contains no stars within SDSS DR12/SEGUE.

More focused efforts collect or derive some combination of \teff, \logg, \feh, \vsini, and \vrad, sometimes for BA stars \citep{degeus1989,royer2007,zorec2012,david2015,draper2018} and more often for FGKM stars \citep{preibisch2002,nordstrom2004,holmberg2009,schroeder2009,martinez-arnaiz2010,ramirez2012,rojas-ayala2012,gaidos2014,gomes2014,newton2014,ramirez2014,astudillo2017,cummings2017,hinkel2017,bailey2018,borosaikia2018,reiners2018,houdebine2019,gomes2021,llorente2021,brown2022,meunier2022a,boesgaard2022,reiners2022,marvin2023,mignon2023,rathsam2023,zuo2024,cifuentes2025}. Several include all spectral types \citep{stassun2019}. These approaches use a variety of photometric and spectroscopic methods to derive basic stellar parameters. In general, the most accurate results perform detailed model atmosphere fits to high signal-to-noise high resolution spectra. All told, we collect more than 100,000 measurements for more than 2000 CDDS stars.

{\bf Operations.} \quad
For stars with multiple measurements of the same parameter, we construct the average (for two measurements) or the median (for three or more measurements). Typically, the differences between multiple measurements are similar to the quoted uncertainties. For the uncertainty in the average or median, we adopt the standard deviation of the average or the interquartile range of the median and add measurement uncertainties in quadrature. Where needed, we discuss the impact of combining different studies into a single average or median. 

Many figures below illustrate the dependence of one stellar parameter with respect to another {\it and} the density of stars across the figure. To encode density, we generally adopt the Python `copper' sequence of black (low density), gold (intermediate), and yellow (high density). In some figures, we use the `Blues', `Greens', or `Reds' sequences where darker (lighter) colors indicate higher (lower) density. In all of these plots, we use density to indicate the relative concentrations; however, the density values themselves have no significance. 

In the discussion that follows, we derive \av, \teff, and \lstar\ with optical and infrared photometry. From \gaia\ (the literature), we compile \nteffg\ (\nteffl) measurements of \teff\ derived from model atmosphere fits. Combined, these data yield \teff's for \nteffgl\ stars, $\sim$ 92\% of the sample. Using tables for main sequence \citep{pecaut2013} or giant \citep{alonso1999,vanbelle2021} stars, optical spectral types provide \teff's for \nteffs\ stars, $\sim$ 94\% of the sample. 

With so many \teff\ measurements from model atmosphere fits, we adopt another approach to derive \av, \teff, and \lstar\ from optical and infrared photometry. For each star, we compile a set of colors, \bt--\vt, B--V, V--I, V--K, J--K, G--K, and BP--RP, that minimizes contamination from dusty debris at longer wavelengths. For an adopted \av\ and extinction curve (see below), matching the observed colors to a set of colors tabulated as a function of \teff\ yields an estimate of \teff\ for each observed color. To enable the best match between observed and tabulated colors, we interpolate in (log~\teff, color) space. We then iterate on \av\ to minimize the differences in \teff\ among the available colors and adopt the median \teff\ as the color \teff. The interquartile range in the set of \teff's establishes the uncertainty in the color \teff.

Once \av\ and \teff\ are set, we use bolometric corrections at G, K, and V to derive \lstar.  As discussed below, the typical ratio between the \teff's and \lstar's derived from this method and those from \gaia, the literature, and spectral types for the same star is almost identical to one with an interquartile range similar to the respective errors of the measurements. Thus, we consider this approach successful. 

For most stars, the full set of colors yields the color \teff. The full range in \teff\ estimates is usually less than 5\% of the median \teff; the interquartile range is roughly half that value. Sometimes \bt--\vt\ or V--I are not available; BP--RP may be compromised by light from a nearby brighter star. Still, the remaining colors are sufficient to derive a color \teff\ with an uncertainty of a few per cent. Luminosity estimates follow a similar pattern. The typical uncertainty for \lstar\ derived from G, K, and V is $\lesssim$ 2\%. If one or two measures are unavailable, we sometimes substitute J or make do with one or two measures. Typical uncertainties remain the same.

An advantage of this approach is the ability to consider various tables of stellar properties commonly used in the literature. As the basis for the algorithm, we adopt (i) the \citet{pecaut2013} tables to relate SpT, \teff, and \logg\ to broadband colors, radii, and masses for main sequence stars, (ii) the \citet{worthey2011} tables to relate \teff, \logg, and V-band bolometric corrections to optical--IR colors for giant stars with \logg\ $\lesssim$ 3.5--4 \citep[see also][]{alonso1999}, and the \citet{mucciarelli2021} expressions to derive \teff\ from BP--RP and G--K for main sequence stars and giants. To compare the first two tables, we derive \logg\ for stars in \citet{pecaut2013} from the listed \mstar\ and \rstar. We interpolate in the \citet{worthey2011} tables to match \logg\ and B--V to \citet{pecaut2013} and then calculate the differences in U--B, V--I, J--K, and V--K as a function of B--V. The typical color difference is $\lesssim$ 0.01~mag for all B--V. These differences grow with decreasing \logg; although the differences remain small ($\lesssim$ 0.02 mag) for U--B and V--I, they reach $\sim$ 0.05~mag for J--K and V--K when \logg\ = 3. 

We repeat this exercise comparing \citet{pecaut2013} \teff\ as a function BP--RP and G--K with \teff\ derived from the polynomial relations in \citet{mucciarelli2021}. For the \citet{mucciarelli2021} main sequence star relations, the median of the ratio $\rm T_{eff, m} / T_{eff, p}$ is 0.99 for BP--RP and 1.00 for G--K with inter-quartile ranges of 0.01. Results with the giant relations are identical for BP--RP and somewhat smaller, 0.98, for G--K. We conclude that the combination of the \citet{worthey2011} and \citet{mucciarelli2021} relations accurately relate the set of CDDS colors to \teff\ for giant stars.

As an additional check, we calculate color differences between the \citet{kh1995} colors for main sequence stars and the adopted color tables. Although the differences for the \citet{kh1995} table are somewhat larger, $\sim$ 0.01--0.02 mag for each B--V, the tabulated values for colors involving broadband LMNQ and IRAS fluxes complement the \gaia\ and WISE colors in \citet{pecaut2013}. These comparisons suggest that the choice of table has a small impact on the derived \av, \teff, and \lstar.

{\bf Interstellar Extinction.}\quad
Modern observations now provide detailed maps of interstellar extinction within a few kpc of the Sun \citep[e.g.,][and references therein]{green2019,lallement2019,edenhofer2024,zucker2025}. In addition to revealing the structure of nearby interstellar clouds, these data demonstrate that the Sun lies within a `Local Bubble' having negligible visual extinction within $\sim$ 100--200~pc \citep[see also][]{frisch1983,sfeir1999,welsh2009,linsky2021,oneill2024,oneill2025}. At larger distances, the optical extinction is patchy, $A_V \lesssim$ 2--3 mag. The densest parts of local star-forming regions have much larger extinctions \citep[e.g.,][]{chapman2009,cao2023}.

We adopt a simple approach to derive \av. For ground-based optical and near-IR data, the \citet{fitzpatrick1999} relation is a popular choice for extinction in the standard Johnson BVRIJHKLM filters \citep[see also][]{cardelli1989,gordon2021,decleir2022}. Recent updates analyze a broad collection of survey data and demonstrate that the extinction law in broadband filters depends on the stellar effective temperature \citep[e.g.,][]{wang2019,li2023,rzhang2023,cao2024}. Nearly all CDDS stars have modest reddening, \av\ $\lesssim$ 1~mag, where the corrections due to stellar effective temperature are small. We adopt 
\av\ = 0.76~$A_G$ = 3.1~E(B--V) = 2.67~E(V--I$_C$) = 2.46~E(BP--RP) = 1.37~E(V--J) = 1.12~E(V--K) = 1.98~E(G--J) = 1.51~E(G--K). When we need a reddening for Tycho photometry, we convert \bt--\vt\ to B--V \citep[][and references therein]{bessell2012} and then correct B--V for reddening.

In addition to deriving \av\ from the algorithm for \teff\ and \lstar, we estimate \av\ from comparisons with the full set of color indices -- \bt--\vt, B--V, BP--RP, V--I$_{\rm C}$, V--J, V--K, and G--K -- appropriate for the \gaia, literature, and spectral type \teff's from \citet{pecaut2013} or \citet{worthey2011}. This exercise yields up to seven independent measures of \av\ for each \teff\ measure. The median of these results are indistnguishable for the results with the iterative algorithm described above.

In addition to deriving \av, we include results from the literature quoted above, including \gaia\ and various \spitz\ and \herschel\ studies. For stars inside the Local Bubble, we adopt \av\ = 0 and confirm that the median extinction for stars with $d \le$ 80~pc is zero with an interquartile range of 0.05~mag. For stars beyond 80~pc, detailed analyses of open cluster color-magnitude diagrams yield estimates and uncertainties for the visual extinction. Some results quote \av; others quote the color excess E(B-V). We include the following estimates with those derived from optical and infrared colors, \av\ (mag) = 
0.18 \citep[$\alpha$ Per;][see also Pinsonneault et al 1998]{boyle2023},
0.03 \citep[Blanco-1;][]{jackson2020},
0.005 \citep[Coma;][]{souto2021},
0.0 \citep[$\eta$ Cha;][]{rugel2018},
0.003 \citep[Hyades;][see also Taylor 2006]{brandner2023b},
0.06 \citep[IC~2391;][]{jackson2020}, 
0.03 \citep[IC~2602;][]{jackson2020},
0.19 \citep[NGC~2232;][]{jackson2020},
0.22 \citep[NGC~2422 (M47);][]{kharchenko2005},
0.06 \citep[NGC~2451a;][]{jackson2020},
0.22 \citep[NGC2451~b;][]{jackson2020},
0.40 \citep[NGC~2516;][]{jackson2020},
0.25 \citep[NGC~2547;][]{jackson2020},
0.30 \citep[Ori OB1a and Ori OB1b;][]{briceno2019},
0.12 \citep[Pleiades;][see also Taylor2008]{brandner2023c},
0.08 \citep[Praesepe;][]{taylor2006}, and
0.20 \citep[$\sigma$ Ori;][]{sherry2008}.

Some young stars have K-band excesses, as measured by the ground-based K--L and K--M colors or the K--W1 and K--W2 colors. We derive  \av\ from observations at I$_{\rm C}$ or J, which minimizes the contribution from circumstellar disk emission \citep[e.g.,][]{kh1990,cieza2005}. The results agree with extinction measurements from B--V and BP--RP. Thus, these stars appear to have little optical emission from accretion.
 
Collectively, we have at least 3 \av\ measurements for each CDDS star. Typically, the range in \av\ estimates is 0.1--0.2 mag. Estimates from color excesses usually agree well with the \gaia\ and cluster results. To derive a single estimate and an associated uncertainty, we calculate the median and the interquartile range. Using the average and standard deviation produces identical results. Sometimes, multiple \av\ estimates have a spread of more than 0.5 mag. Using the \teff, \logg, and \feh\ measurements as a guide, we analyze the outliers and choose the \av\ estimate where the derived \teff\ and \lstar\ are most consistent with the atmospheric parameters.

{\bf Effective Temperature.}\quad
The set of literature studies quoted above includes \teff\ measurements for \nteffl\ CDDS stars. To consolidate these data among the various ground-based surveys, we compare the ratio of each pair of \teff\ estimates as a function of \teff. For stars common to any of two surveys, the maximum ratio of 1.02 corresponds to a temperature difference of $\sim$ 100~K for solar-type stars. Among CDDS stars, the temperature differences and ratios show no trend with \teff. These results are similar to those derived for APOGEE, LAMOST and RAVE \citep{nandakumar2017} or for APOGEE, \gaia--ESO, and GALAH \citep{hegedus2023}. By deriving the average or median of multiple measurements, we minimize differences in \teff\ scales across the compilation.

In addition to \teff\ from \gaia\ and the literature, we convert dereddened broadband colors and adopted SpTs into \teff\ estimates as discussed above. Error estimates assume uncertainties of $\pm$1 spectral subclass for SpT (see below) and the measured uncertainties for colors and extinction estimates. Including the \gaia\ and literature estimates, each CDDS star has at least two \teff\ estimates.

\begin{figure}[t]
\begin{center}
\hspace*{-0.7cm}
\includegraphics[width=2.5in]{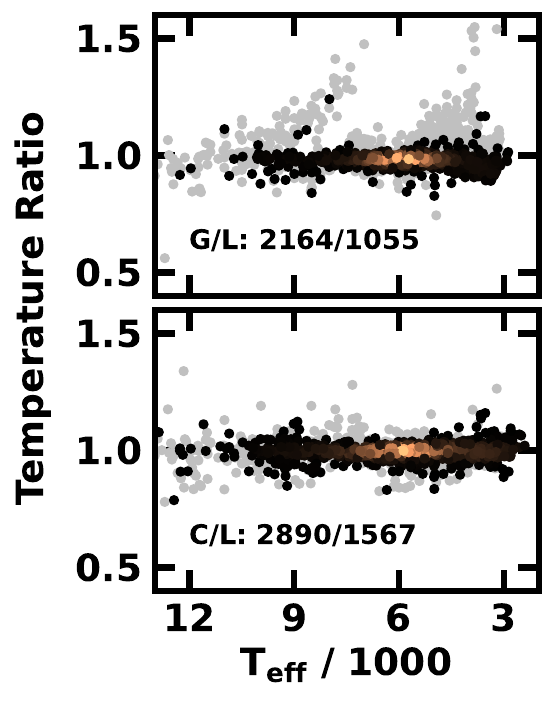}
%\vskip -2ex
\caption{
\label{fig: teffcomp}
Ratio of \teff\ estimates from \gaia\ and the literature (upper panel) and from B--V and the literature (lower panel). In each panel, light grey points plot results for the complete set of common stars. The `copper' sequence of black, gold, and yellow points indicates ratios for stars within 80~pc of the Sun; yellow (black) regions have the highest (lowest) density of points. The numbers in each panel indicate the number of common stars in the full sample and the 80~pc sample. In the upper (lower) panel, the median ratio for all stars is 0.986 (0.998) with an interquartile range of 0.032 (0.027). For the 80~pc sample, the median ratio is 0.983 (upper panel) and 0.999 (lower panel) with an interquartile ranges of 0.018 and 0.021. There is no correlation between the ratios and \teff.
}
\end{center}
\end{figure}

To evaluate the four sets of \teff\ estimates, we calculate the temperature ratio, $r_{T,il}$ = $\rm T_{eff,i} / T_{eff,l}$, where $i$ represents the color, \gaia, or SpT \teff\ estimate and $l$ represents the adopted value from the ground-based literature. With \nteffc\ (color), \nteffg\ (\gaia), \nteffl\ (ground-based literature), and \nteffs\ (spectral types) estimates for \teff, we use the average (or median) ratio and the variance (inter-quartile range) to assess the accuracy of the estimates. 

Figure~\ref{fig: teffcomp} shows temperature ratios for common stars in the \gaia--literature (upper panel) and the color--literature (lower panel) samples. As described in the caption, light (dark) colors indicate results for all CDDS stars (stars within 80~pc). The ratios for the full and 80~pc samples cluster around unity. For the full sample, the median ratios are 0.986 (upper panel) and 1.003 (lower panel) with inter-quartile ranges of 0.031. The spread about unity for the 80~pc samples is smaller: median ratios of 0.983 (upper panel) and 0.998 (lower panel) with interquartile ranges of 0.017 and 0.021. As in \citet{nandakumar2017} and \citet{hegedus2023}, we look for trends in the ratio with \teff; there are none. 

Extinction causes many of the outliers in Figure~\ref{fig: teffcomp}. Sometimes, the \gaia\ analysis appears to overestimate the extinction, which results in higher \teff\ and $r_{T,gl} \approx$ 1.1--1.5  for AF and KM stars. With no extinction correction, the larger B--V and other colors of reddened stars systematically reduce \teff's and generate a set of stars with $r_{T,cl} \approx$ 0.5--0.9. As shown in the lower panel of Figure~\ref{fig: teffcomp}, using reddening-corrected colors eliminates nearly all outliers.

The set of \teff's derived from SpT provide a good substitute for other options. The median \teff\ ratio is 1.000 (1.001) for the full (80~pc) sample. The interquartile ranges of 0.024 (all stars) and 0.021 (stars within 80~pc) are nearly identical to those for the Gaia and color samples. The spectral types derived from \gaia\ XP spectra are responsible for the excellent agreement between the literature and spectral type \teff\ measures. Using spectral types from the literature, the median temperature ratios are still very close to unity, but the interquartile ranges are a factor of two larger. Thus, the spectral types discussed below from \gaia\ spectra are somewhat more robust than those from the literature.

Finally, we compare color \teff's with those derived from \gaia\ and spectral types. For the full sample, color \teff's are typically 1\% (0.5\%) smaller (larger) than \gaia\ (spectral type) \teff's. The differences between color \teff's and spectral type \teff's are independent of distance. At large distances, $\gtrsim$ 80~pc, the color and \gaia\ \teff's are typically identical. The inter-quartile ranges of these comparisons are similar, 0.01--0.02.

{\bf Spectral Types.}\quad
Most of the \spitz\ and \herschel\ studies quoted above draw spectral types from the literature \citep[e.g.,][]{cannon1924,houk1978,hoffleit1991,hamdy1993,gray2003,wright2003,abt2004,gray2006}. To construct a more uniform set of types, we compile types from more recent analyses \citep{zorec2012,david2015,reiners2018,gomes2021,llorente2021,brown2022,meunier2022a}. Most classifications date from the HD catalog and tend to emphasize subclasses 0, 2, 5, and 8. Many are more recent. When a star has more than one classification, we adopt the average or median of multiple measurements. This exercise yields spectral types for \nspectype\ CDDS stars.

To modernize the older classifications, we analyze \ngspec\ \gaia\ XP spectra. These spectra come in two forms: a set of 55 coefficients for each XP spectrum or a more traditional set of fluxes as a function of wavelength. Converting the natural system of 55 coefficients to a fluxed spectrum introduces features that compromise analysis \citep[e.g.,][]{deangeli2023,montegriffo2023,andrae2023a,bhuang2024}. Thus, we analyze the coefficients.

As outlined in \citet{deangeli2023}, the first few coefficients of the BP and RP spectra contain sufficient information to classify most stellar types. We first normalize the XP spectra. Defining $S_i = \Sigma_{n=0}^{54} c_i$, the normalized coefficients for the blue ($b_i$) and red ($r_i$) spectra are
\begin{equation}
\label{eq: gaia1}
b_i, r_i = c_i / S_i
\end{equation}
where $c_i$ is the set of coefficients for the BP or RP spectra supplied by the \gaia\ project. We then define a \gaia\ spectral index from the first six normalized coefficients for the blue spectra:
\begin{equation}
\label{eq: gaia2}
I_B = b_0 - b_1 + b_2 - b_3 + b_4 - b_5 ~ .
\end{equation}
Even-numbered (odd-numbered) coefficients of the XP spectra are usually positive (negative). To add the information contained in these coefficients, the signs in the expression for $I_B$ alternate. 

\begin{figure}[t]
\begin{center}
\hspace*{-0.25cm}
\includegraphics[width=3.5in]{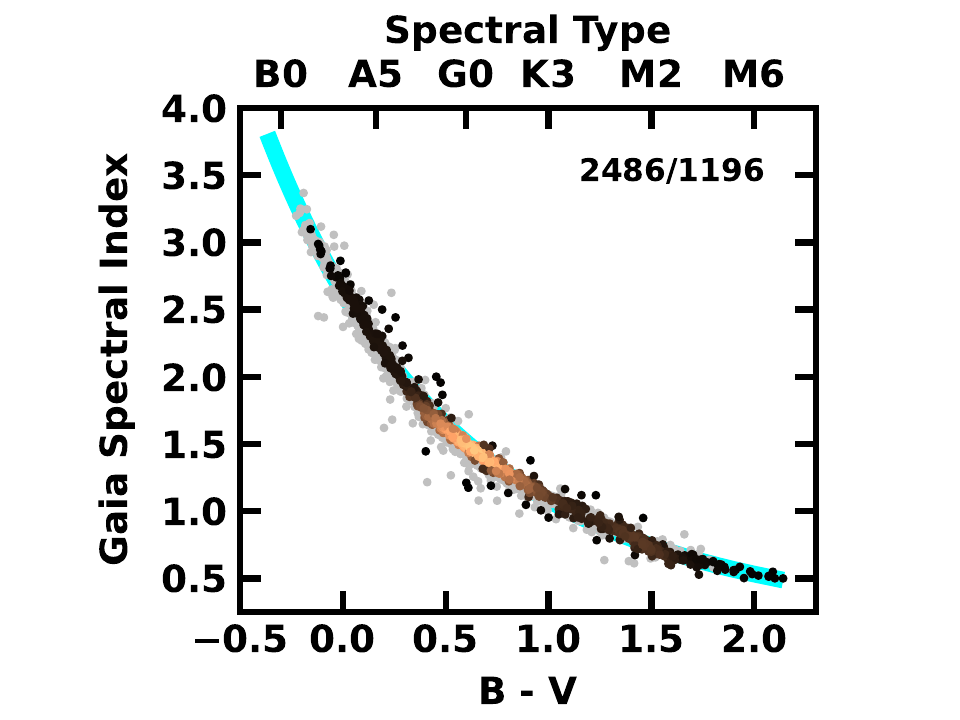}
\caption{
\label{fig: gaiacolor}
Relation between the \gaia\ spectral index $I_B$ and the B--V color. The upper x-axis relates spectral types of main sequence stars to B--V \citep{pecaut2013}. Grey points: all \ngspecg\ stars with \gaia\ spectra, B--V colors, and no saturation issues; black, yellow, and gold points: \ngspecgn\ stars with $d \le$ 80~pc. The cyan curve is the best-fitting polynomial to the 80~pc data (eqs. \ref{eq: bluefit0}--\ref{eq: bluefit1}). The clear correlation between $I_B$ and B--V enables spectral classification with an accuracy of $\pm$ 0.5--0.75 subclass. 
}
\end{center}
\end{figure}

Figure~\ref{fig: gaiacolor} illustrates the relation between the \gaia\ spectral index and the B--V color for the complete set of CDDS stars (light grey symbols) and the 80~pc sample (black, gold, and yellow points). This relation excludes stars with $S_C(G) \gtrsim$ 10, large differences between observed and predicted V and I ($\delta V, I \gtrsim$ 0.1), and large RUWE $\gtrsim$ 3. Aside from a few outliers, there is a clear correlation, where stars with larger $I_B$ have bluer colors. The full sample and the 80~pc sample follow the same relation. 

Although there is also a strong correlation between $I_B$ and the spectral type, the scatter about the relation is much larger than the scatter about the relation between $I_B$ and B--V. Similarly, we consider relations between (i) B--V and an index derived from the red spectra $I_R$, (ii) the \gaia\ BP--RP color and either $I_B$ and $I_R$, and (iii) the V--K and J--K colors and either $I_B$ or $I_R$. In all cases, there is a clear correlation between color or spectral type and \gaia\ spectral index. However, none are as clean as the relation shown in Figure~\ref{fig: gaiacolor}. There is more scatter about the relation and the relation breaks down for spectral types $\gtrsim$ M5--M6 (B--V $\gtrsim$ 2; BP--RP $\gtrsim$ 4.5). 

To make a useful relation between $I_B$ and B--V, we take the logarithm of $I_B$ and derive a simple quadratic:
\begin{equation}
\label{eq: bluefit0}
\rm {log~I_B = 0.41958 - 0.42413 ~ (B-V) + 0.03866 ~ (B-V)^2}
\end{equation}
The small quadratic term allows a better fit for small and large B--V where the correlation displays some curvature. Constructing samples of stars with $d \le$ 20~pc, 40~pc, and 60~pc yields nearly identical results. Solving this equation for B--V yields
\begin{equation}
\label{eq: bluefit1}
\rm{B-V = 5.48544 - \sqrt{19.23779 + 26.85997~log(I_B)}} ~ .
\end{equation}
Adopting an expression between B--V and spectral type \citep[e.g.,][]{pecaut2013} provides a method to infer spectral types from $I_B$. The upper x-axis of Fig.~\ref{fig: gaiacolor} illustrates this relation.

To test eq.~\ref{eq: bluefit1}, we explore results for $\sim$ 1200 stars with $d \le$ 80~pc. For each star with a known spectral type, B--V, and $I_B$, we derive B--V from eq.~\ref{eq: bluefit1}. We then (i) interpolate in the \citet{pecaut2013} table for main sequence stars to infer a spectral type from the measured B--V and the B--V estimated from $I_B$ and (ii) compute the difference in spectral types, $d_i$, where $i$ represents the known spectral type, the type from B--V only, and the type from $I_B$. The median offset (interquartile range) between the \gaia\ and B--V types is 0.05 (0.9) subclass independent of B--V. Compared to literature spectral classifications, the \gaia\ and B--V types are 0.25 subclass later, with interquartile ranges of $\sim$ 0.75 subclass. This comparison demonstrates that the \gaia\ $I_B$ index and the B--V color yield accurate spectral types with an uncertainty of somewhat less than one subclass.

The BP spectra and BP--RP color yield a relation similar to eq.~\ref{eq: bluefit1}:
\begin{equation}
\label{eq: bluefit2}
\rm{BP-RP = 3.13678 - \sqrt{2.94796 + 15.49763~log(I_B)}} ~ .
\end{equation}
As with eq.~\ref{eq: bluefit1}, this relation and \citet{pecaut2013} yield spectral types with a median offset (inter-quartile range) of 0.25 (0.85) of a subclass. 

Both approaches to derive spectral types from $I_B$ fail for types $\gtrsim$ M5--M6. For these late-type stars, $I_B$ changes little with B--V or BP--RP. Small fluctuations in $I_B$ also produce much larger variations in the derived B--V and BP--RP, which creates large uncertainties in the derived spectral type. This behavior also occurs in red \gaia\ spectra and with additional coefficients in the expressions for $I_B$ and $I_R$. We therefore recommend using eqs.~\ref{eq: bluefit1}--\ref{eq: bluefit2} only for stars earlier than $\sim$ M5.

For stars with $d >$ 80~pc, the \gaia\ or B--V spectral types are generally later than the literature spectral types. Stars with negligible (measurable) reddening, \av\ $\lesssim$ 0.01~mag (\av\ $\gtrsim$ 0.1--0.2~mag) are typically $\lesssim$ 0.25 subclass ($\gtrsim$ 1 subclass) later than the literature spectral types. Correcting for extinction is straightforward. The dereddened B--V or BP--RP color yields a corrected $I_B$ and SpT through eq.~\ref{eq: bluefit0} or the equivalent result for BP--RP. As with the 80~pc sample, the corrected spectral types are nearly indistinguishable from the known SpTs or SpTs derived from measured \teff's. Thus, the \gaia\ spectra yield accurate spectral types for stars with a measured B--V or BP--RP.

{\bf Stellar Luminosity.}\quad
With \teff\ estimates for all CDDS stars, combining a bolometric correction appropriate for the measured \teff\ \citep{pecaut2013}; a G, K, or V magnitude; \av; and $d$ yields the bolometric luminosity \lstar. For the brightest stars (V $\lesssim$ 5), we prefer G or V to derive \lstar\ due to saturation issues with 2MASS K. Smaller uncertainties in the extinction favor K over G and V for fainter stars. In all cases, we adopt a \teff\ from those available and derive \lstar\ from the average of the G, K, and V luminosity estimates and the uncertainty in \lstar\ from the dispersion in the average. For nine stars, we derive \lstar\ from a single measurement, either G, K, or V.

Formally, the lack of parallaxes prevent \lstar\ estimates for 32 stars. However, many stars without distances have positions, proper motions, and radial velocities that strongly imply cluster membership (see \S\ref{sec: clusters} below). Adopting the nominal distance of the cluster for these stars yields luminosities within a few percent of cluster stars with identical \teff. Thus, we adopt the association or cluster distance for these stars. In other cases, the \logg\ and \teff\ measures together with data for the stellar activity measures discussed below place stars robustly on the \citet{pecaut2013} main sequence. Using a stellar mass from \citet{pecaut2013} that corresponds to the measured \teff\ and the measured \logg\ yields a luminosity and distance. With these approaches, we derive \lstar\ for \nhrd\ CDDS stars.

\begin{figure}[t]
\begin{center}
\hspace*{-0.5cm}
\includegraphics[width=2.5in]{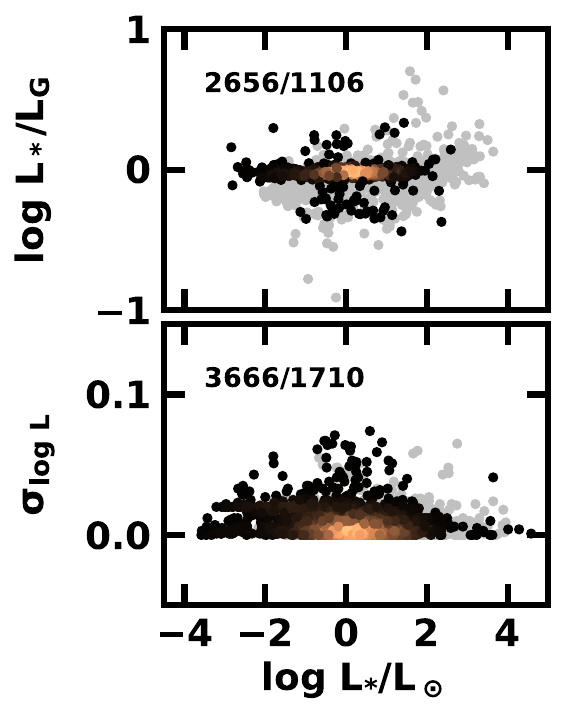}
\caption{
\label{fig: lstarcomp}
Properties of derived \lstar\ for the full sample (light points) and the 80~pc sample (dark points). {\it Upper panel:} logarithms of the ratio of \lstar\ to the \gaia\ luminosity $L_G$ as a function of \lstar. {\it Lower panel:} uncertainty in the logarithm of \lstar\ derived from the G, K, and V luminosities.
}
\end{center}
\end{figure}

Figure~\ref{fig: lstarcomp} summarizes aspects of the \lstar\ estimates. The lower panel shows the dispersion in the logarithm of \lstar, $\sigma_{log~L}$, as a function of \lstar. For the full (80~pc) sample, the median uncertainty in log~\lstar\ is 0.004 (0.005); the interquartile range of the uncertainty is 0.009 (0.010) for the 80~pc (full) sample. Among the main sequence stars, the set of \lstar\ estimates ranges from 15\% below to a factor of 2--3 above the \citet{pecaut2013} main sequence locus. As expected, K-type and M-type stars in the Sco--Cen association typically lie a factor of 5--10 above the main sequence. A small group of GKM giant stars are 10--1000 times more luminous than their main sequence counterparts.

The upper panel of Fig.~\ref{fig: lstarcomp} compares \lstar\ estimates with results from \gaia\ APSIS. We define $\Delta L_{CG}$ = log~\lstar -- log~$L_G$, where $L_G$ is the APSIS estimate. For stars within 80~pc, the APSIS luminosity estimates are less than 2\% larger than \lstar: the median $\Delta L_{CG} = -0.018$ with an interquartile range of 0.020. The median difference grows to 3\% among more distant stars, with an interquartile range of 9\%. 

There are two main sources of large outliers in the upper panel of Fig.~\ref{fig: lstarcomp}. Within the 80~pc sample, unresolved binaries with approximately equal masses distort the APSIS luminosity estimate. We usually correct luminosities based on the measured magnitude difference \citep[e.g.,][]{hummel1995,WDS1997} or the measured masses \citep[e.g.,][]{ORB62001a,ORB62001b,SBC2004,tokovinin2018,torres2021,gaiasb2024} and a mass--luminosity relation \citep{pecaut2013}. Sometimes, we add constraints derived from analyses of specific spectroscopic binaries \citep[e.g.,][]{cameron1981,fekel1988,popper1990,fekel1994,fekel1996,fekel1997,berdyugina1998,torres2002,cakirli2003,fekel2004,tokovinin2006,fuhrmann2008,houdebine2009,raghavan2009,fuhrmann2011,griffin2011,baron2012,helminiak2012,ramirez2012,eker2014,anthonioz2015,fuhrmann2015,hummel2017,kiefer2018,czekala2019,kochukhov2019,piccotti2020,akeson2021,hahlin2021,meunier2022b,muirhead2022,ryabchikova2022,torres2022,gallenne2023,torres2024}. For more distant stars, differences in the adopted extinction estimate lead to differences in the derived luminosities. Typically, \gaia\ derives larger \av\ and a larger luminosity. For distant stars, optical G and V along with 2MASS K yield a good estimate for the reddening; G, K, and V then enable a more accurate estimate for \lstar\ than APSIS. Often, the measured \logg\ helps to choose between possible options for \teff, \av, and \lstar. 

{\bf Surface Gravity.}\quad
After discarding $\sim$ 15 measurements where the surface gravity is a factor of more than two smaller or larger than the expected surface gravity for the measured SpT, \teff, and \lstar\ and deriving the average or median of multiple measurements, the compilation of stellar atmospheric parameters from the literature yields surface gravity measurements for \ngravl\ stars. The \gaia\ DR3 catalog provides another \ngravg\ estimates. Together, the full set of literature and \gaia\ data provides at least one measurement of \logg\ for \ngravt\ stars, $\sim$ 93\% of the full sample.

\begin{figure}[t]
\begin{center}
\hspace*{-0.7cm}
\includegraphics[width=2.75in]{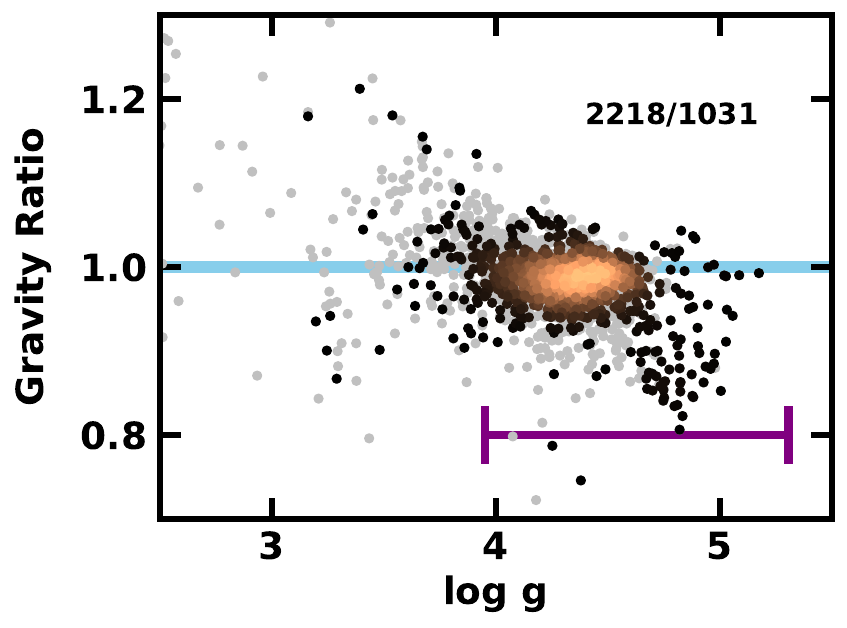}
%\vskip -2ex
\caption{
\label{fig: gravcomp0}
Ratio of gravity estimates from \gaia\ and the literature. Light (dark) points show results for stars with $d >$ 80~pc ($d \le$ 80~pc). The light blue bar indicates a ratio of one. The bar in the lower right indicates the range of gravities for stars on the main sequence, from early OB stars with \logg\ $\sim$ 4 to late M stars with \logg\ $\sim$ 5.3 \citep{pecaut2013}. 
}
\end{center}
\end{figure}

Figure~\ref{fig: gravcomp0} compares the ratio $r_\text{grav}$ of \logg\ values for the \gaia\ estimates relative to the literature estimates ($r_{grav}$ = \logg$_\text{Gaia}$ / \logg$_\text{lit}$). For most stars, the ratios are very close to unity. For the 80~pc (full) sample, the median ratio is 0.986 (0.989) with an inter-quartile range of 0.032 (0.036). Thus the \gaia\ \logg\ values are typically 1\% smaller than the literature estimates. The derived inter-quartile ranges are similar to the typical differences in \logg\ among a set of multiple measurements, $\Delta$~\logg\ $\sim$ 0.05--0.06. Thus, the difference between the \gaia\ and literature \logg\ values is smaller than the typical differences among multiple ground-based literature estimates.

There are two sets of outliers in Fig.~\ref{fig: gravcomp0}. For high gravity M dwarfs, the \gaia\ \logg\ estimates are systematically lower than the literature estimates by 10\% to 20\%. Comparison with the expected \logg\ for M dwarfs on the main sequence indicates that the \gaia\ estimates are smaller than the true gravity. Among some lower gravity stars, \gaia\ \logg\ estimates are up to 20\% larger than the ground-based literature estimates. This group mostly includes pre-main sequence stars and red giant stars, where the literature gravity appears more appropriate given results for \teff\, \lstar, and the activity indicators discussed below. Based on this analysis, we prefer the literature gravity estimates when available.

{\bf Metallicity.} \quad
As with \teff\ and \logg, we extract [Fe/H] data for CDDS stars from \gaia, large ground-based surveys, compilations of published results \citep{gaspar2016,xzhang2023}, and recent, more focused studies on specific types of main sequence stars not included in the \citet{gaspar2016} compilation. We construct two sets of [Fe/H] measurements: a) \nfehg\ stars from \citet{gaspar2016} and b) \nfehl\ stars with multiple measurements from more recent ground-based studies. For the second set, we set [Fe/H] equal to the average (median) for stars with two (three or more) measurements; we adopt the standard deviation or the inter-quartile range as the uncertainty in [Fe/H].

We verify the lack of a large systematic offset between these two sets. For the $\sim$ 1000 stars with two or more measurements, the average adopted uncertainty in [Fe/H] is 0.07$\pm$0.06, which is similar to the quoted uncertainty of a typical measurement. Nearly 97\% of the stars with two or more measurements have an uncertainty $\lesssim$ 0.20. As a function of [Fe/H], there is no trend in the measured uncertainty. The lack of a trend suggests there is a negligible offset among the multiple sets of metallicity measurements.

As another test, we consider the \nfehc\ common sources between the \citet{gaspar2016} compilation and subsequent non-\gaia\ studies. The median offset (inter-quartile range) is $\Delta$\feh\ = 0.00 (0.070). Following \citet{nandakumar2017}, we derive the median offset in bins from \feh\ = $-0.5$ to 0.5 in steps of 0.1 and look for trends in the median with \feh. The minimum (maximum) offset in this range is $\Delta$\feh\ = $-0.01$ for \feh\ = $-0.5$ to $-0.4$ (0.02 for \feh\ = 0.2--0.3). These offsets are less than the typical errors in \feh. 

The combined studies of \citet{nandakumar2017} and \citet{hegedus2023} demonstrate relationships between [Fe/H] measures among the APOGEE, GALAH, Gaia-ESO, LAMOST, and RAVE surveys. For the [Fe/H] range among CDDS stars, the metallicity differences among these surveys is small compared to the uncertainties in [Fe/H].  We also consider offsets between common sources in the \citet{gaspar2016} compilation and these recent, much larger surveys. Although we identify variations in $\Delta$\feh\ with \feh, for the full \citet{gaspar2016} sample with respect to the APOGEE, GALAH, and RAVE samples, the offsets for the subset of CDDS stars in the \citet{gaspar2016} compilation are small, $\sim$ 0.02--0.04 per dex, and less than the typical uncertainty in [Fe/H]. We conclude that the full set of \nfeh\ \feh\ measurements for CDDS stars are on a roughly similar scale. 

To try to supplement these data, we consider several compilations of \feh\ derived from \gaia\ XP spectra \citep{andrae2023a,andrae2023b,xzhang2023}. The median offsets are +0.31 for \gaia\ DR3 metallicities \citep{andrae2023a} and +0.23 for the machine-learning metallicities from \citet{xzhang2023}. Eliminating the OBA stars where analyses of XP spectra have issues reduces the offsets to +0.23 for \gaia\ DR3 and +0.20 for \citet{xzhang2023}. There is no obvious trend of the offset with \feh\ for the \gaia\ metallicities. However, the offsets with the \citet{xzhang2023} metallicities show a clear trend with \feh. Rather than try to place these data on the same scale as the ground-based data, we discard them for this analysis. 

\begin{figure}[t]
\begin{center}
\includegraphics[width=3.5in]{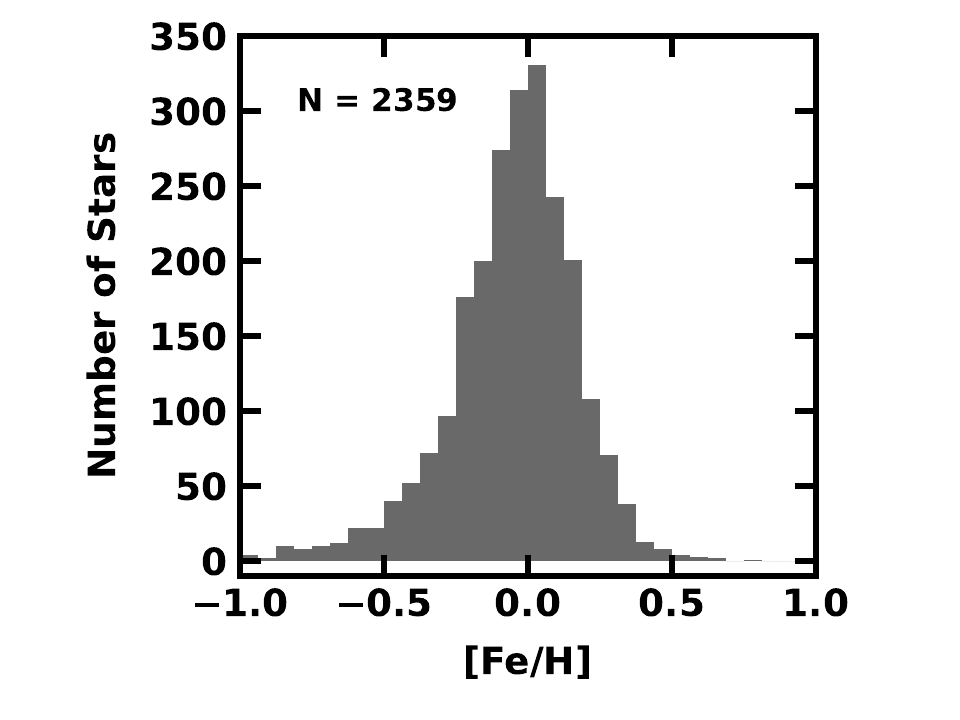}
\vskip -4ex
\caption{
Frequency distribution of \feh\ for the full sample of \nfeh\ CDDS stars with metallicity estimates. The sample has a median \feh\ of $-0.03$ and a full-width at half maximum $\sim$ 0.13.
}
\label{fig: metals}
\end{center}
\end{figure}

The distribution of [Fe/H] among CDDS stars resembles a gaussian with a longer tail to smaller \feh\ than to larger \feh\ (Fig.~\ref{fig: metals}). Although there are a few very low (high) metallicity stars with [Fe/H] = $-2.30$ (0.97), the distribution is highly concentrated about the median \feh\ = $-0.03$. Half (90\%, 98\%) of the sample has \feh\ = $-0.17$ to 0.09 ($-0.48$ to 0.26, $-0.93$ to 0.40). This distribution is similar to that in other collections of nearby stars 
\citep[e.g.,][]{hinkel2014}.

{\bf Lithium.}\quad
With applications in big bang nucleosynthesis and galactic chemical evolution, lithium plays a big role in stellar astrophysics \citep[e.g.,][]{bensby2018,gao2020,romano2021,buder2022}. On the main sequence, the convective atmospheres of late-F and GKM stars draw lithium into the hydrogen-burning core, where it is destroyed \citep[e.g.,][]{pinsonneault1997}. Hence the Li abundances of lower main sequence stars decline with time and may serve as a useful age indicator for CDDS and other main sequence stars \citep[e.g.,][]{sestito2005,jeffries2023,rathsam2023,gutierrezalbarran2024}.  

Published data for the Li abundance generally fall into three categories: the absolute lithium abundance on a log scale where the hydrogen abundance is 12 \citep[A(Li); e.g.,][]{jeffries1999,ford2001,boesgaard2003,lambert2004,dasilva2009,ghezzi2010,pace2012,ramirez2012,delgado2014,delgado2015,lopezvaldivia2015,guiglion2016,cummings2017,bouvier2018,aguileragomez2018,carlos2019,chavero2019,llorente2021,boesgaard2022,hourihane2023,rathsam2023,gutierrezalbarran2024,lubin2024}, the lithium abundance relative to the solar abundance [Li/H] \citep[e.g.,][]{buder2021}, and the equivalent width of the Li~I 670.8~nm absorption line \citep{preibisch2002,pecaut2016,zerjal2019,zerjal2021,luhman2022b,jeffries2023,zerjal2023}. We derive the absolute abundance from the relative abundance: A(Li) = [Li/H] + A(Li)$_\odot$, where A(Li)$_\odot$ = 1.05 is the solar Li abundance \citep{asplund2009}.

Converting Li~I equivalent widths EW(Li) to absolute or relative abundances requires a detailed model atmosphere calculation. Here, we adopt the grid of model atmospheres in \citet{franciosini2022}, which compiles Li~I curves of growth as a function of \teff, \logg, and \feh. For CDDS stars with measured \teff, \logg, \feh, and Li~I equivalent widths, we interpolate in their Table A.1 (for FGK stars) and Table A.3 (for M stars) to derive A(Li). For stars with literature estimates of A(Li) and quoted EW(Li) which we convert to A(Li), we derive a median offset from the model conversion of $\Delta$A(Li) = A(Li)$_{lit}$ - A(Li)$_{EW}$ = $-0.04$ with an interquartile range of 0.09. Thus, the model-conversion of EW(Li) to A(Li) results in a slightly larger Li abundance than published abundances. However, this difference is smaller than the typical uncertainty, $\sim$ 0.1--0.2 dex, of a single A(Li) estimate.

Some studies report \teff\ and Li~I equivalent widths without \logg\ or \feh\ data. As a test for stars in the Hyades and Praesepe \citep{cummings2017}, we match a solar metallicity PARSEC isochrone \citep{ychen2014,jtang2014,ychen2015} with age = 800~Myr to the measured \teff, adopt the corresponding \logg, and estimate A(Li) for $\sim$ 100 stars from observed equivalent widths. For Hyades stars and Praesepe stars, the results are indistinguishable: a median offset of $-$0.04 in A(Li) relative to the values in \citet{cummings2017} and an interquartile range of 0.08. For an ensemble of Pleiades stars from diverse sources, we achieve similar uncertainties using a 100~Myr solar metallicity PARSEC isochrone to set \logg. Finally, a set of 150 stars with membership probability greater than 0.5 in the young cluster 25~Ori from the Gaia--ESO survey \citep{hourihane2023} also yields good agreement between the A(Li) derived in \citet{hourihane2023} and A(Li) derived using a 10~Myr isochrone for \logg: a median offset of $+0.00$ and an interquartile range of 0.12. Thus, this approach provides reasonably accurate Li abundance estimates from a set of measured equivalent widths for ages 10~Myr to 1~Gyr.

For each CDDS star with two (or more) measurements of A(Li), we compute the average (median) value. The range of multiple published measurements for a single CDDS star, $\Delta$A(Li) $\approx$ 0.1--0.2, is generally comparable to the typical error of a single measurement. For stars with upper limits, we adopt the most recent measurement or the measurement from the highest signal-to-noise spectrum (when available).

\begin{figure}[t]
\begin{center}
\includegraphics[width=3.0in]{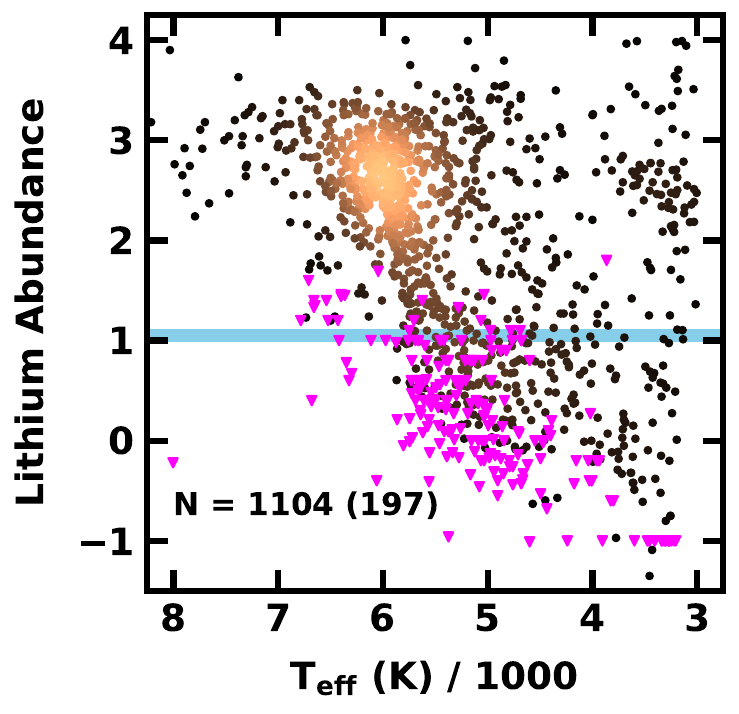}
%\vskip -2ex
\caption{
\label{fig: lithium}
Lithium abundance A(Li) as a function of effective temperature for CDDS stars. Colors of filled circles (\nlith\ points) indicate the relative density, ranging from low (black) to moderate (dark gold) to high (bright yellow). Filled fuchsia triangles indicate upper limits for \nlithl\ stars. The solid horizontal blue line indicates the solar lithium abundance \citep{asplund2009}. CDDS stars follow the general trend of decreasing A(Li) with decreasing \teff, but they display a clear gap at 1.0 $\lesssim$ A(Li) $\lesssim$ 2.0 not observed in much larger samples.
}
\end{center}
\end{figure}

Fig.~\ref{fig: lithium} shows the variation of A(Li) with \teff\ for \nlitht\ CDDS stars. The overall morphology is similar to that in larger compilations of low mass MS stars \citep[e.g.,][and references therein]{llorente2021,rathsam2023,gutierrezalbarran2024}. The upper envelope of the distribution peaks at A(Li) $\approx$ 3.5 for mid F-type stars with \teff\ $\sim$ 6500~K and falls to A(Li) $\approx$ 3 towards hotter and cooler \teff. Stars in the densest part of the ensemble have \teff\ $\approx$ 5000--6500~K and A(Li) $\approx$ 2--3. These stars have abundances 10--100 times larger than the solar abundance. For \teff\ $\approx$ 3500--6000~K, there is a set of stars with A(Li) $\approx$ 0--2 and a few stars with A(Li) $\approx -1$ to 0. Within this \teff\ range, many stars only have upper limits for A(Li).

Compared to the general population of FGKM stars, CDDS stars have larger lithium abundances \citep[e.g.,][]{ramirez2012,llorente2021,rathsam2023,gutierrezalbarran2024}. In a population of cool stars with \teff\ $\lesssim$ 4500--5000~K, A(Li) declines to values comparable to or less than the solar Li abundance on time scales $\lesssim$ 100--200~Myr \citep[e.g.,][]{jeffries2023}. The CDDS K--M dwarfs with A(Li) $\gtrsim$ 1 are likely much younger than typical nearby M dwarfs with ages $\gtrsim$ 1~Gyr and negligible Li abundances. Among a sample of `solar-twin' stars with ages $\tau \approx$ 0.5--11~Gyr \citep{carlos2019,rathsam2023}, few (many) have A(Li) $\gtrsim$ 2 (A(Li) $\lesssim$ 1.5). In the Hyades and Praesepe \citep[$\tau \approx$ 750--800~Myr;][]{brandner2023b}, solar-type stars with \teff\ $\gtrsim$ 5600~K (\teff\ $\lesssim$ 5600~K) have A(Li) $\gtrsim$ 2.0 \citep[A(Li) $\lesssim$ 2;][]{cummings2017}. Many solar-type stars in the CDDS have A(Li) $\gtrsim$ 2 and are thus probably younger than stars in the Hyades. This group includes field stars {\it not} in an association, cluster, or moving group.

{\bf Stellar Rotation.}\quad
Measurements of stellar rotation periods \prot\ require a dedicated facility to cover time scales ranging from a few to a few hundred days. Along with several studies based on \gaia, {\it Kepler}, and {\it TESS} data, programs with small ground-based telescopes provide \prot\ for \nprot\ stars \citep{rebull2004,lawson2005,hartmann2010,delorme2011,cargile2014,kovacs2014,newton2016,rebull2016,astudillo2017,douglas2017,mellon2017,messina2017,newton2017,rebull2017,newton2018,rebull2018,diezalonso2019,douglas2019,kuker2019,schofer2019,bouchard2020,cantomartins2020,gillen2020,magaudda2020,rebull2020,rampalli2021,magaudda2022,nunez2022,petit2022,reiners2022,rebull2022,distefano2023,popinchalk2023,nunez2024,douglas2024,colman2024,campelo2025}. For some stars, we calculate the average (median) of two (three or more) measurements. 

Sources for \vsini\ measurements cover a broad range. In addition to \gaia, we collect data from the APOGEE \citep{abdurrouf2022}, GALAH \citep{buder2021}, and RAVE \citep{steinmetz2020} surveys.  \citet{hoffleit1991} has a good compilation for the brightest stars. Other studies report data for B-type and A-type stars \citep{abt1995,abt2002,royer2002a,royer2002b,royer2007,zorec2012,david2015}, FGK stars \citep{white2007,schroeder2009,rainer2023}, M-type stars \citep{torres2002,jenkins2009,malo2014,houdebine2011,reiners2012,houdebine2015,houdebine2016,reiners2018,schofer2019,reiners2022}, and stars in open clusters \citep{glebocki2000,bailey2018,fritzewski2020,bouma2021}. For all of the \vsini\ data, we eliminate duplicate measurements from multiple compilations and adopt the average (median) of two (three or more) measurements. Often, studies report upper limits for stars with small \vsini. For these stars, we compute the average (or median) for all measurements without an upper limit and adopt the smaller of the average (or median) and the upper limit. Several studies using cross-correlation techniques report \vsini\ = 0 for a dozen stars when the width of the cross-correlation peak is indistinguishable from that of a non-rotating star. We adopt an upper limit of 3 \kms, which is roughly half the median of all \vsini\ upper limits in the CDDS and corresponds to the peak of the \vsini\ distribution in the extensive compilation of \citet{nordstrom2004}. Often these stars have measurements from other sources. We then adopt the smaller of the average of other measurements and the adopted upper limit.

\begin{figure}[t]
\begin{center}
\includegraphics[width=3.0in]{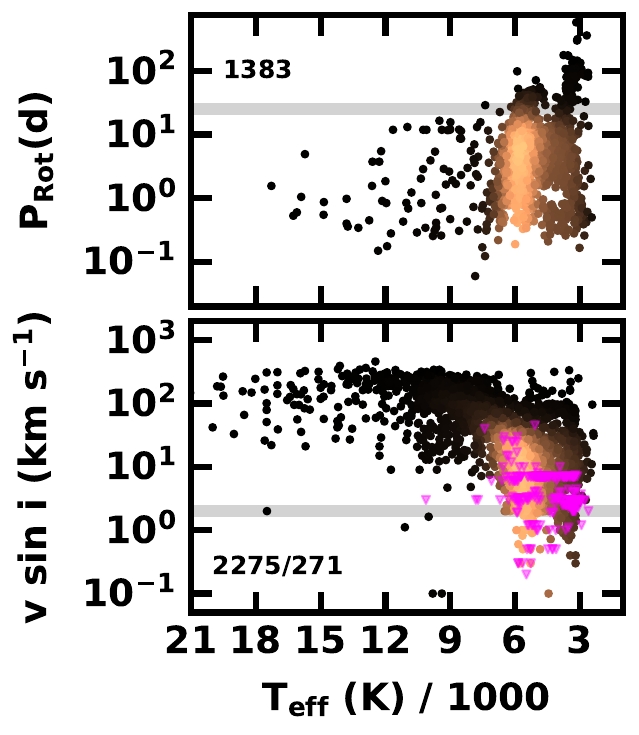}
\caption{
\label{fig: teffRot}
Rotation as a function of \teff. {\it Upper panel:} rotational periods in days for \nprot\ stars. {\it Lower panel:} projected rotational velocities in \kms\ for \nvrot\ stars (black, yellow, gold points) and upper limits for \nvrotl\ stars (fuchsia triangles). In both panels, the horizontal grey band indicates solar values.
}
\end{center}
\end{figure}

Fig.~\ref{fig: teffRot} plots \prot\ (upper panel) and \vsini\ (lower panel) as a function of \teff. Among B-, A-, and early F-type stars, the lack of photospheric features prevents robust measurements for all but a handful of rapidly rotating stars with \prot\ $\lesssim$ 10~days. Late F-type and GK stars have a broad range of rotational periods, \prot\ $\approx$ 0.1--100 days, and a median \prot\ $\approx$ 8~days that is somewhat shorter than the 
typical \prot\ $\sim$ 10~days observed in recent large-scale surveys \citep[e.g.,][]{mcquillan2014,johnstone2021,gruner2023,ye2024}. Nearly 90\% of these stars rotate more rapidly than the Sun. M-type stars display the largest range in periods, with \prot\ $\sim$ 100--300 days for older M stars and \prot\ $\sim$ 0.1--3 days for the youngest M stars. The median rotational period of 5 days for M stars reflects the large frequency of younger stars in the CDDS sample \citep[e.g.,][]{newton2016,newton2017,shan2024}.

Among early-type stars with \teff\ $\gtrsim$ 7500--8000~K, the projected rotational velocity ranges from $\sim$ 30~\kms\ to $\sim$ 300~\kms, with a median value of $\sim$ 100~\kms. Although the sample of B-type stars is small, the lower envelope of \vsini\ appears to fall from $\sim$ 30~\kms\ for the hottest stars with \teff\ $\sim$ 20,000~K to $\sim$ 10~\kms\ for A-type stars with \teff\ $\approx$ 9000~K. Stars with \teff\ $\sim$ 9000~K have a bimodal distribution. A group of older stars has \vsini\ $\sim$ 10--40~\kms\ while the younger stars have \vsini\ $\sim$ 50--300~\kms. 

The ensemble of FGK stars has a very broad \vsini\ distribution, with values ranging from $\sim$ 1~\kms\ to $\sim$ 300~\kms. Although there are many FG stars with \vsini\ $\approx$ 100--300~\kms, most have \vsini\ $\sim$ 3--30~\kms. Throughout this temperature range, the lower envelope of \vsini\ continues to fall with decreasing \teff. Cooler G-type stars have lower values of \vsini\ than warmer F-type stars. Among the G stars, $\sim$ 10 stars have \vsini\ $\lesssim$ 1 \kms\ (or upper limits below 1~\kms); only $\sim$ 4\% have \vsini\ values or upper limits below the Sun's rotational velocity of 2~\kms. There are no F stars with such low \vsini.

K-type and M-type stars generally follow the \vsini\ distribution of G-type stars. The lack of K and M stars in young clusters surveyed by \spitz\ severely limits the population of rapidly-rotating KM stars with \vsini\ $\gtrsim$ 100~\kms. There is also a dearth of K-type stars at $\sim$ 30~\kms. Despite these differences, the KM stars in the CDDS sample have a similar lower limit, $\sim$ 1~\kms, and otherwise fill the space between 1~\kms\ and 100~\kms\ fairly uniformly. 

{\bf Chromospheric and Coronal Emission.}\quad
All late-type stars with convective envelopes have hotter material above the photosphere that generates continuum and line emission from X-ray to radio wavelengths \citep[e.g.,][]{hall2008,testa2015,demoortel2015,degrijs2021}. The European missions \rosat\ and \erosita\ conducted all-sky surveys that provide X-ray luminosities or upper limits for the corona of each CDDS star \citep{aschenbach1981,predehl2006}. Probing chromospheric emission requires ground-based observations, which are not available for every star. Here we focus on emission from the Ca H \& K and IR triplet lines as robust chromospheric activity indicators \citep[e.g.,][]{busa2007,borosaikia2018}. Although H$\alpha$ emission is a common activity indicator for M-type stars \citep[e.g.,][]{souzadossantos2024}, it is often omitted in studies of Ca~II emission in FGK stars. Thus, we do not include it in this compilation.

{\bf Calcium emission.}\quad
Emission in the cores of the Ca~II H \& K absorption lines is a hallmark of chromospheric activity in late F and GKM main sequence stars \citep[e.g.,][]{wilson1963,vaughan1980,henry1996,borosaikia2018}. The $S$ index -- defined as the ratio of the total flux in narrow filters centered on H \& K to the continuum flux outside H \& K -- is $\sim$ 0.1 (10) for stars with weak (strong) chromospheric emission \citep{vaughan1980,noyes1984}. To correct for the dependence of $S$ on spectral type, \citet{middlekoop1982} derived a color-dependent factor $\mathcal{C}(B-V$) = $R_{HK} /S$, where \rhkp\ = $F_{HK} / \sigma {\rm T_{eff}^4}$ is the ratio of the stellar flux in the H \& K passbands to the total stellar flux. \citet{noyes1984} outline a procedure to eliminate the photospheric contribution to \rhk, yielding \rhkp\ as the ratio of the chromospheric flux in H \& K to the total energy output. Most studies follow a variant of the \citet{noyes1984} procedure \citep[e.g.,][]{henry1996,borosaikia2018}; however, some measure the total equivalent width $W_{HK}$ of the H \& K emission features from high resolution spectra. As outlined in \citet{lanzafame2023}, \rhkp\ = $ F_{HK} W_{HK} / \sigma {\rm T_{eff}^4}$. The two approaches -- $S_{HK} \rightarrow$ \rhkp\ and $W_{HK} \rightarrow$ \rhkp\ -- are equivalent.

The success of $S_{HK}$ and \rhkp\ as activity indicators led to similar measurements for the Ca II IR triplet lines at 849.8~nm, 854.2~nm, and 866.2~nm \citep[e.g.,][]{linsky1979,cayrel1983,foing1989,soderblom1993b,busa2007}. Along with theoretical studies \citep[e.g.,][]{chmielewski2000,andretta2005}, these analyses demonstrate that the analogous indices $S_{IRT}$ and \rirtp\ yield similarly good measurements of activity as $S_{HK}$ and \rhkp\ \citep[see also][]{lorenzooliveira2016,martin2017}. These efforts culminated in the release of IR triplet activity indices for two million stars in \gaia\ DR3 \citep{lanzafame2023}.

We compile \rhkp\ from the literature \citep{duncan1991,henry1996,strass2000,wright2004,white2007,schroder2009,gomes2014,astudillo2017,houdebine2017,borosaikia2018,lorenzooliveira2018,gondoin2020b,gomes2021,perdelwitz2021,brown2022,marvin2023,mignon2023}. For each CDDS star, we adopt the average (median) of two (three or more) measurements. Many publications include summaries of previous \rhkp\ measurements. Before performing the median, we remove repeat measurements collected from several sources. Typically, the range in \rhkp\ for a single star is $\sim$ 0.1--0.2, which is similar to the amplitude of the variation in \rhkp\ over a complete stellar cycle \citep[e.g.,][]{borosaikia2018}.

Studies that quote some variant of the $S$ index instead of \rhkp\ generally place their results on the original Mt. Wilson scale \citep[e.g.,][]{wzhang2022}. We convert these $S$ values to \rhkp\ following the procedures described in \citet{noyes1984} and \citet{henry1996}. For stars in common, the derived \rhkp\ values are within 10\% of published results. Overall, we compile \rhkp\ for \nrhk\ stars.

From the \gaia\ data set, we collect the activity index $\alpha$ for \nirt\ stars. The $\alpha$ index measures the strength of the Ca II IR triplet lines relative to a theoretical photospheric spectrum and is related to the emission line equivalent width $W_{IRT}$. For each measured $\alpha$, the photospheric spectrum is matched to the values for \teff, \logg, and metallicity [M/H] derived from analysis of the \gaia\ RVS spectrum. As with the H \& K lines, deriving the IR triplet index \rirtp\ requires a measurement of the stellar flux across the IR triplet lines: \rirtp\ = $\mathcal{F} \alpha$. \citet{lanzafame2023} adopt the theoretical spectrum and derive a function $\mathcal{F}$([M/H], \teff) that depends on the [M/H] and \teff\ derived from \gaia\ spectra. In their analysis, this function is nearly independent of \logg. For each CDDS star with a measured $\alpha$, [M/H], and \teff\ from \gaia\ measurements, we linearly interpolate $\mathcal{F}$ in Table~1 of \citet{lanzafame2023} to generate \rirtp.

Among the set of \nirt\ stars, \nirtp\ have a positive emission line equivalent width ($\alpha > 0$). The rest have extra absorption relative to the matching photospheric spectrum ($\alpha < 0$). Stars with weak chromospheres often produce extra absorption in the Ca II H \& K and the IR triplet lines \citep[e.g.,][and references therein]{andretta2005,linsky2017}. Among the stars with extra absorption, those with the largest $|\alpha|$ are probably the most chromospherically active.

\begin{figure}[t]
\begin{center}
\includegraphics[width=3.0in]{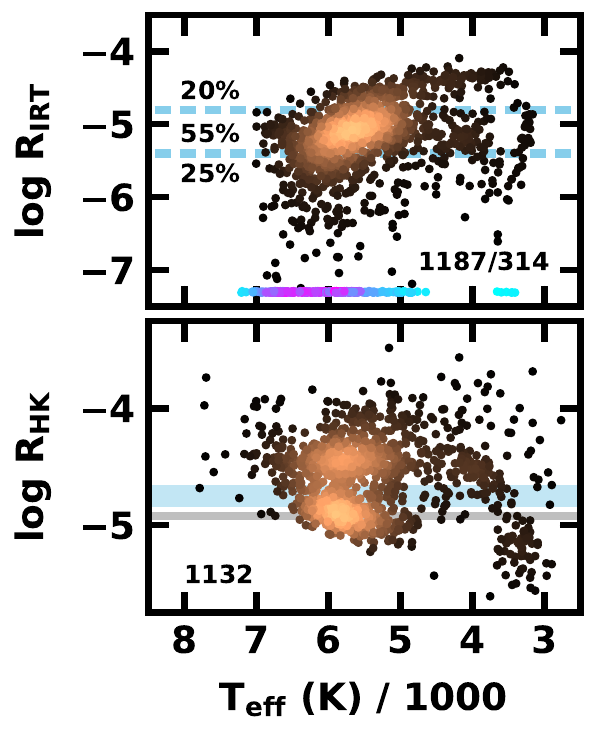}
%\vskip -2ex
\caption{
\label{fig: teffCa}
Activity indices for Ca~II H \& K (\rhkp; lower panel, \nrhk\ stars) and the Ca II infrared triplet (\rirtp; upper panel, \nirt\ stars including \nirtul\ upper limits) as a function of \teff.  {\it Lower panel:} The thin grey bar (thick blue bar) indicates the range of activity for the Sun (the Vaughan--Preston gap). {\it Upper panel:} Points with cyan (low density) to pink (high density) colors at the bottom of the plot show the \nirtul\ stars with $\alpha < 0$. Three points with $\alpha > 0$ have \lrirtp\ $\approx$ $-7.5$ to $-8.3$ and lie below the extent of the plot. Dashed blue lines indicate the boundaries of the high, moderate, and low activity regimes of \citet{lanzafame2023}; the percentage of stars with $\alpha > 0$ in each regime appears at the left edge of the plot.
}
\end{center}
\end{figure}

Fig.~\ref{fig: teffCa} shows the distribution of \rhkp\ (lower panel) and \rirtp\ (upper panel) as a function of \teff. Only \nabrhk\ (\nmrhk) of the \nabspec\ (\nmspec) stars with \teff\ $>$ 6500~K (\teff\ $<$ 4500~K) have measured \rhkp. Most of the F stars are very active (\lrhkp\ $\gtrsim -4.5$). Although a few KM stars are among the most active stars, most have low activity levels (\lrhkp\ $\lesssim -5$). More than half (\nfgkrhk\ out of \nfgkspec) of the stars with \teff\ = 4500--6500~K have measured Ca~II H\&K emission, at levels ranging from minimal activity similar to the Sun \citep[\lrhkp\ $\sim -5$, as illustrated by the thin grey band in the figure;][]{egeland2017,borosaikia2018,dineva2022,zills2024} to maximal activity with \lrhkp\ $\sim -4$. Roughly 32\% (53\%) of the stars lie below (above) the Vaughan--Preston gap, a region with very few stars in the first large-scale surveys of stellar activity among solar-type main sequence stars \citep{vaughan1980,noyes1984}. 

To put these \rhkp\ measurements in context, we compare with the \citet{borosaikia2018} analysis of 4454 late-type stars. In their Figure~3, H \& K emission is rare among middle F-type and earlier stars with \teff\ $\gtrsim$ 6500~K. A few M stars have high activity levels; most have low \rhkp. Compared to the \citet{borosaikia2018} sample, the CDDS has a much larger fraction of GK stars with high activity levels (above the Vaughan-Preston gap) and a much smaller fraction with low activity levels (below the gap). The CDDS GK stars have a somewhat stronger gap, but the sample is much smaller than the \citet{borosaikia2018} sample, where the gap is nearly absent. 

As in the full sample of \gaia\ stars, stars with \teff\ = 3000--7000~K have Ca~II IR triplet indices ranging from a maximum \lrirtp\ $\approx -4$ to a minimum \lrirtp\ $\approx -8$ (Fig.~\ref{fig: teffCa}, upper panel). In the middle of the diagram, \lrirtp\ $\approx -5.4$ to $-4.8$, FG stars produce a dense concentration with two tails: a group of KM stars with similar \lrirtp\ and a group of mostly FG stars with weaker emission (\rirtp\ $\lesssim -5.4$). Above the main body, there is a group of FGK stars with strong IR triplet emission (\lrirtp\ $\approx -4.8$ to $-4.2$.

In addition to the low activity stars with \lrirtp\ $\gtrsim -8.5$, another set of stars has extra Ca II absorption compared to the \gaia\ photospheric comparison spectrum \citep{lanzafame2023}. Based on the photospheric models, these stars have extra chromospheric absorption \citep[see the discussion in][]{linsky2017,carlsson2019}. Following \citet{lanzafame2023}, we interpret this extra absorption as an indicator of a weak chromosphere. Most of the stars in this group are F-type stars, where chromospheric emission is generally weaker than for GK stars.

To compare with the full \gaia\ sample, we follow \citet{lanzafame2023} and define stars with low ($-8.5 \lesssim$ \lrirtp\ $\lesssim -5.4$), moderate (\lrirtp\ = $-5.4$ to $-4.7$), and high (\lrirtp\ $\ge$ $-4.7$) activity levels. Most CDDS stars (55\%) have moderate activity. This group contains the full range of spectral types with Ca II triplet emission, FGKM. Roughly 25\% have low activity. Only a few KM stars have low activity levels. The rest of the sample (20\%) have high activity levels. Most of these are G stars; some are K stars; very few are F or M stars.

Overall, CDDS stars for any \teff\ have much larger activity levels than those in the ground-based \citep[e.g.,][]{borosaikia2018} and the full \gaia\ sample \citep{lanzafame2023}. In the CDDS (full sample), moderate (low) activity stars dominate; high activity stars are common (rare). Although the lowest activity stars in the CDDS and full samples have similar activity levels, CDDS stars with these levels are relatively rare. Similarly, CDDS stars with the highest activity levels are rare in the full ground-based and \gaia\ samples. As discussed above, the \spitz\ and \herschel\ surveys favored younger stars with ages much smaller than the Sun, $\lesssim$ 1~Gyr. Thus, CDDS stars have higher activity levels than stars with ages similar to the Sun's.

{\bf Stellar X-rays}\quad
For this study, we focus on data acquired with \rosat\ \citep{aschenbach1981}, \chandra\ \citep{weisskopf2002}, \xmm\ \citep{jansen2001}, and \erosita\ \citep{predehl2006}. In addition to the \rosat\ all sky survey \citep[RASS;][]{voges1999,boller2016,freund2022}, \rosat\ acquired more sensitive measurements with pointed observations of selected open clusters \citep[e.g.,][and references therein]{micela1999}. \citet{voges2001} describe all of the ROSAT data catalogs; \citet{boller2016} include a link to the catalog of pointed observations accessible in \vizier. \citet{freund2022} associate RASS stellar sources with optical counterparts from \gaia\ eDR3 \citep{fabricius2021,gaiaedr3}. Explicit matches of an eDR3 identifier (or a Tycho-2 identifier) with each RASS source enables a robust association between a stellar X-ray source and its optical counterpart. \citet{freund2022} also use the known maximum in the fractional X-ray luminosity \lxlstar\ to avoid spurious matches. We identify \nrosat\ sources in these ROSAT databases.

Several \chandra\ and \xmm\ programs conducted deep pointed observations of nearby open clusters. We collect \chandra\ data for the Pleiades \citep{daniel2002} and NGC~2516 \citep{damiani2003}. Several other CDDS stars appear in various {\it Chandra Source Catalogs} \citep[original versions:][]{evans2010,wang2016}. These compilations yield \nchandra\ X-ray sources.

The \xmm\ coverage of nearby open clusters is more extensive, with data from the Pleiades \citep{briggs2003}, Praesepe \citep{franciosini2003}, Blanco~1 \citep{pillitteri2004}, NGC~2516 \citep{pillitteri2006}, NGC~2547 \citep{jeffries2006}, and Upper Sco \citep{argiroffi2006}. As \xmm\ moves from one pointed observation to the next, it identifies X-ray sources along its path. We select \nxmm\ CDDS stars from the \citet{freund2018} catalog of stellar counterparts to these sources. 

The first data release of the \erosita\ all-sky survey covers the western galactic hemisphere, $l$ = 180\deg--360\deg\ \citep{merloni2024}. This region includes \nerositap\ out of \ncdds\ CDDS sources. In addition to $\eta$ Cha \citep[16 sources;][]{robrade2022} and the Sco-Cen association \citep[106 sources;][]{schmitt2022}, this release includes roughly half of the CDDS stars in the Hyades along with CDDS stars in Coma, IC~2391, NGC~2232, NGC~2422, NGC~2451, NGC~2516, NGC~2547, Praesepe, and the TW Hya association. However, it does not include $\alpha$~Per, Blanco~1, the Pleiades, and the other half of the CDDS stars in the Hyades, which are in the eastern galactic hemisphere and remain unpublished. Among the \nerositap\ possibilities, the \erosita\ catalogs list absolute flux measurements for \nerosita\ stars and upper limits for the rest.

\begin{figure}[t]
\begin{center}
\includegraphics[width=3.25in]{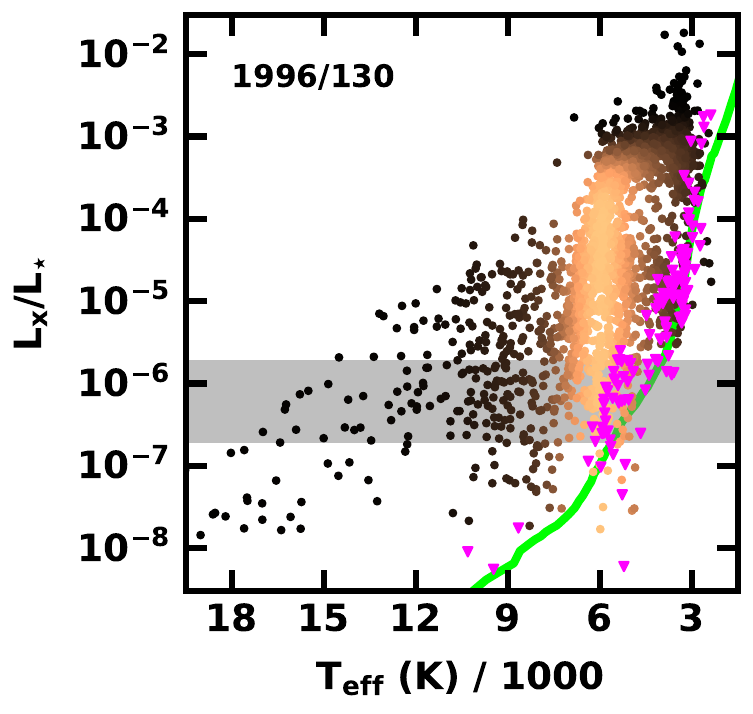}
\vskip -1ex
\caption{
\label{fig: teffxray}
Fractional X-ray luminosity ($L_X/L_{\star}$) as a function of the effective temperature for \nxray\ stars with measured \teff, \lstar, $L_X$. Filled fuchsia triangles indicate upper limits for \nxrayul\ stars with $d \le$ 17~pc. The horizontal grey band shows the range of the fractional X-ray luminosity for the Sun \citep{judge2003}. Adopting the \citet{pecaut2013} \teff-\lstar\ relation for main sequence stars, the green line indicates approximate \erosita\ and ROSAT detection limits at distances of 10--20~pc.
}
\end{center}
\end{figure}

From \rosat, \chandra, \xmm, and \erosita, the CDDS has X-ray fluxes $F_X$ for \nxray\ stars. All fluxes lie within a similar bandpass of $\sim$ 0.1--2~keV. To put the X-ray luminosities of these stars in context, we derive the fractional X-ray luminosity \lxlstar. For each star with an accurate distance $d$, we calculate the average X-ray flux $\langle F_X \rangle$ of the individual measurements from each satellite; we then set $L_X = 4 \pi d^2 \langle F_X \rangle$.  Among stars with robust X-ray fluxes from several satellites, the brightest sources vary by a factor of $\sim$ 2--3. For stars with accurate parallaxes and \lstar, the fractional X-ray luminosity, $L_X$/\lstar, follows.

Fig.~\ref{fig: teffxray} shows how \lxlstar\ depends on \teff. The distribution follows that of all coronal X-ray sources in \erosita: (i) an upper envelope with \lxlstar\ $\lesssim 10^{-5}$ for B-type stars with \teff\ $\approx$ 12,000--20,000~K that gradually rises to \lxlstar\ $\lesssim 10^{-2}$ for KM stars with \teff\ $\approx$ 3000--4000~K and (ii) a lower envelope with \lxlstar\ $\gtrsim 10^{-8}$ to $10^{-7}$ for early-type stars and \lxlstar\ $\gtrsim 10^{-6}$ for late-type stars \citep[see Fig.~11 of][]{freund2024}.  Main sequence stars detected in the ROSAT all sky survey show a similar pattern \citep{freund2022}. Once again, the concentration of stars with \teff\ $\approx$ 4500--7500~K is a feature of the \spitz\ and \herschel\ programs that focus on FGK stars in their searches for debris disks.

As summarized in \citet{freund2022,freund2024}, the maximum in \lxlstar\ reflects `saturation' of the mechanism that generates coronal X-ray emission in the youngest and most rapidly rotating solar-type stars \citep[e.g.,][]{stauffer1994,randich1996,gudel2004,cargile2009,argiroffi2016,fang2018}. Currently, there is no consensus on the origin of this maximum \citep[e.g.,][]{testa2015,demoortel2015}. For FG stars, the lower limit of \lxlstar\ $\sim 10^{-8}$ is characteristic of old, slowly rotating stars \citep[e.g.,][]{gudel2004}, including the Sun \citep{judge2003}. 

The fuchsia triangles in Fig.~\ref{fig: teffxray} show upper limits for \nxrayul\ stars with $d \lesssim$ 17~pc. Most stars in this group hug the lower envelope of fractional X-ray luminosities for stars with \teff\ $\approx$ 3000--9000~K. Stars with X-ray upper limits and $d \gtrsim$ 17~pc have similar \teff\ as those with $d \lesssim$ 17~pc and lie at larger \lxlstar. Two stars with measured \teff\ and \lxlstar\ = $5.1 \times 10^{-10}$ and $2.8 \times 10^{-9}$ are not shown. 

To demonstrate that the lower envelope of X-ray detections for GKM stars is a sensitivity limit, we derive \lxlstar\ using the approximate \rosat\ (\erosita) sensitivity limit, $F_X$ = $1.5 \times 10^{-13}$~erg~cm$^{-2}$~s$^{-1}$ ($F_X = 5 \times 10^{-14}$~erg~cm$^{-2}$~s$^{-1}$), from \citet{freund2022}
% bcb ok?
%(\citet{merloni2024}). 
(\citealt{merloni2024}). 
For the \citet{pecaut2013} locus of main sequence stars in the \teff--\lstar\ plane, the green line in Fig.~\ref{fig: teffxray} shows the expected \teff--\lxlstar\ relation for main sequence stars with $d$ = 17~pc (10~pc) and an X-ray flux at the \erosita\ (\rosat) sensitivity limit. This estimate matches the lower limit of \lxlstar\ for late G and KM stars rather well.
 
{\bf Host Star Properties Summary.}\quad
Table~\ref{tab: stardata} summarizes basic statistics for the properties of CDDS stars as a function of spectral type. As in \S\ref{sec: companions}, we adopt the \citet{pecaut2013} relation between spectral type and effective temperature.  Nearly all stars, $\sim$ 94\% have an accurate spectral type published in the literature or derived from \gaia\ low resolution spectroscopy. Together with published effective temperatures in the literature, from \gaia, and from color indices, the full set of spectral types yields stellar effective temperatures for 100\% of the sample. All of the stars have an accurate K-band magnitude and most have an accurate optical magnitude (either G, V, or $g$), which yields a luminosity derived from G, K, V, or $g$; the optical extinction; and a \gaia\ or \hipparcos\ parallax (or distance estimate based on cluster membership or location on the main sequence). Approximately 80\% (64\%) have a measured \logg\ (metallicity). Among the other stellar properties, 30\% to 38\% of the sample has a measured Li abundance, rotational period, and the two Ca~II chromospheric activity indicators. More than half (59\%) have measurements for \vsini; nearly 55\% have  \lxlstar\ $>$ 0. 

All of these measures have an uneven distribution with respect to SpT, \teff, or \lstar. FG stars have an almost complete set of \logg\ data; A stars lag; KM stars lag more; and B stars lag the most. BM stars have relatively less \feh\ data than AK stars and much less than FG stars. With little or no stellar activity, BA stars have few measurements of \prot, A(Li), \lrhkp, or \lrirtp. For these measures, G stars have a larger fraction of data than FKM stars. The \vsini\ and \lxlstar\ measurements are rather evenly distributed among all spectral types; A stars have the smallest fraction of \lxlstar\ measurements, but are well-represented in the set of \vsini\ measurements.

\begin{deluxetable}{lccccccC}
\tablecolumns{8}
%\tablewidth{10cm}
\tabletypesize{\scriptsize}
\tablecaption{Stellar Data Summary}
\tablehead{
  \colhead{Measure} &
  \colhead{All} &
  \colhead{B} &
  \colhead{A} &
  \colhead{F} &
  \colhead{G} &
  \colhead{K} &
  \colhead{M}
}
\label{tab: stardata}
\startdata
SpT      & 0.937 & 1.000 & 1.000 & 0.972 & 0.960 & 0.859 & 0.855 \\
\teff\   & 1.000 & 1.000 & 1.000 & 1.000 & 1.000 & 1.000 & 1.000 \\
\lstar\  & 1.000 & 1.000 & 1.000 & 1.000 & 1.000 & 1.000 & 1.000 \\
\logg\   & 0.804 & 0.709 & 0.907 & 0.937 & 0.919 & 0.645 & 0.627 \\
{[Fe/H]} & 0.642 & 0.179 & 0.461 & 0.829 & 0.865 & 0.700 & 0.446 \\
{[Li/H]} & 0.301 & 0.006 & 0.052 & 0.367 & 0.504 & 0.363 & 0.282 \\
\prot\   & 0.376 & 0.109 & 0.098 & 0.336 & 0.626 & 0.407 & 0.538 \\
\vsini\  & 0.586 & 0.641 & 0.620 & 0.627 & 0.683 & 0.505 & 0.433 \\
\lxlstar\ & 0.543 & 0.406 & 0.376 & 0.609 & 0.599 & 0.542 & 0.627 \\
Ca II HK & 0.312 & 0.000 & 0.015 & 0.358 & 0.630 & 0.354 & 0.268 \\
Ca II IRT & 0.303 & 0.000 & 0.000 & 0.351 & 0.604 & 0.432 & 0.141 \\
\enddata
\end{deluxetable}

Table~\ref{tab: activity} addresses the frequency of activity indicators -- \lih, \prot, \vsini, \lxlstar, \lrhkp, and \lrirtp\ --  as a function of spectral type. BA stars have more than two measures less than 5\% of the time, but they usually have at least one (71\% for B stars and 72\% for A stars). F-type stars are the most common stars in the CDDS; they often have two or three activity measures (44\%); another 35\% have one or four measures. F stars more often have five or six measures (13\%) than none (8\%).

G-type stars dominate the activity measure statistics. Only 2\% have no measure of activity within the set of six. Among the six measures for G stars, 25\% have four, 22\% have five, 20\% have three, 12\% have two, 11\% have six, and 7\% have one. K-type stars have somewhat worse statistics than F-type stars. More than 90\% of the K stars have at least one measure; two to four measures are most common, followed in frequency by one, five, and six. Although M-type stars rarely have none of the activity indicators, they never have all six. More than 55\% of M stars have two or three indicators; another 35\% have one or four. 

\begin{deluxetable}{lcccccccC}
\tablecolumns{9}
%\tablewidth{10cm}
\tabletypesize{\scriptsize}
\tablecaption{Stellar Activity Statistics}
\tablehead{
  \colhead{SpT} &
  \colhead{All} &
  \colhead{0} &
  \colhead{1} &
  \colhead{2} &
  \colhead{3} &
  \colhead{4} &
  \colhead{5} &
  \colhead{6}
}
\label{tab: activity}
\startdata
B & 357 & 0.232 & 0.409 & 0.322 & 0.036 & 0.000 & 0.000 & 0.000 \\
A & 518 & 0.230 & 0.434 & 0.286 & 0.044 & 0.006 & 0.000 & 0.000 \\
F & 852 & 0.093 & 0.177 & 0.208 & 0.218 & 0.158 & 0.107 & 0.039 \\
G & 681 & 0.063 & 0.057 & 0.106 & 0.186 & 0.247 & 0.214 & 0.126 \\
K & 771 & 0.188 & 0.130 & 0.145 & 0.170 & 0.204 & 0.122 & 0.042 \\
M & 496 & 0.109 & 0.200 & 0.230 & 0.264 & 0.149 & 0.048 & 0.000 \\
\enddata
\end{deluxetable}
In the current paradigm, these activity indicators change as the interior structure of a star evolves \citep[e.g.,][]{johnstone2021,dumont2021a}. For example, the observed \lih\ in low mass stars depends on metallicity and complex interactions between convection, rotation, and various transport processes, including diffusion, meridional circulation, magnetic torque, turbulence, and viscosity \citep[e.g.,][]{deliyannis2000,talon2005,semenova2020,dumont2021b,dumont2023}. The chromospheric properties of FGKM stars, including \lrhkp, \lrirtp, and \lxlstar, depend on the evolution of the magnetic field and the Rossby number, $Ro$, which is the ratio of the rotational period to the convective turnover time, $\tau_c$: $Ro$ = \prot\ / $\tau_c$ \citep[e.g.,][]{mamajek2008,gallet2015,tu2015,johnstone2021}. As \prot, $B$, and $Ro$ decline with time, \lih, \lrhkp, \lrirtp, and \lxlstar\ should follow. 

\begin{figure}[t]
\begin{center}
\includegraphics[width=3.25in]{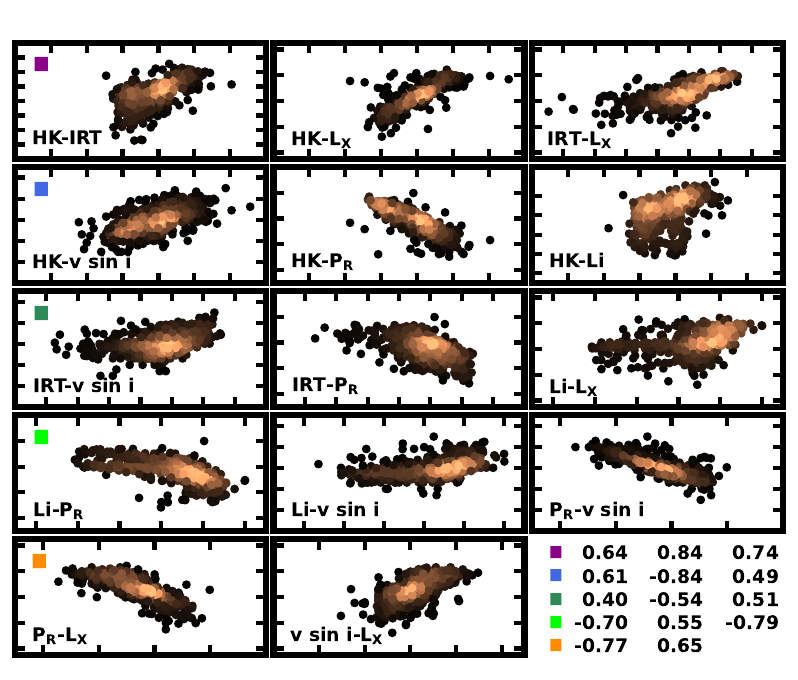}
\vskip -1ex
\caption{
\label{fig: CorrPlot}
Correlations for 14 pairs of activity indices listed in Table~\ref{tab: correlations} (not shown: \rirtp--\lih, which has a negligible correlation coefficient). Text in the lower left corner of each plot indicates the pair of indices; the x-axis (y-axis) plots the first (second) index listed. The lower right corner lists the {\it Spearman} correlation coefficients from Table~\ref{tab: correlations}. For convenience, squares at the start of each row of coefficients correspond to squares in the upper left corners of plots in the first column of the array of plots.
}
\end{center}
\end{figure}

To make an initial analysis of activity indicators among CDDS stars, we consider correlations between each pair of measures. In a future paper, we explore ways to combine these into stellar age estimates. Here, we simply note which measures are more correlated with other measures. For this task, we employ the Pearson, Spearman, and Kendall methods to search for correlations. The number of $(x,y)$ pairs for the correlations ranges from 467 for \rhkp--\lih\ to 859 for \rhkp--\vsini. We quote results for the Pearson and Spearman tests; the Kendall $\tau$ method yields results indistinguishable from the Spearman test. Table~\ref{tab: correlations} summarizes the correlation coefficients and the $p$-values for each pair of variables. Small $p$-values indicate a small probability that the variables are uncorrelated. Fig.~\ref{fig: CorrPlot} shows the nearly complete set of measurements for CDDS stars. With the largest likelihood of no correlation, we do not plot results for \lrirtp\ versus A(Li).

\begin{deluxetable}{lccccC}
\tablecolumns{6}
%\tablewidth{10cm}
\tabletypesize{\footnotesize}
\tablecaption{Correlation Coefficients\tablenotemark{{\footnotesize \rm 1}}}
\tablehead{
  \colhead{Pair} &
  \colhead{$N$} &
  \colhead{~$r_p$} &
  \colhead{$p_p$} &
  \colhead{~$r_S$} &
  \colhead{$p_S$}
}
\label{tab: correlations}
\startdata
\rhk--\rirt\      & 501 & $ 0.58$ & $ 1 \times 10^{-46}$ & $ 0.63$ & $9 \times 10^{-58}$ \\
\rhk--\lxlstar\   & 589 & $ 0.74$ & $ 7 \times 10^{-104}$ & $ 0.84$ & $2 \times 10^{-158}$ \\
\rhk--\lih\       & 467 & $ 0.37$ & $ 8 \times 10^{-17}$ & $ 0.49$ & $3 \times 10^{-29}$ \\
\rhk--\prot\      & 534 & $-0.75$ & $ 1 \times 10^{-98}$ & $-0.84$ & $5 \times 10^{-142}$ \\
\rhk--\vsini\     & 859 & $ 0.58$ & $ 6 \times 10^{-80}$ & $ 0.60$ & $4 \times 10^{-86}$ \\
\rirt--\lxlstar\  & 559 & $ 0.68$ & $ 8 \times 10^{-77}$ & $ 0.74$ & $8 \times 10^{-99}$ \\
\rirt--\lih\      & 469 & $ 0.14$ & $ 2 \times 10^{-3}$ & $ 0.20$ & $1 \times 10^{-05}$ \\
\rirt--\prot\     & 564 & $-0.53$ & $ 7 \times 10^{-43}$ & $-0.54$ & $3 \times 10^{-44}$ \\
\rirt--\vsini\    & 656 & $ 0.38$ & $ 5 \times 10^{-24}$ & $ 0.40$ & $3 \times 10^{-26}$ \\
\lih--\lxlstar\   & 471 & $ 0.41$ & $ 1 \times 10^{-20}$ & $ 0.51$ & $3 \times 10^{-32}$ \\
\lih--\prot\      & 530 & $-0.61$ & $ 3 \times 10^{-54}$ & $-0.70$ & $5 \times 10^{-78}$ \\
\lih--\vsini\     & 627 & $ 0.43$ & $ 3 \times 10^{-30}$ & $ 0.55$ & $1 \times 10^{-50}$ \\
\prot--\lxlstar\  & 586 & $-0.77$ & $ 3 \times 10^{117}$ & $-0.81$ & $6 \times 10^{-137}$ \\
\prot--\vsini\    & 671 & $-0.78$ & $ 3 \times 10^{-141}$ & $-0.79$ & $9 \times 10^{-144}$ \\
\vsini--\lxlstar\ & 682 & $ 0.63$ & $ 1 \times 10^{-77}$ & $ 0.65$ & $5 \times 10^{-83}$ \\
\enddata
\tablenotetext{1}{These results measure correlations between logarithms of each quantity.}
\end{deluxetable}

Aside from the \rirtp--\lih\ pair, all pairs of activity measures show obvious correlations. Among these, the $p$-values do not obviously depend on the number of stars with both activity measures. Two of the three strongest correlations, \rhkp--\prot\ (534) and \prot--\lxlstar\ (586), have rather few data points compared to \rhk--\vsini\ (846), which has the most data points and one of the smaller $p$-values. 

The three strongest correlations involve \rhkp, \prot, and \lxlstar. Among others, \citet{johnstone2021} demonstrates clear correlations between \prot\ and \lxlstar\ with the evolution of the Rossby number, $Ro$. As stars age, \prot\ grows while $\tau_c$ reaches an equilibrium. Thus, $Ro$ steadily increases as \prot\ increases. Increasing $Ro$ leads to a decrease in \lxlstar, far-ultraviolet continuum emission, and other activity indicators. The Ca~II activity indicators clearly share this evolution: \rhkp\ is clearly tied to \lxlstar\ and \prot. Although the Ca~II IR triplet flux is less correlated with \prot, it is strongly correlated with \lxlstar. 

Differences of the $p$-values among the pairs of variables illustrates the fickleness of low mass stars. Among stars with the same mass, \teff, and \lstar, some maintain high rotation rates far longer than the rest of their cohort; others spin down more rapidly \citep[e.g.,][]{johnstone2021,getman2023,santos2025,mathur2025}. Similarly, some stars lose their lithium rapidly while others somehow maintain relatively high lithium abundances compared to the rest of their cohort \citep[e.g.,][]{dumont2021b,rathsam2023,winnick2025}. While there are clear trends in the evolution of stellar activity with age, individuality introduces profound and sometimes large differences from star to star.

Comparing these results with other estimates is challenging. While many studies show a clear correlation between various activity indicators, they generally do not quote a correlation coefficient \citep[e.g.,][]{mamajek2008,dumont2021b,johnstone2021,nunez2024,shan2024}. In general, the correlations look similar to those shown in Fig.~\ref{fig: CorrPlot}. 

\citet{lanzafame2023} consider correlations between the two Ca indices, using unpublished archival data from ESO--FEROS (671 sources) and the extensive \citet{borosaikia2018} compilation (220 sources). With many CDDS stars in \citet{borosaikia2018}, we focus on this comparison. \citet{lanzafame2023} derive a {\it Pearson} linear correlation coefficient $r$ = 0.5 and a relation \lrirtp\ = $a$ \lrhkp\ - $b$, with $a$ = 0.841 and $b$ = 1.661. The CDDS data yield similar results for 501 stars: $r$ = 0.63, $a$ = 0.817, and $b = 1.525$. We also infer correlations between \teff\ and the difference in the two Ca indices, $\delta R_{\rm Ca} =$ \lrirtp\ $-$ \lrhkp: 
\begin{align}
\label{eq: deltaRCa}
\delta R_{\rm Ca} = 
\left\{
    \begin {aligned}
          & -0.547 ~ T_{1000} + 2.374  & ~~T_{1000} \ge 5 \\
          & -0.50 & ~~T_{1000} < 5
    \end {aligned}
\right.
\end{align}
where $T_{1000}$ = \teff/1000. Both correlations have $r$ = 0.57. These results differ from those in \citet{martin2017}, who analyze 2274 observations of 82 stars. 

\citet{freund2025} consider correlations between \lxlstar\ and \rirtp\ for a set of 43,200 stars with reliable measurements of \erosita\ X-ray and \gaia\ DR3 Ca~II IR triplet fluxes. Aside from F-type stars, the smaller set of CDDS stars shows similar trends but with stronger correlations; the Pearson correlation coefficients for CDDS stars are 0.75 (all stars), 0.42 (F), 0.77 (G), 0.86 (K), and 0.74 (M) compared to 0.67 (all), 0.45 (F), 0.69 (G), 0.81 (K), and 0.57 (M) for the \citet{freund2025} sample. Differences could be a result of small number statistics or the overall youth of CDDS stars compared to other stellar samples.

\section{Membership in Associations, Clusters, and Moving Groups}
\label{sec: clusters}

To select and to verify targets for observations of moving groups, open clusters, and stellar associations, \spitz\ programs often relied on published membership lists. For nearby (distant) regions, \hip\ (ground-based) parallaxes and proper motions helped to separate members from the background or foreground. Sometimes, color-magnitude diagrams and stellar activity measures also served to identify cluster stars. 

Today, \gaia\ data enable significant improvements in defining the members of associations, clusters, and moving groups. Among Hyades stars, for example, the typical fractional error for ground-based parallax or proper motion measurements is $\sim$ 10\% \citep[e.g.,][and references therein]{vanaltena1995,zacharias2004}. \hip\ improved on parallax (proper motion) errors by a factor of 2--3 ($\sim$ 10). In contrast, the fractional errors for \gaia\ parallax (proper motion) measurements of Hyades stars are $\sim$ 0.1\% (0.02\%)! Although the fractional errors increase for more distant stars, the parallax uncertainty for Blanco~1 ($d \approx$ 237~pc) is still small, $\sim$ 1\%.

To update membership data for CDDS stars, we rely on recent \gaia-based studies. These analyses adopt different techniques that include some combination of coordinates on the sky, parallax, proper motion, and radial velocity. \gaia\ or ground-based color-magnitude diagrams often add to the analysis. For CDDS stars, all have a measured or adopted parallax, all but three have a proper motion, and \nvrad\ (97\%) have a measured \vrad. When the dynamical data are used to construct galactic coordinates $(X, Y, Z)$ and velocities $(U, V, W)$ relative to the galactic center, analyses must account for uncertainties in the solar motion \citep[e.g.,][]{bobylev2016,mikkola2023}. Among the studies we consult, several are all-sky compilations \citep{cantat2018,dias2021,hunt2023,hunt2024}. Others focus on 10--20 specific, well-known regions \citep[e.g.,][]{liu2019,pang2021,jackson2022}. As outlined below, many analyses focus on 1--3 associations, clusters, or groups. When any of these studies include membership probabilities $p$, we exclude stars with $p < 0.5$.

While the focus of membership analysis is on dynamical properties, we consider comparisons with model stellar isochrones when stars considered members in \spitz\ studies are rejected as members in \gaia\ dynamical studies. Usually, rejected stars lie well outside the volume of {\it bone fide} cluster members. Sometimes, rejected stars lie within the cluster; model stellar isochrones are then helpful when deciding cluster membership. For this task, we use version 1.2S of the PAdova and tRieste Stellar Evolutionary Code \citep[PARSEC;][]{ychen2014,jtang2014,ychen2015}. While other options are available, \citet{brandner2023a,brandner2023b,brandner2023c} matched observations of two CDDS clusters, the Hyades and Pleiades, 
% per referee comment, bcb added
to
PARSEC models. In a detailed study of the Castor system ($\alpha$ Gem AB, a CDDS binary system), \citet{torres2022} matched interferometric measures of mass and radius to PARSEC isochrones. 

We begin the discussion with a few examples of associations, clusters, and moving groups and then continue with summaries of the remaining clusters. For each system, we briefly outline relevant \spitz\ programs, \gaia\ analyses, and the number of CDDS stars considered as dynamical members. After describing dynamical measurements that exclude membership for a set of stars, we consider membership based on stellar activity measures. Combining the activity and dynamical information, we close with a best-estimate of CCDS members in each association, cluster, and moving group.

\subsection{The Pleiades and Praesepe Open Clusters}
\label{sec: PlPr}

\begin{figure}[t]
\begin{center}
\hskip -20ex
\includegraphics[width=4.5in]{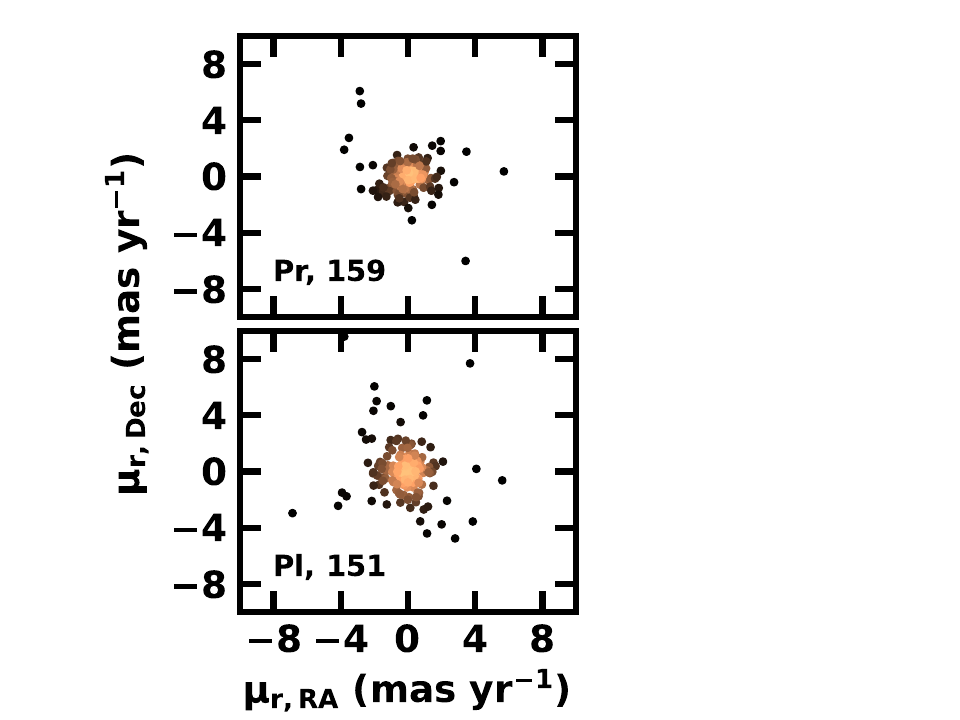}
%\vskip -2ex
\caption{
\label{fig: pmPlPr}
Final proper motion distributions for CDDS stars in Praesepe (upper panel, 160 stars) and the Pleiades (lower panel, 151 stars). Data for each cluster are centered on the median proper motion from \citet{alfonso2023}: ($\mu_{RA}, \mu_{Dec}$) = ($19.95, -45.44$) for the Pleiades and ($-35.93, -12.88$) for Praesepe. The density of members increases from low (black points) to medium (gold points) to high (bright yellow points). 
}
\end{center}
\end{figure}

The Pleiades and Praesepe are good examples of relatively nearby open clusters with well-defined proper motion distributions and accurate parallaxes. Fig.~\ref{fig: pmPlPr} shows the distribution of proper motions for cluster members (defined below), centered on their median proper motions \citep{alfonso2023}. Each has a dense, yellow core of stars close to the median proper motion, a set of gold-colored stars on the outskirts of this core, and a few outliers in black. The Pleiades has a few more outliers than Praesepe, which is probably a result of its smaller distance. 

{\bf Pleiades.}\quad
Several \spitz\ programs targeted the Pleiades \citep{stauffer2005,gorlova2006,meyer2006,su2006,sierchio2010}. Each group selected high probability members based on parallax, proper motion, chromospheric activity, lithium abundance, and position on the color-magnitude diagram \citep[e.g.,][and references therein]{stauffer1987,soderblom1993b,soderblom1993c}. Across the five programs, the CDDS sample contains 165 stars with spectral types from late B to early M.

Four recent studies of the Pleiades with \gaia\ data place new constraints on likely members. \citet[][DR3: 959 stars]{alfonso2023} and \citet[][DR3: 911 stars]{brandner2023c} focus on the core regions. The stars in these surveys have a proper motion footprint similar to the CDDS stars in Fig.~\ref{fig: pmPlPr}. \citet[][DR2: 2281 stars]{lodieu2019b} and \citet[][eDR3: 1273 stars]{heyl2022} cast a much wider net to identify stars in the core and in tidal tails. The combined set of likely members comprises 2313 stars; 146 CDDS stars match entries in all four catalogs. 

Of the 19 non-matches in the CDDS, we restore five where the \gaia\ parallax and proper motion data have large errors and the ground-based parallax, proper motion, and radial velocity data suggest membership. For the 14 other non-matches, some combination of the \gaia, \hip, or ground-based parallaxes and proper motions (and their errors) suggest a star in the foreground, the background, or with a significant velocity relative to Pleiades stars. Some of these stars have strong indicators of youth and thus might be Pleiades-age stars in the vicinity of (but not within) the cluster.

{\bf Praesepe.}\quad
\citet{su2006} selected 5 A-type stars in Praesepe from the literature. To construct a more comprehensive view, \citet{gaspar2009} matched a large set of \spitz\ targets with accurate photometry to high quality proper motion-based membership lists \citep[e.g.,][]{wang1995,kraus2007}. They collected photometry from the literature, constructed a $g-r, r$ color-magnitude diagram, and focused on a set of high probability members based on proper motions and distance (in magnitudes) from an isochrone delinated by the highest probability cluster members. The final target list includes 193 high probability members with optical, near-infrared, and \spitz\ photometry \citep[see also][]{urban2012}; five of these are in \citet{su2006}.

From the historical literature, \citet{lodieu2019b} compiled 2078 unique sources among a list of 6479 across all studies. After including \gaia\ DR2 data and applying detailed kinematic and Bayesian procedures to the full compilation, they produce a set of 2199 candidates. From \gaia\ DR3 data, \citet{alfonso2023} use two separate analyses to construct a set of 744 likely members within a somewhat smaller volume than those of \citet{lodieu2019b}; all but 19 of these are included in the \citet{lodieu2019b} list.

Of the 193 CDDS stars, 146 match candidates in \citet{cantat2018}, \citet{lodieu2019b}, \citet{alfonso2023}, and \citet{hunt2023,hunt2024}. Of the 47 non-matches, 11 (19) are clearly in the foreground (background) of the cluster, closer than (beyond) a few tidal radii from the cluster center. Five stars have distances close to the cluster median, but the proper motions are well outside the cluster limits. Of the remaining stars, (i) three have parallax or proper motion errors too large to make a clear conclusion regarding their membership; (ii) five have parallaxes and proper motions within the bounds of most of the matches; and (iii) another five have no \gaia\ or \hip\ data, have reasonable proper motions from ground-based measurements, and have radial velocities consistent with other cluster members. We assign these last thirteen stars to Praesepe, which brings the number of cluster stars to 159 out of the original 194 cluster members. 

Information on stellar activity is insufficient to promote any other non-matches to cluster membership. Stars obviously in the cluster foreground or background have activity levels that place them with older field stars. Several stars with similar parallax but discrepant proper motion relative to the cluster also have similar activity levels as Praesepe stars, but the proper motions preclude cluster membership. However, all of the 13 non-matches assigned to the cluster and described above have activity levels similar to other Praesepe members with similar spectral types, which supports our assignment of these stars to the cluster. 

\subsection{The Sco--Cen Association}
\label{sec: assoc}

Of the three stellar associations with more than 100 stars in the CDDS, Upper Scorpius (USco) was the most popular \spitz\ target \citep{chen2005a,meyer2006,su2006,carpenter2009b,chen2011,chen2012}. Four programs included stars in Upper Centaurus-Lupus \citep[UCL;][]{meyer2006,su2006,chen2011,chen2012}, while only two focused on Lower Centaurus-Crux \citep[LCC;][]{chen2011,chen2012}. These programs selected stars from ground-based optical \citep{blaauw1964,degeus1989,priebisch1998}, \hip\ \citep{dezeeuw1999}, and X-ray \citep{walter1994,sciortino1998} membership lists. As in \citet{dezeeuw1999}, we use galactic coordinates to make initial assignments of stars to USco, UCL, and LCC. After eliminating stars with distances $d <$ 100~pc, we start with 601 CDDS stars in these associations, 292 in USco, 148 in UCL, and 161 in LCC. 

\begin{figure}[t]
\begin{center}
\vspace*{-0.25cm}
\hspace*{-0.40cm}
\includegraphics[width=4.0in]{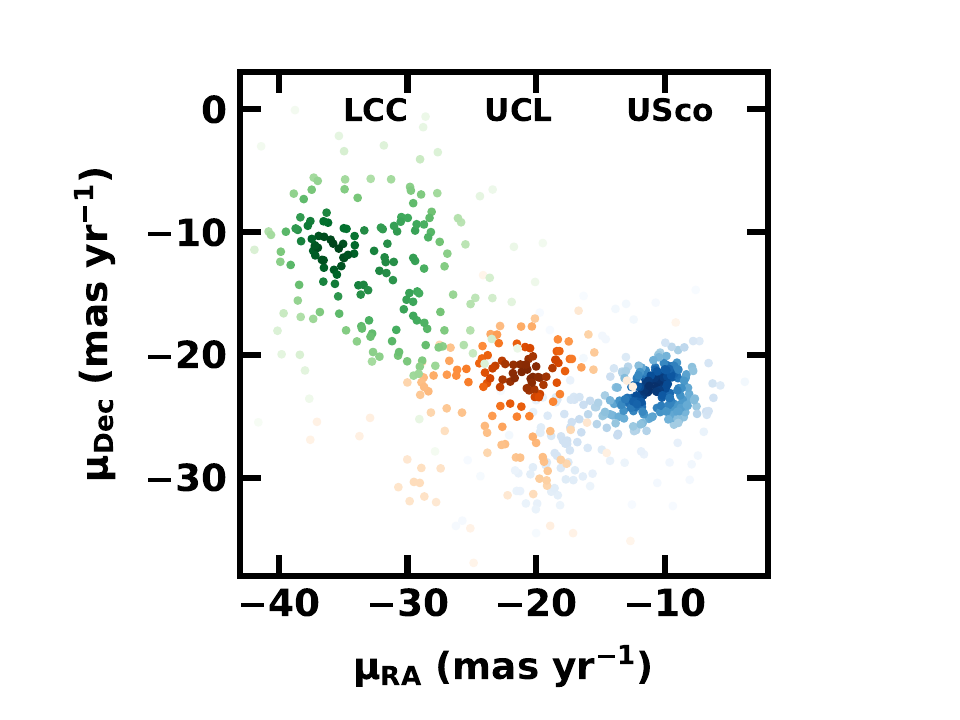}
\vskip -2ex
\caption{
\label{fig: pmLCUCUS}
Proper motion distributions for CDDS stars in LCC (179 stars; upper left, green colors), UCL (140 stars, middle, orange colors) and USco (307 stars; lower right, blue colors). The density of members increases from low (light colors) to high (dark colors). There is some overlap in proper motions between UCL and USco and less overlap between LCC and UCL.
}
\end{center}
\end{figure}

Fig.~\ref{fig: pmLCUCUS} shows the proper motion distributions of the final set of LCC, UCL, and USco members on an expanded scale compared to Fig.~\ref{fig: pmPlPr}. In this space, USco is the most dense, followed by UCL and then LCC. Most stars lie in one of the main concentrations. In addition to some overlap of UCL and USco stars, some stars lie well outside any concentration.  In proper motion space, all three associations are much more extended than either the Pleiades or Praesepe.

As summarized in \citet{luhman2022a}, numerous groups have analyzed \gaia\ data to select young stars in the region that includes USco, UCL, and LCC. Here, we match CDDS stars to candidates identified in \citet{luhman2022a} and several other investigations \citep{goldman2018,damiani2019,miretroig2022,briceno2023,wood2023,ratzenbock2023}. All of these studies, including others listed in \citet{luhman2022a}, apply different techniques to analyze different sets of \gaia\ parallaxes and proper motions (e.g., DR2, eDR3, or DR3) with or without \gaia\ or ground-based radial velocities. While these studies have a set of common candidates, each study has additional candidates not included in the others. Our strategy is to (i) compile a set of CDDS stars within a coordinate range that generously encompasses USco, UCL, and LCL, (ii) match CCDS stars to candidates from one of the seven studies listed above, and (iii) eliminate matches from the original list. We then repeat (ii) and (iii) for each of the remaining catalogs. From the distributions of parallax and proper motion for the matches, we remove outliers from the set of non-matches. This exercise results in 617 matches and 44 non-matches with parallaxes and proper motions similar to the matched stars. A few matches have radial velocities that seem discrepant from the rest of the matches; however, these stars have similar measurements of \teff, \lstar, and \logg\ as the rest of the stars in LCL, UCL, and USco. 

Among the non-matches, we reject several with sky positions that fall well outside the complete collection of members from sources listed above. We then map the distribution of distances and radial velocities for members across the sky and use this map to include or reject the rest of the non-matches. This exercise results in 28 additional matches with distances and radial velocities close to members with similar sky position and proper motions.

As a final check on membership, we compare the stellar activity measurements of the remaining non-matches with the set of matches. Compared to Sco--Cen stars, all but three of the 16 remaining non-matches have the activity levels of much older stars. Two stars with moderate activity levels suggestive of a young star lie either too far north or too far south relative to the association boundaries. However, HD~119022 has strong Ca II H\&K and X-ray emission, lies on the near side of the bulk of the LCC stars, and has a proper motion on the edge of the LCC proper motion distribution. We include this star in LCC.  

The approach outlined above maximizes the number of CDDS stars in the three associations. For each star not included in the original set of assignments to USco, UCL, and LCC or in some other cluster or association, we assign the star to the closest association based on sky position and distance. We then have 645 matches to CDDS stars within or near the galactic coordinate boundaries defined in \citet{dezeeuw1999}: 308 in USco, 140 in UCL, 179 in LCC, 12 in IC~2602 (discussed below), 5 in TW Hya (discussed below), 1 in $\epsilon$ Cha, and 1 in Cha-Near. The full set of 627 Sco--Cen stars includes $\sim$ 25 more stars than the original assignment.

\subsection{The 93 Tau Association}
\label{sec: taurus}

The FEPS program targeted eight stars in the general vicinity of the Taurus--Auriga association \citep{meyer2006}. As noted in the Appendix, we assign V1299~Tau to the field. Of the remainder, five lie within the 93~Tau association, with distances of 104--122~pc \citep{luhman2023a}. Another lies behind 93 Tau at 136~pc and has the low gravity, short \prot, and large \rhkp\ expected for a pre-main sequence star. The final target lies at the appropriate distance but has a much different proper motion. Still, the star has a short rotational period and proper placement in the HR diagram for a young star. We assign five of these stars to 93 Tau and two to the Taurus molecular cloud.

\subsection{Nearby Moving Groups}
\label{sec: groups}

Moving groups, also known as stellar kinematic groups, have a long history \citep[see the discussions in][and references therein]{roman1949,eggen1963,eggen1970,soderblom1993a}. Based primarily on color-magnitude diagrams, proper motions, and parallaxes for the nearest stars, these studies identified groups of stars with similar ages and space motions. The UMa group (also known as the Sirius group), the Hyades group, the Wolf 630 group, and the Local Association (Pleiades moving group) are good examples of moving groups discovered prior to \hip\ \citep[see also][]{montes2001}.

\hip\ parallaxes enabled the discovery of new moving groups and the reexamination of older groups \citep[e.g.,][]{barrado1998,montes2001,zuckerman2001a,zuckerman2001b,zuckerman2001c,king2003,bovy2009,zuckerman2011,zuckerman2013}. While many moving groups consist of stars with ages of 100--300~Myr, others have ages as young as 10--20~Myr. The mix of young and old stars among the complete ensemble of moving groups is an interesting outcome of the \hip\ mission.

Various \spitz\ and \herschel\ programs targeted $\sim$ 210 stars in young moving groups \citep{chen2005b,su2006,rebull2008,plavchan2009,zuckerman2011,donaldson2012,chen2014,thureau2014,vican2016,sibthorpe2018}. Within the CDDS, the AB Dor (47), Tuc--Hor (47), $\beta$ Pic (30 stars), TW Hya (23),  Ursa Major (16), and Columba (16) moving groups are well-represented. There are also stars from the Argus, Her-Lyra, and Octans moving groups. 

\begin{figure}[t]
\begin{center}
\hskip -20ex
\hspace*{1.7cm}\includegraphics[width=3.7in]{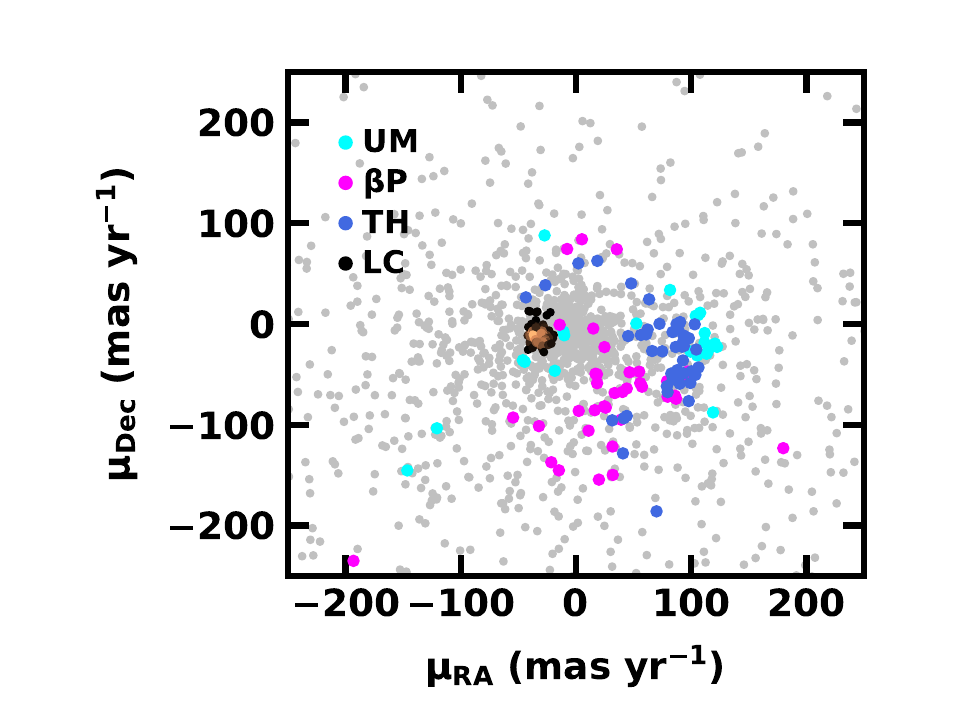}
\vskip -2ex
\caption{
\label{fig: pmMG}
Proper motions for CDDS stars in LCC (black, gold, yellow points as in Fig.~\ref{fig: pmLCUCUS}), $\beta$ Pic (fuchsia points), Tuc--Hor (dark blue points) and Ursa Major (cyan points). Nearby stars in moving groups have a much larger range of proper motions than stars in associations (e.g., LCC) and clusters (e.g., Pleiades and Praesepe; Fig.~\ref{fig: pmPlPr}). 
}
\end{center}
\end{figure}

Recent studies using \gaia\ data suggest 205 CDDS stars as members of the standard moving groups \citep{cantat2018,gagne2018,kounkel2019,miretroig2020,ujjwal2020,hunt2023,popinchalk2023,hunt2024,lee2024}. Of these, most had been listed as moving group members in various \spitz\ and \herschel\ programs. Others are new identifications. \citet{kounkel2019} recognize 11 other CDDS stars as members of 11 previously unknown groups (e.g., Theia 73 and Theia 113). Together, the combined set of old and new candidates includes 259 potential young, nearby moving group members. After analyzing the stellar properties in \S\ref{sec: host-stars}, we reject 46 stars due to (i) more accurate \gaia\ parallaxes and proper motions that result in space velocities inconsistent with other moving group stars, (ii) more accurate \teff, \logg, \lstar\ that do not place the stars on the appropriate PARSEC isochrone for their nominal group, and (iii) new activity indicators that imply much older ages than the ages of other nominal group members. Thus, we assign 213 CDDS stars to moving groups (Table~\ref{tab: clusters}). 

% to justify the numbers above
% 143 original members of moving groups from Spitzer and Herschel
% 181 members of previously known groups using Gaia data (76 in original list -> 105 new, so now the old+new = 248)
% 11 in Kounkel, together with 248 -> 259 !!
% we adopt 213 in moving groups and reject 46
% the 46 rejections have better Gaia parallaxes and proper motions
% 

Fig.~\ref{fig: pmMG} compares the proper motion distributions of several moving groups with measurements for LCC. Instead of the compact structure observed in open clusters and stellar associations, proper motions for moving groups have a wide spread and do not follow a coherent pattern. Isolating these moving groups requires converting parallaxes, proper motions, radial velocities, and positions on the sky to galactic coordinates and finding their common motions. 

\subsection{Other Open Clusters}

{\bf $\alpha$ Per}\quad
As with other \spitz\ programs, \citet[][13 FGK-type stars]{meyer2006} and \citet[][5 A-type stars]{su2006} selected targets from ground-based proper motion data and measurements of stellar activity. Among recent \gaia\ surveys, \citet{cantat2018}, \citet{lodieu2019b}, and \citet{hunt2023,hunt2024} provide extensive membership lists. Thirteen CDDS stars, including one nominal field star, are cluster members. \gaia\ parallaxes for three (two) others place them in the foreground (background) of the cluster. 

{\bf Blanco~1.}\quad
Based on studies published prior to 2004, \citet{stauffer2010} selected 38 members for \spitz\ observations. Subsequent photometric and spectroscopic surveys suggest all 38 are members \citep{mermilliod2008,gonzalez2009}. \citet{su2006} lists HD 225200 as a member, but this star is not considered a member in more recent dynamical studies. \citet{cantat2018}, \citet{jackson2020,jackson2022}, \citet{zhang2020}, \citet{pang2021}, \citet{alfonso2023}, and \citet{hunt2023,hunt2024} analyze \gaia\ data and report new membership lists. 

The 38 CDDS members of Blanco~1 have a tight core in proper motion space, similar to those in the Pleiades and Praesepe (Fig.~\ref{fig: pmPlPr}). Distances range from 218.2~pc to 241.8~pc; with an inter-quartile range of 4.53~pc, the median distance, $d_{med}$ = 234.3 pc, is identical to the \citet{alfonso2023} distance of 236.67$\pm$2.10. The dynamical studies quoted above yield 33 \citep{cantat2018}, 31 \citep{zhang2020}, 33 \citep{pang2021}, 24 \citep{alfonso2023}, 27 \citep{hunt2023}, and 34 \citep{hunt2024} CDDS members. These analyses conclusively reject four stars as members; all are in the cluster background. The remaining 34 are high probability members; all have similar parallaxes, proper motions, and radial velocities. 

{\bf $\eta$ Cha.}\quad
\citet{mamajek1999} discovered this cluster from a combination of ROSAT X-ray observations and \hip\ astrometry. Later, 
\citet{gautier2008} 
% bcb this ok? 
% (\citet{riviere2015}) 
(\citealt{riviere2015}) 
reported \spitz\ (\herschel) observations of the brightest cluster members. Among the 16 CDDS stars in $\eta$~Cha, 15 have \gaia\ parallaxes and proper motions; all but two form a tight cluster with a median distance, $d_{med} \approx$ 98.4~pc, close to the $d$ = 98.2~pc in \citet{hunt2023,hunt2024}. These stars are clearly cluster members \citep{cantat2018,gagne2018,hunt2023,hunt2024}. EN Cha does not have a \gaia\ parallax. The parallax for ET Cha places it slightly behind the cluster, but the measurement has a large error. With strong X-ray luminosities, EN Cha and ET Cha are certain cluster members.

{\bf Coma.}\quad
Within larger \spitz\ surveys, \citet{su2006} (\citet{moor2011a}) assigned five A-type (two F-type) stars to Coma. \citet{urban2012} later reported observations of 83 stars that included those discussed in \citet{su2006}. While all were once considered proper motion members of Coma \citep[e.g.,][]{abad1999,kraus2007}, \gaia\ data place only 18 within the cluster \citep[e.g.,][]{tang2018,tang2019,olivares2023,hunt2024}. These have a median distance, $d_{med}$ = 85.3~pc, close to the nominal distance of 85.8~pc \citep{tang2018,tang2019,olivares2023,hunt2024}. Of the remaining stars, four are well in the foreground; five are 10--20~pc more distant than the nominal distance and have proper motions larger than the median cluster proper motion. The rest lie well beyond 130~pc.

{\bf Hyades.}\quad
Four \spitz\ programs targeted the Hyades, selecting 22 \citep{meyer2006}, 11 \citep{su2006}, and 45 members \citep{cieza2008}. \citet{urban2012} reconsidered the full set of 78 stars. Along with a few Hyads from other programs \citep{chen2005b} and later assignments \citep{chen2014}, the CDDS sample has 84 nominal members. In addition to a comprehensive series of Hyades studies based on \gaia\ DR3 \citep{brandner2023a,brandner2023b,brandner2023c}, \citet{lodieu2019a} combined \gaia\ DR2 data with ground-based infrared parallaxes to investigate the core and tidal tails of the cluster. \citet{oh2020} made a similar study only with \gaia\ data. \citet{hunt2023,hunt2024} included the Hyades in a comprehensive \gaia\ DR3 cluster study. Combined, these studies include 76 CDDS stars; \citet{oh2020} add two others in the tidal tails. For the six non-matches, four are in the foreground of the cluster, one is in the background, and one has an appropriate parallax but inappropriate proper motions.

{\bf IC~2391.}\quad
\citet{siegler2007} cast a wide net for possible IC~2391 members from a set of deep \spitz\ images, ultimately selecting 34 from ground-based proper motion catalogs \citep{monet2003,zacharias2004}. The 34 members have a median distance $d_{med}$ = 150.6~pc close to the $d$ = 150.2~pc of \citet{hunt2023,hunt2024}. Nearly all of the nominal members lie within 4~\masyr\ of the median proper motion in RA and Dec. Recent analyses using \gaia\ data yield 100---376 cluster members \citep{cantat2018,pang2021,nisak2022,jackson2020,jackson2022,hunt2023,hunt2024}. All but two CDDS stars match a cluster member in these surveys. One CDDS star is in the foreground; the other has proper motions inconsistent with other cluster members.

\begin{deluxetable*}{lccccc}
\tablecolumns{6}
%\tablewidth{10cm}
\tabletypesize{\footnotesize}
\tablecaption{Associations, Open Clusters, and Moving Groups$^{\rm a}$}
\tablehead{
  \colhead{Group} &
  \colhead{$N_{SH}$} &
  \colhead{$N_g$} &
  \colhead{Age (Myr)} &
  \colhead{Methods\tablenotemark{\rm \scriptsize b}} &
  \colhead{References} 
}
\label{tab: clusters}
\startdata
AB Dor          & ~47 & ~33 & 149      & iso            & Be15 \\
$\alpha$ Per    & ~18 & ~13 & 85       & Li             & B02, B04a, B19, D21 \\
Argus           & ~~8 & ~14 & 40--60   & iso, Li        & T08, Be15 \\                  
$\beta$ Pic     & ~30 & ~41 & 24       & iso            & Be15 \\
Blanco 1        & ~38 & ~34 & 85--125  & iso, Li        & M07, B19, D21, J22, R22 \\
Carina--Near    & ~~3 & ~~3 & 200      & act, Li        & Z06 \\
Cha--Near       & ~~1 & ~~1 & 10       & act            & Z04, R07\\      
Columba         & ~16 & ~16 & 42       & iso            & Be15 \\
Coma            & ~85 & ~18 & 650      & iso            & S14, D21 \\
CrA             & ~~7 & ~~7 & 4.5      & EB             & G12 \\
$\epsilon$ Cha  & ~~1 & ~~1 & 5        & iso            & DV21 \\
$\eta$ Cha      & ~16 & ~16 & 11       & iso            & Be15 \\
Hyades          & ~84 & ~76 & 600--850 & iso, MSTO, rot & Br15, G18, D21, B23 \\
Her-Lyra        & ~~1 & ~~1 & 210--305 & Li             & E13 \\
IC2391          & ~34 & ~32 & 40--50   & iso, Li        & B99, B04b, B19, D21 \\
IC2602          & ~12 & ~12 & 25-50    & iso, Li        & S97, D10, B19, D21\\
LCC             & 161 & 179 & 15       & iso            & P16 \\
NGC 2232        & 209 & ~54 & 18--38   & iso, Li        & B21, D21, J22, R22 \\
NGC 2422 (M47)  & ~32 & ~28 & 123--195 & iso            & B19, D21, L21 \\
NGC 2451a       & ~70 & ~35 & 35--80   & iso, Li        & H03, B19, D21, J22, R22 \\
NGC 2451b       & 131 & ~78 & 30--50   & iso, Li        & H03, B19, D21, F22, R22 \\
NGC 2516        & ~50 & ~42 & 138--265 & iso, Li        & M93, S02, B19, D21, F22, R22 \\
NGC 2547        & ~30 & ~23 & 25--40   & iso, Li        & J03, O03, J05, B19, D21, F22 \\
Oceanus         & ~~0 & ~~1 & 400--600 & act, iso, rot  & G23 \\
Octans          & ~~1 & ~~1 & 35       & Li             & M15 \\
Ori OB1a        & ~~9 & ~~9 & 10-11    & iso            & Br19, H23 \\
Ori OB1b        & ~~6 & ~~6 & 3--6     & iso            & Br19, H23 \\
Pleiades        & 165 & 151 & 100--160 & iso, MSTO, rot & S98, B04b, D15, G18, B19, D21 \\
Praesepe        & 193 & 159 & 590--760 & iso, MSTO, rot & G18, B19, D21 \\
$\sigma$ Ori    & ~~3 & ~~3 & 3--5  & iso               & O02, O04 \\
Taurus          & ~~7 & ~~2 & 1--3  & iso               & L23 \\
TW Hya          & ~23 & ~22 & 10    & iso               & Be15 \\
Tuc--Hor        & ~46 & ~48 & 45    & iso               & Be15 \\
UCL             & 148 & 140 & 16    & iso               & P16 \\
USco            & 292 & 308 & 10    & iso               & P16 \\
UMa             & ~16 & ~24 & 414   & int               & J15, J17 \\
93~Tau          & ~~0 & ~~5 & 20--50 & iso              & L23 \\
\enddata
\tablenotetext{a}{Age references: 
B99: \citet{barrado1999};
B02: \citet{barrado2002};
B04a: \citet{barrado2004a};
B04b: \citet{barrado2004b}; 
Be15: \citet{bell2015};
B19: \citet{bossini2019};
B21: \citet{binks2021};
B23: \citet{brandner2023b};
Br15: \citet{brandt2015};
Br19: \citet{briceno2019};
D15: \citet{dahm2015}; 
D21: \citet{dias2021};
DV21: \citet{dickson2021};
D10: \citet{dobbie2010};
E13: \citet{eisenbeiss2013};
F22: \citet{franciosini2022};
G12: \citet{gennaro2012}; 
G18: \citet{gossage2018};
G23: \citet{gagne2023};
H03: \citet{hunsch2003};
H23: \citet{hernandez2023};
J22: \citet{jackson2022};
J03: \citet{jeffries2003};
J04: \citet{jeffries2005};
J15: \citet{jones2015};
J17: \citet{jones2017}; 
L21: \citet{lipartito2021};
L23: \citet{luhman2023a};
M93: \citet{meynet1993};
M07: \citet{moraux2007};
M15: \citet{murphy2015}; 
O02: \citet{oliveira2002};
O03: \citet{oliveira2003};
O04: \citet{oliveira2004}; 
P16: \citet{pecaut2016}; 
R07: \citet{rhee2007};
R22: \citet{randich2022};
S14: \citet{silaj2014}; 
S97: \citet{stauffer1997}; 
S98: \citet{stauffer1998};
S02: \citet{sung2002}; 
T08: \citet{torres2008};
Z04: \citet{zucksong2004};
Z06: \citet{zuckerman2006}
}
\tablenotetext{b}{Methods:
act: stellar activity, including X-ray emission;
EB: eclipsing binary;
int: interferometric sizes of A-type stars;
iso: isochronal fitting;
Li: Lithium depletion boundary;
MSTO: main sequence turnoff;
rot: rotation, gyrochronology
}
\end{deluxetable*}

{\bf IC 2602.}\quad
Based on ground-based proper motion catalogs, \citet{rieke2005}, \citet{su2006}, and \citet{meyer2006} targeted 12 stars in this cluster. \gaia\ analyses suggest all twelve are cluster members \citep{cantat2018,pang2021,nisak2022,hunt2023,hunt2024}.

{\bf NGC~2232.}\quad
Starting with deep \spitz\ photometry, \citet{currie2008} selected 526 MIPS 24~\mum\ sources with 2MASS counterparts, matched these with ROSAT X-ray data \citep{voges1999} and optical photometry \citep{claria1972,lyra2006}, and published a catalog of 209 stars. \citet{monroe2010} acquired optical spectra of 23 stars with \spitz\ photometry, concluding that only five are members. From \gaia, the distribution of parallaxes and proper motions consists of three groups: (i) $\sim$ 25 foreground stars with $d$ = 100--300~pc, (ii) a group of $\sim$ 50--60 stars with a small range in proper motions and a median distance $d_{med}$ = 317.4~pc, and (iii) a disperse set of $\sim$ 125 stars with $d \gtrsim$ 360~pc. We identify the middle group as the cluster; the median distance agrees with the $d$ = 318~pc of \citet{hunt2023,hunt2024}. To examine cluster membership in more detail, we consult results from several recent \gaia\ studies \citep{cantat2018,liu2019,jackson2020,pang2021,jackson2022,hunt2023,hunt2024}. This analysis yields 54 clear members and a few outliers in $\mu$ and $\pi$ that might be part of a tidal tail. 

{\bf NGC~2422 (M47).}\quad
Two \spitz\ programs \citep{gorlova2004,rieke2005} published observations for 32 stars with spectral types of K0 and earlier and confirmed membership for all but one from ground-based \citep[e.g.,][]{urban1998,kharchenko2001,zacharias2004} and space-based \citep{hog2000} proper motion catalogs. Using \gaia\ data, all but three of the 32 nominal members form a tight knot in proper motion space within a small range of parallax. The median distance of these stars $d_{med}$ = 468.6~pc agrees with the $d$ = 468.3~pc derived from 510 members in \citet{hunt2023}. From \citet{hunt2023,hunt2024} and other \gaia\ studies \citep{cantat2018,liu2019,pang2021,jackson2022}, at least 28 CDDS stars are members of M47; two others are within 10~pc of the cluster core and almost share the proper motions of cluster members. One candidate lies $\sim$ 30~pc in front of the cluster and has discrepant proper motions. The last candidate is clearly a background star.

{\bf NGC~2451.}\quad
NGC~2451 consists of two clusters, NGC 2451a and NGC 2451b, along the same line of sight. \citet{balog2009} used a combination of optical photometry, high resolution optical spectroscopy, X-ray emission, and deep IRAC and MIPS images to select 70 (131) members in NGC 2451a (NGC2451b). The CDDS contains 42 cluster stars with MIPS data, 22 in NGC 2451a and 20 in NGC 2451b. Most of the rest have IRAC data. The full set of stars has a broad range in $\mu$ and $\pi$. Some stars assigned to NGC~2451a (NGC2451b) have $d \sim$ 190~pc (360~pc); both have a concentrated group with a common proper motion. Consulting several \gaia\ analyses \citep{cantat2018,liu2019,pang2021,jackson2022,hunt2023,hunt2024} results in 35 (78) high probability members in NGC~2451a (NGC~2451b). Of the members in NGC~2451a from \citet{balog2009}, 5 (23) are in the foreground (background); one does not have a \gaia\ parallax. For NGC~2451b, 18 (40) lie in the foreground (background) and one has large \gaia\ parallax errors. The \gaia\ analyses swap several members between the clusters. Among the foreground and background stars, several have parallaxes in the vicinity of one of the clusters but they have discrepant proper motions.

{\bf NGC~2516.}\quad
From ground-based proper motion studies, \citet{rieke2005} observed all 50 CDDS stars in NGC~2516. \gaia\ data show a concentrated set of stars with similar proper motion and a small set of outliers. The median distance of 409~pc for the sample agrees with the 407~pc distance derived from more than 500 stars in \citet{hunt2023}. Recent \gaia\ surveys classify all but eight stars as members \citep{cantat2018,liu2019,pang2021,jackson2022,hunt2023,hunt2024}. Several non-members lie within the cluster but have discrepant proper motions. Two others are clear background stars, while another has poor \gaia\ data.

{\bf NGC~2547.}\quad
\citet{rieke2005} and \citet{gorlova2007} together selected 30 plausible proper motion members for \spitz\ observations.  The \gaia\ proper motion distribution consists of a tight core similar to that observed in the Pleiades and Praesepe (Fig.~\ref{fig: pmPlPr}) with 7--8 outliers. The median distance of stars in the core $d_{med}$ = 384~pc is quite close to the \citet{hunt2023,hunt2024} estimate of 382~pc from analysis of 424 member stars. From a set of detailed \gaia\ analyses, 23 CDDS stars in or close to the proper motion core are cluster members \citep{cantat2018,liu2019,pang2021,jackson2022,hunt2023,hunt2024}. Of the seven non-members, one (three) star(s) is (are) definitely in the foreground (background). For three others, two are slightly in the foreground; and one has the same parallax as member stars, but has discrepant proper motions with respect to to cluster members. 

{\bf Summary.}\quad
To conclude this section, Table~\ref{tab: clusters} summarizes several aspects of the set of stellar associations, open clusters, and moving groups in the CDDS. The columns in the table include the group name, number of group members in the CDDS from \spitz\ and \herschel\ studies ($N_{SH}$), the number of members suggested by the \gaia\ analyses quoted above ($N_g$), the range of cluster ages, codes for the age method(s) used, and the appropriate references. As in published \spitz\ and \herschel\ studies, isochronal methods and the Li-depletion boundary are popular choices to derive cluster ages. The position of the main sequence turn-off and the rotational period distribution are also common methods for estimating ages.  Although the age range for several clusters is a factor of $\sim$ 2, most results from independent approaches agree to within 10\% to 25\%.  

The CDDS contains a representative set of associations and cluster stars with ages ranging from 5--15~Myr (e.g., $\eta$ Cha, $\sigma$ Ori, and the Sco--Cen associations) to close to 1~Gyr (the Hyades and Praesepe). Across this age range, the Sco--Cen associations (10--20~Myr, $\gtrsim$ 600 members), Blanco-1 and the Pleiades ($\sim$ 100~Myr; 34 and 151 members), and the Hyades and Praesepe (600--850~Myr; 76 and 150 members) are particularly well-represented. The clusters IC~2391, IC~2602, NGC~2232, NGC~2451a, NGC~2451b, and NGC~2547 are intermediate in age between Sco--Cen and the Pleiades. Along with the Coma cluster, NGC~2422 and NGC~2416 span the age range between Blanco-1 and Praesepe. 

Moving groups within the CDDS are also well-populated and cover a broad range of ages. Among the younger groups, the CDDS includes most of the stars in Tuc--Hor (45~Myr, 48 members) and $\beta$~Pic (24~Myr, 41 members). With most members of the TW Hya association discovered since \spitz\ and \herschel, the CDDS contains only ten of 66 possible members \citep{luhman2023b}. Although none of the moving groups in the CDDS are as old as the Hyades, AB Dor (149~Myr, 33 members) and UMa (400~Myr, 24 members) provide a useful comparison to the Pleiades and Praesepe.

\section{HR Diagrams}
\label{sec: hrdiagram}

For the past 10--20 years, advances in stellar structure calculations together with results from large astrometric and spectroscopic surveys of field stars and open clusters have greatly improved understanding of the evolution of the chromospheric and photospheric properties of main sequence stars. Recent studies illustrate how rotational evolution and internal mixing processes influence Li and Be depletion \citep[e.g.,][]{dumont2021b} and the decline in chromospheric emission \citep[e.g.,][]{johnstone2021} as a star evolves. Along with precise cluster membership probabilities from \gaia, \citep[e.g.,][]{cantat2020b,dias2021,jackson2022,hunt2024}, new model isochrones provide more accurate estimates of cluster ages as a function of metallicity and other physical parameters \citep[e.g.,][]{bell2015,gossage2018,dias2021}. 

\begin{figure}[t]
\begin{center}
\hspace{-0.5cm}\includegraphics[width=3.0in]{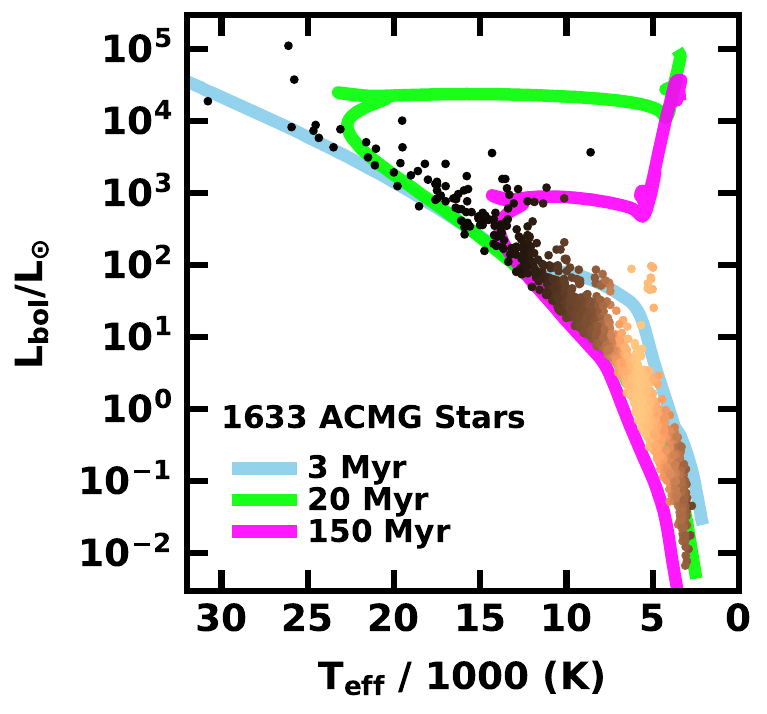}
\vskip -2ex
\caption{
HR diagram for \nchrd\ CDDS stars in stellar associations, clusters, and moving groups (ACMG). The light blue (lime, fuchsia) curves shows the PARSEC 1.2S isochrone for \feh\ = 0 and an age of 3 Myr (20, 150~Myr), displaced by -500~K (0~K, +750~K) for clarity. The lower bound of the two curves for \teff\ $\gtrsim$ 15000~K (light blue) and \teff\ $\lesssim$ 15000~K (fuchsia) marks the approximate zero-age main sequence; all CDDS stars lie above this locus. For cooler stars, the 3~Myr isochrone sets a strong upper limit to the set of young stars contracting to the main sequence. Some cluster stars lie at the main sequence turnoff for ages 20--150~Myr.
\label{fig: clusterhrd}
}
\end{center}
\end{figure}

\begin{figure}[t]
\begin{center}
\hspace{-0.5cm}\includegraphics[width=3.0in]{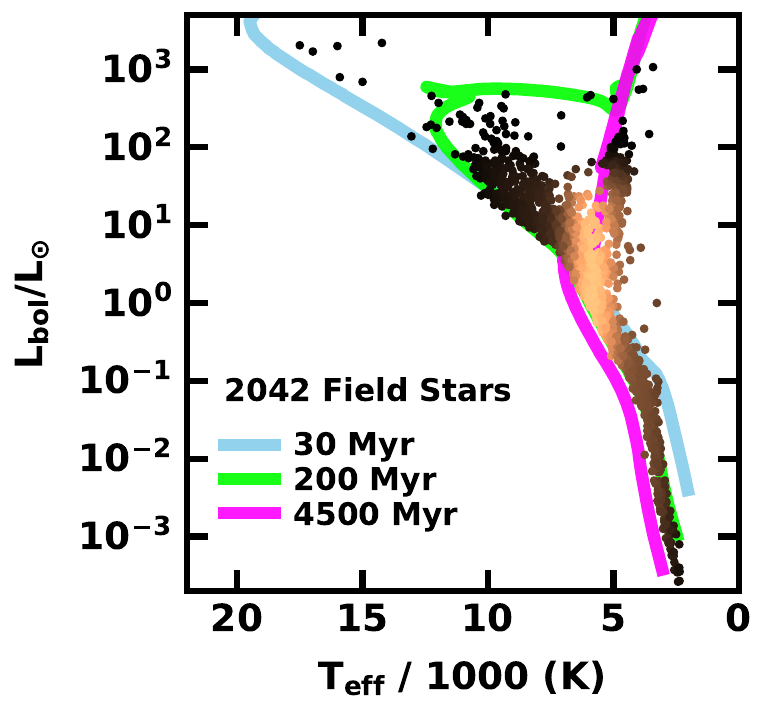}
\vskip -2ex
\caption{
HR diagram for \nfhrd\ CDDS field stars. The light blue (lime, fuchsia) curves shows the PARSEC 1.2S isochrone for \feh\ = 0 and an age of 30~Myr (200~Myr, 4.5~Gyr). The lower bound of the light blue and light green curves for \teff\ $\gtrsim$ 10000~K marks the approximate zero-age main sequence; all CDDS stars lie near or above this locus. Nearly all cooler stars lie near the 200~Myr isochrone; others lie on the giant branch. 
\label{fig: fieldhrd}
}
\end{center}
\end{figure}

Although we defer an extended discussion of age estimates to Paper IV in this series, we conclude our summary of host star properties with HR diagrams for association, cluster, and moving group stars (Fig.~\ref{fig: clusterhrd}) and for field stars (Fig.~\ref{fig: fieldhrd}). For stars in clusters, a combination of the 3~Myr isochrone for massive stars and the 150~Myr isochrone for lower mass stars roughly corresponds to the zero-age main sequence (ZAMS). Among the lower mass stars with \teff\ $\lesssim$ 10,000~K, the 3~Myr (150~Myr) isochrone sets a rough upper bound (strict lower bound) to the observed luminosities. Although most stars lie close to the ZAMS, others are starting to turn off the main sequence and evolve towards the red giant branch. Many of these lie on or in between the turnoffs for the 3~Myr and 150~Myr isochrones. 

For nearly all of the clusters, an age of 3--150~Myr nicely brackets age estimates for the youngest associations \citep[e.g., LCC, UCL, and Upper Sco;][]{pecaut2016} to clusters like the Pleiades \citep[e.g.,][]{stauffer1998,dahm2015,gossage2018,bossini2019,dias2021}. Aside from stars on the main sequence turnoff, most stars in older clusters \citep[e.g.,Hyades and Praesepe;][]{brandt2015,gossage2018,bossini2019,dias2021,brandner2023b} lie on or close to the main sequence. Across all of the clusters, few stars lie on the red giant branch.

The HR diagram for CDDS field stars has a very different morphology (Fig.~\ref{fig: fieldhrd}). Most of this group is bound by the 200~Myr and 4.5~Gyr isochrones. Nearly all of the cool stars with \teff\ $\lesssim$ 5000~K lie on or near the 200~Myr isochrone. Few lie close to the 30~Myr isochrone or the 4.5~Gyr isochrone. Aside from a few stars clearly on the red giant branch, these stars are older than cluster stars of similar \teff, but are much younger than 4.5~Gyr. Among hotter stars with \teff\ = 5000--10000~K, many are either on the main sequence or evolving off the main sequence. The more evolved stars are bound by the 200~Myr and 4.5~Gyr isochrones. Finally, there are a few very hot stars, \teff\ = 12000--20000~K, between the 30~Myr and 200~Myr isochrones that are clearly young. These stars are not quite as young as their cluster counterparts, but they are clearly the youngest stars not obviously associated with  an association, cluster, or moving group.

\section{Discussion}
\label{sec: discussion}

As described in the previous sections, we have assembled from the literature a sample of \ncdds\ stars that has strong constraints on cold circumstellar dust emission from observations with \spitz\ and \herschel. The stars span a range of spectral types (B5--M9), luminosity classes (V--II), and ages ($\gtrsim$ 5--10~Myr). Most stars are in the field (\nfhrd; 56\%); the rest (\nchrd; 44\%) are members of stellar associations, open clusters, and moving groups. For this sample, we compiled optical-infrared photometry, parallaxes, and proper motions (\S\ref{sec: sample}), along with information on planetary and stellar companions (\S\ref{sec: companions}); host star properties (spectral type SpT, effective temperature \teff, extinction \av, luminosity \lstar, gravity \logg, metallicity \feh, Li abundance A(Li), stellar rotation \prot, chromospheric and coronal emission; \S\ref{sec: host-stars}) and their membership in associations, clusters, and moving groups (\S\ref{sec: clusters}). Several tables report the extent of the compilation; Table~\ref{tab: phot-stats} summarizes photometric data; Table~\ref{tab: non-single-stars} outlines the sources of data on stellar multiplicity; Table~\ref{tab: companion stats} lists statistics on companions; Table~\ref{tab: stardata} recaps the availability of host star properties; Table~\ref{tab: activity} reports the frequency of activity measures as a function of spectral type; and Table~\ref{tab: correlations} lists the correlation coefficients for different stellar activity measures within the sample. Table~\ref{tab: binary stars} lists properties of candidate common proper motion companions. 

{\bf Missing Data.}\quad
Despite efforts to construct a complete compilation, some data are inevitably missing. Among the photometric observations, \bvic\ are sometimes unavailable for faint red stars discovered in the 2MASS survey; bright stars saturated in 2MASS and WISE sometimes do not have counterparts in DIRBE or other literature sources. These misses are rare, $\lesssim$ 1\% of the compilation. Deep \bvic\ photometry or shallow \jhks\ photometry could complete these parts of the compilation. Short time series limit the availability of robust rotational periods for many CDDS stars. Analyses of {\it TESS} data and additional ground-based programs could alleviate this deficiency. 

At longer wavelengths, stars are often too faint for W4 and sometimes too faint for MIPS at 70~\mum\ or \herschel\ PACS and SPIRE at any wavelength. \herschel\ generally observed nearby CDDS stars; 75\% (89\%) of the stars with \herschel\ 5$\sigma$ or better detections lie within 50~pc (100~pc). \herschel\ observing time constraints and the large population ($\sim$ 50\%) of CDDS stars with distances $\gtrsim$ 100~pc precluded a complete set of \herschel\ data. Adding these data require a new and more sensitive space-based mission.

On the rare occasion when \gaia\ misses on a parallax or proper motion, the star is near a much brighter star or is a binary with a period comparable to the lifetime of the satellite. The available data in the rest of the compilation is then sufficient to derive a distance. The activity measures are often helpful in assigning a star to a cluster when proper motion is not available. We anticipate that a future \gaia\ data release will include accurate parallaxes and proper motions for these stars.

Missing spectroscopic diagnostics have several causes. Early type stars with radiative envelopes and \teff\ $\gtrsim$ 6500~K comprise $\sim$ 30\% of the CDDS and are often too hot or lack the appropriate absorption features to measure \logg, \feh, \lih, \prot, \rhkp, and \rirtp. Roughly half of the BA and early F stars have measured \vsini\ and \lxlstar. Moderate resolution optical or infrared spectra are sufficient for \vsini. Some of these stars lie in the unpublished half of the \erosita\ survey and will at least have improved upper limits once these \erosita\ data are published.

Among the late F and G stars, nearly all have measured \logg; $\sim$ 85\% have \feh; and $\sim$ 60--70\% have \prot, \vsini, \rhkp\, \rirtp, and \lxlstar. Only $\sim$ 50\% have an \lih\ measurement. The second half of the \erosita\ data release should add \lxlstar\ data for many stars. High resolution echelle data can yield \vsini, \rhkp, \rirtp, and \lih. The next \gaia\ data release should add new measurements of \lrirtp. 

Aside from \lxlstar, KM stars in the CDDS are less likely to have measurements of \logg, abundances, and activity measures than G stars. These stars are often fainter; stronger molecular bands may complicate measures of abundances and activity indicators. The CARMENES project searched many nearby M dwarfs for exoplanets \citep[e.g.,][]{alonso2015,cortes2017}. Future high resolution spectroscopic observations on 4-m class or larger telescopes are required to expand the CDDS collection of M dwarf photospheric and chromospheric properties.

{\bf Frequency of CDDS Stars in Associations, Clusters, and Moving Groups.}\quad
Within the CDDS, stars in open clusters (682) and stellar associations (652) clearly outnumber the stars in moving groups (213). Overall, $\sim$ \fcl\ of CDDS stars lie within an association, cluster, or moving group. There are another 40--50 on the outskirts of a cluster or association but have discrepant proper motions. Several dozen possibly lie within a moving group. 

Compared to the \hipparcos\ data available for \spitz\ and \herschel\ programs, exquisite \gaia\ data enable much more robust estimates of membership probabilities in associations, clusters, and moving groups at much greater distances \citep[e.g.,][]{cantat2018,cantat2020a,dias2021,hunt2023,hunt2024}. For the CDDS, \gaia\ data eliminate many potential members in the Pleiades, Praesepe, and other clusters. However, we now have many more CDDS stars in nearby associations. The collection of stars within moving groups is impressive: \gaia\ confirms most of the moving group members discussed in \spitz\ programs and adds enough new members to replace members lost due to age, distance, and proper motion considerations.

The frequency of cluster stars in the CDDS is much larger than the frequency within 100--200~pc of the Sun. As an example, \citet{hunt2024} identify $\sim$ 2700 cluster members within 100~pc and $\sim$ 19000 within 200~pc. Other analyses yield similar results \citep[e.g.,][]{cantat2018,dias2021,hunt2023}. Within 100~pc, the GCNS contains information on close to 190,000 stars \citep{gaianearby2021}. For 200~pc, we scale the GCNS up by a factor of eight. The frequency of cluster stars is then $\sim$ 1.25\% to 1.5\% within 100--200~pc.  The CDDS contains a factor of 30--35 more stars in clusters than the solar neighborhood.

Although \herschel\ targeted few cluster stars, \spitz\ programs focused on young clusters to have a better chance of detecting IR excesses from debris disks and to have a better estimate of stellar ages from analyses of the full cluster instead of individual field stars (see Table~\ref{tab: clusters}). With $\sim$ \fcl\ of CDDS stars in groups with fairly well-defined ages and $\sim$ \ffld\ in the field, the CDDS sample provides a way to follow the evolution of debris disk emission as a function of environment. When we consider age estimates in Paper IV, we will investigate this issue.

{\bf Potential for Gaia--Spitzer Cross-matching.}\quad
We have applied many \gaia\ analyses of clusters to assign membership of CDDS stars to associations, clusters, and moving groups. In most cases, \gaia\ parallax and proper motion data coupled with \gaia\ or new ground-based radial velocity data eliminate cluster members from \spitz\ catalogs. Sometimes the losses are minimal. In the Pleiades, \gaia\ analyses retain 91.5\% of Pleiads in the combined \spitz\ catalog; in Blanco-1, we retain 90\% of the \spitz\ catalog. For other clusters, losses are severe. In Coma (NGC~2451ab), we retain only 21\% (56\%) of proposed \spitz\ cluster members.

These results suggest that matching \gaia\ cluster members \citep[e.g.,][]{hunt2024} with the full set of \spitz\ photometric data for individual clusters might increase the set of cluster members with high quality \spitz\ photometry. The strategy would be to derive coordinates and photometry for all \spitz\ sources within the nominal boundary of a cluster, match these to \gaia\ DR3 sources to verify the coordinates and to obtain G, BP, and RP for each \spitz\ source, and then derive cluster membership probabilities from the detailed \gaia\ cluster studies \citep[e.g.,][]{hunt2024}. 

As a test of this concept, we consider Blanco-1 which has $\sim$ 800 members from \citet{hunt2024}. Matching these to the SEIP Source List \citep{capak2019} at IPAC recovers 205 matches. While many of these are IRAC measurements, we identify 51 (7) MIPS 24~\mum\ aperture (PSF) measurements. Compared to the 38 measurements reported in \citet{stauffer2010}, this \gaia\ match adds as many as $\sim$ 50\% more 24~\mum\ measurements from Blanco-1 stars not considered members at the time of the \spitz\ observations. Considering the high quality requirements for SEIP, it seems plausible that additional \spitz\ sources might be identified by reducing the \spitz\ images from scratch. We set this possibility aside for another day.

{\bf Applications}\quad
As noted in the Introduction, we currently plan three additional studies with the CDDS compilation. Paper II will explore the magnitude and frequency of IR excess emission as a function of the physical properties of the central star, known companions, and membership in a group. In Paper III, we then search for new planetary and stellar companions with \gaia. Finally, Paper IV will consider age estimates for the full sample of CDDS stars and examine trends in IR excess emission with age for stars in clusters and the field. Aside from these, CDDS can serve as a starting point for many investigations of stellar physics. We discuss several possibilities.

Although we do not compile information on stellar magnetic fields, the connection between magnetic topology and chromospheric and coronal emission is well-established \citep[e.g.,][]{see2016,brown2022,jeffers2023,weber2023}. Recent investigations of magnetic field topology in several CDDS stars -- $\tau$~Boo, 61~Cyg, $\epsilon$~Eri, $\chi$~Dra, $\iota$~Hor, and 18~Sco among others -- illustrate the relation between the topology, chromospheric and coronal emission, stellar activity cycles, and the stellar dynamo \citep[e.g.,][]{metcalfe2010,jeffers2018,jeffers2022,marsden2023,nascimento2023}. With the success of deriving rotational periods and stellar cycles from ground-based and space-based facilities \citep[e.g.,][]{hartmann2010,cargile2014,cantomartins2020,rebull2022,popinchalk2023,douglas2024}, it seems plausible that many CDDS stars with high levels of stellar activity will show long-term activity cycles and merit more detailed measurements of their magnetic field topology.

In contrast to the dynamo-produced magnetic fields in the atmospheres of low mass main sequence stars, magnetic fields detected in B-type or A-type stars are probably fossil fields left over from the original molecular cloud \citep[e.g.,][]{braithwaite2013,schleicher2023}. Although the CDDS contains many apparently pristine B-type and A-type stars, there are several with strong metallic lines (SpT: Am or Bm, e.g., 67 UMa) or with magnetic fields \citep[SpT: Ap or Bp, e.g., $\alpha$ Lyr (Vega), $\alpha$~CMa (Sirius), and $\beta$~UMa (Merak),][]{lignieres2009,petit2011,blazere2015}. Several others, such as $\alpha$ Aql, $\alpha$ Lyr, and $\alpha$ Lac, have detectable rotation periods, \prot\ $\lesssim$ 1 day, perhaps suggestive of features induced by a weak magnetic field. Detailed studies of other CDDS BA (and perhaps early F) stars might similarly reveal weak fields or short periods. Enlarging these samples may shed light on the evolution of magnetic and rotational structures observed in massive stars.

\section{Summary}
\label{sec: summary}

We conclude with a synopsis of the compilation, the main results, and the next steps in this study.

{\bf The CDDS Compilation.}\quad
We compile photometric, dynamical, and spectroscopic
%reordered to match the order discussed:
%dynamical, photometric, and spectroscopic 
data for \ncdds\ stars targeted by \spitz\ and \herschel\ to search for cold circumstellar dust emission from debris disks. We name the compilation the Cold Debris Disk Surveys (CDDS). The photometric data include \gaia\ G, BP, and RP; Tycho B$_{\rm T}$V$_{\rm T}$; Johnson--Cousins BVI$_{\rm C}$; APASS \agri, Pan-STARRS and Skymapper \griz; 2MASS JHK$_{\rm s}$; WISE W1--W4; \spitz\ IRAC B1--B4, IRS 8--25~\mum, and MIPS 24--70~\mum; IRAS 12--100~\mum; \herschel\ PACS 70--160~\mum\ and SPIRE 250--500~\mum; and submm 450--1200~\mum. Figs.~\ref{fig: stats1}--\ref{fig: stats2} and Table~\ref{tab: phot-stats} show that $\gtrsim$ 80\% of CDDS stars have photometry from B to 25~\mum; the fraction of available data declines at longer wavelengths.  Fig.~\ref{fig: gaiabprphist} compares the distribution of BP--RP colors of CDDS stars with the colors of main sequence stars in the GCNS; compared to the local solar neighborhood, the compilation has an overabundance of AFG stars and an underabundance of KM stars. When available, the CDDS includes rotational periods \prot\ measured from photometric time series.

The dynamical data comprise \hipparcos\ and \gaia\ parallaxes and proper motions along with properties of stellar and planetary companions. All but a handful of stars have parallaxes and proper motions. Roughly half (more than 90\%) of the sample lies within 100~pc (300~pc). Fig.~\ref{fig: plx} contrasts the distribution of parallaxes in the CDDS (where the number of stars as a function of distance is approximately $N \propto d$) and the GCNS (where $N \propto d^3$). Approximately 44\% (9\%) of the stars have at least one stellar (planetary) companion.  Figs.~\ref{fig: non-single-stars} and \ref{fig: planets} illustrate the broad semimajor axis distributions of stellar and planetary companions to CDDS primary stars. The fraction of non-single FGKM stars in the CDDS is comparable to the fraction in the general field \citep[e.g.,][]{gao2014,niu2021,susemiehl2022}; the CDDS has a somewhat smaller fraction of non-single BA stars than the field \citep[e.g.,][]{guo2022}. The frequency of planets in the CDDS sample is smaller than the 0.2--1 per star derived for typical field FGKM stars \citep[e.g.,][]{petigura2013,fernandes2019,fulton2021,kaminski2025,mignon2025}. 

The spectroscopic data include \gaia\ XP spectra and literature measurements of spectral type SpT, effective temperature \teff, gravity \logg, metallicity \feh, lithium abundance \lih, projected rotational velocity \vsini, the Ca~II HK and IR triplet activity indicators \rhkp\ and \rirtp, and the 0.1--3~keV X-ray flux $F_X$. Combined with the dynamical and photometric data, the measured effective temperatures yield estimates for interstellar extinction \av\ and stellar luminosity \lstar. Tables~\ref{tab: stardata}--\ref{tab: activity} summarize the statistics of these measurements as a function of SpT. Data for \teff\ and \lstar\ are complete. Approximately 80\% (64\%, 59\%, 54\%) of the stars have measured \logg\ (\feh, \vsini, \lxlstar). Only $\sim$ 50\%  of FGK stars with \teff\ = 4500--6500~K have measured \lih, \prot, \rhkp, and \rirtp. Figs.~\ref{fig: teffcomp}--\ref{fig: teffxray} illustrate (i) the excellent agreement between \teff, \logg, and \lstar\ estimates from \gaia\ and ground-based data, (ii) a gaussian-like \feh\ distribution with a median \feh\ $\approx -0.03$ and a full-width at half maximum of $\sim$ 0.26, and (iii) the extensive range in \lih, \prot, \vsini, \lrhkp, \lrirtp, and \lxlstar\ as a function of \teff.   

Together with detailed studies from the literature, we use the dynamical, photometric, and spectroscopic data to assign CDDS stars to the field (\nfhrd\ stars; 56\% of the sample) or to one of many moving groups, open clusters, or stellar associations (\nchrd\ stars; 44\%; Table~\ref{tab: clusters}). The fraction of CDDS stars within an association, cluster, or moving group is much larger than we would find in a random sample of stars within 100--200~pc, where the fraction is $\sim$ 1\% to 2\% \citep[e.g.,][]{gaianearby2021,hunt2024}.

{\bf Main Results.}\quad
In parallel with an analysis of \gaia\ data for \spitz\ and \herschel\ targets, we identify a set of faint stars with nearly identical parallaxes and proper motions as the main target. Table~\ref{tab: binary stars} in the Appendix lists G and BP--RP colors, separation from the apparent primary, and velocity difference. The magnitude differences suggest most of these candidate proper motion companions are M dwarfs or brown dwarfs. 

Compared to \spitz\ and \herschel\ studies, \gaia\ data subtract stars from the population of cluster stars, add stars to the population of associations, and add {\it and} subtract stars from moving groups. Some clusters (e.g., Pleiades and Praesepe) lose few stars; others (e.g., Coma, NGC2232, and NGC2451a,b) lose many. The number of moving group stars with \spitz\ and \herschel\ data is significant.

The addition of stars to some groups suggests that a full analysis of available \spitz\ photometry within the footprint of \gaia\ cluster boundaries might result in a larger sample of cluster stars with \spitz\ photometry. Illustrations with Blanco-1 and Coma imply increases of 25\% to 50\%. Adding unpublished \spitz\ photometry to systems with few members in Table~\ref{tab: clusters} would enhance the \spitz\ legacy.

Fig.~\ref{fig: CorrPlot} and Table~\ref{tab: correlations} summarize the correlations of pairs of activity indicators. With 467--859 points per correlation, the CDDS provides good statistics for this analysis. The best correlations involve any combination of \lrhkp, \prot, and \lxlstar\ ($p$-values $\lesssim 10^{-137}$ for the probability of no correlation). The strength of most other correlations remains strong, with $p$-values between $10^{-19}$ and $10^{-95}$. The relation between the Ca~II IR triplet and the lithium abundance has the largest probability of no correlation ($p$-value = $10^{-5}$). 

The complete HR diagrams for cluster (Fig.~\ref{fig: clusterhrd}) and field (Fig.~\ref{fig: fieldhrd}) stars indicate that CDDS stars have ages ranging from $\sim$ 5--10~Myr to $\sim$ 5--10~Gyr. Comparisons of \lih, \rhkp, \rirtp, and \lxlstar\ with results for other cluster and field stars in the literature confirm this general conclusion. While the range of ages is a feature for \spitz\ and \herschel\ targets, few of the field stars are considered as young as 50--100~Myr in \spitz\ and \herschel\ publications. Some of the young stars might be in extended halos (instead of tidal tails) surrounding the inner cores of the clusters \citep[e.g.,][]{currie2010,yu2020,pang2022} or stars ejected from the cluster through dynamical interactions with other cluster members \citep[e.g.,][]{leonard1990,kounkel2022,fajrin2025}.

{\bf Next Steps.}\quad
Although the CDDS compilation is extensive, additional data for \logg, \feh\ and \lih, rotation measures (\prot\ and \vsini), and activity indicators (\rhkp, \rirtp, and \lxlstar) would improve understanding of stellar activity evolution and the relationship between stellar activity and the presence of planetary companions and debris disks. Large surveys and focused studies are each responsible for roughly half of the ground-based spectroscopic data in the CDDS. Both approaches are needed to enhance the CDDS.

In paper II of this series, we plan to explore the extent of IR excess emission among CDDS stars. This study will yield the fraction of stars with IR excess as a function of \teff\ and the age and activity indicators. We will also consider the association of IR excess emission with the presence of stellar and planetary companions. 

Paper III will conduct a search for dynamical companions to CDDS stars within \gaia\ DR3. \gaia\ anomalous accelerations have been an important contributor to recent detections of planets and low mass stellar companions \citep[e.g.,][]{currie2023,tobin2024}. In addition to isolating accelerating stars within the CDDS, we plan to consider other options for detecting unresolved companions.

Finally, paper IV will consider approaches for deriving ages of CDDS stars and applying these results to the evolution of debris disk emission as a function of stellar age and other physical properties. This analysis will allow tests of various models for the evolution of debris as a function of stellar age \citep[e.g.,][]{wyatt2007,kb2008,krivov2021,nkb2022}.

{\bf Acknowledgements.}\quad

%We thank M. Geller and anyone else for helpful discussions and comments on the manuscript.
This project was supported by the NASA Exoplanets Research Program, through contract 80NSSC24K0158. We thank the reviewer for a courteous and constructive report.

This work makes use of data from the ESA missions Hipparcos and Gaia\footnote{https://www.cosmos.esa.int/gaia}. \gaia\ data are processed by the Gaia Data Processing and Analysis Consortium\footnote{https://www.cosmos.esa.int/web/gaia/dpac/consortium} (DPAC). 
Funding for the DPAC has been provided by national institutions, in particular the institutions participating in the Gaia Multilateral Agreement. Herschel is an ESA space observatory with science instruments provided by European-led Principal Investigator consortia and with important participation from NASA. 
This research also makes use of data products from (i) the Two Micron All Sky Survey, which is a joint project of the University of Massachusetts and the Infrared Processing and Analysis Center/California Institute of Technology, funded by the National Aeronautics and Space Administration and the National Science Foundation; (ii) the Wide-field Infrared Survey Explorer, which is a joint project of the University of California, Los Angeles, and the Jet Propulsion Laboratory/California Institute of Technology, funded by the National Aeronautics and Space Administration; (iii) the Herschel Space Observatory, which is an ESA space observatory with science instruments provided by European-led Principal Investigator consortia and with important participation from NASA; (iv) AKARI, a JAXA project with the participation of ESA; (v) ISO, an ESA project with instruments funded by ESA Member States (especially the PI countries: France, Germany, the Netherlands and the United Kingdom) and with the participation of ISAS and NASA; and (vi) the Infrared Astronomical Satellite (IRAS), which was was a joint project of the US, UK and the Netherlands. 
We also used the SIMBAD database and VizieR catalogue access tool, CDS, Strasbourg, France \citep[DOI : 10.26093/cds/vizier;][]{vizier2000}; the NASA/IPAC Infrared Science Archive, which is funded by the National Aeronautics and Space Administration and operated by the California Institute of Technology; the Washington Double Star Catalog maintained at the U.S. Naval Observatory; the NASA \nxa, which is operated by the California Institute of Technology, under contract with the National Aeronautics and Space Administration under the Exoplanet Exploration Program; and the WEBDA database, operated at the Department of Theoretical Physics and Astrophysics of the Masaryk University. 
The Pan-STARRS1 Surveys (PS1) and the PS1 public science archive have been made possible through contributions by the Institute for Astronomy, the University of Hawaii, the Pan-STARRS Project Office, the Max-Planck Society and its participating institutes, the Max Planck Institute for Astronomy, Heidelberg and the Max Planck Institute for Extraterrestrial Physics, Garching, The Johns Hopkins University, Durham University, the University of Edinburgh, the Queen's University Belfast, the Harvard-Smithsonian Center for Astrophysics, the Las Cumbres Observatory Global Telescope Network Incorporated, the National Central University of Taiwan, the Space Telescope Science Institute, the National Aeronautics and Space Administration under Grant No. NNX08AR22G issued through the Planetary Science Division of the NASA Science Mission Directorate, the National Science Foundation Grant No. AST–1238877, the University of Maryland, Eotvos Lorand University (ELTE), the Los Alamos National Laboratory, and the Gordon and Betty Moore Foundation.
The national facility capability for SkyMapper has been funded through ARC LIEF grant LE130100104 from the Australian Research Council, awarded to the University of Sydney, the Australian National University, Swinburne University of Technology, the University of Queensland, the University of Western Australia, the University of Melbourne, Curtin University of Technology, Monash University and the Australian Astronomical Observatory. SkyMapper is owned and operated by The Australian National University's Research School of Astronomy and Astrophysics. The survey data were processed and provided by the SkyMapper Team at ANU. The SkyMapper node of the All-Sky Virtual Observatory (ASVO) is hosted at the National Computational Infrastructure (NCI). Development and support of the SkyMapper node of the ASVO has been funded in part by Astronomy Australia Limited (AAL) and the Australian Government through the Commonwealth's Education Investment Fund (EIF) and National Collaborative Research Infrastructure Strategy (NCRIS), particularly the National eResearch Collaboration Tools and Resources (NeCTAR) and the Australian National Data Service Projects (ANDS).

{\bf Data Availability.}\quad
All of the data discussed in this paper are publicly available from the websites noted above and within the literature quoted throughout. We plan to publish the full compilation in parallel with Paper IV of this series.

\appendix

While building this compilation, we encountered several confusing source identifications and derived distances for stars with uncertain or unknown parallaxes. For investigations of stellar or planetary companions, we needed to separate physically associated companions from those along the same line-of-sight. 
% bcb: replaced this text....
% For all CDDS stars, we used \gaia\ parallaxes, proper motions, and (when available) radial velocities to distinguish physical pairs from chance alignments.  Along the way, we identified plausible common proper motion companions not included in the Washington Double Stars catalog. Here, we clarify source identifications, discuss distance estimates, and include a table of candidate proper motion companions derived from \gaia\ data.
% (bcb) ...with this...
For all CDDS stars, we used \gaia\ sky positions, parallaxes, and proper motions to distinguish physical pairs from chance alignments. By gathering sources that lie at a parallax within 3-$\sigma$ of each star and comparing inferred physical separation and relative speed, we identified plausible common-proper motion companions of the CDDS stars. Table~\ref{tab: binary stars} lists some of these companions, chosen as a result of their low relative velocity ($< 2$~km/s at the physical distance inferred from the CDDS stars' parallax), their small physical separation ($< 1,000$~au), their large distance from a known star cluster, and their failure to appear as a known companion in the double star catalogs considered here. 

\begin{deluxetable*}{lcccccccc}
\tablecolumns{9}
%\tablewidth{10cm}
\tabletypesize{\scriptsize}
\tablecaption{Binary and triple star candidates from Gaia astrometry}
\tablehead{
 % \colhead{CDDS #}
  \colhead{\begin{tabular}{c}SIMBAD\\\texttt{main\_id}\end{tabular}} &
  \colhead{\begin{tabular}{c}Gaia DR3\\\texttt{source\_id}\end{tabular}} &
  \colhead{\begin{tabular}{c}G-band\\(mag)\end{tabular}} &
  \colhead{\begin{tabular}{c}BP-RP\\(mag)\end{tabular}} &
  \colhead{\begin{tabular}{c}companion\\\texttt{source\_id}\end{tabular}} &
  \colhead{\begin{tabular}{c}G-band\\(mag)\end{tabular}} &
  \colhead{\begin{tabular}{c}BP-RP\\(mag)\end{tabular}} &
  \colhead{\begin{tabular}{c}sep.\\(au)\end{tabular}} & 
  \colhead{\begin{tabular}{c}vel.\\(km/s)\end{tabular}}
}
\label{tab: binary stars}
\decimals
\startdata
HD 109536    & 6146899611411058304 &  5.067 & 0.313 &  6146899611411103360 & 13.841 & 2.786 &  15.7 & 0.62 \\
G  54-20     & 625740966539620096 & 13.479 & 2.506 &  625740966539596032 & 15.447 & 2.987 &   5.2 & 1.48 \\
* phi For    & 4967177781457918976 &  5.112 & 0.109 &  4967153630858709120 &  7.465 & 0.683 & 391.1 & 0.21 \\
HD  20759    & 4854737122493173760 &  7.595 & 0.628 &  4854737122493187200 & 14.469 &   nan &   3.5 & 1.56 \\
HD  37548    & 2971060398858087168 &  7.451 & 0.764 &  2971060330138611968 & 15.234 & 2.855 &  44.0 & 0.17 \\
HD  40540    & 2889357751382815872 &  7.495 & 0.379 &  2889357751382248704 & 15.269 & 2.944 &  40.6 & 0.52 \\
HD  46190    & 5478200622494775552 &  6.599 & 0.118 &  5478200588135044096 & 14.849 & 2.776 &  82.2 & 0.11 \\
HD  71722    & 5321368860896990720 &  6.044 & 0.080 &  5321369406350033664 & 11.076 & 1.366 & 268.9 & 0.12 \\
HD  71722    & 5321368860896990720 &  6.044 & 0.080 &  5321368860896992000 & 15.644 & 2.437 &   9.9 & 1.06 \\
HD  77052    & 686698960631875968 &  8.708 & 0.814 &  686698891912397696 & 14.130 & 2.545 &  55.4 & 0.23 \\
HD 101259    & 3491774880840574464 &  6.167 & 1.040 &  3491774880840574208 & 15.395 & 1.974 &   8.0 & 0.81 \\
HD 109391    & 3703459387769393920 &  8.485 & 0.601 &  3703459387768058624 & 15.962 &   nan &   3.5 & 1.30 \\
HD 113319    & 1467164096785657088 &  7.367 & 0.820 &  1467167257881586816 & 15.657 & 3.558 & 175.1 & 0.13 \\
HD 150697    & 4326683743996162432 &  7.933 & 0.690 &  4326683743996161408 & 16.884 & 3.242 &  21.3 & 0.31 \\
* b Oph      & 4111251723554095104 &  4.105 & 0.394 &  4111251689120620160 & 14.566 & 3.122 &  18.1 & 1.92 \\
HD 220150    & 1910520582770593792 &  8.231 & 0.548 &  1910520544116627840 & 13.652 & 1.674 &   4.0 & 0.81 \\
HD  33283    & 2955981936912654592 &  7.912 & 0.767 &  2955981696394487296 & 16.972 & 3.194 &  55.7 & 0.33 \\
HD 107148    & 3693358861640279296 &  7.864 & 0.827 &  3693358792919145600 & 17.413 & 0.649 &  35.0 & 1.07 \\
HD 188015    & 2028419118720795392 &  8.078 & 0.848 &  2028419118720966912 & 15.434 & 3.084 &  13.0 & 0.75 \\
BD-20   638  & 5101544806816778752 & 10.194 & 0.815 &  5101544802521433216 & 17.951 & 4.091 &  40.0 & 0.20 \\
HD 279788    & 179245207355823232 & 10.482 & 1.083 &  179244760679229184 & 13.484 & 2.098 &  93.5 & 0.33 \\
HD 285840    & 3410633602770942464 & 10.554 & 1.054 &  3410633980728058624 & 13.448 & 2.242 & 154.2 & 1.42 \\
BD+08   742  & 3292617105588895104 & 10.772 & 0.984 &  3292617041166009216 & 14.703 & 2.741 &  15.7 & 0.18 \\
HD 245567    & 3341551046679446016 &  9.386 & 0.963 &  3341556922194703232 & 13.726 & 2.576 &  49.8 & 1.22 \\
HD  90712    & 5447414606155987200 &  7.375 & 0.758 &  5447414606155987840 & 11.970 & 2.294 &  33.3 & 0.47 \\
HD 157664    & 1638293082964887424 &  7.847 & 0.691 &  1638293082964794880 & 13.831 & 2.317 &  16.5 & 1.80 \\
HD 170778    & 2112083741571086592 &  7.344 & 0.762 &  2112083741570227072 & 12.555 & 2.765 &  20.2 & 1.29 \\
HD 193017    & 4219087666404378624 &  7.148 & 0.705 &  4219087700764115840 & 16.643 & 3.911 &  53.6 & 0.35 \\
BD+22  4792  & 2839360861722307200 & 10.009 & 0.738 &  2839360861721116416 & 13.675 &   nan &   1.4 & 1.31 \\
HD 104731    & 6148648281576023424 &  5.033 & 0.574 &  6148648281576023552 & 10.537 &   nan &   2.2 & 1.24 \\
HD 202884    & 1743504800870418432 &  7.149 & 0.655 &  1743504525992505600 & 15.086 & 2.953 &  89.5 & 0.24 \\
HD  40335    & 3315163867123687808 &  6.890 & 0.298 &  3315169399041560832 &  5.869 & 0.317 & 177.8 & 0.84 \\
HD  45557    & 5481786576589734400 &  5.773 & -0.004 &  5481833477632605056 & 14.185 & 2.382 &  33.2 & 0.34 \\
HD  65517    & 5194852249768916224 &  7.482 & 0.304 &  5194852249769630976 & 12.102 &   nan &   2.2 & 0.63 \\
HD 114905    & 1522893908790675456 &  6.721 & 0.630 &  1522893908790675584 & 12.606 & 1.941 &   6.6 & 0.48 \\
HD  73487    & 664329465285221248 &  9.003 & 1.073 &  664329465285129856 & 16.473 & 2.331 &  32.1 & 0.07 \\
HD 146569    & 6037205940527834624 &  8.243 & 0.165 &  6037206692130852352 & 17.854 & 3.019 &  30.0 & 0.55 \\
* eta Aps    & 5772879737424829184 &  4.835 & 0.325 &  5772878981507554048 & 15.219 & -0.136 &  24.9 & 0.16 \\
HD 102956    & 845335158256083584 &  7.619 & 1.097 &  845335158256818176 & 15.319 & 2.543 &  31.9 & 0.39 \\
HD 108863    & 3952692457305436800 &  7.467 & 1.147 &  3952693281939158144 & 10.004 & 0.725 & 115.4 & 0.17 \\
BD+26  2339  & 4008339427981993088 &  9.927 & 0.592 &  4008339359262516224 & 14.000 & 1.513 &  13.1 & 0.18 \\
BD+28  2091  & 4009747107807825664 &  9.933 & 0.658 &  4009747146463096192 & 15.748 & 2.186 &  22.8 & 0.58 \\
HD 105898    & 4003139184658921472 &  7.264 & 1.059 &  4003139180364904064 & 17.331 & 2.091 &  13.1 & 0.13 \\
HD 107086    & 4008609736043668864 &  7.454 & 0.589 &  4008609731748472192 & 17.080 & 3.368 &  19.0 & 0.58 \\
HD 108956    & 3959079382912168320 &  6.979 & 0.778 &  3959079382911362944 & 14.099 & 2.260 &   7.0 & 1.02 \\
\enddata
\end{deluxetable*}

{\bf Stars with special circumstances.}\quad
Some CDDS stars have no measured parallax. Several others have an interesting history in the literature. Here we outline approaches to derive distances and provide summaries for mis-classified stars.

{\it EN Cha.} Within the uncertainties, EN Cha has the same radial velocity and proper motion as other members of the $\eta$ Cha cluster, but it has no distance. Adopting a distance $d$ = 98.2~pc \citep{hunt2023,hunt2024} yields a similar \lstar\ as other M4--M5 stars in the group.

{\it HIP 59154.} Far away from any association or cluster, this star has an appropriate \lxlstar\ $\sim 5 \times 10^{-4}$ for a moderately young star on the main sequence. The adopted distance is then 27~pc. 

{\it 2MASS J06434532-6424396.} The \logg\ and \vsini\ suggest a star contracting to the main sequence. Adopting \mstar\ = 0.32~\msun\ for \teff\ = 3350~K implies $d$ = 37.7~pc. 

{\it 2MASS J07453282-3751230.} The \teff, \logg, \vsini, and [Li/H] suggest an old M3--M4 star on the main sequence. Adopting \mstar\ = 0.32~\msun\ for \teff\ = 3350~K implies $d$ = 139~pc. 

{\it Pleiades.} The B star Cl* Melotte 22 HII 1432 ($\eta$ Tau) is assigned to the Pleiades, but has large errors in parallax and proper motion. The nominal parallax places it on the near side of the cluster. Within the large uncertainties, the M star Cl* Melotte 22 HCG 0277 (QV Tau) has similar proper motion and radial velocity to $\eta$ Tau; with no distance estimate, we use \logg, \teff, and a mass estimate (0.27~\msun) to derive \lstar\ and a distance of 95~pc. \gaia\ studies place the star in the field, which agrees with our distance estimate. 

{\it Praesepe.} Five stars assigned to Praesepe on dynamical grounds have no distance measurement (Cl* NGC 2632 KW 236, BD+20 2162, Cl* NGC 2632 WJJP 607, BD+19 2076, and Cl* NGC 2632 JC 243; see \S\ref{sec: clusters}). Adopting the median distance of Praesepe stars, $d$ = 183.43~pc, places all five on or close to the 600--700~Myr PARSEC isochrones appropriate for Praesepe \citep{gossage2018,bossini2019,dias2021}. 

{\it Sco--Cen Association.} Several stars (CD$-22^{\deg}$ 11432, HD 128242, 2MASS J15571674-2529192, 2MASS J15572986-2258438, 2MASS J16080157-1927579, 2MASS J16082387-1935518, 2MASS J16082751-1949047, 2MASS J16084309-1900519, and V1154~Sco) have appropriate radial velocities and proper motions for LCC, UCL, or USco, but have no distance estimate. They also have some combination of \logg, \prot, and \lxlstar\ indicative of very young stars. Adopting a median distance to LCC, UCL, or USco places these stars on a 10--20~Myr PARSEC isochrone. As a check, we adopt a mass appropriate for the spectral type and derive \lstar\ (and then a distance similar to the median distance) from \logg\ and \teff. For ScoPMS 13, ScoPMS 23, and ScoPMS 31 (no distance estimates and large proper motion errors), their \logg, \prot, and \lxlstar\ are consistent with other USco stars that have ages of $\sim$ 10~Myr. Adopting the 10~Myr age implies $d$ = 96~pc for ScoPMS~13 and ScoPMS~31 and $d$ = 107~pc for ScoPMS~23. These stars lie in the foreground of most association stars.

{\it TYC~654-1274-1 (also known as 1E0307.4+1424, 1E0307.5+1424, and 2E0307.4+1424).} In SIMBAD, 1E0307.4+1424 has J2000.0 coordinates shifted by $\delta$RA = 1.64~sec and $\delta$Dec = 15.95~arcsec relative to the \gaia\ coordinates for TYC 654-1274-1 (which are almost identical to the 2MASS and WISE coordinates). However, they are the same star; on the Digitized Sky Survey, PanSTARRS DR1, 2MASS, and allWISE available with Aladin Lite, there is no star at the coordinates listed for 1E0307.4+1424, but there is clearly a star at the coordinates for TYC~654-1274-1. Curiously, there is no X-ray source coincident with either coordinate in ROSAT data. 

{\it TIC 454363834 (also known as CHX~18N, CHX~18, and RXJ1111.7-7620).} Like TYC~654-1274-1, this K5 star has two entries in SIMBAD, one for CHX~18N (TIC 454363834) and another for CHX~18 (RXJ1111.7-7620), but the information in the two entries corresponds to the same star. The coordinates for CHX~18 are shifted by $\delta$RA = 3.46~sec and $\delta$Dec = +20.9~arcsec relative to CHX~18N. Although this field is too far south for PAN-STARRS, CHX~18N is visible on Digitized Sky Survey, 2MASS, and allWISE images on Aladin Lite; CHX~18 is not visible. With ROSAT and eROSITA, there is an obvious X-ray source associated with CHX~18N and none associated with the position of CHX~18. 

{\it V1299~Tau.} The negligible reddening, \logg\ = 4.273, and \teff\ = 5775~K place the star on a PARSEC isochrone for an age of $\sim$ 15~Myr and yields \lstar $\approx$ 2~\lsun\ and $d$ = 103.5~pc. Younger isochrones are precluded. Stellar activity diagnostics, \vsini\ $\approx$ 70~\kms\ and \lxlstar\ $\sim 10^{-3}$, are consistent with a young age. Although the star might be in the greater Taurus region \citep{kraus2017}, we assign it to the field.

{\bf Candidate Common Proper Motion Companions.}\quad
To construct a set of candidate common proper motion stars, we consider possible companion stars that satisfy three constraints: (i) the projected separation $s$ at the distance of the nominal CDDS primary star is $s \le s_{max}$ = 1000~au, (ii) the parallax difference between the candidate companion and the primary is zero within the 3-$\sigma$ errors, and (iii) the difference in relative tangential velocities is $\delta v \le$ 2~\kms, where the tangential velocity of each component is $v = 4.74~\mu/\pi$ (units of \kms, mas yr$^{-1}$, and mas, respectively) and $\delta v = |v_1 - v_2|$. As a final cut, we eliminate stars where the parallax error is too large to make a reliable estimate of the projected separation; typically, this constraint precludes companions to stars in the vicinity of associations or clusters with distances larger than $\sim$ 300~pc. Once we have a final set of possible common proper motion companions, we remove those previously reported in the literature (see Table~\ref{tab: non-single-stars}). 

For all of the candidates, uncertainties in the parallax preclude an accurate assessment of the 3D separation. Typical errors in the relative positions are $\lesssim$ 0.1 arcsec, which corresponds to $\sim$ 5~au at the 50~pc median distance of CDDS stars. Parallax uncertainties are $\sim$ 0.2\% (0.5\%) for primary (secondary) stars; at 50~pc, the line-of-sight separation then has a minimum uncertainty of a few thousand au. For orbits randomly oriented in space, large line-of-sight separations and small projected separations are rare. Thus, we only consider the projected separation. 

Table~\ref{tab: binary stars} lists the candidates. The first four columns list the nominal ID of the CDDS primary along with its \gaia\ ID, G-band magnitude, and BP--RP color. The next three columns lists the \gaia\ ID, G, and BP--RP for the proposed companion. The table concludes with the projected separation in au and velocity difference in \kms. The candidates fall into three broad classes. Some have no entries in SIMBAD; others have an entry with only a Gaia DR3 identification, the G-magnitude, and sometimes 2MASS \jhks\ data. Several have information on spectral types that we discuss below. All of the candidates are consistent with main sequence or white dwarf companions to brighter primary stars. Additional information, such as supplementary epochs of imaging, \bvic\ photometry, and optical or infrared spectroscopy would test the possibility that these companions are physical pairs.

%Stellar companions of the CDDS stars may affect inferred physical parameters (if blended) and may have important dynamical impacts on debris disks. In addition to a bibliographic search for companions, we queried the Gaia DR3 archive for sources close to each CDDS member, evaluating whether they might be bound by assessing their relative proper motion. Our search 

{\it $\eta$ Apodis.} The possible companion to this A-type star is a DA white dwarf \citep{jimenez2023}, one of many discovered with \gaia. The stars are not obviously known as a binary pair. 

{\it b Ophiuchi.} With a magnitude difference $\Delta G \approx$ 10.4, this pair consists of an A-type primary and an M-type secondary. The magnitude difference implies a middle M spectral type for the fainter star. Infrared colors (2MASS J17262200-2410487) and G--K paint an inconsistent picture: G--K and H--K suggest a late type star, while J--H is more consistent with an early K star.

{\it HD 33283.} Although the companion is included in \citet{gonzalez2024}, it merits inclusion due to uncertainties in the quoted separation. The primary HD 33283~A is a solar-type star with $d \approx$ 90~pc. The secondary HD~33283~B is listed in SIMBAD and has the G--K and J--K colors of a middle M-type star. \citet{gonzalez2024} show a schematic with a planet orbiting A inside 1~au and the companion star beyond $10^4$~au. With a projected separation of 50--60~au, most of the quoted separation is due to differences in a parallax which has large uncertainties of $\sim$ 5000~au. Optical or infrared spectra would establish the spectral type of component B; IR imaging over several years would constrain the orbit. 

{\it HD~77052.} This G-type star has companion with a magnitude difference $\Delta$G $\approx$ 5.8. For the \citet{pecaut2013} main sequence locus, the G-magnitude difference implies a spectral type, $\approx$ M3~V, that is consistent with the spectral type of M2.5 V listed in SIMBAD. The companion is also listed as TIC 203226434.

{\it HD 104731.} The middle F primary is roughly 5.5~mag brighter than the potential companion at G, which implies a spectral type of M0--M1 for a main sequence companion. The lack of BP--RP precludes a consistency check. Aside from listings in catalogs of nearby stars \citep{gaianearby2021,gondoin2023}, the companion appears in SIMBAD as HD~104731~B without any additional information 

{\it HD 107148.} The early G-type primary is $\sim$ 10\% more luminous than the Sun and approximately 9.5~mag brighter than the white dwarf companion at G \citep{jimenez2023}. Although SIMBAD does not list the white dwarf as a companion, \citet{gonzalez2024} place the star at a separation of $\sim 10^4$~au and note the two planets within 1~au of the primary. The projected separation is only $\sim$ 35~au; the line-of-sight separation has an uncertainty of $\sim 2 \times 10^4$~au. 

{\it HD~202884.} The companion to this F5 star has an M4 spectral type in SIMBAD (TIC 262277113). The G-magnitude difference implies a spectral type of M3--M5 \citep{pecaut2013}. 2MASS data quoted in SIMBAD (2MASS J21184222+085645, TIC 262277113) are consistent with an M4 main sequence star. 

%{\it HD 245567.} This G0 star has a fainter companion with an M3 spectral type - need to check this one more as it is in an association.

%\bibliography{cdds}{}
\bibliographystyle{aasjournalv7}
\bibliography{ms.bbl}

\begin{thebibliography}{}
\expandafter\ifx\csname natexlab\endcsname\relax\def\natexlab#1{#1}\fi
\providecommand{\url}[1]{\href{#1}{#1}}
\providecommand{\dodoi}[1]{doi:~\href{http://doi.org/#1}{\nolinkurl{#1}}}
\providecommand{\doeprint}[1]{\href{http://ascl.net/#1}{\nolinkurl{http://ascl.net/#1}}}
\providecommand{\doarXiv}[1]{\href{https://arxiv.org/abs/#1}{\nolinkurl{https://arxiv.org/abs/#1}}}

% type= article
\bibitem[{C. {Abad} \& B. {Vicente}(1999){Abad} \& {Vicente}}]{abad1999}
{Abad}, C., \& {Vicente}, B. 1999, \bibinfo{title}{{An astrometric catalogue
  for the area of Coma Berenices},} \aaps, 136, 307,
  \dodoi{10.1051/aas:1999216}

% type= article
\bibitem[{ {Abdurro'uf} {et~al.}(2022){Abdurro'uf}, {Accetta}, {Aerts}, {Silva
  Aguirre}, {Ahumada}, {Ajgaonkar}, {Filiz Ak}, {Alam}, {Allende Prieto},
  {Almeida}, {Anders}, {Anderson}, {Andrews}, {Anguiano}, {Aquino-Ort{\'\i}z},
  {Arag{\'o}n-Salamanca}, {Argudo-Fern{\'a}ndez}, {Ata}, {Aubert},
  {Avila-Reese}, {Badenes}, {Barb{\'a}}, {Barger}, {Barrera-Ballesteros},
  {Beaton}, {Beers}, {Belfiore}, {Bender}, {Bernardi}, {Bershady}, {Beutler},
  {Bidin}, {Bird}, {Bizyaev}, {Blanc}, {Blanton}, {Boardman}, {Bolton},
  {Boquien}, {Borissova}, {Bovy}, {Brandt}, {Brown}, {Brownstein}, {Brusa},
  {Buchner}, {Bundy}, {Burchett}, {Bureau}, {Burgasser}, {Cabang}, {Campbell},
  {Cappellari}, {Carlberg}, {Wanderley}, {Carrera}, {Cash}, {Chen}, {Chen},
  {Cherinka}, {Chiappini}, {Choi}, {Chojnowski}, {Chung}, {Clerc}, {Cohen},
  {Comerford}, {Comparat}, {da Costa}, {Covey}, {Crane}, {Cruz-Gonzalez},
  {Culhane}, {Cunha}, {Dai}, {Damke}, {Darling}, {Davidson}, {Davies},
  {Dawson}, {De Lee}, {Diamond-Stanic}, {Cano-D{\'\i}az}, {S{\'a}nchez},
  {Donor}, {Duckworth}, {Dwelly}, {Eisenstein}, {Elsworth}, {Emsellem},
  {Eracleous}, {Escoffier}, {Fan}, {Farr}, {Feng}, {Fern{\'a}ndez-Trincado},
  {Feuillet}, {Filipp}, {Fillingham}, {Frinchaboy}, {Fromenteau}, {Galbany},
  {Garc{\'\i}a}, {Garc{\'\i}a-Hern{\'a}ndez}, {Ge}, {Geisler}, {Gelfand},
  {G{\'e}ron}, {Gibson}, {Goddy}, {Godoy-Rivera}, {Grabowski}, {Green},
  {Greener}, {Grier}, {Griffith}, {Guo}, {Guy}, {Hadjara}, {Harding},
  {Hasselquist}, {Hayes}, {Hearty}, {Hern{\'a}ndez}, {Hill}, {Hogg},
  {Holtzman}, {Horta}, {Hsieh}, {Hsu}, {Hsu}, {Huber}, {Huertas-Company},
  {Hutchinson}, {Hwang}, {Ibarra-Medel}, {Chitham}, {Ilha}, {Imig}, {Jaekle},
  {Jayasinghe}, {Ji}, {Johnson}, {Jones}, {J{\"o}nsson}, {Katkov}, {Khalatyan},
  {Kinemuchi}, {Kisku}, {Knapen}, {Kneib}, {Kollmeier}, {Kong}, {Kounkel},
  {Kreckel}, {Krishnarao}, {Lacerna}, {Lane}, {Langgin}, {Lavender}, {Law},
  {Lazarz}, {Leung}, {Leung}, {Lewis}, {Li}, {Li}, {Lian}, {Liang}, {Lin},
  {Lin}, {Lin}, {Lintott}, {Long}, {Longa-Pe{\~n}a}, {L{\'o}pez-Cob{\'a}},
  {Lu}, {Lundgren}, {Luo}, {Mackereth}, {de la Macorra}, {Mahadevan},
  {Majewski}, {Manchado}, {Mandeville}, {Maraston}, {Margalef-Bentabol},
  {Masseron}, {Masters}, {Mathur}, {McDermid}, {Mckay}, {Merloni},
  {Merrifield}, {Meszaros}, {Miglio}, {Di Mille}, {Minniti}, {Minsley},
  {Monachesi}, {Moon}, {Mosser}, {Mulchaey}, {Muna}, {Mu{\~n}oz}, {Myers},
  {Myers}, {Nadathur}, {Nair}, {Nandra}, {Neumann}, {Newman}, {Nidever},
  {Nikakhtar}, {Nitschelm}, {O'Connell}, {Garma-Oehmichen}, {Luan Souza de
  Oliveira}, {Olney}, {Oravetz}, {Ortigoza-Urdaneta}, {Osorio}, {Otter},
  {Pace}, {Padilla}, {Pan}, {Pan}, {Parikh}, {Parker}, {Peirani}, {Pe{\~n}a
  Ram{\'\i}rez}, {Penny}, {Percival}, {Perez-Fournon}, {Pinsonneault},
  {Poidevin}, {Poovelil}, {Price-Whelan}, {B{\'a}rbara de Andrade Queiroz},
  {Raddick}, {Ray}, {Rembold}, {Riddle}, {Riffel}, {Riffel}, {Rix}, {Robin},
  {Rodr{\'\i}guez-Puebla}, {Roman-Lopes}, {Rom{\'a}n-Z{\'u}{\~n}iga}, {Rose},
  {Ross}, {Rossi}, {Rubin}, {Salvato}, {S{\'a}nchez}, {S{\'a}nchez-Gallego},
  {Sanderson}, {Santana Rojas}, {Sarceno}, {Sarmiento}, {Sayres}, {Sazonova},
  {Schaefer}, {Schiavon}, {Schlegel}, {Schneider}, {Schultheis}, {Schwope},
  {Serenelli}, {Serna}, {Shao}, {Shapiro}, {Sharma}, {Shen}, {Shetrone}, {Shu},
  {Simon}, {Skrutskie}, {Smethurst}, {Smith}, {Sobeck}, {Spoo}, {Sprague},
  {Stark}, {Stassun}, {Steinmetz}, {Stello}, {Stone-Martinez},
  {Storchi-Bergmann}, {Stringfellow}, {Stutz}, {Su}, {Taghizadeh-Popp},
  {Talbot}, {Tayar}, {Telles}, {Teske}, {Thakar}, {Theissen}, {Tkachenko},
  {Thomas}, {Tojeiro}, {Hernandez Toledo}, {Troup}, {Trump}, {Trussler},
  {Turner}, {Tuttle}, {Unda-Sanzana}, {V{\'a}zquez-Mata}, {Valentini},
  {Valenzuela}, {Vargas-Gonz{\'a}lez}, {Vargas-Maga{\~n}a}, {Alfaro},
  {Villanova}, {Vincenzo}, {Wake}, {Warfield}, {Washington}, {Weaver},
  {Weijmans}, {Weinberg}, {Weiss}, {Westfall}, {Wild}, {Wilde}, {Wilson},
  {Wilson}, {Wilson}, {Wolf}, {Wood-Vasey}, {Yan}, {Zamora}, {Zasowski},
  {Zhang}, {Zhao}, {Zheng}, {Zheng}, \& {Zhu}}]{abdurrouf2022}
{Abdurro'uf}, {Accetta}, K., {Aerts}, C., {et~al.} 2022, \bibinfo{title}{{The
  Seventeenth Data Release of the Sloan Digital Sky Surveys: Complete Release
  of MaNGA, MaStar, and APOGEE-2 Data},} \apjs, 259, 35,
  \dodoi{10.3847/1538-4365/ac4414}

% type= article
\bibitem[{H.~A. {Abt}(2004){Abt}}]{abt2004}
{Abt}, H.~A. 2004, \bibinfo{title}{{Spectral Classification of Stars in A
  Supplement to the Bright Star Catalogue},} \apjs, 155, 175,
  \dodoi{10.1086/423803}

% type= article
\bibitem[{H.~A. {Abt} {et~al.}(2002){Abt}, {Levato}, \& {Grosso}}]{abt2002}
{Abt}, H.~A., {Levato}, H., \& {Grosso}, M. 2002, \bibinfo{title}{{Rotational
  Velocities of B Stars},} \apj, 573, 359, \dodoi{10.1086/340590}

% type= article
\bibitem[{H.~A. {Abt} \& N.~I. {Morrell}(1995){Abt} \& {Morrell}}]{abt1995}
{Abt}, H.~A., \& {Morrell}, N.~I. 1995, \bibinfo{title}{{The Relation between
  Rotational Velocities and Spectral Peculiarities among A-Type Stars},} \apjs,
  99, 135, \dodoi{10.1086/192182}

% type= article
\bibitem[{C. {Aguilera-G{\'o}mez} {et~al.}(2018){Aguilera-G{\'o}mez},
  {Ram{\'\i}rez}, \& {Chanam{\'e}}}]{aguileragomez2018}
{Aguilera-G{\'o}mez}, C., {Ram{\'\i}rez}, I., \& {Chanam{\'e}}, J. 2018,
  \bibinfo{title}{{Lithium abundance patterns of late-F stars: an in-depth
  analysis of the lithium desert},} \aap, 614, A55,
  \dodoi{10.1051/0004-6361/201732209}

% type= article
\bibitem[{R. {Akeson} {et~al.}(2021){Akeson}, {Beichman}, {Kervella},
  {Fomalont}, \& {Benedict}}]{akeson2021}
{Akeson}, R., {Beichman}, C., {Kervella}, P., {Fomalont}, E., \& {Benedict},
  G.~F. 2021, \bibinfo{title}{{Precision Millimeter Astrometry of the
  {\ensuremath{\alpha}} Centauri AB System},} \aj, 162, 14,
  \dodoi{10.3847/1538-3881/abfaff}

% type= article
\bibitem[{J. {Alfonso} \& A. {Garc{\'\i}a-Varela}(2023){Alfonso} \&
  {Garc{\'\i}a-Varela}}]{alfonso2023}
{Alfonso}, J., \& {Garc{\'\i}a-Varela}, A. 2023, \bibinfo{title}{{A Gaia
  astrometric view of the open clusters Pleiades, Praesepe, and Blanco 1},}
  \aap, 677, A163, \dodoi{10.1051/0004-6361/202346569}

% type= article
\bibitem[{A. {Alonso} {et~al.}(1999){Alonso}, {Arribas}, \&
  {Mart{\'\i}nez-Roger}}]{alonso1999}
{Alonso}, A., {Arribas}, S., \& {Mart{\'\i}nez-Roger}, C. 1999,
  \bibinfo{title}{{The effective temperature scale of giant stars (F0-K5). II.
  Empirical calibration of T$_{eff}$ versus colours and [Fe/H]},} \aaps, 140,
  261, \dodoi{10.1051/aas:1999521}

% type= article
\bibitem[{F.~J. {Alonso-Floriano} {et~al.}(2015){Alonso-Floriano}, {Morales},
  {Caballero}, {Montes}, {Klutsch}, {Mundt}, {Cort{\'e}s-Contreras}, {Ribas},
  {Reiners}, {Amado}, {Quirrenbach}, \& {Jeffers}}]{alonso2015}
{Alonso-Floriano}, F.~J., {Morales}, J.~C., {Caballero}, J.~A., {et~al.} 2015,
  \bibinfo{title}{{CARMENES input catalogue of M dwarfs. I. Low-resolution
  spectroscopy with CAFOS},} \aap, 577, A128,
  \dodoi{10.1051/0004-6361/201525803}

% type= article
\bibitem[{R. {Andrae} {et~al.}(2023{\natexlab{a}}){Andrae}, {Rix}, \&
  {Chandra}}]{andrae2023b}
{Andrae}, R., {Rix}, H.-W., \& {Chandra}, V. 2023{\natexlab{a}},
  \bibinfo{title}{{Robust Data-driven Metallicities for 175 Million Stars from
  Gaia XP Spectra},} \apjs, 267, 8, \dodoi{10.3847/1538-4365/acd53e}

% type= article
\bibitem[{R. {Andrae} {et~al.}(2023{\natexlab{b}}){Andrae}, {Fouesneau},
  {Sordo}, {Bailer-Jones}, {Dharmawardena}, {Rybizki}, {De Angeli},
  {Lindstr{\o}m}, {Marshall}, {Drimmel}, {Korn}, {Soubiran}, {Brouillet},
  {Casamiquela}, {Rix}, {Abreu Aramburu}, {{\'A}lvarez}, {Bakker},
  {Bellas-Velidis}, {Bijaoui}, {Brugaletta}, {Burlacu}, {Carballo}, {Chaoul},
  {Chiavassa}, {Contursi}, {Cooper}, {Creevey}, {Dafonte}, {Dapergolas}, {de
  Laverny}, {Delchambre}, {Demouchy}, {Edvardsson}, {Fr{\'e}mat}, {Garabato},
  {Garc{\'\i}a-Lario}, {Garc{\'\i}a-Torres}, {Gavel}, {Gomez},
  {Gonz{\'a}lez-Santamar{\'\i}a}, {Hatzidimitriou}, {Heiter}, {Jean-Antoine
  Piccolo}, {Kontizas}, {Kordopatis}, {Lanzafame}, {Lebreton}, {Licata},
  {Livanou}, {Lobel}, {Lorca}, {Magdaleno Romeo}, {Manteiga}, {Marocco},
  {Mary}, {Nicolas}, {Ordenovic}, {Pailler}, {Palicio}, {Pallas-Quintela},
  {Panem}, {Pichon}, {Poggio}, {Recio-Blanco}, {Riclet}, {Robin},
  {Santove{\~n}a}, {Sarro}, {Schultheis}, {Segol}, {Silvelo}, {Slezak},
  {Smart}, {S{\"u}veges}, {Th{\'e}venin}, {Torralba Elipe}, {Ulla}, {Utrilla},
  {Vallenari}, {van Dillen}, {Zhao}, \& {Zorec}}]{andrae2023a}
{Andrae}, R., {Fouesneau}, M., {Sordo}, R., {et~al.} 2023{\natexlab{b}},
  \bibinfo{title}{{Gaia Data Release 3. Analysis of the Gaia BP/RP spectra
  using the General Stellar Parameterizer from Photometry},} \aap, 674, A27,
  \dodoi{10.1051/0004-6361/202243462}

% type= article
\bibitem[{V. {Andretta} {et~al.}(2005){Andretta}, {Bus{\`a}}, {Gomez}, \&
  {Terranegra}}]{andretta2005}
{Andretta}, V., {Bus{\`a}}, I., {Gomez}, M.~T., \& {Terranegra}, L. 2005,
  \bibinfo{title}{{The Ca II Infrared Triplet as a stellar activity diagnostic
  . I. Non-LTE photospheric profiles and definition of the R$_{IRT}$
  indicator},} \aap, 430, 669, \dodoi{10.1051/0004-6361:20041745}

% type= article
\bibitem[{F. {Anthonioz} {et~al.}(2015){Anthonioz}, {M{\'e}nard}, {Pinte}, {Le
  Bouquin}, {Benisty}, {Thi}, {Absil}, {Duch{\^e}ne}, {Augereau}, {Berger},
  {Casassus}, {Duvert}, {Lazareff}, {Malbet}, {Millan-Gabet}, {Schreiber},
  {Traub}, \& {Zins}}]{anthonioz2015}
{Anthonioz}, F., {M{\'e}nard}, F., {Pinte}, C., {et~al.} 2015,
  \bibinfo{title}{{The VLTI/PIONIER near-infrared interferometric survey of
  southern T Tauri stars. I. First results},} \aap, 574, A41,
  \dodoi{10.1051/0004-6361/201424520}

% type= article
\bibitem[{C. {Argiroffi} {et~al.}(2016){Argiroffi}, {Caramazza}, {Micela},
  {Sciortino}, {Moraux}, {Bouvier}, \& {Flaccomio}}]{argiroffi2016}
{Argiroffi}, C., {Caramazza}, M., {Micela}, G., {et~al.} 2016,
  \bibinfo{title}{{Supersaturation and activity-rotation relation in PMS stars:
  the young cluster h Persei},} \aap, 589, A113,
  \dodoi{10.1051/0004-6361/201526539}

% type= article
\bibitem[{C. {Argiroffi} {et~al.}(2006){Argiroffi}, {Favata}, {Flaccomio},
  {Maggio}, {Micela}, {Peres}, \& {Sciortino}}]{argiroffi2006}
{Argiroffi}, C., {Favata}, F., {Flaccomio}, E., {et~al.} 2006,
  \bibinfo{title}{{XMM-Newton survey of two upper Scorpius regions},} \aap,
  459, 199, \dodoi{10.1051/0004-6361:20065674}

% type= article
\bibitem[{B. {Aschenbach} {et~al.}(1981){Aschenbach}, {Br{\"a}uninger},
  {Briel}, {Brinkmann}, {Fink}, {Heinecke}, {Hippmann}, {Kettenring},
  {Metzner}, {Ondrusch}, {Pfeffermann}, {Predehl}, {Reger}, {Stephan},
  {Tr{\"u}mper}, \& {Zimmermann}}]{aschenbach1981}
{Aschenbach}, B., {Br{\"a}uninger}, H., {Briel}, U., {et~al.} 1981,
  \bibinfo{title}{{The ROSAT Mission},} \ssr, 30, 569,
  \dodoi{10.1007/BF01246075}

% type= article
\bibitem[{M. {Asplund} {et~al.}(2009){Asplund}, {Grevesse}, {Sauval}, \&
  {Scott}}]{asplund2009}
{Asplund}, M., {Grevesse}, N., {Sauval}, A.~J., \& {Scott}, P. 2009,
  \bibinfo{title}{{The Chemical Composition of the Sun},} \araa, 47, 481,
  \dodoi{10.1146/annurev.astro.46.060407.145222}

% type= article
\bibitem[{N. {Astudillo-Defru} {et~al.}(2017){Astudillo-Defru}, {Delfosse},
  {Bonfils}, {Forveille}, {Lovis}, \& {Rameau}}]{astudillo2017}
{Astudillo-Defru}, N., {Delfosse}, X., {Bonfils}, X., {et~al.} 2017,
  \bibinfo{title}{{Magnetic activity in the HARPS M dwarf sample. The
  rotation-activity relationship for very low-mass stars through R'$_{HK}$},}
  \aap, 600, A13, \dodoi{10.1051/0004-6361/201527078}

% type= article
\bibitem[{H.~H. {Aumann} {et~al.}(1984){Aumann}, {Gillett}, {Beichman}, {de
  Jong}, {Houck}, {Low}, {Neugebauer}, {Walker}, \& {Wesselius}}]{aumann84}
{Aumann}, H.~H., {Gillett}, F.~C., {Beichman}, C.~A., {et~al.} 1984,
  \bibinfo{title}{{Discovery of a shell around alpha Lyrae.},} \apjl, 278, L23,
  \dodoi{10.1086/184214}

% type= article
\bibitem[{C.~A.~L. {Bailer-Jones} {et~al.}(2021){Bailer-Jones}, {Rybizki},
  {Fouesneau}, {Demleitner}, \& {Andrae}}]{bailerjones2021}
{Bailer-Jones}, C.~A.~L., {Rybizki}, J., {Fouesneau}, M., {Demleitner}, M., \&
  {Andrae}, R. 2021, \bibinfo{title}{{Estimating Distances from Parallaxes. V.
  Geometric and Photogeometric Distances to 1.47 Billion Stars in Gaia Early
  Data Release 3},} \aj, 161, 147, \dodoi{10.3847/1538-3881/abd806}

% type= article
\bibitem[{C.~A.~L. {Bailer-Jones} {et~al.}(2013){Bailer-Jones}, {Andrae},
  {Arcay}, {Astraatmadja}, {Bellas-Velidis}, {Berihuete}, {Bijaoui},
  {Carri{\'o}n}, {Dafonte}, {Damerdji}, {Dapergolas}, {de Laverny},
  {Delchambre}, {Drazinos}, {Drimmel}, {Fr{\'e}mat}, {Fustes},
  {Garc{\'\i}a-Torres}, {Gu{\'e}d{\'e}}, {Heiter}, {Janotto}, {Karampelas},
  {Kim}, {Knude}, {Kolka}, {Kontizas}, {Kontizas}, {Korn}, {Lanzafame},
  {Lebreton}, {Lindstr{\o}m}, {Liu}, {Livanou}, {Lobel}, {Manteiga},
  {Martayan}, {Ordenovic}, {Pichon}, {Recio-Blanco}, {Rocca-Volmerange},
  {Sarro}, {Smith}, {Sordo}, {Soubiran}, {Surdej}, {Th{\'e}venin},
  {Tsalmantza}, {Vallenari}, \& {Zorec}}]{bailerjones2013}
{Bailer-Jones}, C.~A.~L., {Andrae}, R., {Arcay}, B., {et~al.} 2013,
  \bibinfo{title}{{The Gaia astrophysical parameters inference system (Apsis).
  Pre-launch description},} \aap, 559, A74, \dodoi{10.1051/0004-6361/201322344}

% type= article
\bibitem[{J.~I. {Bailey} {et~al.}(2018){Bailey}, {Mateo}, {White}, {Shectman},
  \& {Crane}}]{bailey2018}
{Bailey}, J.~I., {Mateo}, M., {White}, R.~J., {Shectman}, S.~A., \& {Crane},
  J.~D. 2018, \bibinfo{title}{{Radial velocity variability and stellar
  properties of FGK stars in the cores of NGC 2516 and NGC 2422},} \mnras, 475,
  1609, \dodoi{10.1093/mnras/stx3266}

% type= article
\bibitem[{Z. {Balog} {et~al.}(2009){Balog}, {Kiss}, {Vink{\'o}}, {Rieke},
  {Muzerolle}, {G{\'a}sp{\'a}r}, {Young}, \& {Gorlova}}]{balog2009}
{Balog}, Z., {Kiss}, L.~L., {Vink{\'o}}, J., {et~al.} 2009,
  \bibinfo{title}{{Spitzer/IRAC-MIPS Survey of NGC 2451A AND B: Debris Disks at
  50-80 Million Years},} \apj, 698, 1989, \dodoi{10.1088/0004-637X/698/2/1989}

% type= article
\bibitem[{S.~A. {Barenfeld} {et~al.}(2016){Barenfeld}, {Carpenter}, {Ricci}, \&
  {Isella}}]{barenfield2016}
{Barenfeld}, S.~A., {Carpenter}, J.~M., {Ricci}, L., \& {Isella}, A. 2016,
  \bibinfo{title}{{ALMA Observations of Circumstellar Disks in the Upper
  Scorpius OB Association},} \apj, 827, 142,
  \dodoi{10.3847/0004-637X/827/2/142}

% type= article
\bibitem[{F. {Baron} {et~al.}(2012){Baron}, {Monnier}, {Pedretti}, {Zhao},
  {Schaefer}, {Parks}, {Che}, {Thureau}, {ten Brummelaar}, {McAlister},
  {Ridgway}, {Farrington}, {Sturmann}, {Sturmann}, \& {Turner}}]{baron2012}
{Baron}, F., {Monnier}, J.~D., {Pedretti}, E., {et~al.} 2012,
  \bibinfo{title}{{Imaging the Algol Triple System in the H Band with the CHARA
  Interferometer},} \apj, 752, 20, \dodoi{10.1088/0004-637X/752/1/20}

% type= article
\bibitem[{D. {Barrado y Navascues}(1998){Barrado y Navascues}}]{barrado1998}
{Barrado y Navascues}, D. 1998, \bibinfo{title}{{The Castor moving group. The
  age of Fomalhaut and VEGA},} \aap, 339, 831,
  \dodoi{10.48550/arXiv.astro-ph/9905243}

% type= inproceedings
\bibitem[{D. {Barrado Y Navascu{\'e}s}(2004){Barrado Y
  Navascu{\'e}s}}]{barrado2004a}
{Barrado Y Navascu{\'e}s}, D. 2004, \bibinfo{title}{{The stellar population in
  the field of the {\ensuremath{\alpha}} Persei cluster},} in ESA Special
  Publication, Vol. 538, Stellar Structure and Habitable Planet Finding, ed.
  F.~{Favata}, S.~{Aigrain}, \& A.~{Wilson} (ESA Publications Division),
  269--271

% type= article
\bibitem[{D. {Barrado y Navascu{\'e}s} {et~al.}(2002){Barrado y Navascu{\'e}s},
  {Bouvier}, {Stauffer}, {Lodieu}, \& {McCaughrean}}]{barrado2002}
{Barrado y Navascu{\'e}s}, D., {Bouvier}, J., {Stauffer}, J.~R., {Lodieu}, N.,
  \& {McCaughrean}, M.~J. 2002, \bibinfo{title}{{A substellar mass function for
  Alpha Persei},} \aap, 395, 813, \dodoi{10.1051/0004-6361:20021262}

% type= article
\bibitem[{D. {Barrado y Navascu{\'e}s} {et~al.}(2004){Barrado y Navascu{\'e}s},
  {Stauffer}, \& {Jayawardhana}}]{barrado2004b}
{Barrado y Navascu{\'e}s}, D., {Stauffer}, J.~R., \& {Jayawardhana}, R. 2004,
  \bibinfo{title}{{Spectroscopy of Very Low Mass Stars and Brown Dwarfs in IC
  2391: Lithium Depletion and H{\ensuremath{\alpha}} Emission},} \apj, 614,
  386, \dodoi{10.1086/423485}

% type= article
\bibitem[{D. {Barrado y Navascu{\'e}s} {et~al.}(1999){Barrado y Navascu{\'e}s},
  {Stauffer}, \& {Patten}}]{barrado1999}
{Barrado y Navascu{\'e}s}, D., {Stauffer}, J.~R., \& {Patten}, B.~M. 1999,
  \bibinfo{title}{{The Lithium-Depletion Boundary and the Age of the Young Open
  Cluster IC 2391},} \apjl, 522, L53, \dodoi{10.1086/312212}

% type= article
\bibitem[{C.~A. {Beichman} {et~al.}(2005){Beichman}, {Bryden}, {Rieke},
  {Stansberry}, {Trilling}, {Stapelfeldt}, {Werner}, {Engelbracht}, {Blaylock},
  {Gordon}, {Chen}, {Su}, \& {Hines}}]{beichman2005}
{Beichman}, C.~A., {Bryden}, G., {Rieke}, G.~H., {et~al.} 2005,
  \bibinfo{title}{{Planets and Infrared Excesses: Preliminary Results from a
  Spitzer MIPS Survey of Solar-Type Stars},} \apj, 622, 1160,
  \dodoi{10.1086/428115}

% type= article
\bibitem[{C.~A. {Beichman} {et~al.}(2006){Beichman}, {Tanner}, {Bryden},
  {Stapelfeldt}, {Werner}, {Rieke}, {Trilling}, {Lawler}, \&
  {Gautier}}]{beichman2006}
{Beichman}, C.~A., {Tanner}, A., {Bryden}, G., {et~al.} 2006,
  \bibinfo{title}{{IRS Spectra of Solar-Type Stars: A Search for Asteroid Belt
  Analogs},} \apj, 639, 1166, \dodoi{10.1086/499424}

% type= book
\bibitem[{C.~A. {Beichmann}(1985){Beichmann}}]{beichman1985}
{Beichmann}, C.~A. 1985, {Infrared Astronomical Satellite (IRAS) catalogs and
  atlases. Explanatory supplement} (Jet Propulsion Laboratory, Pasadena, CA
  USA)

% type= article
\bibitem[{C.~P.~M. {Bell} {et~al.}(2015){Bell}, {Mamajek}, \&
  {Naylor}}]{bell2015}
{Bell}, C. P.~M., {Mamajek}, E.~E., \& {Naylor}, T. 2015, \bibinfo{title}{{A
  self-consistent, absolute isochronal age scale for young moving groups in the
  solar neighbourhood},} \mnras, 454, 593, \dodoi{10.1093/mnras/stv1981}

% type= article
\bibitem[{T. {Bensby} \& K. {Lind}(2018){Bensby} \& {Lind}}]{bensby2018}
{Bensby}, T., \& {Lind}, K. 2018, \bibinfo{title}{{Exploring the production and
  depletion of lithium in the Milky Way stellar disk},} \aap, 615, A151,
  \dodoi{10.1051/0004-6361/201833118}

% type= article
\bibitem[{S.~V. {Berdyugina} {et~al.}(1998){Berdyugina}, {Jankov}, {Ilyin},
  {Tuominen}, \& {Fekel}}]{berdyugina1998}
{Berdyugina}, S.~V., {Jankov}, S., {Ilyin}, I., {Tuominen}, I., \& {Fekel},
  F.~C. 1998, \bibinfo{title}{{The active RS Canum Venaticorum binary II
  Pegasi. I. Stellar and orbital parameters},} \aap, 334, 863

% type= article
\bibitem[{M.~S. {Bessel}(1990){Bessel}}]{bessell1990}
{Bessel}, M.~S. 1990, \bibinfo{title}{{BVRI photometry of the Gliese catalogue
  stars.},} \aaps, 83, 357

% type= article
\bibitem[{M. {Bessell} \& S. {Murphy}(2012){Bessell} \& {Murphy}}]{bessell2012}
{Bessell}, M., \& {Murphy}, S. 2012, \bibinfo{title}{{Spectrophotometric
  Libraries, Revised Photonic Passbands, and Zero Points for UBVRI, Hipparcos,
  and Tycho Photometry},} \pasp, 124, 140, \dodoi{10.1086/664083}

% type= article
\bibitem[{A.~S. {Binks} {et~al.}(2021){Binks}, {Jeffries}, {Jackson},
  {Franciosini}, {Sacco}, {Bayo}, {Magrini}, {Randich}, {Arancibia-Silva},
  {Bergemann}, {Bragaglia}, {Gilmore}, {Gonneau}, {Hourihane}, {Jofr{\'e}},
  {Korn}, {Morbidelli}, {Prisinzano}, {Worley}, \& {Zaggia}}]{binks2021}
{Binks}, A.~S., {Jeffries}, R.~D., {Jackson}, R.~J., {et~al.} 2021,
  \bibinfo{title}{{The Gaia-ESO survey: a lithium depletion boundary age for
  NGC 2232},} \mnras, 505, 1280, \dodoi{10.1093/mnras/stab1351}

% type= article
\bibitem[{A. {Blaauw}(1964){Blaauw}}]{blaauw1964}
{Blaauw}, A. 1964, \bibinfo{title}{{The O Associations in the Solar
  Neighborhood},} \araa, 2, 213, \dodoi{10.1146/annurev.aa.02.090164.001241}

% type= inproceedings
\bibitem[{A. {Blaz{\`e}re} {et~al.}(2015){Blaz{\`e}re}, {Petit},
  {Ligni{\`e}res}, {Auri{\`e}re}, {Ballot}, {B{\"o}hm}, {Folsom}, {Ariste}, \&
  {Wade}}]{blazere2015}
{Blaz{\`e}re}, A., {Petit}, P., {Ligni{\`e}res}, F., {et~al.} 2015,
  \bibinfo{title}{{Ultra-weak magnetic fields in Am stars: {\ensuremath{\beta}}
  UMa and {\ensuremath{\theta}} Leo},} in IAU Symposium, Vol. 305, Polarimetry,
  ed. K.~N. {Nagendra}, S.~{Bagnulo}, R.~{Centeno}, \& M.~{Jes{\'u}s
  Mart{\'\i}nez Gonz{\'a}lez}, 67--72, \dodoi{10.1017/S1743921315004536}

% type= article
\bibitem[{V.~V. {Bobylev} \& A.~T. {Bajkova}(2016){Bobylev} \&
  {Bajkova}}]{bobylev2016}
{Bobylev}, V.~V., \& {Bajkova}, A.~T. 2016, \bibinfo{title}{{Kinematic analysis
  of solar-neighborhood stars based on RAVE4 data},} Astronomy Letters, 42, 90,
  \dodoi{10.1134/S1063773716020018}

% type= article
\bibitem[{A.~M. {Boesgaard} {et~al.}(2003){Boesgaard}, {Armengaud}, \&
  {King}}]{boesgaard2003}
{Boesgaard}, A.~M., {Armengaud}, E., \& {King}, J.~R. 2003,
  \bibinfo{title}{{Beryllium Abundances in F and G Dwarfs in the Coma Cluster
  and the Ursa Major Moving Group from Keck HIRES Observations},} \apj, 583,
  955, \dodoi{10.1086/345410}

% type= article
\bibitem[{A.~M. {Boesgaard} {et~al.}(2022){Boesgaard}, {Deliyannis}, {Lum}, \&
  {Chontos}}]{boesgaard2022}
{Boesgaard}, A.~M., {Deliyannis}, C.~P., {Lum}, M.~G., \& {Chontos}, A. 2022,
  \bibinfo{title}{{Lithium and Beryllium in One-solar-mass Stars},} \apj, 941,
  21, \dodoi{10.3847/1538-4357/ac9625}

% type= article
\bibitem[{T. {Boller} {et~al.}(2016){Boller}, {Freyberg}, {Tr{\"u}mper},
  {Haberl}, {Voges}, \& {Nandra}}]{boller2016}
{Boller}, T., {Freyberg}, M.~J., {Tr{\"u}mper}, J., {et~al.} 2016,
  \bibinfo{title}{{Second ROSAT all-sky survey (2RXS) source catalogue},} \aap,
  588, A103, \dodoi{10.1051/0004-6361/201525648}

% type= article
\bibitem[{A. {Bonsor} {et~al.}(2014){Bonsor}, {Kennedy}, {Wyatt}, {Johnson}, \&
  {Sibthorpe}}]{bonsor2014}
{Bonsor}, A., {Kennedy}, G.~M., {Wyatt}, M.~C., {Johnson}, J.~A., \&
  {Sibthorpe}, B. 2014, \bibinfo{title}{{Herschel observations of debris discs
  orbiting planet-hosting subgiants},} \mnras, 437, 3288,
  \dodoi{10.1093/mnras/stt2128}

% type= article
\bibitem[{M. {Booth} {et~al.}(2021){Booth}, {Schulz}, {Krivov}, {Marino},
  {Pearce}, \& {Launhardt}}]{booth2021}
{Booth}, M., {Schulz}, M., {Krivov}, A.~V., {et~al.} 2021,
  \bibinfo{title}{{Resolving the outer ring of HD 38206 using ALMA and
  constraining limits on planets in the system},} \mnras, 500, 1604,
  \dodoi{10.1093/mnras/staa3362}

% type= article
\bibitem[{S. {Boro Saikia} {et~al.}(2018){Boro Saikia}, {Marvin}, {Jeffers},
  {Reiners}, {Cameron}, {Marsden}, {Petit}, {Warnecke}, \&
  {Yadav}}]{borosaikia2018}
{Boro Saikia}, S., {Marvin}, C.~J., {Jeffers}, S.~V., {et~al.} 2018,
  \bibinfo{title}{{Chromospheric activity catalogue of 4454 cool stars.
  Questioning the active branch of stellar activity cycles},} \aap, 616, A108,
  \dodoi{10.1051/0004-6361/201629518}

% type= inproceedings
\bibitem[{J. {Borsenberger} {et~al.}(2006){Borsenberger}, {de Batz},
  {Derriere}, {Mamon}, {Omont}, {Paturel}, {Simon}, \&
  {Vaughin}}]{borsenberger2006}
{Borsenberger}, J., {de Batz}, B., {Derriere}, S., {et~al.} 2006,
  \bibinfo{title}{{DENIS, a European DEep Near Infrared Survey of the Southern
  Sky},} in Visions for Infrared Astronomy, Instrumentation, Mesure,
  M{\'e}trologie, ed. V.~{Coud{\'e} du Foresto}, D.~{Rouan}, \& G.~{Rousset},
  135--138

% type= article
\bibitem[{D. {Bossini} {et~al.}(2019){Bossini}, {Vallenari}, {Bragaglia},
  {Cantat-Gaudin}, {Sordo}, {Balaguer-N{\'u}{\~n}ez}, {Jordi}, {Moitinho},
  {Soubiran}, {Casamiquela}, {Carrera}, \& {Heiter}}]{bossini2019}
{Bossini}, D., {Vallenari}, A., {Bragaglia}, A., {et~al.} 2019,
  \bibinfo{title}{{Age determination for 269 Gaia DR2 open clusters},} \aap,
  623, A108, \dodoi{10.1051/0004-6361/201834693}

% type= article
\bibitem[{K. {Bouchaud} {et~al.}(2020){Bouchaud}, {Domiciano de Souza},
  {Rieutord}, {Reese}, \& {Kervella}}]{bouchard2020}
{Bouchaud}, K., {Domiciano de Souza}, A., {Rieutord}, M., {Reese}, D.~R., \&
  {Kervella}, P. 2020, \bibinfo{title}{{A realistic two-dimensional model of
  Altair},} \aap, 633, A78, \dodoi{10.1051/0004-6361/201936830}

% type= article
\bibitem[{L.~G. {Bouma} {et~al.}(2021){Bouma}, {Curtis}, {Hartman}, {Winn}, \&
  {Bakos}}]{bouma2021}
{Bouma}, L.~G., {Curtis}, J.~L., {Hartman}, J.~D., {Winn}, J.~N., \& {Bakos},
  G.~{\'A}. 2021, \bibinfo{title}{{Rotation and Lithium Confirmation of a 500
  pc Halo for the Open Cluster NGC 2516},} \aj, 162, 197,
  \dodoi{10.3847/1538-3881/ac18cd}

% type= article
\bibitem[{J. {Bouvier} {et~al.}(2018){Bouvier}, {Barrado}, {Moraux},
  {Stauffer}, {Rebull}, {Hillenbrand}, {Bayo}, {Boisse}, {Bouy}, {DiFolco},
  {Lillo-Box}, \& {Morales Calder{\'o}n}}]{bouvier2018}
{Bouvier}, J., {Barrado}, D., {Moraux}, E., {et~al.} 2018, \bibinfo{title}{{The
  lithium-rotation connection in the 125 Myr-old Pleiades cluster},} \aap, 613,
  A63, \dodoi{10.1051/0004-6361/201731881}

% type= article
\bibitem[{J. {Bovy} {et~al.}(2009){Bovy}, {Hogg}, \& {Roweis}}]{bovy2009}
{Bovy}, J., {Hogg}, D.~W., \& {Roweis}, S.~T. 2009, \bibinfo{title}{{The
  Velocity Distribution of Nearby Stars from Hipparcos Data. I. The
  Significance of the Moving Groups},} \apj, 700, 1794,
  \dodoi{10.1088/0004-637X/700/2/1794}

% type= article
\bibitem[{A.~W. {Boyle} \& L.~G. {Bouma}(2023){Boyle} \& {Bouma}}]{boyle2023}
{Boyle}, A.~W., \& {Bouma}, L.~G. 2023, \bibinfo{title}{{Stellar Rotation and
  Structure of the {\ensuremath{\alpha}} Persei Complex: When Does
  Gyrochronology Start to Work?},} \aj, 166, 14,
  \dodoi{10.3847/1538-3881/acd3e8}

% type= article
\bibitem[{J. {Braithwaite} \& M. {Cantiello}(2013){Braithwaite} \&
  {Cantiello}}]{braithwaite2013}
{Braithwaite}, J., \& {Cantiello}, M. 2013, \bibinfo{title}{{Weak magnetic
  fields in early-type stars: failed fossils},} \mnras, 428, 2789,
  \dodoi{10.1093/mnras/sts109}

% type= article
\bibitem[{W. {Brandner} {et~al.}(2023{\natexlab{a}}){Brandner},
  {Calissendorff}, \& {Kopytova}}]{brandner2023b}
{Brandner}, W., {Calissendorff}, P., \& {Kopytova}, T. 2023{\natexlab{a}},
  \bibinfo{title}{{Astrophysical Properties of 600 Bona Fide Single Stars in
  the Hyades Open Cluster},} \aj, 165, 108, \dodoi{10.3847/1538-3881/acb208}

% type= article
\bibitem[{W. {Brandner} {et~al.}(2023{\natexlab{b}}){Brandner},
  {Calissendorff}, \& {Kopytova}}]{brandner2023c}
{Brandner}, W., {Calissendorff}, P., \& {Kopytova}, T. 2023{\natexlab{b}},
  \bibinfo{title}{{Benchmarking Gaia DR3 Apsis with the Hyades and Pleiades
  open clusters},} \aap, 677, A162, \dodoi{10.1051/0004-6361/202346790}

% type= article
\bibitem[{W. {Brandner} {et~al.}(2023{\natexlab{c}}){Brandner},
  {Calissendorff}, \& {Kopytova}}]{brandner2023a}
{Brandner}, W., {Calissendorff}, P., \& {Kopytova}, T. 2023{\natexlab{c}},
  \bibinfo{title}{{Benchmarking MESA isochrones against the Hyades single star
  sequence},} \mnras, 518, 662, \dodoi{10.1093/mnras/stac2247}

% type= article
\bibitem[{T.~D. {Brandt} \& C.~X. {Huang}(2015){Brandt} \&
  {Huang}}]{brandt2015}
{Brandt}, T.~D., \& {Huang}, C.~X. 2015, \bibinfo{title}{{Bayesian Ages for
  Early-type Stars from Isochrones Including Rotation, and a Possible Old Age
  for the Hyades},} \apj, 807, 58, \dodoi{10.1088/0004-637X/807/1/58}

% type= article
\bibitem[{C. {Brice{\~n}o} {et~al.}(2019){Brice{\~n}o}, {Calvet},
  {Hern{\'a}ndez}, {Vivas}, {Mateu}, {Downes}, {Loerincs}, {P{\'e}rez-Blanco},
  {Berlind}, {Espaillat}, {Allen}, {Hartmann}, {Mateo}, \&
  {Bailey}}]{briceno2019}
{Brice{\~n}o}, C., {Calvet}, N., {Hern{\'a}ndez}, J., {et~al.} 2019,
  \bibinfo{title}{{The CIDA Variability Survey of Orion OB1. II. Demographics
  of the Young, Low-mass Stellar Populations},} \aj, 157, 85,
  \dodoi{10.3847/1538-3881/aaf79b}

% type= article
\bibitem[{G. {Brice{\~n}o-Morales} \& J.
  {Chanam{\'e}}(2023){Brice{\~n}o-Morales} \& {Chanam{\'e}}}]{briceno2023}
{Brice{\~n}o-Morales}, G., \& {Chanam{\'e}}, J. 2023,
  \bibinfo{title}{{Substructure, supernovae, and a time-resolved star formation
  history for Upper Scorpius},} \mnras, 522, 1288,
  \dodoi{10.1093/mnras/stad608}

% type= article
\bibitem[{K.~R. {Briggs} \& J.~P. {Pye}(2003){Briggs} \& {Pye}}]{briggs2003}
{Briggs}, K.~R., \& {Pye}, J.~P. 2003, \bibinfo{title}{{XMM-Newton and the
  Pleiades - I. Bright coronal sources and the X-ray emission from
  intermediate-type stars},} \mnras, 345, 714,
  \dodoi{10.1046/j.1365-8711.2003.06991.x}

% type= article
\bibitem[{E.~L. {Brown} {et~al.}(2022){Brown}, {Jeffers}, {Marsden}, {Morin},
  {Boro Saikia}, {Petit}, {Jardine}, {See}, {Vidotto}, {Mengel}, {Dahlkemper},
  \& {the BCool Collaboration}}]{brown2022}
{Brown}, E.~L., {Jeffers}, S.~V., {Marsden}, S.~C., {et~al.} 2022,
  \bibinfo{title}{{Linking chromospheric activity and magnetic field properties
  for late-type dwarf stars},} \mnras, 514, 4300,
  \dodoi{10.1093/mnras/stac1291}

% type= article
\bibitem[{G. {Bryden} {et~al.}(2006){Bryden}, {Beichman}, {Trilling}, {Rieke},
  {Holmes}, {Lawler}, {Stapelfeldt}, {Werner}, {Gautier}, {Blaylock}, {Gordon},
  {Stansberry}, \& {Su}}]{bryden2006}
{Bryden}, G., {Beichman}, C.~A., {Trilling}, D.~E., {et~al.} 2006,
  \bibinfo{title}{{Frequency of Debris Disks around Solar-Type Stars: First
  Results from a Spitzer MIPS Survey},} \apj, 636, 1098, \dodoi{10.1086/498093}

% type= article
\bibitem[{G. {Bryden} {et~al.}(2009){Bryden}, {Beichman}, {Carpenter}, {Rieke},
  {Stapelfeldt}, {Werner}, {Tanner}, {Lawler}, {Wyatt}, {Trilling}, {Su},
  {Blaylock}, \& {Stansberry}}]{bryden2009}
{Bryden}, G., {Beichman}, C.~A., {Carpenter}, J.~M., {et~al.} 2009,
  \bibinfo{title}{{Planets and Debris Disks: Results from a Spitzer/MIPS Search
  for Infrared Excess},} \apj, 705, 1226, \dodoi{10.1088/0004-637X/705/2/1226}

% type= article
\bibitem[{S. {Buder} {et~al.}(2021){Buder}, {Sharma}, {Kos}, {Amarsi},
  {Nordlander}, {Lind}, {Martell}, {Asplund}, {Bland-Hawthorn}, {Casey}, {de
  Silva}, {D'Orazi}, {Freeman}, {Hayden}, {Lewis}, {Lin}, {Schlesinger},
  {Simpson}, {Stello}, {Zucker}, {Zwitter}, {Beeson}, {Buck}, {Casagrande},
  {Clark}, {Cotar}, {da Costa}, {de Grijs}, {Feuillet}, {Horner}, {Kafle},
  {Khanna}, {Kobayashi}, {Liu}, {Montet}, {Nandakumar}, {Nataf}, {Ness},
  {Spina}, {Tepper-Garc{\'\i}a}, {Ting}, {Traven}, {Vogrincic}, {Wittenmyer},
  {Wyse}, {Zerjal}, \& {Galah Collaboration}}]{buder2021}
{Buder}, S., {Sharma}, S., {Kos}, J., {et~al.} 2021, \bibinfo{title}{{The
  GALAH+ survey: Third data release},} \mnras, 506, 150,
  \dodoi{10.1093/mnras/stab1242}

% type= article
\bibitem[{S. {Buder} {et~al.}(2022){Buder}, {Lind}, {Ness}, {Feuillet},
  {Horta}, {Monty}, {Buck}, {Nordlander}, {Bland-Hawthorn}, {Casey}, {de
  Silva}, {D'Orazi}, {Freeman}, {Hayden}, {Kos}, {Martell}, {Lewis}, {Lin},
  {Schlesinger}, {Sharma}, {Simpson}, {Stello}, {Zucker}, {Zwitter},
  {Ciuc{\u{a}}}, {Horner}, {Kobayashi}, {Ting}, {Wyse}, \& {Wyse}}]{buder2022}
{Buder}, S., {Lind}, K., {Ness}, M.~K., {et~al.} 2022, \bibinfo{title}{{The
  GALAH Survey: chemical tagging and chrono-chemodynamics of accreted halo
  stars with GALAH+ DR3 and Gaia eDR3},} \mnras, 510, 2407,
  \dodoi{10.1093/mnras/stab3504}

% type= article
\bibitem[{I. {Bus{\`a}} {et~al.}(2007){Bus{\`a}}, {Aznar Cuadrado},
  {Terranegra}, {Andretta}, \& {Gomez}}]{busa2007}
{Bus{\`a}}, I., {Aznar Cuadrado}, R., {Terranegra}, L., {Andretta}, V., \&
  {Gomez}, M.~T. 2007, \bibinfo{title}{{The Ca II infrared triplet as a stellar
  activity diagnostic. II. Test and calibration with high resolution
  observations},} \aap, 466, 1089, \dodoi{10.1051/0004-6361:20065588}

% type= article
\bibitem[{O. {Cakirli} {et~al.}(2003){Cakirli}, {Ibanoglu}, {Frasca}, \&
  {Catalano}}]{cakirli2003}
{Cakirli}, O., {Ibanoglu}, C., {Frasca}, A., \& {Catalano}, S. 2003,
  \bibinfo{title}{{H{\ensuremath{\alpha}} variations of the RS CVn type binary
  ER Vulpeculae},} \aap, 400, 257, \dodoi{10.1051/0004-6361:20021885}

% type= article
\bibitem[{N. {Calvet} \& E. {Gullbring}(1998){Calvet} \&
  {Gullbring}}]{calvet1998}
{Calvet}, N., \& {Gullbring}, E. 1998, \bibinfo{title}{{The Structure and
  Emission of the Accretion Shock in T Tauri Stars},} \apj, 509, 802,
  \dodoi{10.1086/306527}

% type= article
\bibitem[{A.~C. {Cameron} {et~al.}(1981){Cameron}, {Hearnshaw}, \&
  {Austin}}]{cameron1981}
{Cameron}, A.~C., {Hearnshaw}, J.~B., \& {Austin}, R.~R.~D. 1981,
  \bibinfo{title}{{The southern RS CVn binary, HD 5303.},} \mnras, 197, 769,
  \dodoi{10.1093/mnras/197.3.769}

% type= article
\bibitem[{J.~P.~S. {Campelo} {et~al.}(2025){Campelo}, {Canto Martins},
  {Le{\~a}o}, {Fontinele}, {Gomes}, {Messias}, {Janot-Pacheco}, {Almeida},
  {Brito}, {Ferreira Lopes}, \& {De Medeiros}}]{campelo2025}
{Campelo}, J.~P.~S., {Canto Martins}, B.~L., {Le{\~a}o}, I.~C., {et~al.} 2025,
  \bibinfo{title}{{Rotation Signatures of TESS B-type Stars: Enlarging the
  Sample},} \apj, 989, 177, \dodoi{10.3847/1538-4357/adef2f}

% type= book
\bibitem[{A.~J. {Cannon} \& E.~C. {Pickering}(1924){Cannon} \&
  {Pickering}}]{cannon1924}
{Cannon}, A.~J., \& {Pickering}, E.~C. 1924, {Henry Draper (HD) catalog and HD
  extension} (Harvard College Observatory, Cambridge, MA USA)

% type= article
\bibitem[{T. {Cantat-Gaudin} \& F. {Anders}(2020){Cantat-Gaudin} \&
  {Anders}}]{cantat2020a}
{Cantat-Gaudin}, T., \& {Anders}, F. 2020, \bibinfo{title}{{Clusters and
  mirages: cataloguing stellar aggregates in the Milky Way},} \aap, 633, A99,
  \dodoi{10.1051/0004-6361/201936691}

% type= article
\bibitem[{T. {Cantat-Gaudin} \& T.~D. {Brandt}(2021){Cantat-Gaudin} \&
  {Brandt}}]{cantat2021}
{Cantat-Gaudin}, T., \& {Brandt}, T.~D. 2021, \bibinfo{title}{{Characterizing
  and correcting the proper motion bias of the bright Gaia EDR3 sources},}
  \aap, 649, A124, \dodoi{10.1051/0004-6361/202140807}

% type= article
\bibitem[{T. {Cantat-Gaudin} {et~al.}(2018){Cantat-Gaudin}, {Jordi},
  {Vallenari}, {Bragaglia}, {Balaguer-N{\'u}{\~n}ez}, {Soubiran}, {Bossini},
  {Moitinho}, {Castro-Ginard}, {Krone-Martins}, {Casamiquela}, {Sordo}, \&
  {Carrera}}]{cantat2018}
{Cantat-Gaudin}, T., {Jordi}, C., {Vallenari}, A., {et~al.} 2018,
  \bibinfo{title}{{A Gaia DR2 view of the open cluster population in the Milky
  Way},} \aap, 618, A93, \dodoi{10.1051/0004-6361/201833476}

% type= article
\bibitem[{T. {Cantat-Gaudin} {et~al.}(2020){Cantat-Gaudin}, {Anders},
  {Castro-Ginard}, {Jordi}, {Romero-G{\'o}mez}, {Soubiran}, {Casamiquela},
  {Tarricq}, {Moitinho}, {Vallenari}, {Bragaglia}, {Krone-Martins}, \&
  {Kounkel}}]{cantat2020b}
{Cantat-Gaudin}, T., {Anders}, F., {Castro-Ginard}, A., {et~al.} 2020,
  \bibinfo{title}{{Painting a portrait of the Galactic disc with its stellar
  clusters},} \aap, 640, A1, \dodoi{10.1051/0004-6361/202038192}

% type= article
\bibitem[{B.~L. {Canto Martins} {et~al.}(2020){Canto Martins}, {Gomes},
  {Messias}, {de Lira}, {Le{\~a}o}, {Almeida}, {Teixeira}, {das Chagas},
  {Bravo}, {Bewketu Belete}, \& {De Medeiros}}]{cantomartins2020}
{Canto Martins}, B.~L., {Gomes}, R.~L., {Messias}, Y.~S., {et~al.} 2020,
  \bibinfo{title}{{A Search for Rotation Periods in 1000 TESS Objects of
  Interest},} \apjs, 250, 20, \dodoi{10.3847/1538-4365/aba73f}

% type= article
\bibitem[{Z. {Cao} {et~al.}(2024){Cao}, {Jiang}, {Wang}, \& {Li}}]{cao2024}
{Cao}, Z., {Jiang}, B., {Wang}, S., \& {Li}, J. 2024,
  \bibinfo{title}{{Extinction of Taurus, Orion, Perseus and California
  Molecular Clouds Based on the LAMOST, 2MASS, and Gaia Surveys II: The
  Extinction Law},} arXiv e-prints, arXiv:2410.02731,
  \dodoi{10.48550/arXiv.2410.02731}

% type= article
\bibitem[{Z. {Cao} {et~al.}(2023){Cao}, {Jiang}, {Zhao}, \& {Sun}}]{cao2023}
{Cao}, Z., {Jiang}, B., {Zhao}, H., \& {Sun}, M. 2023,
  \bibinfo{title}{{Extinction of Taurus, Orion, Perseus, and California
  Molecular Clouds Based on the LAMOST, 2MASS, and Gaia Surveys. I. 3D
  Extinction and Structure},} \apj, 945, 132, \dodoi{10.3847/1538-4357/acbbc7}

% type= misc
\bibitem[{P. {Capak}(2019){Capak}}]{capak2019}
{Capak}, P. 2019, {Spitzer Enhanced Imaging Products (SEIP) Source List},, NASA
  IPAC DataSet, IRSA3 \dodoi{10.26131/IRSA3}

% type= article
\bibitem[{J.~A. {Cardelli} {et~al.}(1989){Cardelli}, {Clayton}, \&
  {Mathis}}]{cardelli1989}
{Cardelli}, J.~A., {Clayton}, G.~C., \& {Mathis}, J.~S. 1989,
  \bibinfo{title}{{The Relationship between Infrared, Optical, and Ultraviolet
  Extinction},} \apj, 345, 245, \dodoi{10.1086/167900}

% type= article
\bibitem[{P.~A. {Cargile} {et~al.}(2014){Cargile}, {James}, {Pepper}, {Kuhn},
  {Siverd}, \& {Stassun}}]{cargile2014}
{Cargile}, P.~A., {James}, D.~J., {Pepper}, J., {et~al.} 2014,
  \bibinfo{title}{{Evaluating Gyrochronology on the Zero-age-main-sequence:
  Rotation Periods in the Southern Open Cluster Blanco 1 from the KELT-South
  Survey},} \apj, 782, 29, \dodoi{10.1088/0004-637X/782/1/29}

% type= article
\bibitem[{P.~A. {Cargile} {et~al.}(2009){Cargile}, {James}, \&
  {Platais}}]{cargile2009}
{Cargile}, P.~A., {James}, D.~J., \& {Platais}, I. 2009, \bibinfo{title}{{A New
  X-Ray Analysis of the Open Cluster Blanco 1 Using Wide-Field BVI$_{c}$
  Photometric and Proper Motion Surveys},} \aj, 137, 3230,
  \dodoi{10.1088/0004-6256/137/2/3230}

% type= article
\bibitem[{M. {Carlos} {et~al.}(2019){Carlos}, {Mel{\'e}ndez}, {Spina}, {dos
  Santos}, {Bedell}, {Ramirez}, {Asplund}, {Bean}, {Yong}, {Yana Galarza}, \&
  {Alves-Brito}}]{carlos2019}
{Carlos}, M., {Mel{\'e}ndez}, J., {Spina}, L., {et~al.} 2019,
  \bibinfo{title}{{The Li-age correlation: the Sun is unusually Li deficient
  for its age},} \mnras, 485, 4052, \dodoi{10.1093/mnras/stz681}

% type= article
\bibitem[{M. {Carlsson} {et~al.}(2019){Carlsson}, {De Pontieu}, \&
  {Hansteen}}]{carlsson2019}
{Carlsson}, M., {De Pontieu}, B., \& {Hansteen}, V.~H. 2019,
  \bibinfo{title}{{New View of the Solar Chromosphere},} \araa, 57, 189,
  \dodoi{10.1146/annurev-astro-081817-052044}

% type= article
\bibitem[{J.~M. {Carpenter} {et~al.}(2025){Carpenter}, {Esplin}, {Luhman},
  {Mamajek}, \& {Andrews}}]{carpenter2025}
{Carpenter}, J.~M., {Esplin}, T.~L., {Luhman}, K.~L., {Mamajek}, E.~E., \&
  {Andrews}, S.~M. 2025, \bibinfo{title}{{Extending the ALMA Census of
  Circumstellar Disks in the Upper Scorpius OB Association},} \apj, 978, 117,
  \dodoi{10.3847/1538-4357/ad8ebc}

% type= article
\bibitem[{J.~M. {Carpenter} {et~al.}(2009){Carpenter}, {Mamajek},
  {Hillenbrand}, \& {Meyer}}]{carpenter2009b}
{Carpenter}, J.~M., {Mamajek}, E.~E., {Hillenbrand}, L.~A., \& {Meyer}, M.~R.
  2009, \bibinfo{title}{{XXXXX Upper Sco},} \apj, 705, 1646,
  \dodoi{10.1088/0004-637X/705/2/1646}

% type= article
\bibitem[{J.~M. {Carpenter} {et~al.}(2014){Carpenter}, {Ricci}, \&
  {Isella}}]{carpenter2014}
{Carpenter}, J.~M., {Ricci}, L., \& {Isella}, A. 2014, \bibinfo{title}{{An ALMA
  Continuum Survey of Circumstellar Disks in the Upper Scorpius OB
  Association},} \apj, 787, 42, \dodoi{10.1088/0004-637X/787/1/42}

% type= article
\bibitem[{J.~M. {Carpenter} {et~al.}(2008){Carpenter}, {Bouwman},
  {Silverstone}, {Kim}, {Stauffer}, {Cohen}, {Hines}, {Meyer}, \&
  {Crockett}}]{carpenter2008}
{Carpenter}, J.~M., {Bouwman}, J., {Silverstone}, M.~D., {et~al.} 2008,
  \bibinfo{title}{{The Formation and Evolution of Planetary Systems:
  Description of the Spitzer Legacy Science Database},} \apjs, 179, 423,
  \dodoi{10.1086/592274}

% type= article
\bibitem[{R. {Cayrel} {et~al.}(1983){Cayrel}, {de Strobel}, {Campbell}, {Mein},
  {Mein}, \& {Dumont}}]{cayrel1983}
{Cayrel}, R., {de Strobel}, G.~C., {Campbell}, B., {et~al.} 1983,
  \bibinfo{title}{{Evidence of high chromospheric activity in Hyades dwarfs
  from spectroscopic observations.},} \aap, 123, 89

% type= article
\bibitem[{K.~C. {Chambers} {et~al.}(2016){Chambers}, {Magnier}, {Metcalfe},
  {Flewelling}, {Huber}, {Waters}, {Denneau}, {Draper}, {Farrow}, {Finkbeiner},
  {Holmberg}, {Koppenhoefer}, {Price}, {Rest}, {Saglia}, {Schlafly}, {Smartt},
  {Sweeney}, {Wainscoat}, {Burgett}, {Chastel}, {Grav}, {Heasley}, {Hodapp},
  {Jedicke}, {Kaiser}, {Kudritzki}, {Luppino}, {Lupton}, {Monet}, {Morgan},
  {Onaka}, {Shiao}, {Stubbs}, {Tonry}, {White}, {Ba{\~n}ados}, {Bell},
  {Bender}, {Bernard}, {Boegner}, {Boffi}, {Botticella}, {Calamida},
  {Casertano}, {Chen}, {Chen}, {Cole}, {Deacon}, {Frenk}, {Fitzsimmons},
  {Gezari}, {Gibbs}, {Goessl}, {Goggia}, {Gourgue}, {Goldman}, {Grant},
  {Grebel}, {Hambly}, {Hasinger}, {Heavens}, {Heckman}, {Henderson}, {Henning},
  {Holman}, {Hopp}, {Ip}, {Isani}, {Jackson}, {Keyes}, {Koekemoer}, {Kotak},
  {Le}, {Liska}, {Long}, {Lucey}, {Liu}, {Martin}, {Masci}, {McLean}, {Mindel},
  {Misra}, {Morganson}, {Murphy}, {Obaika}, {Narayan}, {Nieto-Santisteban},
  {Norberg}, {Peacock}, {Pier}, {Postman}, {Primak}, {Rae}, {Rai}, {Riess},
  {Riffeser}, {Rix}, {R{\"o}ser}, {Russel}, {Rutz}, {Schilbach}, {Schultz},
  {Scolnic}, {Strolger}, {Szalay}, {Seitz}, {Small}, {Smith}, {Soderblom},
  {Taylor}, {Thomson}, {Taylor}, {Thakar}, {Thiel}, {Thilker}, {Unger},
  {Urata}, {Valenti}, {Wagner}, {Walder}, {Walter}, {Watters}, {Werner},
  {Wood-Vasey}, \& {Wyse}}]{chambers2016}
{Chambers}, K.~C., {Magnier}, E.~A., {Metcalfe}, N., {et~al.} 2016,
  \bibinfo{title}{{The Pan-STARRS1 Surveys},} arXiv e-prints, arXiv:1612.05560,
  \dodoi{10.48550/arXiv.1612.05560}

% type= article
\bibitem[{N.~L. {Chapman} {et~al.}(2009){Chapman}, {Mundy}, {Lai}, \&
  {Evans}}]{chapman2009}
{Chapman}, N.~L., {Mundy}, L.~G., {Lai}, S.-P., \& {Evans}, Neal~J., I. 2009,
  \bibinfo{title}{{The Mid-Infrared Extinction Law in the Ophiuchus, Perseus,
  and Serpens Molecular Clouds},} \apj, 690, 496,
  \dodoi{10.1088/0004-637X/690/1/496}

% type= article
\bibitem[{C. {Chavero} {et~al.}(2019){Chavero}, {de la Reza}, {Ghezzi},
  {Llorente de Andr{\'e}s}, {Pereira}, {Giuppone}, \&
  {Pinz{\'o}n}}]{chavero2019}
{Chavero}, C., {de la Reza}, R., {Ghezzi}, L., {et~al.} 2019,
  \bibinfo{title}{{Emerging trends in metallicity and lithium properties of
  debris disc stars},} \mnras, 487, 3162, \dodoi{10.1093/mnras/stz1496}

% type= article
\bibitem[{C.~H. {Chen} {et~al.}(2005{\natexlab{a}}){Chen}, {Jura}, {Gordon}, \&
  {Blaylock}}]{chen2005a}
{Chen}, C.~H., {Jura}, M., {Gordon}, K.~D., \& {Blaylock}, M.
  2005{\natexlab{a}}, \bibinfo{title}{{A Spitzer Study of Dusty Disks in the
  Scorpius-Centaurus OB Association},} \apj, 623, 493, \dodoi{10.1086/428607}

% type= article
\bibitem[{C.~H. {Chen} {et~al.}(2011){Chen}, {Mamajek}, {Bitner}, {Pecaut},
  {Su}, \& {Weinberger}}]{chen2011}
{Chen}, C.~H., {Mamajek}, E.~E., {Bitner}, M.~A., {et~al.} 2011,
  \bibinfo{title}{{A Magellan MIKE and Spitzer MIPS Study of 1.5-1.0 M $_{sun}$
  Stars in Scorpius-Centaurus},} \apj, 738, 122,
  \dodoi{10.1088/0004-637X/738/2/122}

% type= article
\bibitem[{C.~H. {Chen} {et~al.}(2014){Chen}, {Mittal}, {Kuchner}, {Forrest},
  {Lisse}, {Manoj}, {Sargent}, \& {Watson}}]{chen2014}
{Chen}, C.~H., {Mittal}, T., {Kuchner}, M., {et~al.} 2014, \bibinfo{title}{{The
  Spitzer Infrared Spectrograph Debris Disk Catalog. I. Continuum Analysis of
  Unresolved Targets},} \apjs, 211, 25, \dodoi{10.1088/0067-0049/211/2/25}

% type= article
\bibitem[{C.~H. {Chen} {et~al.}(2012){Chen}, {Pecaut}, {Mamajek}, {Su}, \&
  {Bitner}}]{chen2012}
{Chen}, C.~H., {Pecaut}, M., {Mamajek}, E.~E., {Su}, K. Y.~L., \& {Bitner}, M.
  2012, \bibinfo{title}{{A Spitzer MIPS Study of 2.5-2.0 M
  $_{{\ensuremath{\odot}}}$ Stars in Scorpius-Centaurus},} \apj, 756, 133,
  \dodoi{10.1088/0004-637X/756/2/133}

% type= article
\bibitem[{C.~H. {Chen} {et~al.}(2009){Chen}, {Sheehan}, {Watson}, {Manoj}, \&
  {Najita}}]{chen2009}
{Chen}, C.~H., {Sheehan}, P., {Watson}, D.~M., {Manoj}, P., \& {Najita}, J.~R.
  2009, \bibinfo{title}{{Solar System Analogs Around IRAS-Discovered Debris
  Disks},} \apj, 701, 1367, \dodoi{10.1088/0004-637X/701/2/1367}

% type= article
\bibitem[{C.~H. {Chen} {et~al.}(2020){Chen}, {Su}, \& {Xu}}]{chen2020}
{Chen}, C.~H., {Su}, K. Y.~L., \& {Xu}, S. 2020, \bibinfo{title}{{Spitzer's
  debris disk legacy from main-sequence stars to white dwarfs},} Nature
  Astronomy, 4, 328, \dodoi{10.1038/s41550-020-1067-6}

% type= article
\bibitem[{C.~H. {Chen} {et~al.}(2005{\natexlab{b}}){Chen}, {Patten}, {Werner},
  {Dowell}, {Stapelfeldt}, {Song}, {Stauffer}, {Blaylock}, {Gordon}, \&
  {Krause}}]{chen2005b}
{Chen}, C.~H., {Patten}, B.~M., {Werner}, M.~W., {et~al.} 2005{\natexlab{b}},
  \bibinfo{title}{{A Spitzer Study of Dusty Disks around Nearby, Young Stars},}
  \apj, 634, 1372, \dodoi{10.1086/497124}

% type= article
\bibitem[{C.~H. {Chen} {et~al.}(2006){Chen}, {Sargent}, {Bohac}, {Kim},
  {Leibensperger}, {Jura}, {Najita}, {Forrest}, {Watson}, {Sloan}, \&
  {Keller}}]{chen2006}
{Chen}, C.~H., {Sargent}, B.~A., {Bohac}, C., {et~al.} 2006,
  \bibinfo{title}{{Spitzer IRS Spectroscopy of IRAS-discovered Debris Disks},}
  \apjs, 166, 351, \dodoi{10.1086/505751}

% type= article
\bibitem[{Y. {Chen} {et~al.}(2015){Chen}, {Bressan}, {Girardi}, {Marigo},
  {Kong}, \& {Lanza}}]{ychen2015}
{Chen}, Y., {Bressan}, A., {Girardi}, L., {et~al.} 2015,
  \bibinfo{title}{{PARSEC evolutionary tracks of massive stars up to 350
  M$_{{\ensuremath{\odot}}}$ at metallicities 0.0001 {\ensuremath{\leq}} Z
  {\ensuremath{\leq}} 0.04},} \mnras, 452, 1068, \dodoi{10.1093/mnras/stv1281}

% type= article
\bibitem[{Y. {Chen} {et~al.}(2014){Chen}, {Girardi}, {Bressan}, {Marigo},
  {Barbieri}, \& {Kong}}]{ychen2014}
{Chen}, Y., {Girardi}, L., {Bressan}, A., {et~al.} 2014,
  \bibinfo{title}{{Improving PARSEC models for very low mass stars},} \mnras,
  444, 2525, \dodoi{10.1093/mnras/stu1605}

% type= article
\bibitem[{Y. {Chmielewski}(2000){Chmielewski}}]{chmielewski2000}
{Chmielewski}, Y. 2000, \bibinfo{title}{{The infrared triplet lines of ionized
  calcium as a diagnostic tool for F, G, K-type stellar atmospheres},} \aap,
  353, 666

% type= article
\bibitem[{L.~A. {Cieza} {et~al.}(2008){Cieza}, {Cochran}, \&
  {Augereau}}]{cieza2008}
{Cieza}, L.~A., {Cochran}, W.~D., \& {Augereau}, J.-C. 2008,
  \bibinfo{title}{{Spitzer Observations of the Hyades: Circumstellar Debris
  Disks at 625 Myr of Age},} \apj, 679, 720, \dodoi{10.1086/586887}

% type= article
\bibitem[{L.~A. {Cieza} {et~al.}(2005){Cieza}, {Kessler-Silacci}, {Jaffe},
  {Harvey}, \& {Evans}}]{cieza2005}
{Cieza}, L.~A., {Kessler-Silacci}, J.~E., {Jaffe}, D.~T., {Harvey}, P.~M., \&
  {Evans}, Neal~J., I. 2005, \bibinfo{title}{{Evidence for J- and H-Band Excess
  in Classical T Tauri Stars and the Implications for Disk Structure and
  Estimated Ages},} \apj, 635, 422, \dodoi{10.1086/497325}

% type= article
\bibitem[{L.~A. {Cieza} {et~al.}(2013){Cieza}, {Olofsson}, {Harvey}, {Evans},
  {Najita}, {Henning}, {Mer{\'\i}n}, {Liebhart}, {G{\"u}del}, {Augereau}, \&
  {Pinte}}]{cieza2013}
{Cieza}, L.~A., {Olofsson}, J., {Harvey}, P.~M., {et~al.} 2013,
  \bibinfo{title}{{The Herschel DIGIT Survey of Weak-line T Tauri Stars:
  Implications for Disk Evolution and Dissipation},} \apj, 762, 100,
  \dodoi{10.1088/0004-637X/762/2/100}

% type= article
\bibitem[{C. {Cifuentes} {et~al.}(2025){Cifuentes}, {Caballero},
  {Gonz{\'a}lez-Payo}, {Amado}, {B{\'e}jar}, {Burgasser},
  {Cort{\'e}s-Contreras}, {Lodieu}, {Montes}, {Quirrenbach}, {Reiners},
  {Ribas}, {Sanz-Forcada}, {Seifert}, \& {Zapatero Osorio}}]{cifuentes2025}
{Cifuentes}, C., {Caballero}, J.~A., {Gonz{\'a}lez-Payo}, J., {et~al.} 2025,
  \bibinfo{title}{{CARMENES input catalogue of M dwarfs: IX. Multiplicity from
  close spectroscopic binaries to ultra-wide systems},} \aap, 693, A228,
  \dodoi{10.1051/0004-6361/202452527}

% type= article
\bibitem[{J.~J. {Claria}(1972){Claria}}]{claria1972}
{Claria}, J.~J. 1972, \bibinfo{title}{{Photometric study of the open cluster
  NGC 2232.},} \aap, 19, 303

% type= article
\bibitem[{I.~L. {Colman} {et~al.}(2024){Colman}, {Angus}, {David}, {Curtis},
  {Hattori}, \& {Lu}}]{colman2024}
{Colman}, I.~L., {Angus}, R., {David}, T., {et~al.} 2024,
  \bibinfo{title}{{Methods for the Detection of Stellar Rotation Periods in
  Individual TESS Sectors and Results from the Prime Mission},} \aj, 167, 189,
  \dodoi{10.3847/1538-3881/ad2c86}

% type= article
\bibitem[{A.~C.~M. {Correia} {et~al.}(2005){Correia}, {Udry}, {Mayor},
  {Laskar}, {Naef}, {Pepe}, {Queloz}, \& {Santos}}]{correia2005}
{Correia}, A.~C.~M., {Udry}, S., {Mayor}, M., {et~al.} 2005,
  \bibinfo{title}{{The CORALIE survey for southern extra-solar planets. XIII. A
  pair of planets around HD{\,}202206 or a circumbinary planet?},} \aap, 440,
  751, \dodoi{10.1051/0004-6361:20042376}

% type= article
\bibitem[{M. {Cort{\'e}s-Contreras} {et~al.}(2017){Cort{\'e}s-Contreras},
  {B{\'e}jar}, {Caballero}, {Gauza}, {Montes}, {Alonso-Floriano}, {Jeffers},
  {Morales}, {Reiners}, {Ribas}, {Sch{\"o}fer}, {Quirrenbach}, {Amado},
  {Mundt}, \& {Seifert}}]{cortes2017}
{Cort{\'e}s-Contreras}, M., {B{\'e}jar}, V.~J.~S., {Caballero}, J.~A., {et~al.}
  2017, \bibinfo{title}{{CARMENES input catalogue of M dwarfs. II.
  High-resolution imaging with FastCam},} \aap, 597, A47,
  \dodoi{10.1051/0004-6361/201629056}

% type= article
\bibitem[{E. {Costa} {et~al.}(2006){Costa}, {M{\'e}ndez}, {Jao}, {Henry},
  {Subasavage}, \& {Ianna}}]{costa2006}
{Costa}, E., {M{\'e}ndez}, R.~A., {Jao}, W.~C., {et~al.} 2006,
  \bibinfo{title}{{The Solar Neighborhood. XVI. Parallaxes from CTIOPI: Final
  Results from the 1.5 m Telescope Program},} \aj, 132, 1234,
  \dodoi{10.1086/505706}

% type= article
\bibitem[{O.~L. {Creevey} {et~al.}(2023){Creevey}, {Sordo}, {Pailler},
  {Fr{\'e}mat}, {Heiter}, {Th{\'e}venin}, {Andrae}, {Fouesneau}, {Lobel},
  {Bailer-Jones}, {Garabato}, {Bellas-Velidis}, {Brugaletta}, {Lorca},
  {Ordenovic}, {Palicio}, {Sarro}, {Delchambre}, {Drimmel}, {Rybizki},
  {Torralba Elipe}, {Korn}, {Recio-Blanco}, {Schultheis}, {De Angeli},
  {Montegriffo}, {Abreu Aramburu}, {Accart}, {{\'A}lvarez}, {Bakker},
  {Brouillet}, {Burlacu}, {Carballo}, {Casamiquela}, {Chiavassa}, {Contursi},
  {Cooper}, {Dafonte}, {Dapergolas}, {de Laverny}, {Dharmawardena},
  {Edvardsson}, {Le Fustec}, {Garc{\'\i}a-Lario}, {Garc{\'\i}a-Torres},
  {Gomez}, {Gonz{\'a}lez-Santamar{\'\i}a}, {Hatzidimitriou}, {Jean-Antoine
  Piccolo}, {Kontiza}, {Kordopatis}, {Lanzafame}, {Lebreton}, {Licata},
  {Lindstr{\o}m}, {Livanou}, {Magdaleno Romeo}, {Manteiga}, {Marocco},
  {Marshall}, {Mary}, {Nicolas}, {Pallas-Quintela}, {Panem}, {Pichon},
  {Poggio}, {Riclet}, {Robin}, {Santove{\~n}a}, {Silvelo}, {Slezak}, {Smart},
  {Soubiran}, {S{\"u}veges}, {Ulla}, {Utrilla}, {Vallenari}, {Zhao}, {Zorec},
  {Barrado}, {Bijaoui}, {Bouret}, {Blomme}, {Brott}, {Cassisi}, {Kochukhov},
  {Martayan}, {Shulyak}, \& {Silvester}}]{creevey2023}
{Creevey}, O.~L., {Sordo}, R., {Pailler}, F., {et~al.} 2023,
  \bibinfo{title}{{Gaia Data Release 3. Astrophysical parameters inference
  system (Apsis). I. Methods and content overview},} \aap, 674, A26,
  \dodoi{10.1051/0004-6361/202243688}

% type= article
\bibitem[{J.~D. {Cummings} {et~al.}(2017){Cummings}, {Deliyannis}, {Maderak},
  \& {Steinhauer}}]{cummings2017}
{Cummings}, J.~D., {Deliyannis}, C.~P., {Maderak}, R.~M., \& {Steinhauer}, A.
  2017, \bibinfo{title}{{WIYN Open Cluster Study. LXXV. Testing the Metallicity
  Dependence of Stellar Lithium Depletion Using Hyades-aged Clusters. I. Hyades
  and Praesepe},} \aj, 153, 128, \dodoi{10.3847/1538-3881/aa5b86}

% type= article
\bibitem[{T. {Currie} {et~al.}(2008){Currie}, {Plavchan}, \&
  {Kenyon}}]{currie2008}
{Currie}, T., {Plavchan}, P., \& {Kenyon}, S.~J. 2008, \bibinfo{title}{{A
  Spitzer Study of Debris Disks in the Young Nearby Cluster NGC 2232: Icy
  Planets Are Common around \raisebox{-0.5ex}\textasciitilde1.5-3
  M$_{{\ensuremath{\odot}}}$ Stars},} \apj, 688, 597, \dodoi{10.1086/591842}

% type= article
\bibitem[{T. {Currie} {et~al.}(2010){Currie}, {Hernandez}, {Irwin}, {Kenyon},
  {Tokarz}, {Balog}, {Bragg}, {Berlind}, \& {Calkins}}]{currie2010}
{Currie}, T., {Hernandez}, J., {Irwin}, J., {et~al.} 2010, \bibinfo{title}{{The
  Stellar Population of h and {\ensuremath{\chi}} Persei: Cluster Properties,
  Membership, and the Intrinsic Colors and Temperatures of Stars},} \apjs, 186,
  191, \dodoi{10.1088/0067-0049/186/2/191}

% type= article
\bibitem[{T. {Currie} {et~al.}(2023){Currie}, {Brandt}, {Brandt}, {Lacy},
  {Burrows}, {Guyon}, {Tamura}, {Liu}, {Sagynbayeva}, {Tobin}, {Chilcote},
  {Groff}, {Marois}, {Thompson}, {Murphy}, {Kuzuhara}, {Lawson}, {Lozi}, {Deo},
  {Vievard}, {Skaf}, {Uyama}, {Jovanovic}, {Martinache}, {Kasdin}, {Kudo},
  {McElwain}, {Janson}, {Wisniewski}, {Hodapp}, {Nishikawa}, {He{\l}miniak},
  {Kwon}, \& {Hayashi}}]{currie2023}
{Currie}, T., {Brandt}, G.~M., {Brandt}, T.~D., {et~al.} 2023,
  \bibinfo{title}{{Direct imaging and astrometric detection of a gas giant
  planet orbiting an accelerating star},} Science, 380, 198,
  \dodoi{10.1126/science.abo6192}

% type= book
\bibitem[{R.~M. {Cutri} {et~al.}(2003){Cutri}, {Skrutskie}, {van Dyk},
  {Beichman}, {Carpenter}, {Chester}, {Cambresy}, {Evans}, {Fowler}, {Gizis},
  {Howard}, {Huchra}, {Jarrett}, {Kopan}, {Kirkpatrick}, {Light}, {Marsh},
  {McCallon}, {Schneider}, {Stiening}, {Sykes}, {Weinberg}, {Wheaton},
  {Wheelock}, \& {Zacarias}}]{cutri2003}
{Cutri}, R.~M., {Skrutskie}, M.~F., {van Dyk}, S., {et~al.} 2003, {2MASS All
  Sky Catalog of point sources.} (NASA/IPAC INfrared Science Archive)

% type= misc
\bibitem[{R.~M. {Cutri} {et~al.}(2013){Cutri}, {Wright}, {Conrow}, {Fowler},
  {Eisenhardt}, {Grillmair}, {Kirkpatrick}, {Masci}, {McCallon}, {Wheelock},
  {Fajardo-Acosta}, {Yan}, {Benford}, {Harbut}, {Jarrett}, {Lake}, {Leisawitz},
  {Ressler}, {Stanford}, {Tsai}, {Liu}, {Helou}, {Mainzer}, {Gettings},
  {Gonzalez}, {Hoffman}, {Marsh}, {Padgett}, {Skrutskie}, {Beck}, {Papin}, \&
  {Wittman}}]{cutri2013}
{Cutri}, R.~M., {Wright}, E.~L., {Conrow}, T., {et~al.} 2013, {Explanatory
  Supplement to the AllWISE Data Release Products},, Explanatory Supplement to
  the AllWISE Data Release Products, by R. M. Cutri et al. IPAC, Caltech

% type= article
\bibitem[{I. {Czekala} {et~al.}(2019){Czekala}, {Chiang}, {Andrews}, {Jensen},
  {Torres}, {Wilner}, {Stassun}, \& {Macintosh}}]{czekala2019}
{Czekala}, I., {Chiang}, E., {Andrews}, S.~M., {et~al.} 2019,
  \bibinfo{title}{{The Degree of Alignment between Circumbinary Disks and Their
  Binary Hosts},} \apj, 883, 22, \dodoi{10.3847/1538-4357/ab287b}

% type= article
\bibitem[{L. {da Silva} {et~al.}(2009){da Silva}, {Torres}, {de La Reza},
  {Quast}, {Melo}, \& {Sterzik}}]{dasilva2009}
{da Silva}, L., {Torres}, C.~A.~O., {de La Reza}, R., {et~al.} 2009,
  \bibinfo{title}{{Search for associations containing young stars (SACY). III.
  Ages and Li abundances},} \aap, 508, 833, \dodoi{10.1051/0004-6361/200911736}

% type= article
\bibitem[{S.~E. {Dahm}(2015){Dahm}}]{dahm2015}
{Dahm}, S.~E. 2015, \bibinfo{title}{{Reexamining the Lithium Depletion Boundary
  in the Pleiades and the Inferred Age of the Cluster},} \apj, 813, 108,
  \dodoi{10.1088/0004-637X/813/2/108}

% type= article
\bibitem[{F. {Damiani} {et~al.}(2003){Damiani}, {Flaccomio}, {Micela},
  {Sciortino}, {Harnden}, {Murray}, {Wolk}, \& {Jeffries}}]{damiani2003}
{Damiani}, F., {Flaccomio}, E., {Micela}, G., {et~al.} 2003,
  \bibinfo{title}{{Chandra X-Ray Observations of the Young Open Cluster NGC
  2516},} \apj, 588, 1009, \dodoi{10.1086/374214}

% type= article
\bibitem[{F. {Damiani} {et~al.}(2019){Damiani}, {Prisinzano}, {Pillitteri},
  {Micela}, \& {Sciortino}}]{damiani2019}
{Damiani}, F., {Prisinzano}, L., {Pillitteri}, I., {Micela}, G., \&
  {Sciortino}, S. 2019, \bibinfo{title}{{Stellar population of Sco OB2 revealed
  by Gaia DR2 data},} \aap, 623, A112, \dodoi{10.1051/0004-6361/201833994}

% type= article
\bibitem[{K.~J. {Daniel} {et~al.}(2002){Daniel}, {Linsky}, \&
  {Gagn{\'e}}}]{daniel2002}
{Daniel}, K.~J., {Linsky}, J.~L., \& {Gagn{\'e}}, M. 2002,
  \bibinfo{title}{{Chandra Observations of the Pleiades Open Cluster: X-Ray
  Emission from Late B- to Early F-Type Binaries},} \apj, 578, 486,
  \dodoi{10.1086/340553}

% type= article
\bibitem[{T.~J. {David} \& L.~A. {Hillenbrand}(2015){David} \&
  {Hillenbrand}}]{david2015}
{David}, T.~J., \& {Hillenbrand}, L.~A. 2015, \bibinfo{title}{{The Ages of
  Early-type Stars: Str{\"o}mgren Photometric Methods Calibrated, Validated,
  Tested, and Applied to Hosts and Prospective Hosts of Directly Imaged
  Exoplanets},} \apj, 804, 146, \dodoi{10.1088/0004-637X/804/2/146}

% type= article
\bibitem[{F. {De Angeli} {et~al.}(2023){De Angeli}, {Weiler}, {Montegriffo},
  {Evans}, {Riello}, {Andrae}, {Carrasco}, {Busso}, {Burgess}, {Cacciari},
  {Davidson}, {Harrison}, {Hodgkin}, {Jordi}, {Osborne}, {Pancino},
  {Altavilla}, {Barstow}, {Bailer-Jones}, {Bellazzini}, {Brown}, {Castellani},
  {Cowell}, {Delchambre}, {De Luise}, {Diener}, {Fabricius}, {Fouesneau},
  {Fr{\'e}mat}, {Gilmore}, {Giuffrida}, {Hambly}, {Hidalgo}, {Holland},
  {Kostrzewa-Rutkowska}, {van Leeuwen}, {Lobel}, {Marinoni}, {Miller},
  {Pagani}, {Palaversa}, {Piersimoni}, {Pulone}, {Ragaini}, {Rainer},
  {Richards}, {Rixon}, {Ruz-Mieres}, {Sanna}, {Sarro}, {Rowell}, {Sordo},
  {Walton}, \& {Yoldas}}]{deangeli2023}
{De Angeli}, F., {Weiler}, M., {Montegriffo}, P., {et~al.} 2023,
  \bibinfo{title}{{Gaia Data Release 3. Processing and validation of BP/RP
  low-resolution spectral data},} \aap, 674, A2,
  \dodoi{10.1051/0004-6361/202243680}

% type= article
\bibitem[{E.~J. {de Geus} {et~al.}(1989){de Geus}, {de Zeeuw}, \&
  {Lub}}]{degeus1989}
{de Geus}, E.~J., {de Zeeuw}, P.~T., \& {Lub}, J. 1989,
  \bibinfo{title}{{Physical parameters of stars in the Scorpio-Centaurus OB
  association.},} \aap, 216, 44

% type= article
\bibitem[{R. {de Grijs} \& D. {Kamath}(2021){de Grijs} \&
  {Kamath}}]{degrijs2021}
{de Grijs}, R., \& {Kamath}, D. 2021, \bibinfo{title}{{Stellar Chromospheric
  Variability},} Universe, 7, 440, \dodoi{10.3390/universe7110440}

% type= article
\bibitem[{I. {De Moortel} \& P. {Browning}(2015){De Moortel} \&
  {Browning}}]{demoortel2015}
{De Moortel}, I., \& {Browning}, P. 2015, \bibinfo{title}{{Recent advances in
  coronal heating},} Philosophical Transactions of the Royal Society of London
  Series A, 373, 20140269, \dodoi{10.1098/rsta.2014.0269}

% type= article
\bibitem[{P.~T. {de Zeeuw} {et~al.}(1999){de Zeeuw}, {Hoogerwerf}, {de
  Bruijne}, {Brown}, \& {Blaauw}}]{dezeeuw1999}
{de Zeeuw}, P.~T., {Hoogerwerf}, R., {de Bruijne}, J.~H.~J., {Brown}, A.~G.~A.,
  \& {Blaauw}, A. 1999, \bibinfo{title}{{A HIPPARCOS Census of the Nearby OB
  Associations},} \aj, 117, 354, \dodoi{10.1086/300682}

% type= article
\bibitem[{M. {Decleir} {et~al.}(2022){Decleir}, {Gordon}, {Andrews}, {Clayton},
  {Cushing}, {Misselt}, {Pendleton}, {Rayner}, {Vacca}, \&
  {Whittet}}]{decleir2022}
{Decleir}, M., {Gordon}, K.~D., {Andrews}, J.~E., {et~al.} 2022,
  \bibinfo{title}{{SpeX Near-infrared Spectroscopic Extinction Curves in the
  Milky Way},} \apj, 930, 15, \dodoi{10.3847/1538-4357/ac5dbe}

% type= article
\bibitem[{E. {Delgado Mena} {et~al.}(2014){Delgado Mena}, {Israelian},
  {Gonz{\'a}lez Hern{\'a}ndez}, {Sousa}, {Mortier}, {Santos}, {Adibekyan},
  {Fernandes}, {Rebolo}, {Udry}, \& {Mayor}}]{delgado2014}
{Delgado Mena}, E., {Israelian}, G., {Gonz{\'a}lez Hern{\'a}ndez}, J.~I.,
  {et~al.} 2014, \bibinfo{title}{{Li depletion in solar analogues with
  exoplanets. Extending the sample},} \aap, 562, A92,
  \dodoi{10.1051/0004-6361/201321493}

% type= article
\bibitem[{E. {Delgado Mena} {et~al.}(2015){Delgado Mena}, {Bertr{\'a}n de Lis},
  {Adibekyan}, {Sousa}, {Figueira}, {Mortier}, {Gonz{\'a}lez Hern{\'a}ndez},
  {Tsantaki}, {Israelian}, \& {Santos}}]{delgado2015}
{Delgado Mena}, E., {Bertr{\'a}n de Lis}, S., {Adibekyan}, V.~Z., {et~al.}
  2015, \bibinfo{title}{{Li abundances in F stars: planets, rotation, and
  Galactic evolution},} \aap, 576, A69, \dodoi{10.1051/0004-6361/201425433}

% type= inproceedings
\bibitem[{C.~P. {Deliyannis} {et~al.}(2000){Deliyannis}, {Pinsonneault}, \&
  {Charbonnel}}]{deliyannis2000}
{Deliyannis}, C.~P., {Pinsonneault}, M.~H., \& {Charbonnel}, C. 2000,
  \bibinfo{title}{{Sinks of Light Elements in Stars - Part I (Invited Paper)},}
  in IAU Symposium, Vol. 198, The Light Elements and their Evolution, ed.
  L.~{da Silva}, R.~{de Medeiros}, \& M.~{Spite}, 61

% type= article
\bibitem[{P. {Delorme} {et~al.}(2011){Delorme}, {Collier Cameron}, {Hebb},
  {Rostron}, {Lister}, {Norton}, {Pollacco}, \& {West}}]{delorme2011}
{Delorme}, P., {Collier Cameron}, A., {Hebb}, L., {et~al.} 2011,
  \bibinfo{title}{{Stellar rotation in the Hyades and Praesepe: gyrochronology
  and braking time-scale},} \mnras, 413, 2218,
  \dodoi{10.1111/j.1365-2966.2011.18299.x}

% type= article
\bibitem[{P. {Delorme} {et~al.}(2013){Delorme}, {Gagn{\'e}}, {Girard},
  {Lagrange}, {Chauvin}, {Naud}, {Lafreni{\`e}re}, {Doyon}, {Riedel},
  {Bonnefoy}, \& {Malo}}]{delorme2013}
{Delorme}, P., {Gagn{\'e}}, J., {Girard}, J.~H., {et~al.} 2013,
  \bibinfo{title}{{Direct-imaging discovery of a 12-14 Jupiter-mass object
  orbiting a young binary system of very low-mass stars},} \aap, 553, L5,
  \dodoi{10.1051/0004-6361/201321169}

% type= misc
\bibitem[{C. {Denis}(2005){Denis}}]{denis2005}
{Denis}, C. 2005, {VizieR Online Data Catalog: The DENIS database (DENIS
  Consortium, 2005)},, VizieR On-line Data Catalog: B/denis. Originally
  published in: The DENIS consortium (2005)

% type= article
\bibitem[{W.~S. {Dias} {et~al.}(2021){Dias}, {Monteiro}, {Moitinho},
  {L{\'e}pine}, {Carraro}, {Paunzen}, {Alessi}, \& {Villela}}]{dias2021}
{Dias}, W.~S., {Monteiro}, H., {Moitinho}, A., {et~al.} 2021,
  \bibinfo{title}{{Updated parameters of 1743 open clusters based on Gaia
  DR2},} \mnras, 504, 356, \dodoi{10.1093/mnras/stab770}

% type= article
\bibitem[{D.~A. {Dickson-Vandervelde} {et~al.}(2021){Dickson-Vandervelde},
  {Wilson}, \& {Kastner}}]{dickson2021}
{Dickson-Vandervelde}, D.~A., {Wilson}, E.~C., \& {Kastner}, J.~H. 2021,
  \bibinfo{title}{{Gaia-based Isochronal, Kinematic, and Spatial Analysis of
  the ɛ Cha Association},} \aj, 161, 87, \dodoi{10.3847/1538-3881/abd0fd}

% type= article
\bibitem[{E. {D{\'\i}ez Alonso} {et~al.}(2019){D{\'\i}ez Alonso}, {Caballero},
  {Montes}, {de Cos Juez}, {Dreizler}, {Dubois}, {Jeffers}, {Lalitha}, {Naves},
  {Reiners}, {Ribas}, {Vanaverbeke}, {Amado}, {B{\'e}jar},
  {Cort{\'e}s-Contreras}, {Herrero}, {Hidalgo}, {K{\"u}rster}, {Logie},
  {Quirrenbach}, {Rau}, {Seifert}, {Sch{\"o}fer}, \& {Tal-Or}}]{diezalonso2019}
{D{\'\i}ez Alonso}, E., {Caballero}, J.~A., {Montes}, D., {et~al.} 2019,
  \bibinfo{title}{{CARMENES input catalogue of M dwarfs. IV. New rotation
  periods from photometric time series},} \aap, 621, A126,
  \dodoi{10.1051/0004-6361/201833316}

% type= article
\bibitem[{E. {Dineva} {et~al.}(2022){Dineva}, {Pearson}, {Ilyin}, {Verma},
  {Diercke}, {Strassmeier}, \& {Denker}}]{dineva2022}
{Dineva}, E., {Pearson}, J., {Ilyin}, I., {et~al.} 2022,
  \bibinfo{title}{{Characterization of chromospheric activity based on
  Sun‑as‑a‑star spectral and disk‑resolved activity indices},}
  Astronomische Nachrichten, 343, e23996, \dodoi{10.1002/asna.20223996}

% type= article
\bibitem[{E. {Distefano} {et~al.}(2023){Distefano}, {Lanzafame}, {Brugaletta},
  {Holl}, {Lanza}, {Messina}, {Pagano}, {Audard}, {Jevardat de Fombelle},
  {Lecoeur-Taibi}, {Mowlavi}, {Nienartowicz}, {Rimoldini}, {Evans}, {Riello},
  {Garc{\'\i}a-Lario}, {Gavras}, \& {Eyer}}]{distefano2023}
{Distefano}, E., {Lanzafame}, A.~C., {Brugaletta}, E., {et~al.} 2023,
  \bibinfo{title}{{Gaia Data Release 3. Rotational modulation and patterns of
  colour variation in solar-like variables},} \aap, 674, A20,
  \dodoi{10.1051/0004-6361/202244178}

% type= article
\bibitem[{J.~A. {Dittmann} {et~al.}(2014){Dittmann}, {Irwin}, {Charbonneau}, \&
  {Berta-Thompson}}]{dittman2014}
{Dittmann}, J.~A., {Irwin}, J.~M., {Charbonneau}, D., \& {Berta-Thompson},
  Z.~K. 2014, \bibinfo{title}{{Trigonometric Parallaxes for 1507 Nearby
  Mid-to-late M Dwarfs},} \apj, 784, 156, \dodoi{10.1088/0004-637X/784/2/156}

% type= article
\bibitem[{J.~D. {do Nascimento} {et~al.}(2023){do Nascimento}, {Barnes},
  {Saar}, {de Mello}, {Hall}, {Anthony}, {de Almeida}, {Velloso}, {da Costa},
  {Petit}, {Strugarek}, {Wargelin}, {Castro}, {Strassmeier}, \&
  {Brun}}]{nascimento2023}
{do Nascimento}, J.~D., {Barnes}, S.~A., {Saar}, S.~H., {et~al.} 2023,
  \bibinfo{title}{{A Hale-like Cycle in the Solar Twin 18 Scorpii},} \apj, 958,
  57, \dodoi{10.3847/1538-4357/acfc1a}

% type= article
\bibitem[{P.~D. {Dobbie} {et~al.}(2010){Dobbie}, {Lodieu}, \&
  {Sharp}}]{dobbie2010}
{Dobbie}, P.~D., {Lodieu}, N., \& {Sharp}, R.~G. 2010, \bibinfo{title}{{IC
  2602: a lithium depletion boundary age and new candidate low-mass stellar
  members},} \mnras, 409, 1002, \dodoi{10.1111/j.1365-2966.2010.17355.x}

% type= article
\bibitem[{S.~E. {Dodson-Robinson} {et~al.}(2016){Dodson-Robinson}, {Su},
  {Bryden}, {Harvey}, \& {Green}}]{dodrob2016}
{Dodson-Robinson}, S.~E., {Su}, K. Y.~L., {Bryden}, G., {Harvey}, P., \&
  {Green}, J.~D. 2016, \bibinfo{title}{{Herschel Observations and Updated
  Spectral Energy Distributions of Five Sunlike Stars with Debris Disks},}
  \apj, 833, 183, \dodoi{10.3847/1538-4357/833/2/183}

% type= article
\bibitem[{J.~K. {Donaldson} {et~al.}(2012){Donaldson}, {Roberge}, {Chen},
  {Augereau}, {Dent}, {Eiroa}, {Krivov}, {Mathews}, {Meeus}, {M{\'e}nard},
  {Riviere-Marichalar}, \& {Sandell}}]{donaldson2012}
{Donaldson}, J.~K., {Roberge}, A., {Chen}, C.~H., {et~al.} 2012,
  \bibinfo{title}{{Herschel PACS Observations and Modeling of Debris Disks in
  the Tucana-Horologium Association},} \apj, 753, 147,
  \dodoi{10.1088/0004-637X/753/2/147}

% type= article
\bibitem[{S.~T. {Douglas} {et~al.}(2017){Douglas}, {Ag{\"u}eros}, {Covey}, \&
  {Kraus}}]{douglas2017}
{Douglas}, S.~T., {Ag{\"u}eros}, M.~A., {Covey}, K.~R., \& {Kraus}, A. 2017,
  \bibinfo{title}{{Poking the Beehive from Space: K2 Rotation Periods for
  Praesepe},} \apj, 842, 83, \dodoi{10.3847/1538-4357/aa6e52}

% type= article
\bibitem[{S.~T. {Douglas} {et~al.}(2024){Douglas}, {Cargile}, {Matt},
  {Breimann}, {P{\'e}rez Ch{\'a}vez}, {Huang}, {Wright}, \&
  {Zhou}}]{douglas2024}
{Douglas}, S.~T., {Cargile}, P.~A., {Matt}, S.~P., {et~al.} 2024,
  \bibinfo{title}{{Constraining Stellar Rotation at the Zero-age Main Sequence
  with TESS},} \apj, 962, 16, \dodoi{10.3847/1538-4357/ad0fe3}

% type= article
\bibitem[{S.~T. {Douglas} {et~al.}(2019){Douglas}, {Curtis}, {Ag{\"u}eros},
  {Cargile}, {Brewer}, {Meibom}, \& {Jansen}}]{douglas2019}
{Douglas}, S.~T., {Curtis}, J.~L., {Ag{\"u}eros}, M.~A., {et~al.} 2019,
  \bibinfo{title}{{K2 Rotation Periods for Low-mass Hyads and a Quantitative
  Comparison of the Distribution of Slow Rotators in the Hyades and Praesepe},}
  \apj, 879, 100, \dodoi{10.3847/1538-4357/ab2468}

% type= article
\bibitem[{Z.~H. {Draper} {et~al.}(2018){Draper}, {Matthews}, {Venn}, {Lambert},
  {Kennedy}, \& {Sitnova}}]{draper2018}
{Draper}, Z.~H., {Matthews}, B., {Venn}, K., {et~al.} 2018,
  \bibinfo{title}{{A-type Stellar Abundances: A Corollary to Herschel
  Observations of Debris Disks},} \apj, 857, 93,
  \dodoi{10.3847/1538-4357/aab1fd}

% type= article
\bibitem[{J.~R. {Ducati} {et~al.}(2001){Ducati}, {Bevilacqua}, {Rembold}, \&
  {Ribeiro}}]{ducati2001}
{Ducati}, J.~R., {Bevilacqua}, C.~M., {Rembold}, S.~B., \& {Ribeiro}, D. 2001,
  \bibinfo{title}{{Intrinsic Colors of Stars in the Near-Infrared},} \apj, 558,
  309, \dodoi{10.1086/322439}

% type= article
\bibitem[{T. {Dumont}(2023){Dumont}}]{dumont2023}
{Dumont}, T. 2023, \bibinfo{title}{{Angular momentum and lithium transport from
  main sequence to sub-giant and red giant low-mass stars},} \aap, 677, A119,
  \dodoi{10.1051/0004-6361/202346915}

% type= article
\bibitem[{T. {Dumont} {et~al.}(2021{\natexlab{a}}){Dumont}, {Charbonnel},
  {Palacios}, \& {Borisov}}]{dumont2021a}
{Dumont}, T., {Charbonnel}, C., {Palacios}, A., \& {Borisov}, S.
  2021{\natexlab{a}}, \bibinfo{title}{{Lithium depletion and angular momentum
  transport in F-type and G-type stars in Galactic open clusters},} \aap, 654,
  A46, \dodoi{10.1051/0004-6361/202141094}

% type= article
\bibitem[{T. {Dumont} {et~al.}(2021{\natexlab{b}}){Dumont}, {Palacios},
  {Charbonnel}, {Richard}, {Amard}, {Augustson}, \& {Mathis}}]{dumont2021b}
{Dumont}, T., {Palacios}, A., {Charbonnel}, C., {et~al.} 2021{\natexlab{b}},
  \bibinfo{title}{{Lithium depletion and angular momentum transport in
  solar-type stars},} \aap, 646, A48, \dodoi{10.1051/0004-6361/202039515}

% type= article
\bibitem[{D.~K. {Duncan} {et~al.}(1991){Duncan}, {Vaughan}, {Wilson},
  {Preston}, {Frazer}, {Lanning}, {Misch}, {Mueller}, {Soyumer}, {Woodard},
  {Baliunas}, {Noyes}, {Hartmann}, {Porter}, {Zwaan}, {Middelkoop}, {Rutten},
  \& {Mihalas}}]{duncan1991}
{Duncan}, D.~K., {Vaughan}, A.~H., {Wilson}, O.~C., {et~al.} 1991,
  \bibinfo{title}{{CA II H and K Measurements Made at Mount Wilson Observatory,
  1966--1983},} \apjs, 76, 383, \dodoi{10.1086/191572}

% type= article
\bibitem[{G. {Edenhofer} {et~al.}(2024){Edenhofer}, {Zucker}, {Frank},
  {Saydjari}, {Speagle}, {Finkbeiner}, \& {En{\ss}lin}}]{edenhofer2024}
{Edenhofer}, G., {Zucker}, C., {Frank}, P., {et~al.} 2024, \bibinfo{title}{{A
  parsec-scale Galactic 3D dust map out to 1.25 kpc from the Sun},} \aap, 685,
  A82, \dodoi{10.1051/0004-6361/202347628}

% type= article
\bibitem[{M.~P. {Egan} \& S.~D. {Price}(1996){Egan} \& {Price}}]{egan1996}
{Egan}, M.~P., \& {Price}, S.~D. 1996, \bibinfo{title}{{The MSX Infrared
  Astometric Catalog},} \aj, 112, 2862, \dodoi{10.1086/118227}

% type= article
\bibitem[{R. {Egeland} {et~al.}(2017){Egeland}, {Soon}, {Baliunas}, {Hall},
  {Pevtsov}, \& {Bertello}}]{egeland2017}
{Egeland}, R., {Soon}, W., {Baliunas}, S., {et~al.} 2017, \bibinfo{title}{{The
  Mount Wilson Observatory S-index of the Sun},} \apj, 835, 25,
  \dodoi{10.3847/1538-4357/835/1/25}

% type= article
\bibitem[{O.~J. {Eggen}(1963){Eggen}}]{eggen1963}
{Eggen}, O.~J. 1963, \bibinfo{title}{{Luminosities, colors, and motions of the
  brightest A-type stars},} \aj, 68, 697, \dodoi{10.1086/109198}

% type= article
\bibitem[{O.~J. {Eggen}(1970){Eggen}}]{eggen1970}
{Eggen}, O.~J. 1970, \bibinfo{title}{{Stellar kinematics and evolution},}
  Vistas in Astronomy, 12, 367, \dodoi{10.1016/0083-6656(70)90049-8}

% type= article
\bibitem[{C. {Eiroa} {et~al.}(2013){Eiroa}, {Marshall}, {Mora}, {Montesinos},
  {Absil}, {Augereau}, {Bayo}, {Bryden}, {Danchi}, {del Burgo}, {Ertel},
  {Fridlund}, {Heras}, {Krivov}, {Launhardt}, {Liseau}, {L{\"o}hne},
  {Maldonado}, {Pilbratt}, {Roberge}, {Rodmann}, {Sanz-Forcada}, {Solano},
  {Stapelfeldt}, {Th{\'e}bault}, {Wolf}, {Ardila}, {Ar{\'e}valo}, {Beichmann},
  {Faramaz}, {Gonz{\'a}lez-Garc{\'{\i}}a}, {Guti{\'e}rrez}, {Lebreton},
  {Mart{\'{\i}}nez-Arn{\'a}iz}, {Meeus}, {Montes}, {Olofsson}, {Su}, {White},
  {Barrado}, {Fukagawa}, {Gr{\"u}n}, {Kamp}, {Lorente}, {Morbidelli},
  {M{\"u}ller}, {Mutschke}, {Nakagawa}, {Ribas}, \& {Walker}}]{eiroa2013}
{Eiroa}, C., {Marshall}, J.~P., {Mora}, A., {et~al.} 2013,
  \bibinfo{title}{{DUst around NEarby Stars. The survey observational
  results},} \aap, 555, A11, \dodoi{10.1051/0004-6361/201321050}

% type= article
\bibitem[{T. {Eisenbeiss} {et~al.}(2013){Eisenbeiss}, {Ammler-von Eiff},
  {Roell}, {Mugrauer}, {Adam}, {Neuh{\"a}user}, {Schmidt}, \&
  {Bedalov}}]{eisenbeiss2013}
{Eisenbeiss}, T., {Ammler-von Eiff}, M., {Roell}, T., {et~al.} 2013,
  \bibinfo{title}{{The Hercules-Lyra association revisited. New age estimation
  and multiplicity study},} \aap, 556, A53, \dodoi{10.1051/0004-6361/201118362}

% type= article
\bibitem[{Z. {Eker} {et~al.}(2014){Eker}, {Bilir}, {Soydugan},
  {G{\"o}k{\c{c}}e}, {Soydugan}, {T{\"u}ys{\"u}z}, {{\c{S}}eny{\"u}z}, \&
  {Demircan}}]{eker2014}
{Eker}, Z., {Bilir}, S., {Soydugan}, F., {et~al.} 2014, \bibinfo{title}{{The
  Catalogue of Stellar Parameters from the Detached Double-Lined Eclipsing
  Binaries in the Milky Way},} \pasa, 31, e024, \dodoi{10.1017/pasa.2014.17}

% type= article
\bibitem[{D. {Engels} {et~al.}(1981){Engels}, {Sherwood}, {Wamsteker}, \&
  {Schultz}}]{engels1981}
{Engels}, D., {Sherwood}, W.~A., {Wamsteker}, W., \& {Schultz}, G.~V. 1981,
  \bibinfo{title}{{Infrared observations of southern bright stars.},} \aaps,
  45, 5

% type= article
\bibitem[{D.~W. {Evans} {et~al.}(2018){Evans}, {Riello}, {De Angeli},
  {Carrasco}, {Montegriffo}, {Fabricius}, {Jordi}, {Palaversa}, {Diener},
  {Busso}, {Cacciari}, {van Leeuwen}, {Burgess}, {Davidson}, {Harrison},
  {Hodgkin}, {Pancino}, {Richards}, {Altavilla}, {Balaguer-N{\'u}{\~n}ez},
  {Barstow}, {Bellazzini}, {Brown}, {Castellani}, {Cocozza}, {De Luise},
  {Delgado}, {Ducourant}, {Galleti}, {Gilmore}, {Giuffrida}, {Holl}, {Kewley},
  {Koposov}, {Marinoni}, {Marrese}, {Osborne}, {Piersimoni}, {Portell},
  {Pulone}, {Ragaini}, {Sanna}, {Terrett}, {Walton}, {Wevers}, \&
  {Wyrzykowski}}]{evans2018}
{Evans}, D.~W., {Riello}, M., {De Angeli}, F., {et~al.} 2018,
  \bibinfo{title}{{Gaia Data Release 2. Photometric content and validation},}
  \aap, 616, A4, \dodoi{10.1051/0004-6361/201832756}

% type= article
\bibitem[{I.~N. {Evans} {et~al.}(2010){Evans}, {Primini}, {Glotfelty},
  {Anderson}, {Bonaventura}, {Chen}, {Davis}, {Doe}, {Evans}, {Fabbiano},
  {Galle}, {Gibbs}, {Grier}, {Hain}, {Hall}, {Harbo}, {He}, {Houck},
  {Karovska}, {Kashyap}, {Lauer}, {McCollough}, {McDowell}, {Miller},
  {Mitschang}, {Morgan}, {Mossman}, {Nichols}, {Nowak}, {Plummer}, {Refsdal},
  {Rots}, {Siemiginowska}, {Sundheim}, {Tibbetts}, {Van Stone}, {Winkelman}, \&
  {Zografou}}]{evans2010}
{Evans}, I.~N., {Primini}, F.~A., {Glotfelty}, K.~J., {et~al.} 2010,
  \bibinfo{title}{{The Chandra Source Catalog},} \apjs, 189, 37,
  \dodoi{10.1088/0067-0049/189/1/37}

% type= article
\bibitem[{C. {Fabricius} {et~al.}(2021){Fabricius}, {Luri}, {Arenou},
  {Babusiaux}, {Helmi}, {Muraveva}, {Reyl{\'e}}, {Spoto}, {Vallenari},
  {Antoja}, {Balbinot}, {Barache}, {Bauchet}, {Bragaglia}, {Busonero},
  {Cantat-Gaudin}, {Carrasco}, {Diakit{\'e}}, {Fabrizio}, {Figueras},
  {Garcia-Gutierrez}, {Garofalo}, {Jordi}, {Kervella}, {Khanna}, {Leclerc},
  {Licata}, {Lambert}, {Marrese}, {Masip}, {Ramos}, {Robichon}, {Robin},
  {Romero-G{\'o}mez}, {Rubele}, \& {Weiler}}]{fabricius2021}
{Fabricius}, C., {Luri}, X., {Arenou}, F., {et~al.} 2021, \bibinfo{title}{{Gaia
  Early Data Release 3. Catalogue validation},} \aap, 649, A5,
  \dodoi{10.1051/0004-6361/202039834}

% type= article
\bibitem[{M. {Fajrin} {et~al.}(2025){Fajrin}, {Armstrong}, {Tan}, {Farias}, \&
  {Eyer}}]{fajrin2025}
{Fajrin}, M., {Armstrong}, J.~J., {Tan}, J.~C., {Farias}, J.~P., \& {Eyer}, L.
  2025, \bibinfo{title}{{Low-mass runaways from the Orion Nebula Cluster -
  kinematic age constraints on star cluster formation},} \mnras, 537, 1320,
  \dodoi{10.1093/mnras/staf074}

% type= article
\bibitem[{X.-S. {Fang} {et~al.}(2018){Fang}, {Zhao}, {Zhao}, \& {Bharat
  Kumar}}]{fang2018}
{Fang}, X.-S., {Zhao}, G., {Zhao}, J.-K., \& {Bharat Kumar}, Y. 2018,
  \bibinfo{title}{{Stellar activity with LAMOST - II. Chromospheric activity in
  open clusters},} \mnras, 476, 908, \dodoi{10.1093/mnras/sty212}

% type= article
\bibitem[{V. {Faramaz} {et~al.}(2019){Faramaz}, {Krist}, {Stapelfeldt},
  {Bryden}, {Mamajek}, {Matr{\`a}}, {Booth}, {Flaherty}, {Hales}, {Hughes},
  {Bayo}, {Casassus}, {Cuadra}, {Olofsson}, {Su}, \& {Wilner}}]{faramaz2019}
{Faramaz}, V., {Krist}, J., {Stapelfeldt}, K.~R., {et~al.} 2019,
  \bibinfo{title}{{From Scattered-light to Millimeter Emission: A Comprehensive
  View of the Gigayear-old System of HD 202628 and its Eccentric Debris Ring},}
  \aj, 158, 162, \dodoi{10.3847/1538-3881/ab3ec1}

% type= article
\bibitem[{V. {Faramaz} {et~al.}(2021){Faramaz}, {Marino}, {Booth}, {Matr{\`a}},
  {Mamajek}, {Bryden}, {Stapelfeldt}, {Casassus}, {Cuadra}, {Hales}, \&
  {Zurlo}}]{faramaz2021}
{Faramaz}, V., {Marino}, S., {Booth}, M., {et~al.} 2021, \bibinfo{title}{{A
  Detailed Characterization of HR 8799's Debris Disk with ALMA in Band 7},}
  \aj, 161, 271, \dodoi{10.3847/1538-3881/abf4e0}

% type= article
\bibitem[{J. {Farihi}(2016){Farihi}}]{farihi16}
{Farihi}, J. 2016, \bibinfo{title}{{Circumstellar debris and pollution at white
  dwarf stars},} \nar, 71, 9, \dodoi{10.1016/j.newar.2016.03.001}

% type= article
\bibitem[{G.~G. {Fazio} {et~al.}(2004){Fazio}, {Hora}, {Allen}, {Ashby},
  {Barmby}, {Deutsch}, {Huang}, {Kleiner}, {Marengo}, {Megeath}, {Melnick},
  {Pahre}, {Patten}, {Polizotti}, {Smith}, {Taylor}, {Wang}, {Willner},
  {Hoffmann}, {Pipher}, {Forrest}, {McMurty}, {McCreight}, {McKelvey},
  {McMurray}, {Koch}, {Moseley}, {Arendt}, {Mentzell}, {Marx}, {Losch},
  {Mayman}, {Eichhorn}, {Krebs}, {Jhabvala}, {Gezari}, {Fixsen}, {Flores},
  {Shakoorzadeh}, {Jungo}, {Hakun}, {Workman}, {Karpati}, {Kichak}, {Whitley},
  {Mann}, {Tollestrup}, {Eisenhardt}, {Stern}, {Gorjian}, {Bhattacharya},
  {Carey}, {Nelson}, {Glaccum}, {Lacy}, {Lowrance}, {Laine}, {Reach},
  {Stauffer}, {Surace}, {Wilson}, {Wright}, {Hoffman}, {Domingo}, \&
  {Cohen}}]{fazio2004}
{Fazio}, G.~G., {Hora}, J.~L., {Allen}, L.~E., {et~al.} 2004,
  \bibinfo{title}{{The Infrared Array Camera (IRAC) for the Spitzer Space
  Telescope},} \apjs, 154, 10, \dodoi{10.1086/422843}

% type= article
\bibitem[{F.~C. {Fekel}(1996){Fekel}}]{fekel1996}
{Fekel}, F.~C. 1996, \bibinfo{title}{{Chromospherically Active Stars. XV. HD
  8357=AR Piscium, an Extremely Active RS CVn System},} \aj, 112, 269,
  \dodoi{10.1086/118010}

% type= article
\bibitem[{F.~C. {Fekel}(1997){Fekel}}]{fekel1997}
{Fekel}, F.~C. 1997, \bibinfo{title}{{Chromospherically Active Stars. XVI. The
  Double-Lined Binary 42 Capricorni},} \aj, 114, 2747, \dodoi{10.1086/118683}

% type= article
\bibitem[{F.~C. {Fekel} {et~al.}(1994){Fekel}, {Dadonas}, {Sperauskas},
  {Vaccaro}, \& {Patterson}}]{fekel1994}
{Fekel}, F.~C., {Dadonas}, V., {Sperauskas}, J., {Vaccaro}, T.~R., \&
  {Patterson}, L.~R. 1994, \bibinfo{title}{{Chromospherically Active Stars.
  XIII. HD 30957: A Double Lined K Dwarf Binary},} \aj, 108, 1936,
  \dodoi{10.1086/117207}

% type= article
\bibitem[{F.~C. {Fekel} {et~al.}(1988){Fekel}, {Gillies}, {Africano}, \&
  {Quigley}}]{fekel1988}
{Fekel}, F.~C., {Gillies}, K., {Africano}, J., \& {Quigley}, R. 1988,
  \bibinfo{title}{{Chromospherically Active Stars. V. HD 91816=LR Hya: A
  Double-Lined BY Draconis Type Binary},} \aj, 96, 1426, \dodoi{10.1086/114893}

% type= article
\bibitem[{F.~C. {Fekel} {et~al.}(2004){Fekel}, {Henry}, \&
  {Alston}}]{fekel2004}
{Fekel}, F.~C., {Henry}, G.~W., \& {Alston}, F.~M. 2004,
  \bibinfo{title}{{Chromospherically Active Stars. XXII. HD 18955, A Massive K
  Dwarf Binary},} \aj, 127, 2303, \dodoi{10.1086/382718}

% type= article
\bibitem[{R.~B. {Fernandes} {et~al.}(2019){Fernandes}, {Mulders}, {Pascucci},
  {Mordasini}, \& {Emsenhuber}}]{fernandes2019}
{Fernandes}, R.~B., {Mulders}, G.~D., {Pascucci}, I., {Mordasini}, C., \&
  {Emsenhuber}, A. 2019, \bibinfo{title}{{Hints for a Turnover at the Snow Line
  in the Giant Planet Occurrence Rate},} \apj, 874, 81,
  \dodoi{10.3847/1538-4357/ab0300}

% type= article
\bibitem[{E.~L. {Fitzpatrick}(1999){Fitzpatrick}}]{fitzpatrick1999}
{Fitzpatrick}, E.~L. 1999, \bibinfo{title}{{Correcting for the Effects of
  Interstellar Extinction},} \pasp, 111, 63, \dodoi{10.1086/316293}

% type= article
\bibitem[{B.~H. {Foing} {et~al.}(1989){Foing}, {Crivellari}, {Vladilo},
  {Rebolo}, \& {Beckman}}]{foing1989}
{Foing}, B.~H., {Crivellari}, L., {Vladilo}, G., {Rebolo}, R., \& {Beckman},
  J.~E. 1989, \bibinfo{title}{{Chromospheres of late-type active and quiescent
  dwarfs. II. an activity index derived from profiles of the CA II lambda 8498
  A and lambda 8542 A triplet lines.},} \aaps, 80, 189

% type= article
\bibitem[{K.~B. {Follette} {et~al.}(2017){Follette}, {Rameau}, {Dong}, {Pueyo},
  {Close}, {Duch{\^e}ne}, {Fung}, {Leonard}, {Macintosh}, {Males}, {Marois},
  {Millar-Blanchaer}, {Morzinski}, {Mullen}, {Perrin}, {Spiro}, {Wang},
  {Ammons}, {Bailey}, {Barman}, {Bulger}, {Chilcote}, {Cotten}, {De Rosa},
  {Doyon}, {Fitzgerald}, {Goodsell}, {Graham}, {Greenbaum}, {Hibon}, {Hung},
  {Ingraham}, {Kalas}, {Konopacky}, {Larkin}, {Maire}, {Marchis}, {Metchev},
  {Nielsen}, {Oppenheimer}, {Palmer}, {Patience}, {Poyneer}, {Rajan},
  {Rantakyr{\"o}}, {Savransky}, {Schneider}, {Sivaramakrishnan}, {Song},
  {Soummer}, {Thomas}, {Vega}, {Wallace}, {Ward-Duong}, {Wiktorowicz}, \&
  {Wolff}}]{follette2017}
{Follette}, K.~B., {Rameau}, J., {Dong}, R., {et~al.} 2017,
  \bibinfo{title}{{Complex Spiral Structure in the HD 100546 Transitional Disk
  as Revealed by GPI and MagAO},} \aj, 153, 264,
  \dodoi{10.3847/1538-3881/aa6d85}

% type= article
\bibitem[{A. {Ford} {et~al.}(2001){Ford}, {Jeffries}, {James}, \&
  {Barnes}}]{ford2001}
{Ford}, A., {Jeffries}, R.~D., {James}, D.~J., \& {Barnes}, J.~R. 2001,
  \bibinfo{title}{{Lithium in the Coma Berenices open cluster},} \aap, 369,
  871, \dodoi{10.1051/0004-6361:20010235}

% type= article
\bibitem[{M. {Fouesneau} {et~al.}(2023){Fouesneau}, {Fr{\'e}mat}, {Andrae},
  {Korn}, {Soubiran}, {Kordopatis}, {Vallenari}, {Heiter}, {Creevey}, {Sarro},
  {de Laverny}, {Lanzafame}, {Lobel}, {Sordo}, {Rybizki}, {Slezak},
  {{\'A}lvarez}, {Drimmel}, {Garabato}, {Delchambre}, {Bailer-Jones},
  {Hatzidimitriou}, {Lorca}, {Le Fustec}, {Pailler}, {Mary}, {Robin},
  {Utrilla}, {Abreu Aramburu}, {Bakker}, {Bellas-Velidis}, {Bijaoui}, {Blomme},
  {Bouret}, {Brouillet}, {Brugaletta}, {Burlacu}, {Carballo}, {Casamiquela},
  {Chaoul}, {Chiavassa}, {Contursi}, {Cooper}, {Dafonte}, {Demouchy},
  {Dharmawardena}, {Garc{\'\i}a-Lario}, {Garc{\'\i}a-Torres}, {Gomez},
  {Gonz{\'a}lez-Santamar{\'\i}a}, {Jean-Antoine Piccolo}, {Kontizas},
  {Lebreton}, {Licata}, {Lindstr{\o}m}, {Livanou}, {Magdaleno Romeo},
  {Manteiga}, {Marocco}, {Martayan}, {Marshall}, {Nicolas}, {Ordenovic},
  {Palicio}, {Pallas-Quintela}, {Pichon}, {Poggio}, {Recio-Blanco}, {Riclet},
  {Santove{\~n}a}, {Schultheis}, {Segol}, {Silvelo}, {Smart}, {S{\"u}veges},
  {Th{\'e}venin}, {Torralba Elipe}, {Ulla}, {van Dillen}, {Zhao}, \&
  {Zorec}}]{fouesneau2023}
{Fouesneau}, M., {Fr{\'e}mat}, Y., {Andrae}, R., {et~al.} 2023,
  \bibinfo{title}{{Gaia Data Release 3. Apsis. II. Stellar parameters},} \aap,
  674, A28, \dodoi{10.1051/0004-6361/202243919}

% type= article
\bibitem[{E. {Franciosini} {et~al.}(2003){Franciosini}, {Randich}, \&
  {Pallavicini}}]{franciosini2003}
{Franciosini}, E., {Randich}, S., \& {Pallavicini}, R. 2003,
  \bibinfo{title}{{Is Praesepe really different from the coeval Hyades cluster?
  The XMM-Newton view},} \aap, 405, 551, \dodoi{10.1051/0004-6361:20030623}

% type= article
\bibitem[{E. {Franciosini} {et~al.}(2022){Franciosini}, {Randich}, {de
  Laverny}, {Biazzo}, {Feuillet}, {Frasca}, {Lind}, {Prisinzano},
  {Tautvaisiene}, {Lanzafame}, {Smiljanic}, {Gonneau}, {Magrini}, {Pancino},
  {Guiglion}, {Sacco}, {Sanna}, {Gilmore}, {Bonifacio}, {Jeffries}, {Micela},
  {Prusti}, {Alfaro}, {Bensby}, {Bragaglia}, {Fran{\c{c}}ois}, {Korn}, {Van
  Eck}, {Bayo}, {Bergemann}, {Carraro}, {Heiter}, {Hourihane}, {Jofr{\'e}},
  {Lewis}, {Martayan}, {Monaco}, {Morbidelli}, {Worley}, \&
  {Zaggia}}]{franciosini2022}
{Franciosini}, E., {Randich}, S., {de Laverny}, P., {et~al.} 2022,
  \bibinfo{title}{{The Gaia-ESO Survey: Lithium measurements and new curves of
  growth},} \aap, 668, A49, \dodoi{10.1051/0004-6361/202244854}

% type= article
\bibitem[{S. {Freund} {et~al.}(2025){Freund}, {Czesla}, {Fuhrmeister},
  {Predehl}, {Robrade}, {Schneider}, \& {Schmitt}}]{freund2025}
{Freund}, S., {Czesla}, S., {Fuhrmeister}, B., {et~al.} 2025,
  \bibinfo{title}{{The stellar corona-chromosphere connection: A comprehensive
  study of X-ray and Ca II IRT fluxes from eROSITA and Gaia},} \aap, 697, A230,
  \dodoi{10.1051/0004-6361/202451421}

% type= article
\bibitem[{S. {Freund} {et~al.}(2022){Freund}, {Czesla}, {Robrade}, {Schneider},
  \& {Schmitt}}]{freund2022}
{Freund}, S., {Czesla}, S., {Robrade}, J., {Schneider}, P.~C., \& {Schmitt},
  J.~H.~M.~M. 2022, \bibinfo{title}{{The stellar content of the ROSAT all-sky
  survey},} \aap, 664, A105, \dodoi{10.1051/0004-6361/202142573}

% type= article
\bibitem[{S. {Freund} {et~al.}(2018){Freund}, {Robrade}, {Schneider}, \&
  {Schmitt}}]{freund2018}
{Freund}, S., {Robrade}, J., {Schneider}, P.~C., \& {Schmitt}, J.~H.~M.~M.
  2018, \bibinfo{title}{{The stellar content of the XMM-Newton slew survey},}
  \aap, 614, A125, \dodoi{10.1051/0004-6361/201732009}

% type= article
\bibitem[{S. {Freund} {et~al.}(2024){Freund}, {Czesla}, {Predehl}, {Robrade},
  {Salvato}, {Schneider}, {Starck}, {Wolf}, \& {Schmitt}}]{freund2024}
{Freund}, S., {Czesla}, S., {Predehl}, P., {et~al.} 2024, \bibinfo{title}{{The
  SRG/eROSITA all-sky survey. Identifying the coronal content with HamStar},}
  \aap, 684, A121, \dodoi{10.1051/0004-6361/202348278}

% type= article
\bibitem[{P.~C. {Frisch} \& D.~G. {York}(1983){Frisch} \& {York}}]{frisch1983}
{Frisch}, P.~C., \& {York}, D.~G. 1983, \bibinfo{title}{{Synthesis maps of
  ultraviolet observations of neutral interstellar gas.},} \apjl, 271, L59,
  \dodoi{10.1086/184095}

% type= article
\bibitem[{D.~J. {Fritzewski} {et~al.}(2020){Fritzewski}, {Barnes}, {James}, \&
  {Strassmeier}}]{fritzewski2020}
{Fritzewski}, D.~J., {Barnes}, S.~A., {James}, D.~J., \& {Strassmeier}, K.~G.
  2020, \bibinfo{title}{{The rotation period distribution of the rich
  Pleiades-age southern open cluster NGC 2516. Existence of a representative
  zero-age main sequence distribution},} \aap, 641, A51,
  \dodoi{10.1051/0004-6361/201936860}

% type= article
\bibitem[{K. {Fuhrmann}(2008){Fuhrmann}}]{fuhrmann2008}
{Fuhrmann}, K. 2008, \bibinfo{title}{{Nearby stars of the Galactic disc and
  halo - IV},} \mnras, 384, 173, \dodoi{10.1111/j.1365-2966.2007.12671.x}

% type= article
\bibitem[{K. {Fuhrmann} \& R. {Chini}(2015){Fuhrmann} \&
  {Chini}}]{fuhrmann2015}
{Fuhrmann}, K., \& {Chini}, R. 2015, \bibinfo{title}{{Multiplicity among F-type
  stars. II.},} \apj, 809, 107, \dodoi{10.1088/0004-637X/809/1/107}

% type= article
\bibitem[{K. {Fuhrmann} {et~al.}(2011){Fuhrmann}, {Chini}, {Hoffmeister},
  {Lemke}, {Murphy}, {Seifert}, \& {Stahl}}]{fuhrmann2011}
{Fuhrmann}, K., {Chini}, R., {Hoffmeister}, V.~H., {et~al.} 2011,
  \bibinfo{title}{{BESO {\'e}chelle spectroscopy of solar-type stars at Cerro
  Armazones},} \mnras, 411, 2311, \dodoi{10.1111/j.1365-2966.2010.17850.x}

% type= article
\bibitem[{B.~J. {Fulton} {et~al.}(2021){Fulton}, {Rosenthal}, {Hirsch},
  {Isaacson}, {Howard}, {Dedrick}, {Sherstyuk}, {Blunt}, {Petigura}, {Knutson},
  {Behmard}, {Chontos}, {Crepp}, {Crossfield}, {Dalba}, {Fischer}, {Henry},
  {Kane}, {Kosiarek}, {Marcy}, {Rubenzahl}, {Weiss}, \& {Wright}}]{fulton2021}
{Fulton}, B.~J., {Rosenthal}, L.~J., {Hirsch}, L.~A., {et~al.} 2021,
  \bibinfo{title}{{California Legacy Survey. II. Occurrence of Giant Planets
  beyond the Ice Line},} \apjs, 255, 14, \dodoi{10.3847/1538-4365/abfcc1}

% type= article
\bibitem[{E. {Furlan} {et~al.}(2006){Furlan}, {Hartmann}, {Calvet},
  {D'Alessio}, {Franco-Hern{\'a}ndez}, {Forrest}, {Watson}, {Uchida},
  {Sargent}, {Green}, {Keller}, \& {Herter}}]{furlan2006}
{Furlan}, E., {Hartmann}, L., {Calvet}, N., {et~al.} 2006, \bibinfo{title}{{A
  Survey and Analysis of Spitzer Infrared Spectrograph Spectra of T Tauri Stars
  in Taurus},} \apjs, 165, 568, \dodoi{10.1086/505468}

% type= article
\bibitem[{J. {Gagn{\'e}} {et~al.}(2023){Gagn{\'e}}, {Moranta}, {Faherty},
  {Kiman}, {Couture}, {Larochelle}, {Popinchalk}, \& {Morrone}}]{gagne2023}
{Gagn{\'e}}, J., {Moranta}, L., {Faherty}, J.~K., {et~al.} 2023,
  \bibinfo{title}{{The Oceanus Moving Group: A New 500 Myr Old Host for the
  Nearest Brown Dwarf},} \apj, 945, 119, \dodoi{10.3847/1538-4357/acb8b7}

% type= article
\bibitem[{J. {Gagn{\'e}} {et~al.}(2018){Gagn{\'e}}, {Mamajek}, {Malo},
  {Riedel}, {Rodriguez}, {Lafreni{\`e}re}, {Faherty}, {Roy-Loubier}, {Pueyo},
  {Robin}, \& {Doyon}}]{gagne2018}
{Gagn{\'e}}, J., {Mamajek}, E.~E., {Malo}, L., {et~al.} 2018,
  \bibinfo{title}{{BANYAN. XI. The BANYAN {\ensuremath{\Sigma}} Multivariate
  Bayesian Algorithm to Identify Members of Young Associations with 150 pc},}
  \apj, 856, 23, \dodoi{10.3847/1538-4357/aaae09}

% type= article
\bibitem[{ {Gaia Collaboration} {et~al.}(2021{\natexlab{a}}){Gaia
  Collaboration}, {Smart}, {Sarro}, {Rybizki}, {Reyl{\'e}}, {Robin}, {Hambly},
  {Abbas}, {Barstow}, {de Bruijne}, {Bucciarelli}, {Carrasco}, {Cooper},
  {Hodgkin}, {Masana}, {Michalik}, {Sahlmann}, {Sozzetti}, {Brown},
  {Vallenari}, {Prusti}, {Babusiaux}, {Biermann}, {Creevey}, {Evans}, {Eyer},
  {Hutton}, {Jansen}, {Jordi}, {Klioner}, {Lammers}, {Lindegren}, {Luri},
  {Mignard}, {Panem}, {Pourbaix}, {Randich}, {Sartoretti}, {Soubiran},
  {Walton}, {Arenou}, {Bailer-Jones}, {Bastian}, {Cropper}, {Drimmel}, {Katz},
  {Lattanzi}, {van Leeuwen}, {Bakker}, {Casta{\~n}eda}, {De Angeli},
  {Ducourant}, {Fabricius}, {Fouesneau}, {Fr{\'e}mat}, {Guerra}, {Guerrier},
  {Guiraud}, {Jean-Antoine Piccolo}, {Messineo}, {Mowlavi}, {Nicolas},
  {Nienartowicz}, {Pailler}, {Panuzzo}, {Riclet}, {Roux}, {Seabroke}, {Sordo},
  {Tanga}, {Th{\'e}venin}, {Gracia-Abril}, {Portell}, {Teyssier}, {Altmann},
  {Andrae}, {Bellas-Velidis}, {Benson}, {Berthier}, {Blomme}, {Brugaletta},
  {Burgess}, {Busso}, {Carry}, {Cellino}, {Cheek}, {Clementini}, {Damerdji},
  {Davidson}, {Delchambre}, {Dell'Oro}, {Fern{\'a}ndez-Hern{\'a}ndez},
  {Galluccio}, {Garc{\'\i}a-Lario}, {Garcia-Reinaldos},
  {Gonz{\'a}lez-N{\'u}{\~n}ez}, {Gosset}, {Haigron}, {Halbwachs}, {Harrison},
  {Hatzidimitriou}, {Heiter}, {Hern{\'a}ndez}, {Hestroffer}, {Holl},
  {Jan{\ss}en}, {Jevardat de Fombelle}, {Jordan}, {Krone-Martins}, {Lanzafame},
  {L{\"o}ffler}, {Lorca}, {Manteiga}, {Marchal}, {Marrese}, {Moitinho}, {Mora},
  {Muinonen}, {Osborne}, {Pancino}, {Pauwels}, {Recio-Blanco}, {Richards},
  {Riello}, {Rimoldini}, {Roegiers}, {Siopis}, {Smith}, {Ulla}, {Utrilla}, {van
  Leeuwen}, {van Reeven}, {Abreu Aramburu}, {Accart}, {Aerts}, {Aguado},
  {Ajaj}, {Altavilla}, {{\'A}lvarez}, {{\'A}lvarez Cid-Fuentes}, {Alves},
  {Anderson}, {Anglada Varela}, {Antoja}, {Audard}, {Baines}, {Baker},
  {Balaguer-N{\'u}{\~n}ez}, {Balbinot}, {Balog}, {Barache}, {Barbato},
  {Barros}, {Bartolom{\'e}}, {Bassilana}, {Bauchet}, {Baudesson-Stella},
  {Becciani}, {Bellazzini}, {Bernet}, {Bertone}, {Bianchi}, {Blanco-Cuaresma},
  {Boch}, {Bombrun}, {Bossini}, {Bouquillon}, {Bragaglia}, {Bramante},
  {Breedt}, {Bressan}, {Brouillet}, {Burlacu}, {Busonero}, {Butkevich},
  {Buzzi}, {Caffau}, {Cancelliere}, {C{\'a}novas}, {Cantat-Gaudin}, {Carballo},
  {Carlucci}, {Carnerero}, {Casamiquela}, {Castellani}, {Castro-Ginard},
  {Castro Sampol}, {Chaoul}, {Charlot}, {Chemin}, {Chiavassa}, {Cioni},
  {Comoretto}, {Cornez}, {Cowell}, {Crifo}, {Crosta}, {Crowley}, {Dafonte},
  {Dapergolas}, {David}, {David}, {de Laverny}, {De Luise}, {De March}, {De
  Ridder}, {de Souza}, {de Teodoro}, {de Torres}, {del Peloso}, {del Pozo},
  {Delgado}, {Delgado}, {Delisle}, {Di Matteo}, {Diakite}, {Diener},
  {Distefano}, {Dolding}, {Eappachen}, {Edvardsson}, {Enke}, {Esquej}, {Fabre},
  {Fabrizio}, {Faigler}, {Fedorets}, {Fernique}, {Fienga}, {Figueras},
  {Fouron}, {Fragkoudi}, {Fraile}, {Franke}, {Gai}, {Garabato},
  {Garcia-Gutierrez}, {Garc{\'\i}a-Torres}, {Garofalo}, {Gavras}, {Gerlach},
  {Geyer}, {Giacobbe}, {Gilmore}, {Girona}, {Giuffrida}, {Gomel}, {Gomez},
  {Gonzalez-Santamaria}, {Gonz{\'a}lez-Vidal}, {Granvik},
  {Guti{\'e}rrez-S{\'a}nchez}, {Guy}, {Hauser}, {Haywood}, {Helmi}, {Hidalgo},
  {Hilger}, {H{\l}adczuk}, {Hobbs}, {Holland}, {Huckle}, {Jasniewicz},
  {Jonker}, {Juaristi Campillo}, {Julbe}, {Karbevska}, {Kervella}, {Khanna},
  {Kochoska}, {Kontizas}, {Kordopatis}, {Korn}, {Kostrzewa-Rutkowska},
  {Kruszy{\'n}ska}, {Lambert}, {Lanza}, {Lasne}, {Le Campion}, {Le Fustec},
  {Lebreton}, {Lebzelter}, {Leccia}, {Leclerc}, {Lecoeur-Taibi}, {Liao},
  {Licata}, {Lindstr{\o}m}, {Lister}, {Livanou}, {Lobel}, {Madrero Pardo},
  {Managau}, {Mann}, {Marchant}, {Marconi}, {Marcos Santos}, {Marinoni},
  {Marocco}, {Marshall}, {Martin Polo}, {Mart{\'\i}n-Fleitas}, {Masip},
  {Massari}, {Mastrobuono-Battisti}, {Mazeh}, {McMillan}, {Messina}, {Millar},
  {Mints}, {Molina}, {Molinaro}, {Moln{\'a}r}, {Montegriffo}, {Mor},
  {Morbidelli}, {Morel}, {Morris}, {Mulone}, {Munoz}, {Muraveva}, {Murphy},
  {Musella}, {Noval}, {Ord{\'e}novic}, {Orr{\`u}}, {Osinde}, {Pagani},
  {Pagano}, {Palaversa}, {Palicio}, {Panahi}, {Pawlak}, {Pe{\~n}alosa
  Esteller}, {Penttil{\"a}}, {Piersimoni}, {Pineau}, {Plachy}, {Plum},
  {Poggio}, {Poretti}, {Poujoulet}, {Prsa}, {Pulone}, {Racero}, {Ragaini},
  {Rainer}, {Raiteri}, {Rambaux}, {Ramos}, {Ramos-Lerate}, {Re Fiorentin},
  {Regibo}, {Ripepi}, {Riva}, {Rixon}, {Robichon}, {Robin}, {Roelens},
  {Rohrbasser}, {Romero-G{\'o}mez}, {Rowell}, {Royer}, {Rybicki}, {Sadowski},
  {Sagrist{\`a} Sell{\'e}s}, {Salgado}, {Salguero}, {Samaras}, {Sanchez
  Gimenez}, {Sanna}, {Santove{\~n}a}, {Sarasso}, {Schultheis}, {Sciacca},
  {Segol}, {Segovia}, {S{\'e}gransan}, {Semeux}, {Shahaf}, {Siddiqui},
  {Siebert}, {Siltala}, {Slezak}, {Solano}, {Solitro}, {Souami}, {Souchay},
  {Spagna}, {Spoto}, {Steele}, {Steidelm{\"u}ller}, {Stephenson},
  {S{\"u}veges}, {Szabados}, {Szegedi-Elek}, {Taris}, {Tauran}, {Taylor},
  {Teixeira}, {Thuillot}, {Tonello}, {Torra}, {Torra}, {Turon}, {Unger},
  {Vaillant}, {van Dillen}, {Vanel}, {Vecchiato}, {Viala}, {Vicente},
  {Voutsinas}, {Weiler}, {Wevers}, {Wyrzykowski}, {Yoldas}, {Yvard}, {Zhao},
  {Zorec}, {Zucker}, {Zurbach}, \& {Zwitter}}]{gaianearby2021}
{Gaia Collaboration}, {Smart}, R.~L., {Sarro}, L.~M., {et~al.}
  2021{\natexlab{a}}, \bibinfo{title}{{Gaia Early Data Release 3. The Gaia
  Catalogue of Nearby Stars},} \aap, 649, A6,
  \dodoi{10.1051/0004-6361/202039498}

% type= article
\bibitem[{ {Gaia Collaboration} {et~al.}(2021{\natexlab{b}}){Gaia
  Collaboration}, {Brown}, {Vallenari}, {Prusti}, {de Bruijne}, {Babusiaux},
  {Biermann}, {Creevey}, {Evans}, {Eyer}, {Hutton}, {Jansen}, {Jordi},
  {Klioner}, {Lammers}, {Lindegren}, {Luri}, {Mignard}, {Panem}, {Pourbaix},
  {Randich}, {Sartoretti}, {Soubiran}, {Walton}, {Arenou}, {Bailer-Jones},
  {Bastian}, {Cropper}, {Drimmel}, {Katz}, {Lattanzi}, {van Leeuwen}, {Bakker},
  {Cacciari}, {Casta{\~n}eda}, {De Angeli}, {Ducourant}, {Fabricius},
  {Fouesneau}, {Fr{\'e}mat}, {Guerra}, {Guerrier}, {Guiraud}, {Jean-Antoine
  Piccolo}, {Masana}, {Messineo}, {Mowlavi}, {Nicolas}, {Nienartowicz},
  {Pailler}, {Panuzzo}, {Riclet}, {Roux}, {Seabroke}, {Sordo}, {Tanga},
  {Th{\'e}venin}, {Gracia-Abril}, {Portell}, {Teyssier}, {Altmann}, {Andrae},
  {Bellas-Velidis}, {Benson}, {Berthier}, {Blomme}, {Brugaletta}, {Burgess},
  {Busso}, {Carry}, {Cellino}, {Cheek}, {Clementini}, {Damerdji}, {Davidson},
  {Delchambre}, {Dell'Oro}, {Fern{\'a}ndez-Hern{\'a}ndez}, {Galluccio},
  {Garc{\'\i}a-Lario}, {Garcia-Reinaldos}, {Gonz{\'a}lez-N{\'u}{\~n}ez},
  {Gosset}, {Haigron}, {Halbwachs}, {Hambly}, {Harrison}, {Hatzidimitriou},
  {Heiter}, {Hern{\'a}ndez}, {Hestroffer}, {Hodgkin}, {Holl}, {Jan{\ss}en},
  {Jevardat de Fombelle}, {Jordan}, {Krone-Martins}, {Lanzafame},
  {L{\"o}ffler}, {Lorca}, {Manteiga}, {Marchal}, {Marrese}, {Moitinho}, {Mora},
  {Muinonen}, {Osborne}, {Pancino}, {Pauwels}, {Petit}, {Recio-Blanco},
  {Richards}, {Riello}, {Rimoldini}, {Robin}, {Roegiers}, {Rybizki}, {Sarro},
  {Siopis}, {Smith}, {Sozzetti}, {Ulla}, {Utrilla}, {van Leeuwen}, {van
  Reeven}, {Abbas}, {Abreu Aramburu}, {Accart}, {Aerts}, {Aguado}, {Ajaj},
  {Altavilla}, {{\'A}lvarez}, {{\'A}lvarez Cid-Fuentes}, {Alves}, {Anderson},
  {Anglada Varela}, {Antoja}, {Audard}, {Baines}, {Baker},
  {Balaguer-N{\'u}{\~n}ez}, {Balbinot}, {Balog}, {Barache}, {Barbato},
  {Barros}, {Barstow}, {Bartolom{\'e}}, {Bassilana}, {Bauchet},
  {Baudesson-Stella}, {Becciani}, {Bellazzini}, {Bernet}, {Bertone}, {Bianchi},
  {Blanco-Cuaresma}, {Boch}, {Bombrun}, {Bossini}, {Bouquillon}, {Bragaglia},
  {Bramante}, {Breedt}, {Bressan}, {Brouillet}, {Bucciarelli}, {Burlacu},
  {Busonero}, {Butkevich}, {Buzzi}, {Caffau}, {Cancelliere}, {C{\'a}novas},
  {Cantat-Gaudin}, {Carballo}, {Carlucci}, {Carnerero}, {Carrasco},
  {Casamiquela}, {Castellani}, {Castro-Ginard}, {Castro Sampol}, {Chaoul},
  {Charlot}, {Chemin}, {Chiavassa}, {Cioni}, {Comoretto}, {Cooper}, {Cornez},
  {Cowell}, {Crifo}, {Crosta}, {Crowley}, {Dafonte}, {Dapergolas}, {David},
  {David}, {de Laverny}, {De Luise}, {De March}, {De Ridder}, {de Souza}, {de
  Teodoro}, {de Torres}, {del Peloso}, {del Pozo}, {Delbo}, {Delgado},
  {Delgado}, {Delisle}, {Di Matteo}, {Diakite}, {Diener}, {Distefano},
  {Dolding}, {Eappachen}, {Edvardsson}, {Enke}, {Esquej}, {Fabre}, {Fabrizio},
  {Faigler}, {Fedorets}, {Fernique}, {Fienga}, {Figueras}, {Fouron},
  {Fragkoudi}, {Fraile}, {Franke}, {Gai}, {Garabato}, {Garcia-Gutierrez},
  {Garc{\'\i}a-Torres}, {Garofalo}, {Gavras}, {Gerlach}, {Geyer}, {Giacobbe},
  {Gilmore}, {Girona}, {Giuffrida}, {Gomel}, {Gomez}, {Gonzalez-Santamaria},
  {Gonz{\'a}lez-Vidal}, {Granvik}, {Guti{\'e}rrez-S{\'a}nchez}, {Guy},
  {Hauser}, {Haywood}, {Helmi}, {Hidalgo}, {Hilger}, {H{\l}adczuk}, {Hobbs},
  {Holland}, {Huckle}, {Jasniewicz}, {Jonker}, {Juaristi Campillo}, {Julbe},
  {Karbevska}, {Kervella}, {Khanna}, {Kochoska}, {Kontizas}, {Kordopatis},
  {Korn}, {Kostrzewa-Rutkowska}, {Kruszy{\'n}ska}, {Lambert}, {Lanza}, {Lasne},
  {Le Campion}, {Le Fustec}, {Lebreton}, {Lebzelter}, {Leccia}, {Leclerc},
  {Lecoeur-Taibi}, {Liao}, {Licata}, {Lindstr{\o}m}, {Lister}, {Livanou},
  {Lobel}, {Madrero Pardo}, {Managau}, {Mann}, {Marchant}, {Marconi}, {Marcos
  Santos}, {Marinoni}, {Marocco}, {Marshall}, {Martin Polo},
  {Mart{\'\i}n-Fleitas}, {Masip}, {Massari}, {Mastrobuono-Battisti}, {Mazeh},
  {McMillan}, {Messina}, {Michalik}, {Millar}, {Mints}, {Molina}, {Molinaro},
  {Moln{\'a}r}, {Montegriffo}, {Mor}, {Morbidelli}, {Morel}, {Morris},
  {Mulone}, {Munoz}, {Muraveva}, {Murphy}, {Musella}, {Noval}, {Ord{\'e}novic},
  {Orr{\`u}}, {Osinde}, {Pagani}, {Pagano}, {Palaversa}, {Palicio}, {Panahi},
  {Pawlak}, {Pe{\~n}alosa Esteller}, {Penttil{\"a}}, {Piersimoni}, {Pineau},
  {Plachy}, {Plum}, {Poggio}, {Poretti}, {Poujoulet}, {Prsa}, {Pulone},
  {Racero}, {Ragaini}, {Rainer}, {Raiteri}, {Rambaux}, {Ramos}, {Ramos-Lerate},
  {Re Fiorentin}, {Regibo}, {Reyl{\'e}}, {Ripepi}, {Riva}, {Rixon}, {Robichon},
  {Robin}, {Roelens}, {Rohrbasser}, {Romero-G{\'o}mez}, {Rowell}, {Royer},
  {Rybicki}, {Sadowski}, {Sagrist{\`a} Sell{\'e}s}, {Sahlmann}, {Salgado},
  {Salguero}, {Samaras}, {Sanchez Gimenez}, {Sanna}, {Santove{\~n}a},
  {Sarasso}, {Schultheis}, {Sciacca}, {Segol}, {Segovia}, {S{\'e}gransan},
  {Semeux}, {Shahaf}, {Siddiqui}, {Siebert}, {Siltala}, {Slezak}, {Smart},
  {Solano}, {Solitro}, {Souami}, {Souchay}, {Spagna}, {Spoto}, {Steele},
  {Steidelm{\"u}ller}, {Stephenson}, {S{\"u}veges}, {Szabados}, {Szegedi-Elek},
  {Taris}, {Tauran}, {Taylor}, {Teixeira}, {Thuillot}, {Tonello}, {Torra},
  {Torra}, {Turon}, {Unger}, {Vaillant}, {van Dillen}, {Vanel}, {Vecchiato},
  {Viala}, {Vicente}, {Voutsinas}, {Weiler}, {Wevers}, {Wyrzykowski}, {Yoldas},
  {Yvard}, {Zhao}, {Zorec}, {Zucker}, {Zurbach}, \& {Zwitter}}]{gaiaedr3}
{Gaia Collaboration}, {Brown}, A.~G.~A., {Vallenari}, A., {et~al.}
  2021{\natexlab{b}}, \bibinfo{title}{{Gaia Early Data Release 3. Summary of
  the contents and survey properties},} \aap, 649, A1,
  \dodoi{10.1051/0004-6361/202039657}

% type= article
\bibitem[{ {Gaia Collaboration} {et~al.}(2023{\natexlab{a}}){Gaia
  Collaboration}, {Vallenari}, {Brown}, {Prusti}, {de Bruijne}, {Arenou},
  {Babusiaux}, {Biermann}, {Creevey}, {Ducourant}, {Evans}, {Eyer}, {Guerra},
  {Hutton}, {Jordi}, {Klioner}, {Lammers}, {Lindegren}, {Luri}, {Mignard},
  {Panem}, {Pourbaix}, {Randich}, {Sartoretti}, {Soubiran}, {Tanga}, {Walton},
  {Bailer-Jones}, {Bastian}, {Drimmel}, {Jansen}, {Katz}, {Lattanzi}, {van
  Leeuwen}, {Bakker}, {Cacciari}, {Casta{\~n}eda}, {De Angeli}, {Fabricius},
  {Fouesneau}, {Fr{\'e}mat}, {Galluccio}, {Guerrier}, {Heiter}, {Masana},
  {Messineo}, {Mowlavi}, {Nicolas}, {Nienartowicz}, {Pailler}, {Panuzzo},
  {Riclet}, {Roux}, {Seabroke}, {Sordo}, {Th{\'e}venin}, {Gracia-Abril},
  {Portell}, {Teyssier}, {Altmann}, {Andrae}, {Audard}, {Bellas-Velidis},
  {Benson}, {Berthier}, {Blomme}, {Burgess}, {Busonero}, {Busso},
  {C{\'a}novas}, {Carry}, {Cellino}, {Cheek}, {Clementini}, {Damerdji},
  {Davidson}, {de Teodoro}, {Nu{\~n}ez Campos}, {Delchambre}, {Dell'Oro},
  {Esquej}, {Fern{\'a}ndez-Hern{\'a}ndez}, {Fraile}, {Garabato},
  {Garc{\'\i}a-Lario}, {Gosset}, {Haigron}, {Halbwachs}, {Hambly}, {Harrison},
  {Hern{\'a}ndez}, {Hestroffer}, {Hodgkin}, {Holl}, {Jan{\ss}en}, {Jevardat de
  Fombelle}, {Jordan}, {Krone-Martins}, {Lanzafame}, {L{\"o}ffler}, {Marchal},
  {Marrese}, {Moitinho}, {Muinonen}, {Osborne}, {Pancino}, {Pauwels},
  {Recio-Blanco}, {Reyl{\'e}}, {Riello}, {Rimoldini}, {Roegiers}, {Rybizki},
  {Sarro}, {Siopis}, {Smith}, {Sozzetti}, {Utrilla}, {van Leeuwen}, {Abbas},
  {{\'A}brah{\'a}m}, {Abreu Aramburu}, {Aerts}, {Aguado}, {Ajaj},
  {Aldea-Montero}, {Altavilla}, {{\'A}lvarez}, {Alves}, {Anders}, {Anderson},
  {Anglada Varela}, {Antoja}, {Baines}, {Baker}, {Balaguer-N{\'u}{\~n}ez},
  {Balbinot}, {Balog}, {Barache}, {Barbato}, {Barros}, {Barstow},
  {Bartolom{\'e}}, {Bassilana}, {Bauchet}, {Becciani}, {Bellazzini},
  {Berihuete}, {Bernet}, {Bertone}, {Bianchi}, {Binnenfeld}, {Blanco-Cuaresma},
  {Blazere}, {Boch}, {Bombrun}, {Bossini}, {Bouquillon}, {Bragaglia},
  {Bramante}, {Breedt}, {Bressan}, {Brouillet}, {Brugaletta}, {Bucciarelli},
  {Burlacu}, {Butkevich}, {Buzzi}, {Caffau}, {Cancelliere}, {Cantat-Gaudin},
  {Carballo}, {Carlucci}, {Carnerero}, {Carrasco}, {Casamiquela}, {Castellani},
  {Castro-Ginard}, {Chaoul}, {Charlot}, {Chemin}, {Chiaramida}, {Chiavassa},
  {Chornay}, {Comoretto}, {Contursi}, {Cooper}, {Cornez}, {Cowell}, {Crifo},
  {Cropper}, {Crosta}, {Crowley}, {Dafonte}, {Dapergolas}, {David}, {David},
  {de Laverny}, {De Luise}, {De March}, {De Ridder}, {de Souza}, {de Torres},
  {del Peloso}, {del Pozo}, {Delbo}, {Delgado}, {Delisle}, {Demouchy},
  {Dharmawardena}, {Di Matteo}, {Diakite}, {Diener}, {Distefano}, {Dolding},
  {Edvardsson}, {Enke}, {Fabre}, {Fabrizio}, {Faigler}, {Fedorets}, {Fernique},
  {Fienga}, {Figueras}, {Fournier}, {Fouron}, {Fragkoudi}, {Gai},
  {Garcia-Gutierrez}, {Garcia-Reinaldos}, {Garc{\'\i}a-Torres}, {Garofalo},
  {Gavel}, {Gavras}, {Gerlach}, {Geyer}, {Giacobbe}, {Gilmore}, {Girona},
  {Giuffrida}, {Gomel}, {Gomez}, {Gonz{\'a}lez-N{\'u}{\~n}ez},
  {Gonz{\'a}lez-Santamar{\'\i}a}, {Gonz{\'a}lez-Vidal}, {Granvik}, {Guillout},
  {Guiraud}, {Guti{\'e}rrez-S{\'a}nchez}, {Guy}, {Hatzidimitriou}, {Hauser},
  {Haywood}, {Helmer}, {Helmi}, {Sarmiento}, {Hidalgo}, {Hilger},
  {H{\l}adczuk}, {Hobbs}, {Holland}, {Huckle}, {Jardine}, {Jasniewicz},
  {Jean-Antoine Piccolo}, {Jim{\'e}nez-Arranz}, {Jorissen}, {Juaristi
  Campillo}, {Julbe}, {Karbevska}, {Kervella}, {Khanna}, {Kontizas},
  {Kordopatis}, {Korn}, {K{\'o}sp{\'a}l}, {Kostrzewa-Rutkowska},
  {Kruszy{\'n}ska}, {Kun}, {Laizeau}, {Lambert}, {Lanza}, {Lasne}, {Le
  Campion}, {Lebreton}, {Lebzelter}, {Leccia}, {Leclerc}, {Lecoeur-Taibi},
  {Liao}, {Licata}, {Lindstr{\o}m}, {Lister}, {Livanou}, {Lobel}, {Lorca},
  {Loup}, {Madrero Pardo}, {Magdaleno Romeo}, {Managau}, {Mann}, {Manteiga},
  {Marchant}, {Marconi}, {Marcos}, {Marcos Santos}, {Mar{\'\i}n Pina},
  {Marinoni}, {Marocco}, {Marshall}, {Martin Polo}, {Mart{\'\i}n-Fleitas},
  {Marton}, {Mary}, {Masip}, {Massari}, {Mastrobuono-Battisti}, {Mazeh},
  {McMillan}, {Messina}, {Michalik}, {Millar}, {Mints}, {Molina}, {Molinaro},
  {Moln{\'a}r}, {Monari}, {Mongui{\'o}}, {Montegriffo}, {Montero}, {Mor},
  {Mora}, {Morbidelli}, {Morel}, {Morris}, {Muraveva}, {Murphy}, {Musella},
  {Nagy}, {Noval}, {Oca{\~n}a}, {Ogden}, {Ordenovic}, {Osinde}, {Pagani},
  {Pagano}, {Palaversa}, {Palicio}, {Pallas-Quintela}, {Panahi},
  {Payne-Wardenaar}, {Pe{\~n}alosa Esteller}, {Penttil{\"a}}, {Pichon},
  {Piersimoni}, {Pineau}, {Plachy}, {Plum}, {Poggio}, {Prsa}, {Pulone},
  {Racero}, {Ragaini}, {Rainer}, {Raiteri}, {Rambaux}, {Ramos}, {Ramos-Lerate},
  {Re Fiorentin}, {Regibo}, {Richards}, {Rios Diaz}, {Ripepi}, {Riva}, {Rix},
  {Rixon}, {Robichon}, {Robin}, {Robin}, {Roelens}, {Rogues}, {Rohrbasser},
  {Romero-G{\'o}mez}, {Rowell}, {Royer}, {Ruz Mieres}, {Rybicki}, {Sadowski},
  {S{\'a}ez N{\'u}{\~n}ez}, {Sagrist{\`a} Sell{\'e}s}, {Sahlmann}, {Salguero},
  {Samaras}, {Sanchez Gimenez}, {Sanna}, {Santove{\~n}a}, {Sarasso},
  {Schultheis}, {Sciacca}, {Segol}, {Segovia}, {S{\'e}gransan}, {Semeux},
  {Shahaf}, {Siddiqui}, {Siebert}, {Siltala}, {Silvelo}, {Slezak}, {Slezak},
  {Smart}, {Snaith}, {Solano}, {Solitro}, {Souami}, {Souchay}, {Spagna},
  {Spina}, {Spoto}, {Steele}, {Steidelm{\"u}ller}, {Stephenson}, {S{\"u}veges},
  {Surdej}, {Szabados}, {Szegedi-Elek}, {Taris}, {Taylor}, {Teixeira},
  {Tolomei}, {Tonello}, {Torra}, {Torra}, {Torralba Elipe}, {Trabucchi},
  {Tsounis}, {Turon}, {Ulla}, {Unger}, {Vaillant}, {van Dillen}, {van Reeven},
  {Vanel}, {Vecchiato}, {Viala}, {Vicente}, {Voutsinas}, {Weiler}, {Wevers},
  {Wyrzykowski}, {Yoldas}, {Yvard}, {Zhao}, {Zorec}, {Zucker}, \&
  {Zwitter}}]{gaiadr3}
{Gaia Collaboration}, {Vallenari}, A., {Brown}, A.~G.~A., {et~al.}
  2023{\natexlab{a}}, \bibinfo{title}{{Gaia Data Release 3. Summary of the
  content and survey properties},} \aap, 674, A1,
  \dodoi{10.1051/0004-6361/202243940}

% type= article
\bibitem[{ {Gaia Collaboration} {et~al.}(2023{\natexlab{b}}){Gaia
  Collaboration}, {Montegriffo}, {Bellazzini}, {De Angeli}, {Andrae},
  {Barstow}, {Bossini}, {Bragaglia}, {Burgess}, {Cacciari}, {Carrasco},
  {Chornay}, {Delchambre}, {Evans}, {Fouesneau}, {Fr{\'e}mat}, {Garabato},
  {Jordi}, {Manteiga}, {Massari}, {Palaversa}, {Pancino}, {Riello}, {Ruz
  Mieres}, {Sanna}, {Santove{\~n}a}, {Sordo}, {Vallenari}, {Walton}, {Brown},
  {Prusti}, {de Bruijne}, {Arenou}, {Babusiaux}, {Biermann}, {Creevey},
  {Ducourant}, {Eyer}, {Guerra}, {Hutton}, {Klioner}, {Lammers}, {Lindegren},
  {Luri}, {Mignard}, {Panem}, {Pourbaix}, {Randich}, {Sartoretti}, {Soubiran},
  {Tanga}, {Bailer-Jones}, {Bastian}, {Drimmel}, {Jansen}, {Katz}, {Lattanzi},
  {van Leeuwen}, {Bakker}, {Casta{\~n}eda}, {Fabricius}, {Galluccio},
  {Guerrier}, {Heiter}, {Masana}, {Messineo}, {Mowlavi}, {Nicolas},
  {Nienartowicz}, {Pailler}, {Panuzzo}, {Riclet}, {Roux}, {Seabroke},
  {Th{\'e}venin}, {Gracia-Abril}, {Portell}, {Teyssier}, {Altmann}, {Audard},
  {Bellas-Velidis}, {Benson}, {Berthier}, {Blomme}, {Busonero}, {Busso},
  {C{\'a}novas}, {Carry}, {Cellino}, {Cheek}, {Clementini}, {Damerdji},
  {Davidson}, {de Teodoro}, {Nu{\~n}ez Campos}, {Dell'Oro}, {Esquej},
  {Fern{\'a}ndez-Hern{\'a}ndez}, {Fraile}, {Garc{\'\i}a-Lario}, {Gosset},
  {Haigron}, {Halbwachs}, {Hambly}, {Harrison}, {Hern{\'a}ndez}, {Hestroffer},
  {Hodgkin}, {Holl}, {Jan{\ss}en}, {Jevardat de Fombelle}, {Jordan},
  {Krone-Martins}, {Lanzafame}, {L{\"o}ffler}, {Marchal}, {Marrese},
  {Moitinho}, {Muinonen}, {Osborne}, {Pauwels}, {Recio-Blanco}, {Reyl{\'e}},
  {Rimoldini}, {Roegiers}, {Rybizki}, {Sarro}, {Siopis}, {Smith}, {Sozzetti},
  {Utrilla}, {van Leeuwen}, {Abbas}, {{\'A}brah{\'a}m}, {Abreu Aramburu},
  {Aerts}, {Aguado}, {Ajaj}, {Aldea-Montero}, {Altavilla}, {{\'A}lvarez},
  {Alves}, {Anderson}, {Anglada Varela}, {Antoja}, {Baines}, {Baker},
  {Balaguer-N{\'u}{\~n}ez}, {Balbinot}, {Balog}, {Barache}, {Barbato},
  {Barros}, {Bartolom{\'e}}, {Bassilana}, {Bauchet}, {Becciani}, {Berihuete},
  {Bernet}, {Bertone}, {Bianchi}, {Binnenfeld}, {Blanco-Cuaresma}, {Boch},
  {Bombrun}, {Bouquillon}, {Bramante}, {Breedt}, {Bressan}, {Brouillet},
  {Brugaletta}, {Bucciarelli}, {Burlacu}, {Butkevich}, {Buzzi}, {Caffau},
  {Cancelliere}, {Cantat-Gaudin}, {Carballo}, {Carlucci}, {Carnerero},
  {Casamiquela}, {Castellani}, {Castro-Ginard}, {Chaoul}, {Charlot}, {Chemin},
  {Chiaramida}, {Chiavassa}, {Comoretto}, {Contursi}, {Cooper}, {Cornez},
  {Cowell}, {Crifo}, {Cropper}, {Crosta}, {Crowley}, {Dafonte}, {Dapergolas},
  {David}, {de Laverny}, {De Luise}, {De March}, {De Ridder}, {de Souza}, {de
  Torres}, {del Peloso}, {del Pozo}, {Delbo}, {Delgado}, {Delisle}, {Demouchy},
  {Dharmawardena}, {Diakite}, {Diener}, {Distefano}, {Dolding}, {Enke},
  {Fabre}, {Fabrizio}, {Faigler}, {Fedorets}, {Fernique}, {Figueras},
  {Fournier}, {Fouron}, {Fragkoudi}, {Gai}, {Garcia-Gutierrez},
  {Garcia-Reinaldos}, {Garc{\'\i}a-Torres}, {Garofalo}, {Gavel}, {Gavras},
  {Gerlach}, {Geyer}, {Giacobbe}, {Gilmore}, {Girona}, {Giuffrida}, {Gomel},
  {Gomez}, {Gonz{\'a}lez-N{\'u}{\~n}ez}, {Gonz{\'a}lez-Santamar{\'\i}a},
  {Gonz{\'a}lez-Vidal}, {Granvik}, {Guillout}, {Guiraud},
  {Guti{\'e}rrez-S{\'a}nchez}, {Guy}, {Hatzidimitriou}, {Hauser}, {Haywood},
  {Helmer}, {Helmi}, {Sarmiento}, {Hidalgo}, {H{\l}adczuk}, {Hobbs}, {Holland},
  {Huckle}, {Jardine}, {Jasniewicz}, {Jean-Antoine Piccolo},
  {Jim{\'e}nez-Arranz}, {Juaristi Campillo}, {Julbe}, {Karbevska}, {Kervella},
  {Khanna}, {Kordopatis}, {Korn}, {K{\'o}sp{\'a}l}, {Kostrzewa-Rutkowska},
  {Kruszy{\'n}ska}, {Kun}, {Laizeau}, {Lambert}, {Lanza}, {Lasne}, {Le
  Campion}, {Lebreton}, {Lebzelter}, {Leccia}, {Leclerc}, {Lecoeur-Taibi},
  {Liao}, {Licata}, {Lindstr{\'o}m}, {Lister}, {Livanou}, {Lobel}, {Lorca},
  {Loup}, {Madrero Pardo}, {Magdaleno Romeo}, {Managau}, {Mann}, {Marchant},
  {Marconi}, {Marcos}, {Marcos Santos}, {Mar{\'\i}n Pina}, {Marinoni},
  {Marocco}, {Marshall}, {Martin Polo}, {Mart{\'\i}n-Fleitas}, {Marton},
  {Mary}, {Masip}, {Mastrobuono-Battisti}, {Mazeh}, {McMillan}, {Messina},
  {Michalik}, {Millar}, {Mints}, {Molina}, {Molinaro}, {Moln{\'a}r}, {Monari},
  {Mongui{\'o}}, {Montero}, {Mor}, {Mora}, {Morbidelli}, {Morel}, {Morris},
  {Muraveva}, {Murphy}, {Musella}, {Nagy}, {Noval}, {Oca{\~n}a}, {Ogden},
  {Ordenovic}, {Osinde}, {Pagani}, {Pagano}, {Palicio}, {Pallas-Quintela},
  {Panahi}, {Payne-Wardenaar}, {Pe{\~n}alosa Esteller}, {Penttil{\"a}},
  {Pichon}, {Piersimoni}, {Pineau}, {Plachy}, {Plum}, {Poggio}, {Prsa},
  {Pulone}, {Racero}, {Ragaini}, {Rainer}, {Raiteri}, {Ramos}, {Ramos-Lerate},
  {Re Fiorentin}, {Regibo}, {Richards}, {Rios Diaz}, {Ripepi}, {Riva}, {Rix},
  {Rixon}, {Robichon}, {Robin}, {Robin}, {Roelens}, {Rogues}, {Rohrbasser},
  {Romero-G{\'o}mez}, {Rowell}, {Royer}, {Rybicki}, {Sadowski}, {S{\'a}ez
  N{\'u}{\~n}ez}, {Sagrist{\`a} Sell{\'e}s}, {Sahlmann}, {Salguero}, {Samaras},
  {Sanchez Gimenez}, {Sarasso}, {Schultheis}, {Sciacca}, {Segol}, {Segovia},
  {S{\'e}gransan}, {Semeux}, {Shahaf}, {Siddiqui}, {Siebert}, {Siltala},
  {Silvelo}, {Slezak}, {Slezak}, {Smart}, {Snaith}, {Solano}, {Solitro},
  {Souami}, {Souchay}, {Spagna}, {Spina}, {Spoto}, {Steele},
  {Steidelm{\"u}ller}, {Stephenson}, {S{\"u}veges}, {Surdej}, {Szabados},
  {Szegedi-Elek}, {Taris}, {Taylor}, {Teixeira}, {Tolomei}, {Tonello}, {Torra},
  {Torra}, {Torralba Elipe}, {Trabucchi}, {Tsounis}, {Turon}, {Ulla}, {Unger},
  {Vaillant}, {van Dillen}, {van Reeven}, {Vanel}, {Vecchiato}, {Viala},
  {Vicente}, {Voutsinas}, {Wevers}, {Wyrzykowski}, {Yoldas}, {Yvard}, {Zhao},
  {Zorec}, {Zucker}, \& {Zwitter}}]{gaiaphot2023}
{Gaia Collaboration}, {Montegriffo}, P., {Bellazzini}, M., {et~al.}
  2023{\natexlab{b}}, \bibinfo{title}{{Gaia Data Release 3. The Galaxy in your
  preferred colours: Synthetic photometry from Gaia low-resolution spectra},}
  \aap, 674, A33, \dodoi{10.1051/0004-6361/202243709}

% type= article
\bibitem[{ {Gaia Collaboration} {et~al.}(2023{\natexlab{c}}){Gaia
  Collaboration}, {Arenou}, {Babusiaux}, {Barstow}, {Faigler}, {Jorissen},
  {Kervella}, {Mazeh}, {Mowlavi}, {Panuzzo}, {Sahlmann}, {Shahaf}, {Sozzetti},
  {Bauchet}, {Damerdji}, {Gavras}, {Giacobbe}, {Gosset}, {Halbwachs}, {Holl},
  {Lattanzi}, {Leclerc}, {Morel}, {Pourbaix}, {Re Fiorentin}, {Sadowski},
  {S{\'e}gransan}, {Siopis}, {Teyssier}, {Zwitter}, {Planquart}, {Brown},
  {Vallenari}, {Prusti}, {de Bruijne}, {Biermann}, {Creevey}, {Ducourant},
  {Evans}, {Eyer}, {Guerra}, {Hutton}, {Jordi}, {Klioner}, {Lammers},
  {Lindegren}, {Luri}, {Mignard}, {Panem}, {Randich}, {Sartoretti}, {Soubiran},
  {Tanga}, {Walton}, {Bailer-Jones}, {Bastian}, {Drimmel}, {Jansen}, {Katz},
  {van Leeuwen}, {Bakker}, {Cacciari}, {Casta{\~n}eda}, {De Angeli},
  {Fabricius}, {Fouesneau}, {Fr{\'e}mat}, {Galluccio}, {Guerrier}, {Heiter},
  {Masana}, {Messineo}, {Nicolas}, {Nienartowicz}, {Pailler}, {Riclet}, {Roux},
  {Seabroke}, {Sordo}, {Th{\'e}venin}, {Gracia-Abril}, {Portell}, {Altmann},
  {Andrae}, {Audard}, {Bellas-Velidis}, {Benson}, {Berthier}, {Blomme},
  {Burgess}, {Busonero}, {Busso}, {C{\'a}novas}, {Carry}, {Cellino}, {Cheek},
  {Clementini}, {Davidson}, {de Teodoro}, {Nu{\~n}ez Campos}, {Delchambre},
  {Dell'Oro}, {Esquej}, {Fern{\'a}ndez-Hern{\'a}ndez}, {Fraile}, {Garabato},
  {Garc{\'\i}a-Lario}, {Haigron}, {Hambly}, {Harrison}, {Hern{\'a}ndez},
  {Hestroffer}, {Hodgkin}, {Jan{\ss}en}, {Jevardat de Fombelle}, {Jordan},
  {Krone-Martins}, {Lanzafame}, {L{\"o}ffler}, {Marchal}, {Marrese},
  {Moitinho}, {Muinonen}, {Osborne}, {Pancino}, {Pauwels}, {Recio-Blanco},
  {Reyl{\'e}}, {Riello}, {Rimoldini}, {Roegiers}, {Rybizki}, {Sarro}, {Smith},
  {Utrilla}, {van Leeuwen}, {Abbas}, {{\'A}brah{\'a}m}, {Abreu Aramburu},
  {Aerts}, {Aguado}, {Ajaj}, {Aldea-Montero}, {Altavilla}, {{\'A}lvarez},
  {Alves}, {Anders}, {Anderson}, {Anglada Varela}, {Antoja}, {Baines}, {Baker},
  {Balaguer-N{\'u}{\~n}ez}, {Balbinot}, {Balog}, {Barache}, {Barbato},
  {Barros}, {Bartolom{\'e}}, {Bassilana}, {Becciani}, {Bellazzini},
  {Berihuete}, {Bernet}, {Bertone}, {Bianchi}, {Binnenfeld}, {Blanco-Cuaresma},
  {Blazere}, {Boch}, {Bombrun}, {Bossini}, {Bouquillon}, {Bragaglia},
  {Bramante}, {Breedt}, {Bressan}, {Brouillet}, {Brugaletta}, {Bucciarelli},
  {Burlacu}, {Butkevich}, {Buzzi}, {Caffau}, {Cancelliere}, {Cantat-Gaudin},
  {Carballo}, {Carlucci}, {Carnerero}, {Carrasco}, {Casamiquela}, {Castellani},
  {Castro-Ginard}, {Chaoul}, {Charlot}, {Chemin}, {Chiaramida}, {Chiavassa},
  {Chornay}, \& {Comoretto}}]{gaiamulti2023}
{Gaia Collaboration}, {Arenou}, F., {Babusiaux}, C., {et~al.}
  2023{\natexlab{c}}, \bibinfo{title}{{Gaia Data Release 3. Stellar
  multiplicity, a teaser for the hidden treasure},} \aap, 674, A34,
  \dodoi{10.1051/0004-6361/202243782}

% type= article
\bibitem[{E. {Gaidos} {et~al.}(2014){Gaidos}, {Mann}, {L{\'e}pine}, {Buccino},
  {James}, {Ansdell}, {Petrucci}, {Mauas}, \& {Hilton}}]{gaidos2014}
{Gaidos}, E., {Mann}, A.~W., {L{\'e}pine}, S., {et~al.} 2014,
  \bibinfo{title}{{Trumpeting M dwarfs with CONCH-SHELL: a catalogue of nearby
  cool host-stars for habitable exoplanets and life},} \mnras, 443, 2561,
  \dodoi{10.1093/mnras/stu1313}

% type= article
\bibitem[{A. {Gallenne} {et~al.}(2023){Gallenne}, {M{\'e}rand}, {Kervella},
  {Graczyk}, {Pietrzy{\'n}ski}, {Gieren}, \& {Pilecki}}]{gallenne2023}
{Gallenne}, A., {M{\'e}rand}, A., {Kervella}, P., {et~al.} 2023,
  \bibinfo{title}{{The Araucaria project: High-precision orbital parallaxes and
  masses of binary stars. I. VLTI/GRAVITY observations of ten double-lined
  spectroscopic binaries},} \aap, 672, A119,
  \dodoi{10.1051/0004-6361/202245712}

% type= article
\bibitem[{F. {Gallet} \& J. {Bouvier}(2015){Gallet} \& {Bouvier}}]{gallet2015}
{Gallet}, F., \& {Bouvier}, J. 2015, \bibinfo{title}{{Improved angular momentum
  evolution model for solar-like stars. II. Exploring the mass dependence},}
  \aap, 577, A98, \dodoi{10.1051/0004-6361/201525660}

% type= article
\bibitem[{S. {Gao} {et~al.}(2014){Gao}, {Liu}, {Zhang}, {Justham}, {Deng}, \&
  {Yang}}]{gao2014}
{Gao}, S., {Liu}, C., {Zhang}, X., {et~al.} 2014, \bibinfo{title}{{The Binarity
  of Milky Way F,G,K Stars as a Function of Effective Temperature and
  Metallicity},} \apjl, 788, L37, \dodoi{10.1088/2041-8205/788/2/L37}

% type= article
\bibitem[{X. {Gao} {et~al.}(2020){Gao}, {Lind}, {Amarsi}, {Buder},
  {Bland-Hawthorn}, {Campbell}, {Asplund}, {Casey}, {de Silva}, {Freeman},
  {Hayden}, {Lewis}, {Martell}, {Simpson}, {Sharma}, {Zucker}, {Zwitter},
  {Horner}, {Munari}, {Nordlander}, {Stello}, {Ting}, {Traven}, {Wittenmyer},
  \& {GALAH Collaboration}}]{gao2020}
{Gao}, X., {Lind}, K., {Amarsi}, A.~M., {et~al.} 2020, \bibinfo{title}{{The
  GALAH survey: a new constraint on cosmological lithium and Galactic lithium
  evolution from warm dwarf stars},} \mnras, 497, L30,
  \dodoi{10.1093/mnrasl/slaa109}

% type= article
\bibitem[{A. {G{\'a}sp{\'a}r} {et~al.}(2016){G{\'a}sp{\'a}r}, {Rieke}, \&
  {Ballering}}]{gaspar2016}
{G{\'a}sp{\'a}r}, A., {Rieke}, G.~H., \& {Ballering}, N. 2016,
  \bibinfo{title}{{The Correlation between Metallicity and Debris Disk Mass},}
  \apj, 826, 171, \dodoi{10.3847/0004-637X/826/2/171}

% type= article
\bibitem[{A. {G{\'a}sp{\'a}r} {et~al.}(2009){G{\'a}sp{\'a}r}, {Rieke}, {Su},
  {Balog}, {Trilling}, {Muzzerole}, {Apai}, \& {Kelly}}]{gaspar2009}
{G{\'a}sp{\'a}r}, A., {Rieke}, G.~H., {Su}, K.~Y.~L., {et~al.} 2009,
  \bibinfo{title}{{The Low Level of Debris Disk Activity at the Time of the
  Late Heavy Bombardment: A Spitzer Study of Praesepe},} \apj, 697, 1578,
  \dodoi{10.1088/0004-637X/697/2/1578}

% type= article
\bibitem[{I. {Gautier} {et~al.}(2008){Gautier}, {Rebull}, {Stapelfeldt}, \&
  {Mainzer}}]{gautier2008}
{Gautier}, Thomas.~N., I., {Rebull}, L.~M., {Stapelfeldt}, K.~R., \& {Mainzer},
  A. 2008, \bibinfo{title}{{Spitzer-MIPS Observations of the
  {\ensuremath{\eta}} Chamaeleontis Young Association},} \apj, 683, 813,
  \dodoi{10.1086/589708}

% type= article
\bibitem[{I. {Gautier} {et~al.}(2007){Gautier}, {Rieke}, {Stansberry},
  {Bryden}, {Stapelfeldt}, {Werner}, {Beichman}, {Chen}, {Su}, {Trilling},
  {Patten}, \& {Roellig}}]{gautier2007}
{Gautier}, Thomas~N., I., {Rieke}, G.~H., {Stansberry}, J., {et~al.} 2007,
  \bibinfo{title}{{Far-Infrared Properties of M Dwarfs},} \apj, 667, 527,
  \dodoi{10.1086/520667}

% type= article
\bibitem[{M. {Gennaro} {et~al.}(2012){Gennaro}, {Prada Moroni}, \&
  {Tognelli}}]{gennaro2012}
{Gennaro}, M., {Prada Moroni}, P.~G., \& {Tognelli}, E. 2012,
  \bibinfo{title}{{Testing pre-main-sequence models: the power of a Bayesian
  approach},} \mnras, 420, 986, \dodoi{10.1111/j.1365-2966.2011.19945.x}

% type= article
\bibitem[{K.~V. {Getman} {et~al.}(2023){Getman}, {Feigelson}, \&
  {Garmire}}]{getman2023}
{Getman}, K.~V., {Feigelson}, E.~D., \& {Garmire}, G.~P. 2023,
  \bibinfo{title}{{Magnetic Activity-Rotation-Age-Mass Relations in
  Late-pre-main-sequence Stars},} \apj, 952, 63,
  \dodoi{10.3847/1538-4357/acd690}

% type= book
\bibitem[{D.~Y. {Gezari} {et~al.}(1993){Gezari}, {Schmitz}, {Pitts}, \&
  {Mead}}]{gezari1993}
{Gezari}, D.~Y., {Schmitz}, M., {Pitts}, P.~S., \& {Mead}, J.~M. 1993, {Catalog
  of infrared observations} (NASA, Greenbelt, MD USA)

% type= article
\bibitem[{L. {Ghezzi} {et~al.}(2010){Ghezzi}, {Cunha}, {Smith}, \& {de la
  Reza}}]{ghezzi2010}
{Ghezzi}, L., {Cunha}, K., {Smith}, V.~V., \& {de la Reza}, R. 2010,
  \bibinfo{title}{{Lithium Abundances in a Sample of Planet-hosting Dwarfs},}
  \apj, 724, 154, \dodoi{10.1088/0004-637X/724/1/154}

% type= article
\bibitem[{L. {Ghezzi} {et~al.}(2018){Ghezzi}, {Montet}, \&
  {Johnson}}]{ghezzi2018}
{Ghezzi}, L., {Montet}, B.~T., \& {Johnson}, J.~A. 2018,
  \bibinfo{title}{{Retired A Stars Revisited: An Updated Giant Planet
  Occurrence Rate as a Function of Stellar Metallicity and Mass},} \apj, 860,
  109, \dodoi{10.3847/1538-4357/aac37c}

% type= article
\bibitem[{E. {Gillen} {et~al.}(2020){Gillen}, {Briegal}, {Hodgkin},
  {Foreman-Mackey}, {Van Leeuwen}, {Jackman}, {McCormac}, {West}, {Queloz},
  {Bayliss}, {Goad}, {Watson}, {Wheatley}, {Belardi}, {Burleigh}, {Casewell},
  {Jenkins}, {Raynard}, {Smith}, {Tilbrook}, \& {Vines}}]{gillen2020}
{Gillen}, E., {Briegal}, J.~T., {Hodgkin}, S.~T., {et~al.} 2020,
  \bibinfo{title}{{NGTS clusters survey - I. Rotation in the young benchmark
  open cluster Blanco 1},} \mnras, 492, 1008, \dodoi{10.1093/mnras/stz3251}

% type= article
\bibitem[{M. {Gillon} {et~al.}(2016){Gillon}, {Jehin}, {Lederer}, {Delrez}, {de
  Wit}, {Burdanov}, {Van Grootel}, {Burgasser}, {Triaud}, {Opitom}, {Demory},
  {Sahu}, {Bardalez Gagliuffi}, {Magain}, \& {Queloz}}]{gillon2016}
{Gillon}, M., {Jehin}, E., {Lederer}, S.~M., {et~al.} 2016,
  \bibinfo{title}{{Temperate Earth-sized planets transiting a nearby ultracool
  dwarf star},} \nat, 533, 221, \dodoi{10.1038/nature17448}

% type= article
\bibitem[{R. {Glebocki} {et~al.}(2000){Glebocki}, {Gnacinski}, \&
  {Stawikowski}}]{glebocki2000}
{Glebocki}, R., {Gnacinski}, P., \& {Stawikowski}, A. 2000,
  \bibinfo{title}{{Catalog of Projected Rotational Velocities},} \actaa, 50,
  509

% type= article
\bibitem[{W. {Gliese}(1969){Gliese}}]{gliese1969}
{Gliese}, W. 1969, \bibinfo{title}{{Catalogue of Nearby Stars. Edition 1969},}
  Veroeffentlichungen des Astronomischen Rechen-Instituts Heidelberg, 22, 1

% type= misc
\bibitem[{W. {Gliese} \& H. {Jahrei{\ss}}(1991){Gliese} \&
  {Jahrei{\ss}}}]{gliese1991}
{Gliese}, W., \& {Jahrei{\ss}}, H. 1991, {Preliminary Version of the Third
  Catalogue of Nearby Stars},, On: The Astronomical Data Center CD-ROM:
  Selected Astronomical Catalogs, Vol. I; L.E. Brotzmann, S.E. Gesser (eds.),
  NASA/Astronomical Data Center, Goddard Space Flight Center, Greenbelt, MD

% type= article
\bibitem[{B. {Goldman} {et~al.}(2018){Goldman}, {R{\"o}ser}, {Schilbach},
  {Mo{\'o}r}, \& {Henning}}]{goldman2018}
{Goldman}, B., {R{\"o}ser}, S., {Schilbach}, E., {Mo{\'o}r}, A.~C., \&
  {Henning}, T. 2018, \bibinfo{title}{{A Large Moving Group within the Lower
  Centaurus Crux Association},} \apj, 868, 32, \dodoi{10.3847/1538-4357/aae64c}

% type= article
\bibitem[{D.~A. {Golimowski} {et~al.}(2006){Golimowski}, {Ardila}, {Krist},
  {Clampin}, {Ford}, {Illingworth}, {Bartko}, {Ben{\'\i}tez}, {Blakeslee},
  {Bouwens}, {Bradley}, {Broadhurst}, {Brown}, {Burrows}, {Cheng}, {Cross},
  {Demarco}, {Feldman}, {Franx}, {Goto}, {Gronwall}, {Hartig}, {Holden},
  {Homeier}, {Infante}, {Jee}, {Kimble}, {Lesser}, {Martel}, {Mei},
  {Menanteau}, {Meurer}, {Miley}, {Motta}, {Postman}, {Rosati}, {Sirianni},
  {Sparks}, {Tran}, {Tsvetanov}, {White}, {Zheng}, \& {Zirm}}]{golimowski06}
{Golimowski}, D.~A., {Ardila}, D.~R., {Krist}, J.~E., {et~al.} 2006,
  \bibinfo{title}{{Hubble Space Telescope ACS Multiband Coronagraphic Imaging
  of the Debris Disk around {\ensuremath{\beta}} Pictoris},} \aj, 131, 3109,
  \dodoi{10.1086/503801}

% type= article
\bibitem[{A. {Golovin} {et~al.}(2023){Golovin}, {Reffert}, {Just}, {Jordan},
  {Vani}, \& {Jahrei{\ss}}}]{gondoin2023}
{Golovin}, A., {Reffert}, S., {Just}, A., {et~al.} 2023, \bibinfo{title}{{The
  Fifth Catalogue of Nearby Stars (CNS5)},} \aap, 670, A19,
  \dodoi{10.1051/0004-6361/202244250}

% type= article
\bibitem[{J. {Gomes da Silva} {et~al.}(2014){Gomes da Silva}, {Santos},
  {Boisse}, {Dumusque}, \& {Lovis}}]{gomes2014}
{Gomes da Silva}, J., {Santos}, N.~C., {Boisse}, I., {Dumusque}, X., \&
  {Lovis}, C. 2014, \bibinfo{title}{{On the long-term correlation between the
  flux in the Ca ii H \& K and H{\ensuremath{\alpha}} lines for FGK stars},}
  \aap, 566, A66, \dodoi{10.1051/0004-6361/201322697}

% type= article
\bibitem[{J. {Gomes da Silva} {et~al.}(2021){Gomes da Silva}, {Santos},
  {Adibekyan}, {Sousa}, {Campante}, {Figueira}, {Bossini}, {Delgado-Mena},
  {Monteiro}, {de Laverny}, {Recio-Blanco}, \& {Lovis}}]{gomes2021}
{Gomes da Silva}, J., {Santos}, N.~C., {Adibekyan}, V., {et~al.} 2021,
  \bibinfo{title}{{Stellar chromospheric activity of 1674 FGK stars from the
  AMBRE-HARPS sample. I. A catalogue of homogeneous chromospheric activity},}
  \aap, 646, A77, \dodoi{10.1051/0004-6361/202039765}

% type= article
\bibitem[{P. {Gondoin}(2020){Gondoin}}]{gondoin2020b}
{Gondoin}, P. 2020, \bibinfo{title}{{Chromospheric activity of nearby Sun-like
  stars. R$^{'}$$_{HK}$ index signature of a recent burst of star formation},}
  \aap, 641, A110, \dodoi{10.1051/0004-6361/202038291}

% type= article
\bibitem[{J.~F. {Gonz{\'a}lez} \& H. {Levato}(2009){Gonz{\'a}lez} \&
  {Levato}}]{gonzalez2009}
{Gonz{\'a}lez}, J.~F., \& {Levato}, H. 2009, \bibinfo{title}{{Spectroscopic
  study of the open cluster Blanco 1},} \aap, 507, 541,
  \dodoi{10.1051/0004-6361/200912772}

% type= article
\bibitem[{J. {Gonz{\'a}lez-Payo} {et~al.}(2024){Gonz{\'a}lez-Payo},
  {Caballero}, {Gorgas}, {Cort{\'e}s-Contreras}, {G{\'a}lvez-Ortiz}, \&
  {Cifuentes}}]{gonzalez2024}
{Gonz{\'a}lez-Payo}, J., {Caballero}, J.~A., {Gorgas}, J., {et~al.} 2024,
  \bibinfo{title}{{Multiplicity of stars with planets in the solar
  neighbourhood},} \aap, 689, A302, \dodoi{10.1051/0004-6361/202450048}

% type= article
\bibitem[{K.~D. {Gordon} {et~al.}(2021){Gordon}, {Misselt}, {Bouwman},
  {Clayton}, {Decleir}, {Hines}, {Pendleton}, {Rieke}, {Smith}, \&
  {Whittet}}]{gordon2021}
{Gordon}, K.~D., {Misselt}, K.~A., {Bouwman}, J., {et~al.} 2021,
  \bibinfo{title}{{Milky Way Mid-Infrared Spitzer Spectroscopic Extinction
  Curves: Continuum and Silicate Features},} \apj, 916, 33,
  \dodoi{10.3847/1538-4357/ac00b7}

% type= article
\bibitem[{N. {Gorlova} {et~al.}(2007){Gorlova}, {Balog}, {Rieke}, {Muzerolle},
  {Su}, {Ivanov}, \& {Young}}]{gorlova2007}
{Gorlova}, N., {Balog}, Z., {Rieke}, G.~H., {et~al.} 2007,
  \bibinfo{title}{{Debris Disks in NGC 2547},} \apj, 670, 516,
  \dodoi{10.1086/521671}

% type= article
\bibitem[{N. {Gorlova} {et~al.}(2006){Gorlova}, {Rieke}, {Muzerolle},
  {Stauffer}, {Siegler}, {Young}, \& {Stansberry}}]{gorlova2006}
{Gorlova}, N., {Rieke}, G.~H., {Muzerolle}, J., {et~al.} 2006,
  \bibinfo{title}{{Spitzer 24 {$\mu$}m Survey of Debris Disks in the
  Pleiades},} \apj, 649, 1028, \dodoi{10.1086/506373}

% type= article
\bibitem[{N. {Gorlova} {et~al.}(2004){Gorlova}, {Padgett}, {Rieke},
  {Muzerolle}, {Morrison}, {Gordon}, {Engelbracht}, {Hines}, {Hinz},
  {Noriega-Crespo}, {Rebull}, {Stansberry}, {Stapelfeldt}, {Su}, \&
  {Young}}]{gorlova2004}
{Gorlova}, N., {Padgett}, D.~L., {Rieke}, G.~H., {et~al.} 2004,
  \bibinfo{title}{{New Debris-Disk Candidates: 24 Micron Stellar Excesses at
  100 Million years},} \apjs, 154, 448, \dodoi{10.1086/422822}

% type= article
\bibitem[{S. {Gossage} {et~al.}(2018){Gossage}, {Conroy}, {Dotter}, {Choi},
  {Rosenfield}, {Cargile}, \& {Dolphin}}]{gossage2018}
{Gossage}, S., {Conroy}, C., {Dotter}, A., {et~al.} 2018, \bibinfo{title}{{Age
  Determinations of the Hyades, Praesepe, and Pleiades via MESA Models with
  Rotation},} \apj, 863, 67, \dodoi{10.3847/1538-4357/aad0a0}

% type= article
\bibitem[{E. {Gosset} {et~al.}(2024){Gosset}, {Damerdji}, {Morel},
  {Delchambre}, {Halbwachs}, {Sadowski}, {Pourbaix}, {Sozzetti}, {Panuzzo}, \&
  {Arenou}}]{gaiasb2024}
{Gosset}, E., {Damerdji}, Y., {Morel}, T., {et~al.} 2024, \bibinfo{title}{{Gaia
  Data Release 3: spectroscopic binary-star orbital solutions and the SB1
  processing chain},} arXiv e-prints, arXiv:2410.14372,
  \dodoi{10.48550/arXiv.2410.14372}

% type= article
\bibitem[{R.~O. {Gray} {et~al.}(2006){Gray}, {Corbally}, {Garrison},
  {McFadden}, {Bubar}, {McGahee}, {O'Donoghue}, \& {Knox}}]{gray2006}
{Gray}, R.~O., {Corbally}, C.~J., {Garrison}, R.~F., {et~al.} 2006,
  \bibinfo{title}{{Contributions to the Nearby Stars (NStars) Project:
  Spectroscopy of Stars Earlier than M0 within 40 pc-The Southern Sample},}
  \aj, 132, 161, \dodoi{10.1086/504637}

% type= article
\bibitem[{R.~O. {Gray} {et~al.}(2003){Gray}, {Corbally}, {Garrison},
  {McFadden}, \& {Robinson}}]{gray2003}
{Gray}, R.~O., {Corbally}, C.~J., {Garrison}, R.~F., {McFadden}, M.~T., \&
  {Robinson}, P.~E. 2003, \bibinfo{title}{{Contributions to the Nearby Stars
  (NStars) Project: Spectroscopy of Stars Earlier than M0 within 40 Parsecs:
  The Northern Sample. I.},} \aj, 126, 2048, \dodoi{10.1086/378365}

% type= article
\bibitem[{J.~S. {Greaves} {et~al.}(2004){Greaves}, {Wyatt}, {Holland}, \&
  {Dent}}]{greaves2004}
{Greaves}, J.~S., {Wyatt}, M.~C., {Holland}, W.~S., \& {Dent}, W.~R.~F. 2004,
  \bibinfo{title}{{The debris disc around {\ensuremath{\tau}} Ceti: a massive
  analogue to the Kuiper Belt},} \mnras, 351, L54,
  \dodoi{10.1111/j.1365-2966.2004.07957.x}

% type= article
\bibitem[{J.~S. {Greaves} {et~al.}(1998){Greaves}, {Holland},
  {Moriarty-Schieven}, {Jenness}, {Dent}, {Zuckerman}, {McCarthy}, {Webb},
  {Butner}, {Gear}, \& {Walker}}]{greaves1998}
{Greaves}, J.~S., {Holland}, W.~S., {Moriarty-Schieven}, G., {et~al.} 1998,
  \bibinfo{title}{{A Dust Ring around ɛ Eridani: Analog to the Young Solar
  System},} \apjl, 506, L133, \dodoi{10.1086/311652}

% type= article
\bibitem[{G.~M. {Green} {et~al.}(2019){Green}, {Schlafly}, {Zucker}, {Speagle},
  \& {Finkbeiner}}]{green2019}
{Green}, G.~M., {Schlafly}, E., {Zucker}, C., {Speagle}, J.~S., \&
  {Finkbeiner}, D. 2019, \bibinfo{title}{{A 3D Dust Map Based on Gaia,
  Pan-STARRS 1, and 2MASS},} \apj, 887, 93, \dodoi{10.3847/1538-4357/ab5362}

% type= article
\bibitem[{M.~J. {Griffin} {et~al.}(2010){Griffin}, {Abergel}, {Abreu}, {Ade},
  {Andr{\'e}}, {Augueres}, {Babbedge}, {Bae}, {Baillie}, {Baluteau}, {Barlow},
  {Bendo}, {Benielli}, {Bock}, {Bonhomme}, {Brisbin}, {Brockley-Blatt},
  {Caldwell}, {Cara}, {Castro-Rodriguez}, {Cerulli}, {Chanial}, {Chen},
  {Clark}, {Clements}, {Clerc}, {Coker}, {Communal}, {Conversi}, {Cox},
  {Crumb}, {Cunningham}, {Daly}, {Davis}, {de Antoni}, {Delderfield}, {Devin},
  {di Giorgio}, {Didschuns}, {Dohlen}, {Donati}, {Dowell}, {Dowell}, {Duband},
  {Dumaye}, {Emery}, {Ferlet}, {Ferrand}, {Fontignie}, {Fox}, {Franceschini},
  {Frerking}, {Fulton}, {Garcia}, {Gastaud}, {Gear}, {Glenn}, {Goizel},
  {Griffin}, {Grundy}, {Guest}, {Guillemet}, {Hargrave}, {Harwit}, {Hastings},
  {Hatziminaoglou}, {Herman}, {Hinde}, {Hristov}, {Huang}, {Imhof}, {Isaak},
  {Israelsson}, {Ivison}, {Jennings}, {Kiernan}, {King}, {Lange}, {Latter},
  {Laurent}, {Laurent}, {Leeks}, {Lellouch}, {Levenson}, {Li}, {Li},
  {Lilienthal}, {Lim}, {Liu}, {Lu}, {Madden}, {Mainetti}, {Marliani}, {McKay},
  {Mercier}, {Molinari}, {Morris}, {Moseley}, {Mulder}, {Mur}, {Naylor},
  {Nguyen}, {O'Halloran}, {Oliver}, {Olofsson}, {Olofsson}, {Orfei}, {Page},
  {Pain}, {Panuzzo}, {Papageorgiou}, {Parks}, {Parr-Burman}, {Pearce},
  {Pearson}, {P{\'e}rez-Fournon}, {Pinsard}, {Pisano}, {Podosek}, {Pohlen},
  {Polehampton}, {Pouliquen}, {Rigopoulou}, {Rizzo}, {Roseboom}, {Roussel},
  {Rowan-Robinson}, {Rownd}, {Saraceno}, {Sauvage}, {Savage}, {Savini},
  {Sawyer}, {Scharmberg}, {Schmitt}, {Schneider}, {Schulz}, {Schwartz},
  {Shafer}, {Shupe}, {Sibthorpe}, {Sidher}, {Smith}, {Smith}, {Smith},
  {Spencer}, {Stobie}, {Sudiwala}, {Sukhatme}, {Surace}, {Stevens}, {Swinyard},
  {Trichas}, {Tourette}, {Triou}, {Tseng}, {Tucker}, {Turner}, {Vaccari},
  {Valtchanov}, {Vigroux}, {Virique}, {Voellmer}, {Walker}, {Ward}, {Waskett},
  {Weilert}, {Wesson}, {White}, {Whitehouse}, {Wilson}, {Winter}, {Woodcraft},
  {Wright}, {Xu}, {Zavagno}, {Zemcov}, {Zhang}, \& {Zonca}}]{griffin2010}
{Griffin}, M.~J., {Abergel}, A., {Abreu}, A., {et~al.} 2010,
  \bibinfo{title}{{The Herschel-SPIRE instrument and its in-flight
  performance},} \aap, 518, L3, \dodoi{10.1051/0004-6361/201014519}

% type= article
\bibitem[{R.~E.~M. {Griffin} \& R.~F. {Griffin}(2011){Griffin} \&
  {Griffin}}]{griffin2011}
{Griffin}, R.~E.~M., \& {Griffin}, R.~F. 2011, \bibinfo{title}{{Composite
  spectra: XVII. 12 Comae, a member of the Coma open cluster},} Astronomische
  Nachrichten, 332, 105, \dodoi{10.1002/asna.201011514}

% type= article
\bibitem[{M.~A.~T. {Groenewegen}(2021){Groenewegen}}]{groenewegen2021}
{Groenewegen}, M.~A.~T. 2021, \bibinfo{title}{{The parallax zero-point offset
  from Gaia EDR3 data},} \aap, 654, A20, \dodoi{10.1051/0004-6361/202140862}

% type= article
\bibitem[{D. {Gruner} {et~al.}(2023){Gruner}, {Barnes}, \&
  {Weingrill}}]{gruner2023}
{Gruner}, D., {Barnes}, S.~A., \& {Weingrill}, J. 2023, \bibinfo{title}{{New
  insights into the rotational evolution of near-solar age stars from the open
  cluster M 67},} \aap, 672, A159, \dodoi{10.1051/0004-6361/202345942}

% type= article
\bibitem[{M. {G{\"u}del}(2004){G{\"u}del}}]{gudel2004}
{G{\"u}del}, M. 2004, \bibinfo{title}{{X-ray astronomy of stellar coronae},}
  \aapr, 12, 71, \dodoi{10.1007/s00159-004-0023-2}

% type= article
\bibitem[{G. {Guiglion} {et~al.}(2016){Guiglion}, {de Laverny}, {Recio-Blanco},
  {Worley}, {De Pascale}, {Masseron}, {Prantzos}, \&
  {Mikolaitis}}]{guiglion2016}
{Guiglion}, G., {de Laverny}, P., {Recio-Blanco}, A., {et~al.} 2016,
  \bibinfo{title}{{The AMBRE project: Constraining the lithium evolution in the
  Milky Way},} \aap, 595, A18, \dodoi{10.1051/0004-6361/201628919}

% type= article
\bibitem[{Y. {Guo} {et~al.}(2022){Guo}, {Liu}, {Wang}, {Wang}, {Zhang}, {Ji},
  {Han}, \& {Chen}}]{guo2022}
{Guo}, Y., {Liu}, C., {Wang}, L., {et~al.} 2022, \bibinfo{title}{{The
  statistical properties of early-type stars from LAMOST DR8},} \aap, 667, A44,
  \dodoi{10.1051/0004-6361/202244300}

% type= article
\bibitem[{M.~L. {Guti{\'e}rrez Albarr{\'a}n} {et~al.}(2024){Guti{\'e}rrez
  Albarr{\'a}n}, {Montes}, {Tabernero}, {Gonz{\'a}lez Hern{\'a}ndez}, {Marfil},
  {Frasca}, {Lanzafame}, {Klutsch}, {Franciosini}, {Randich}, {Smiljanic},
  {Korn}, {Gilmore}, {Alfaro}, {Bensby}, {Biazzo}, {Casey}, {Carraro},
  {Damiani}, {Feltzing}, {Fran{\c{c}}ois}, {Jim{\'e}nez Esteban}, {Magrini},
  {Morbidelli}, {Prisinzano}, {Prusti}, {Worley}, \&
  {Zaggia}}]{gutierrezalbarran2024}
{Guti{\'e}rrez Albarr{\'a}n}, M.~L., {Montes}, D., {Tabernero}, H.~M., {et~al.}
  2024, \bibinfo{title}{{The Gaia-ESO Survey: Calibrating the lithium-age
  relation with open clusters and associations. II. Expanded cluster sample and
  final membership selection},} \aap, 685, A83,
  \dodoi{10.1051/0004-6361/202348438}

% type= article
\bibitem[{A. {Hahlin} {et~al.}(2021){Hahlin}, {Kochukhov}, {Alecian}, {Morin},
  \& {BinaMIcS Collaboration}}]{hahlin2021}
{Hahlin}, A., {Kochukhov}, O., {Alecian}, E., {Morin}, J., \& {BinaMIcS
  Collaboration}. 2021, \bibinfo{title}{{Magnetic field of the eclipsing binary
  UV Piscium},} \aap, 650, A197, \dodoi{10.1051/0004-6361/202140832}

% type= article
\bibitem[{J.-L. {Halbwachs} {et~al.}(2023){Halbwachs}, {Pourbaix}, {Arenou},
  {Galluccio}, {Guillout}, {Bauchet}, {Marchal}, {Sadowski}, \&
  {Teyssier}}]{gaiabin2023}
{Halbwachs}, J.-L., {Pourbaix}, D., {Arenou}, F., {et~al.} 2023,
  \bibinfo{title}{{Gaia Data Release 3. Astrometric binary star processing},}
  \aap, 674, A9, \dodoi{10.1051/0004-6361/202243969}

% type= article
\bibitem[{A.~S. {Hales} {et~al.}(2022){Hales}, {Marino}, {Sheehan}, {Ulloa},
  {P{\'e}rez}, {Matr{\`a}}, {Kral}, {Wyatt}, {Dent}, \&
  {Carpenter}}]{hales2022}
{Hales}, A.~S., {Marino}, S., {Sheehan}, P.~D., {et~al.} 2022,
  \bibinfo{title}{{ALMA Observations of the HD 110058 Debris Disk},} \apj, 940,
  161, \dodoi{10.3847/1538-4357/ac9cd3}

% type= article
\bibitem[{J.~C. {Hall}(2008){Hall}}]{hall2008}
{Hall}, J.~C. 2008, \bibinfo{title}{{Stellar Chromospheric Activity},} Living
  Reviews in Solar Physics, 5, 2, \dodoi{10.12942/lrsp-2008-2}

% type= article
\bibitem[{M.~A. {Hamdy} {et~al.}(1993){Hamdy}, {Abo Elazm}, \&
  {Saad}}]{hamdy1993}
{Hamdy}, M.~A., {Abo Elazm}, M.~S., \& {Saad}, S.~M. 1993, \bibinfo{title}{{A
  Catalogue of Spectral Classification and Photometric Data of B-Type Stars},}
  \apss, 203, 53, \dodoi{10.1007/BF00659414}

% type= article
\bibitem[{P. {Hartigan} {et~al.}(1989){Hartigan}, {Hartmann}, {Kenyon},
  {Hewett}, \& {Stauffer}}]{hartigan1989}
{Hartigan}, P., {Hartmann}, L., {Kenyon}, S., {Hewett}, R., \& {Stauffer}, J.
  1989, \bibinfo{title}{{How to Unveil a T Tauri Star},} \apjs, 70, 899,
  \dodoi{10.1086/191361}

% type= article
\bibitem[{W.~I. {Hartkopf} {et~al.}(2001){Hartkopf}, {Mason}, \&
  {Worley}}]{ORB62001a}
{Hartkopf}, W.~I., {Mason}, B.~D., \& {Worley}, C.~E. 2001,
  \bibinfo{title}{{The 2001 US Naval Observatory Double Star CD-ROM. II. The
  Fifth Catalog of Orbits of Visual Binary Stars},} \aj, 122, 3472,
  \dodoi{10.1086/323921}

% type= article
\bibitem[{J.~D. {Hartman} {et~al.}(2010){Hartman}, {Bakos}, {Kov{\'a}cs}, \&
  {Noyes}}]{hartmann2010}
{Hartman}, J.~D., {Bakos}, G.~{\'A}., {Kov{\'a}cs}, G., \& {Noyes}, R.~W. 2010,
  \bibinfo{title}{{A large sample of photometric rotation periods for FGK
  Pleiades stars},} \mnras, 408, 475, \dodoi{10.1111/j.1365-2966.2010.17147.x}

% type= article
\bibitem[{L. {Hartmann} {et~al.}(2016){Hartmann}, {Herczeg}, \&
  {Calvet}}]{hartmann2016}
{Hartmann}, L., {Herczeg}, G., \& {Calvet}, N. 2016, \bibinfo{title}{{Accretion
  onto Pre-Main-Sequence Stars},} \araa, 54, 135,
  \dodoi{10.1146/annurev-astro-081915-023347}

% type= article
\bibitem[{V. {Heged{\H{u}}s} {et~al.}(2023){Heged{\H{u}}s}, {M{\'e}sz{\'a}ros},
  {Jofr{\'e}}, {Stringfellow}, {Feuillet}, {Garc{\'\i}a-Hern{\'a}ndez},
  {Nitschelm}, \& {Zamora}}]{hegedus2023}
{Heged{\H{u}}s}, V., {M{\'e}sz{\'a}ros}, S., {Jofr{\'e}}, P., {et~al.} 2023,
  \bibinfo{title}{{Comparative analysis of atmospheric parameters from
  high-resolution spectroscopic sky surveys: APOGEE, GALAH, Gaia-ESO},} \aap,
  670, A107, \dodoi{10.1051/0004-6361/202244813}

% type= article
\bibitem[{K.~G. {He{\l}miniak} {et~al.}(2012){He{\l}miniak}, {Konacki},
  {Muterspaugh}, {Browne}, {Howard}, \& {Kulkarni}}]{helminiak2012}
{He{\l}miniak}, K.~G., {Konacki}, M., {Muterspaugh}, M.~W., {et~al.} 2012,
  \bibinfo{title}{{New high-precision orbital and physical parameters of the
  double-lined low-mass spectroscopic binary BY Draconis},} \mnras, 419, 1285,
  \dodoi{10.1111/j.1365-2966.2011.19785.x}

% type= article
\bibitem[{A. {Henden} \& U. {Munari}(2014){Henden} \& {Munari}}]{hendon2014}
{Henden}, A., \& {Munari}, U. 2014, \bibinfo{title}{{The APASS all-sky,
  multi-epoch BVgri photometric survey},} Contributions of the Astronomical
  Observatory Skalnate Pleso, 43, 518

% type= inproceedings
\bibitem[{A.~A. {Henden} {et~al.}(2018){Henden}, {Levine}, {Terrell}, {Welch},
  {Munari}, \& {Kloppenborg}}]{hendon2018}
{Henden}, A.~A., {Levine}, S., {Terrell}, D., {et~al.} 2018,
  \bibinfo{title}{{APASS Data Release 10},} in American Astronomical Society
  Meeting Abstracts, Vol. 232, American Astronomical Society Meeting Abstracts
  \#232, 223.06

% type= article
\bibitem[{S. {Hengst} {et~al.}(2017){Hengst}, {Marshall}, {Horner}, \&
  {Marsden}}]{hengst2017}
{Hengst}, S., {Marshall}, J.~P., {Horner}, J., \& {Marsden}, S.~C. 2017,
  \bibinfo{title}{{A Herschel resolved debris disc around HD 105211},} \mnras,
  468, 4725, \dodoi{10.1093/mnras/stx753}

% type= article
\bibitem[{T.~J. {Henry} {et~al.}(1996){Henry}, {Soderblom}, {Donahue}, \&
  {Baliunas}}]{henry1996}
{Henry}, T.~J., {Soderblom}, D.~R., {Donahue}, R.~A., \& {Baliunas}, S.~L.
  1996, \bibinfo{title}{{A Survey of Ca II H and K Chromospheric Emission in
  Southern Solar-Type Stars},} \aj, 111, 439, \dodoi{10.1086/117796}

% type= article
\bibitem[{T.~J. {Henry} {et~al.}(2018){Henry}, {Jao}, {Winters}, {Dieterich},
  {Finch}, {Ianna}, {Riedel}, {Silverstein}, {Subasavage}, \&
  {Vrijmoet}}]{henry2018}
{Henry}, T.~J., {Jao}, W.-C., {Winters}, J.~G., {et~al.} 2018,
  \bibinfo{title}{{The Solar Neighborhood XLIV: RECONS Discoveries within 10
  parsecs},} \aj, 155, 265, \dodoi{10.3847/1538-3881/aac262}

% type= article
\bibitem[{G.~J. {Herczeg} {et~al.}(2023){Herczeg}, {Chen}, {Donati}, {Dupree},
  {Walter}, {Hillenbrand}, {Johns-Krull}, {Manara}, {G{\"u}nther}, {Fang},
  {Schneider}, {Valenti}, {Alencar}, {Venuti}, {Alcal{\'a}}, {Frasca},
  {Arulanantham}, {Linsky}, {Bouvier}, {Brickhouse}, {Calvet}, {Espaillat},
  {Campbell-White}, {Carpenter}, {Chang}, {Cruz}, {Dahm}, {Eisl{\"o}ffel},
  {Edwards}, {Fischer}, {Guo}, {Henning}, {Ji}, {Jose}, {Kastner}, {Launhardt},
  {Principe}, {Robinson}, {Serna}, {Siwak}, {Sterzik}, \&
  {Takasao}}]{herczeg2023}
{Herczeg}, G.~J., {Chen}, Y., {Donati}, J.-F., {et~al.} 2023,
  \bibinfo{title}{{Twenty-five Years of Accretion onto the Classical T Tauri
  Star TW Hya},} \apj, 956, 102, \dodoi{10.3847/1538-4357/acf468}

% type= article
\bibitem[{J. {Hern{\'a}ndez} {et~al.}(2006){Hern{\'a}ndez}, {Brice{\~n}o},
  {Calvet}, {Hartmann}, {Muzerolle}, \& {Quintero}}]{hernandez2006}
{Hern{\'a}ndez}, J., {Brice{\~n}o}, C., {Calvet}, N., {et~al.} 2006,
  \bibinfo{title}{{Spitzer Observations of the Orion OB1 Association:
  Second-Generation Dust Disks at 5-10 Myr},} \apj, 652, 472,
  \dodoi{10.1086/507942}

% type= article
\bibitem[{J. {Hern{\'a}ndez} {et~al.}(2007){Hern{\'a}ndez}, {Hartmann},
  {Megeath}, {Gutermuth}, {Muzerolle}, {Calvet}, {Vivas}, {Brice{\~n}o},
  {Allen}, {Stauffer}, {Young}, \& {Fazio}}]{hernandez2007}
{Hern{\'a}ndez}, J., {Hartmann}, L., {Megeath}, T., {et~al.} 2007,
  \bibinfo{title}{{A Spitzer Space Telescope Study of Disks in the Young
  {$\sigma$} Orionis Cluster},} \apj, 662, 1067, \dodoi{10.1086/513735}

% type= article
\bibitem[{J. {Hern{\'a}ndez} {et~al.}(2023){Hern{\'a}ndez}, {Zamudio},
  {Brice{\~n}o}, {Calvet}, {Zhu}, {Yuan}, {Liu}, {Manzo-Mart{\'\i}nez},
  {Rom{\'a}n-Z{\'u}{\~n}iga}, {Serna}, {Mauc{\'o}}, \& {Adame}}]{hernandez2023}
{Hern{\'a}ndez}, J., {Zamudio}, L.~F., {Brice{\~n}o}, C., {et~al.} 2023,
  \bibinfo{title}{{A LAMOST Spectroscopic Study of T Tauri Stars in the Orion
  OB1a Subassociation},} \aj, 165, 205, \dodoi{10.3847/1538-3881/acc467}

% type= article
\bibitem[{J. {Heyl} {et~al.}(2022){Heyl}, {Caiazzo}, \& {Richer}}]{heyl2022}
{Heyl}, J., {Caiazzo}, I., \& {Richer}, H.~B. 2022,
  \bibinfo{title}{{Reconstructing the Pleiades with Gaia EDR3},} \apj, 926,
  132, \dodoi{10.3847/1538-4357/ac45fc}

% type= article
\bibitem[{L.~A. {Hillenbrand} {et~al.}(2008){Hillenbrand}, {Carpenter}, {Kim},
  {Meyer}, {Backman}, {Moro-Mart{\'{\i}}n}, {Hollenbach}, {Hines}, {Pascucci},
  \& {Bouwman}}]{hillen2008}
{Hillenbrand}, L.~A., {Carpenter}, J.~M., {Kim}, J.~S., {et~al.} 2008,
  \bibinfo{title}{{The Complete Census of 70 {$\mu$}m-bright Debris Disks
  within ``the Formation and Evolution of Planetary Systems'' Spitzer Legacy
  Survey of Sun-like Stars},} \apj, 677, 630, \dodoi{10.1086/529027}

% type= article
\bibitem[{N.~R. {Hinkel} {et~al.}(2014){Hinkel}, {Timmes}, {Young}, {Pagano},
  \& {Turnbull}}]{hinkel2014}
{Hinkel}, N.~R., {Timmes}, F.~X., {Young}, P.~A., {Pagano}, M.~D., \&
  {Turnbull}, M.~C. 2014, \bibinfo{title}{{Stellar Abundances in the Solar
  Neighborhood: The Hypatia Catalog},} \aj, 148, 54,
  \dodoi{10.1088/0004-6256/148/3/54}

% type= article
\bibitem[{N.~R. {Hinkel} {et~al.}(2017){Hinkel}, {Mamajek}, {Turnbull}, {Osby},
  {Shkolnik}, {Smith}, {Klimasewski}, {Somers}, \& {Desch}}]{hinkel2017}
{Hinkel}, N.~R., {Mamajek}, E.~E., {Turnbull}, M.~C., {et~al.} 2017,
  \bibinfo{title}{{A Catalog of Stellar Unified Properties (CATSUP) for 951
  FGK-Stars within 30 pc},} \apj, 848, 34, \dodoi{10.3847/1538-4357/aa8b0f}

% type= book
\bibitem[{D. {Hoffleit} \& C. {Jaschek}(1991){Hoffleit} \&
  {Jaschek}}]{hoffleit1991}
{Hoffleit}, D., \& {Jaschek}, C. 1991, {The Bright star catalogue} (Yale
  University Observatory, New Haven, CT USA)

% type= article
\bibitem[{E. {H{\o}g} {et~al.}(1997){H{\o}g}, {B{\"a}ssgen}, {Bastian},
  {Egret}, {Fabricius}, {Gro{\ss}mann}, {Halbwachs}, {Makarov}, {Perryman},
  {Schwekendiek}, {Wagner}, \& {Wicenec}}]{hog1997}
{H{\o}g}, E., {B{\"a}ssgen}, G., {Bastian}, U., {et~al.} 1997,
  \bibinfo{title}{{The TYCHO Catalogue},} \aap, 323, L57

% type= article
\bibitem[{E. {H{\o}g} {et~al.}(2000){H{\o}g}, {Fabricius}, {Makarov}, {Urban},
  {Corbin}, {Wycoff}, {Bastian}, {Schwekendiek}, \& {Wicenec}}]{hog2000}
{H{\o}g}, E., {Fabricius}, C., {Makarov}, V.~V., {et~al.} 2000,
  \bibinfo{title}{{The Tycho-2 catalogue of the 2.5 million brightest stars},}
  \aap, 355, L27

% type= article
\bibitem[{W.~S. {Holland} {et~al.}(2017){Holland}, {Matthews}, {Kennedy},
  {Greaves}, {Wyatt}, {Booth}, {Bastien}, {Bryden}, {Butner}, {Chen},
  {Chrysostomou}, {Davies}, {Dent}, {Di Francesco}, {Duch{\^e}ne}, {Gibb},
  {Friberg}, {Ivison}, {Jenness}, {Kavelaars}, {Lawler}, {Lestrade},
  {Marshall}, {Moro-Martin}, {Pani{\'c}}, {Phillips}, {Serjeant}, {Schieven},
  {Sibthorpe}, {Vican}, {Ward-Thompson}, {van der Werf}, {White}, {Wilner}, \&
  {Zuckerman}}]{holland2017}
{Holland}, W.~S., {Matthews}, B.~C., {Kennedy}, G.~M., {et~al.} 2017,
  \bibinfo{title}{{SONS: The JCMT legacy survey of debris discs in the
  submillimetre},} \mnras, 470, 3606, \dodoi{10.1093/mnras/stx1378}

% type= article
\bibitem[{J. {Holmberg} {et~al.}(2009){Holmberg}, {Nordstr{\"o}m}, \&
  {Andersen}}]{holmberg2009}
{Holmberg}, J., {Nordstr{\"o}m}, B., \& {Andersen}, J. 2009,
  \bibinfo{title}{{The Geneva-Copenhagen survey of the solar neighbourhood.
  III. Improved distances, ages, and kinematics},} \aap, 501, 941,
  \dodoi{10.1051/0004-6361/200811191}

% type= article
\bibitem[{J.~R. {Houck} {et~al.}(2004){Houck}, {Roellig}, {van Cleve},
  {Forrest}, {Herter}, {Lawrence}, {Matthews}, {Reitsema}, {Soifer}, {Watson},
  {Weedman}, {Huisjen}, {Troeltzsch}, {Barry}, {Bernard-Salas}, {Blacken},
  {Brandl}, {Charmandaris}, {Devost}, {Gull}, {Hall}, {Henderson}, {Higdon},
  {Pirger}, {Schoenwald}, {Sloan}, {Uchida}, {Appleton}, {Armus}, {Burgdorf},
  {Fajardo-Acosta}, {Grillmair}, {Ingalls}, {Morris}, \& {Teplitz}}]{houck2004}
{Houck}, J.~R., {Roellig}, T.~L., {van Cleve}, J., {et~al.} 2004,
  \bibinfo{title}{{The Infrared Spectrograph (IRS) on the Spitzer Space
  Telescope},} \apjs, 154, 18, \dodoi{10.1086/423134}

% type= article
\bibitem[{E.~R. {Houdebine}(2009){Houdebine}}]{houdebine2009}
{Houdebine}, E.~R. 2009, \bibinfo{title}{{Observation and modelling of
  main-sequence star chromospheres - XII. Two-component model chromospheres for
  five active dM1e stars},} \mnras, 397, 2133,
  \dodoi{10.1111/j.1365-2966.2009.15112.x}

% type= article
\bibitem[{E.~R. {Houdebine}(2011){Houdebine}}]{houdebine2011}
{Houdebine}, E.~R. 2011, \bibinfo{title}{{Observation and modelling of
  main-sequence star chromospheres - XVI. Rotation of dK5 stars},} \mnras, 416,
  2233, \dodoi{10.1111/j.1365-2966.2011.19199.x}

% type= article
\bibitem[{E.~R. {Houdebine} \& D.~J. {Mullan}(2015){Houdebine} \&
  {Mullan}}]{houdebine2015}
{Houdebine}, E.~R., \& {Mullan}, D.~J. 2015, \bibinfo{title}{{Dynamics of
  Rotation in M Dwarfs: Indications for a Change in the Dynamo Regime in Stars
  at the Onset of Complete Convection},} \apj, 801, 106,
  \dodoi{10.1088/0004-637X/801/2/106}

% type= article
\bibitem[{E.~R. {Houdebine} {et~al.}(2017){Houdebine}, {Mullan}, {Bercu},
  {Paletou}, \& {Gebran}}]{houdebine2017}
{Houdebine}, E.~R., {Mullan}, D.~J., {Bercu}, B., {Paletou}, F., \& {Gebran},
  M. 2017, \bibinfo{title}{{The Rotation-Activity Correlations in K and M
  Dwarfs. II. New Constraints on the Dynamo Mechanisms in Late-K and M Dwarfs
  Before and At the Transition to Complete Convection},} \apj, 837, 96,
  \dodoi{10.3847/1538-4357/aa5cad}

% type= article
\bibitem[{{\'E}.~R. {Houdebine} {et~al.}(2019){Houdebine}, {Mullan}, {Doyle},
  {de La Vieuville}, {Butler}, \& {Paletou}}]{houdebine2019}
{Houdebine}, {\'E}.~R., {Mullan}, D.~J., {Doyle}, J.~G., {et~al.} 2019,
  \bibinfo{title}{{The Mass-Activity Relationships in M and K Dwarfs. I.
  Stellar Parameters of Our Sample of M and K Dwarfs},} \aj, 158, 56,
  \dodoi{10.3847/1538-3881/ab23fe}

% type= article
\bibitem[{E.~R. {Houdebine} {et~al.}(2016){Houdebine}, {Mullan}, {Paletou}, \&
  {Gebran}}]{houdebine2016}
{Houdebine}, E.~R., {Mullan}, D.~J., {Paletou}, F., \& {Gebran}, M. 2016,
  \bibinfo{title}{{Rotation-Activity Correlations in K and M Dwarfs. I. Stellar
  Parameters and Compilations of v sin I and P/sin I for a Large Sample of
  Late-K and M Dwarfs},} \apj, 822, 97, \dodoi{10.3847/0004-637X/822/2/97}

% type= book
\bibitem[{N. {Houk}(1978){Houk}}]{houk1978}
{Houk}, N. 1978, {Michigan catalogue of two-dimensional spectral types for the
  HD stars} (Deartment of Astronomy, University of Michigan, Ann Arbor, MI USA)

% type= article
\bibitem[{A. {Hourihane} {et~al.}(2023){Hourihane}, {Fran{\c{c}}ois}, {Worley},
  {Magrini}, {Gonneau}, {Casey}, {Gilmore}, {Randich}, {Sacco}, {Recio-Blanco},
  {Korn}, {Allende Prieto}, {Smiljanic}, {Blomme}, {Bragaglia}, {Walton}, {Van
  Eck}, {Bensby}, {Lanzafame}, {Frasca}, {Franciosini}, {Damiani}, {Lind},
  {Bergemann}, {Bonifacio}, {Hill}, {Lobel}, {Montes}, {Feuillet},
  {Tautvaisiene}, {Guiglion}, {Tabernero}, {Gonz{\'a}lez Hern{\'a}ndez},
  {Gebran}, {Van der Swaelmen}, {Mikolaitis}, {Daflon}, {Merle}, {Morel},
  {Lewis}, {Gonz{\'a}lez Solares}, {Murphy}, {Jeffries}, {Jackson}, {Feltzing},
  {Prusti}, {Carraro}, {Biazzo}, {Prisinzano}, {Jofr{\'e}}, {Zaggia},
  {Drazdauskas}, {Stonkut{\'e}}, {Marfil}, {Jim{\'e}nez-Esteban}, {Mahy},
  {Guti{\'e}rrez Albarr{\'a}n}, {Berlanas}, {Santos}, {Morbidelli}, {Spina}, \&
  {Minkeviciute}}]{hourihane2023}
{Hourihane}, A., {Fran{\c{c}}ois}, P., {Worley}, C.~C., {et~al.} 2023,
  \bibinfo{title}{{The Gaia-ESO Survey: Homogenisation of stellar parameters
  and elemental abundances},} \aap, 676, A129,
  \dodoi{10.1051/0004-6361/202345910}

% type= article
\bibitem[{B. {Huang} {et~al.}(2024){Huang}, {Yuan}, {Xiang}, {Huang}, {Xiao},
  {Xu}, {Zhang}, {Yang}, {Niu}, \& {Gu}}]{bhuang2024}
{Huang}, B., {Yuan}, H., {Xiang}, M., {et~al.} 2024, \bibinfo{title}{{A
  Comprehensive Correction of the Gaia DR3 XP Spectra},} \apjs, 271, 13,
  \dodoi{10.3847/1538-4365/ad18b1}

% type= article
\bibitem[{A.~M. {Hughes} {et~al.}(2018){Hughes}, {Duch{\^e}ne}, \&
  {Matthews}}]{hughes18}
{Hughes}, A.~M., {Duch{\^e}ne}, G., \& {Matthews}, B.~C. 2018,
  \bibinfo{title}{{Debris Disks: Structure, Composition, and Variability},}
  \araa, 56, 541, \dodoi{10.1146/annurev-astro-081817-052035}

% type= article
\bibitem[{C.~A. {Hummel} {et~al.}(1995){Hummel}, {Armstrong}, {Buscher},
  {Mozurkewich}, {Quirrenbach}, \& {Vivekanand}}]{hummel1995}
{Hummel}, C.~A., {Armstrong}, J.~T., {Buscher}, D.~F., {et~al.} 1995,
  \bibinfo{title}{{Orbits of Small Angular Scale Binaries Resolved with the
  Mark III Interferometer},} \aj, 110, 376, \dodoi{10.1086/117528}

% type= article
\bibitem[{C.~A. {Hummel} {et~al.}(2017){Hummel}, {Monnier}, {Roettenbacher},
  {Torres}, {Henry}, {Korhonen}, {Beasley}, {Schaefer}, {Turner}, {Ten
  Brummelaar}, {Farrington}, {Sturmann}, {Sturmann}, {Baron}, \&
  {Kraus}}]{hummel2017}
{Hummel}, C.~A., {Monnier}, J.~D., {Roettenbacher}, R.~M., {et~al.} 2017,
  \bibinfo{title}{{Orbital Elements and Stellar Parameters of the Active Binary
  UX Arietis},} \apj, 844, 115, \dodoi{10.3847/1538-4357/aa7b87}

% type= article
\bibitem[{M. {H{\"u}nsch} {et~al.}(2003){H{\"u}nsch}, {Weidner}, \&
  {Schmitt}}]{hunsch2003}
{H{\"u}nsch}, M., {Weidner}, C., \& {Schmitt}, J.~H.~M.~M. 2003,
  \bibinfo{title}{{An X-ray study of the open clusters NGC 2451 A and B},}
  \aap, 402, 571, \dodoi{10.1051/0004-6361:20030268}

% type= article
\bibitem[{E.~L. {Hunt} \& S. {Reffert}(2023){Hunt} \& {Reffert}}]{hunt2023}
{Hunt}, E.~L., \& {Reffert}, S. 2023, \bibinfo{title}{{Improving the open
  cluster census. II. An all-sky cluster catalogue with Gaia DR3},} \aap, 673,
  A114, \dodoi{10.1051/0004-6361/202346285}

% type= article
\bibitem[{E.~L. {Hunt} \& S. {Reffert}(2024){Hunt} \& {Reffert}}]{hunt2024}
{Hunt}, E.~L., \& {Reffert}, S. 2024, \bibinfo{title}{{Improving the open
  cluster census. III. Using cluster masses, radii, and dynamics to create a
  cleaned open cluster catalogue},} \aap, 686, A42,
  \dodoi{10.1051/0004-6361/202348662}

% type= article
\bibitem[{R.~J. {Jackson} {et~al.}(2020){Jackson}, {Jeffries}, {Wright},
  {Randich}, {Sacco}, {Pancino}, {Cantat-Gaudin}, {Gilmore}, {Vallenari},
  {Bensby}, {Bayo}, {Costado}, {Franciosini}, {Gonneau}, {Hourihane}, {Lewis},
  {Monaco}, {Morbidelli}, \& {Worley}}]{jackson2020}
{Jackson}, R.~J., {Jeffries}, R.~D., {Wright}, N.~J., {et~al.} 2020,
  \bibinfo{title}{{The Gaia-ESO Survey: membership probabilities for stars in
  32 open clusters from 3D kinematics},} \mnras, 496, 4701,
  \dodoi{10.1093/mnras/staa1749}

% type= article
\bibitem[{R.~J. {Jackson} {et~al.}(2022){Jackson}, {Jeffries}, {Wright},
  {Randich}, {Sacco}, {Bragaglia}, {Hourihane}, {Tognelli}, {Degl'Innocenti},
  {Prada Moroni}, {Gilmore}, {Bensby}, {Pancino}, {Smiljanic}, {Bergemann},
  {Carraro}, {Franciosini}, {Gonneau}, {Jofr{\'e}}, {Lewis}, {Magrini},
  {Morbidelli}, {Prisinzano}, {Worley}, {Zaggia}, {Tautvaisiene},
  {Guti{\'e}rrez Albarr{\'a}n}, {Montes}, \&
  {Jim{\'e}nez-Esteban}}]{jackson2022}
{Jackson}, R.~J., {Jeffries}, R.~D., {Wright}, N.~J., {et~al.} 2022,
  \bibinfo{title}{{The Gaia-ESO Survey: Membership probabilities for stars in
  63 open and 7 globular clusters from 3D kinematics},} \mnras, 509, 1664,
  \dodoi{10.1093/mnras/stab3032}

% type= article
\bibitem[{H. {Jang-Condell} {et~al.}(2015){Jang-Condell}, {Chen}, {Mittal},
  {Manoj}, {Watson}, {Lisse}, {Nesvold}, \& {Kuchner}}]{jangcondell2015}
{Jang-Condell}, H., {Chen}, C.~H., {Mittal}, T., {et~al.} 2015,
  \bibinfo{title}{{Spitzer IRS Spectra of Debris Disks in the
  Scorpius-Centaurus OB Association},} \apj, 808, 167,
  \dodoi{10.1088/0004-637X/808/2/167}

% type= article
\bibitem[{F. {Jansen} {et~al.}(2001){Jansen}, {Lumb}, {Altieri}, {Clavel},
  {Ehle}, {Erd}, {Gabriel}, {Guainazzi}, {Gondoin}, {Much}, {Munoz}, {Santos},
  {Schartel}, {Texier}, \& {Vacanti}}]{jansen2001}
{Jansen}, F., {Lumb}, D., {Altieri}, B., {et~al.} 2001,
  \bibinfo{title}{{XMM-Newton observatory. I. The spacecraft and operations},}
  \aap, 365, L1, \dodoi{10.1051/0004-6361:20000036}

% type= article
\bibitem[{M. {Janson} {et~al.}(2021){Janson}, {Gratton}, {Rodet}, {Vigan},
  {Bonnefoy}, {Delorme}, {Mamajek}, {Reffert}, {Stock}, {Marleau}, {Langlois},
  {Chauvin}, {Desidera}, {Ringqvist}, {Mayer}, {Viswanath}, {Squicciarini},
  {Meyer}, {Samland}, {Petrus}, {Helled}, {Kenworthy}, {Quanz}, {Biller},
  {Henning}, {Mesa}, {Engler}, \& {Carson}}]{janson2021}
{Janson}, M., {Gratton}, R., {Rodet}, L., {et~al.} 2021, \bibinfo{title}{{A
  wide-orbit giant planet in the high-mass b Centauri binary system},} \nat,
  600, 231, \dodoi{10.1038/s41586-021-04124-8}

% type= article
\bibitem[{S.~V. {Jeffers} {et~al.}(2023){Jeffers}, {Kiefer}, \&
  {Metcalfe}}]{jeffers2023}
{Jeffers}, S.~V., {Kiefer}, R., \& {Metcalfe}, T.~S. 2023,
  \bibinfo{title}{{Stellar Activity Cycles},} \ssr, 219, 54,
  \dodoi{10.1007/s11214-023-01000-x}

% type= article
\bibitem[{S.~V. {Jeffers} {et~al.}(2018){Jeffers}, {Mengel}, {Moutou},
  {Marsden}, {Barnes}, {Jardine}, {Petit}, {Schmitt}, {See}, {Vidotto}, \&
  {BCool Collaboration}}]{jeffers2018}
{Jeffers}, S.~V., {Mengel}, M., {Moutou}, C., {et~al.} 2018,
  \bibinfo{title}{{The relation between stellar magnetic field geometry and
  chromospheric activity cycles - II The rapid 120-day magnetic cycle of
  {\ensuremath{\tau}} Bootis},} \mnras, 479, 5266,
  \dodoi{10.1093/mnras/sty1717}

% type= article
\bibitem[{S.~V. {Jeffers} {et~al.}(2022){Jeffers}, {Cameron}, {Marsden}, {Boro
  Saikia}, {Folsom}, {Jardine}, {Morin}, {Petit}, {See}, {Vidotto}, {Wolter},
  \& {Mittag}}]{jeffers2022}
{Jeffers}, S.~V., {Cameron}, R.~H., {Marsden}, S.~C., {et~al.} 2022,
  \bibinfo{title}{{The crucial role of surface magnetic fields for stellar
  dynamos: ϵ Eridani, 61 Cygni A, and the Sun},} \aap, 661, A152,
  \dodoi{10.1051/0004-6361/202142202}

% type= article
\bibitem[{R.~D. {Jeffries}(1999){Jeffries}}]{jeffries1999}
{Jeffries}, R.~D. 1999, \bibinfo{title}{{Lithium in the low-mass stars of the
  Coma Berenices open cluster},} \mnras, 304, 821,
  \dodoi{10.1046/j.1365-8711.1999.02347.x}

% type= article
\bibitem[{R.~D. {Jeffries} {et~al.}(2006){Jeffries}, {Evans}, {Pye}, \&
  {Briggs}}]{jeffries2006}
{Jeffries}, R.~D., {Evans}, P.~A., {Pye}, J.~P., \& {Briggs}, K.~R. 2006,
  \bibinfo{title}{{An XMM-Newton observation of the young open cluster NGC
  2547: coronal activity at 30 Myr},} \mnras, 367, 781,
  \dodoi{10.1111/j.1365-2966.2005.09988.x}

% type= article
\bibitem[{R.~D. {Jeffries} \& J.~M. {Oliveira}(2005){Jeffries} \&
  {Oliveira}}]{jeffries2005}
{Jeffries}, R.~D., \& {Oliveira}, J.~M. 2005, \bibinfo{title}{{The lithium
  depletion boundary in NGC 2547 as a test of pre-main-sequence evolutionary
  models},} \mnras, 358, 13, \dodoi{10.1111/j.1365-2966.2005.08820.x}

% type= article
\bibitem[{R.~D. {Jeffries} {et~al.}(2003){Jeffries}, {Oliveira}, {Barrado y
  Navascu{\'e}s}, \& {Stauffer}}]{jeffries2003}
{Jeffries}, R.~D., {Oliveira}, J.~M., {Barrado y Navascu{\'e}s}, D., \&
  {Stauffer}, J.~R. 2003, \bibinfo{title}{{Cool stars in NGC 2547 and
  pre-main-sequence lithium depletion},} \mnras, 343, 1271,
  \dodoi{10.1046/j.1365-8711.2003.06768.x}

% type= article
\bibitem[{R.~D. {Jeffries} {et~al.}(2023){Jeffries}, {Jackson}, {Wright},
  {Weaver}, {Gilmore}, {Randich}, {Bragaglia}, {Korn}, {Smiljanic}, {Biazzo},
  {Casey}, {Frasca}, {Gonneau}, {Guiglion}, {Morbidelli}, {Prisinzano},
  {Sacco}, {Tautvaisien{\.{e}}}, {Worley}, \& {Zaggia}}]{jeffries2023}
{Jeffries}, R.~D., {Jackson}, R.~J., {Wright}, N.~J., {et~al.} 2023,
  \bibinfo{title}{{The Gaia-ESO Survey: empirical estimates of stellar ages
  from lithium equivalent widths (EAGLES)},} \mnras, 523, 802,
  \dodoi{10.1093/mnras/stad1293}

% type= article
\bibitem[{J.~S. {Jenkins} {et~al.}(2009){Jenkins}, {Ramsey}, {Jones},
  {Pavlenko}, {Gallardo}, {Barnes}, \& {Pinfield}}]{jenkins2009}
{Jenkins}, J.~S., {Ramsey}, L.~W., {Jones}, H.~R.~A., {et~al.} 2009,
  \bibinfo{title}{{Rotational Velocities for M Dwarfs},} \apj, 704, 975,
  \dodoi{10.1088/0004-637X/704/2/975}

% type= article
\bibitem[{F.~M. {Jim{\'e}nez-Esteban} {et~al.}(2023){Jim{\'e}nez-Esteban},
  {Torres}, {Rebassa-Mansergas}, {Cruz}, {Murillo-Ojeda}, {Solano}, {Rodrigo},
  \& {Camisassa}}]{jimenez2023}
{Jim{\'e}nez-Esteban}, F.~M., {Torres}, S., {Rebassa-Mansergas}, A., {et~al.}
  2023, \bibinfo{title}{{Spectral classification of the 100 pc white dwarf
  population from Gaia-DR3 and the virtual observatory},} \mnras, 518, 5106,
  \dodoi{10.1093/mnras/stac3382}

% type= article
\bibitem[{C.~P. {Johnstone} {et~al.}(2021){Johnstone}, {Bartel}, \&
  {G{\"u}del}}]{johnstone2021}
{Johnstone}, C.~P., {Bartel}, M., \& {G{\"u}del}, M. 2021, \bibinfo{title}{{The
  active lives of stars: A complete description of the rotation and XUV
  evolution of F, G, K, and M dwarfs},} \aap, 649, A96,
  \dodoi{10.1051/0004-6361/202038407}

% type= inproceedings
\bibitem[{J. {Jones} {et~al.}(2017){Jones}, {White}, {Boyajian}, {Schaefer},
  {Baines}, {Ireland}, {Quinn}, \& {CHARA Team}}]{jones2017}
{Jones}, J., {White}, R.~J., {Boyajian}, T.~S., {et~al.} 2017,
  \bibinfo{title}{{The Ages of A-Stars: Interferometric Observations of Our
  Brightest Neighbors},} in American Astronomical Society Meeting Abstracts,
  Vol. 229, American Astronomical Society Meeting Abstracts \#229, 131.05

% type= article
\bibitem[{J. {Jones} {et~al.}(2015){Jones}, {White}, {Boyajian}, {Schaefer},
  {Baines}, {Ireland}, {Patience}, {ten Brummelaar}, {McAlister}, {Ridgway},
  {Sturmann}, {Sturmann}, {Turner}, {Farrington}, \& {Goldfinger}}]{jones2015}
{Jones}, J., {White}, R.~J., {Boyajian}, T., {et~al.} 2015,
  \bibinfo{title}{{The Ages of A-Stars. I. Interferometric Observations and Age
  Estimates for Stars in the Ursa Major Moving Group},} \apj, 813, 58,
  \dodoi{10.1088/0004-637X/813/1/58}

% type= article
\bibitem[{N.~K. {Jones} {et~al.}(2025){Jones}, {Wang}, {Nielsen}, {De Rosa},
  {Peck}, {Roberson}, {Ruffio}, {Xuan}, {Macintosh}, {Ammons}, {Bailey},
  {Barman}, {Bulger}, {Chiang}, {Chilcote}, {Duch{\^e}ne}, {Esposito},
  {Fitzgerald}, {Follette}, {Goodsell}, {Graham}, {Greenbaum}, {Hibon},
  {Ingraham}, {Kalas}, {Konopacky}, {Liu}, {Marchis}, {Maire}, {Marois},
  {Matthews}, {Mawet}, {Metchev}, {Millar-Blanchaer}, {Oppenheimer}, {Palmer},
  {Patience}, {Perrin}, {Poyneer}, {Pueyo}, {Rajan}, {Rameau}, {Rantakyr{\"o}},
  {Ren}, {Sanghi}, {Savransky}, {Schneider}, {Sivaramakrishnan}, {Smith},
  {Song}, {Soummer}, {Thomas}, {Ward-Duong}, \& {Wolff}}]{jones2025}
{Jones}, N.~K., {Wang}, J.~J., {Nielsen}, E.~L., {et~al.} 2025,
  \bibinfo{title}{{HD 143811 AB b: A Directly Imaged Planet Orbiting a
  Spectroscopic Binary in Sco-Cen},} \apjl, 995, L41,
  \dodoi{10.3847/2041-8213/ae2007}

% type= article
\bibitem[{J.~K. {J{\o}rgensen} {et~al.}(2006){J{\o}rgensen}, {Harvey}, {Evans},
  {Huard}, {Allen}, {Porras}, {Blake}, {Bourke}, {Chapman}, {Cieza}, {Koerner},
  {Lai}, {Mundy}, {Myers}, {Padgett}, {Rebull}, {Sargent}, {Spiesman},
  {Stapelfeldt}, {van Dishoeck}, {Wahhaj}, \& {Young}}]{jorgensen2006}
{J{\o}rgensen}, J.~K., {Harvey}, P.~M., {Evans}, Neal~J., I., {et~al.} 2006,
  \bibinfo{title}{{The Spitzer c2d Survey of Large, Nearby, Interstellar
  Clouds. III. Perseus Observed with IRAC},} \apj, 645, 1246,
  \dodoi{10.1086/504373}

% type= article
\bibitem[{P.~G. {Judge} {et~al.}(2003){Judge}, {Solomon}, \&
  {Ayres}}]{judge2003}
{Judge}, P.~G., {Solomon}, S.~C., \& {Ayres}, T.~R. 2003, \bibinfo{title}{{An
  Estimate of the Sun's ROSAT-PSPC X-Ray Luminosities Using SNOE-SXP
  Measurements},} \apj, 593, 534, \dodoi{10.1086/376405}

% type= article
\bibitem[{M. {Jura} {et~al.}(2004){Jura}, {Chen}, {Furlan}, {Green}, {Sargent},
  {Forrest}, {Watson}, {Barry}, {Hall}, {Herter}, {Houck}, {Sloan}, {Uchida},
  {D'Alessio}, {Brandl}, {Keller}, {Kemper}, {Morris}, {Najita}, {Calvet},
  {Hartmann}, \& {Myers}}]{jura2004}
{Jura}, M., {Chen}, C.~H., {Furlan}, E., {et~al.} 2004,
  \bibinfo{title}{{Mid-Infrared Spectra of Dust Debris around Main-Sequence
  Stars},} \apjs, 154, 453, \dodoi{10.1086/422975}

% type= inproceedings
\bibitem[{N. {Kaiser} {et~al.}(2010){Kaiser}, {Burgett}, {Chambers}, {Denneau},
  {Heasley}, {Jedicke}, {Magnier}, {Morgan}, {Onaka}, \& {Tonry}}]{kaiser2010}
{Kaiser}, N., {Burgett}, W., {Chambers}, K., {et~al.} 2010,
  \bibinfo{title}{{The Pan-STARRS wide-field optical/NIR imaging survey},} in
  Society of Photo-Optical Instrumentation Engineers (SPIE) Conference Series,
  Vol. 7733, Ground-based and Airborne Telescopes III, ed. L.~M. {Stepp},
  R.~{Gilmozzi}, \& H.~J. {Hall}, 77330E, \dodoi{10.1117/12.859188}

% type= article
\bibitem[{P. {Kalas} {et~al.}(2005){Kalas}, {Graham}, \& {Clampin}}]{kalas05}
{Kalas}, P., {Graham}, J.~R., \& {Clampin}, M. 2005, \bibinfo{title}{{A
  planetary system as the origin of structure in Fomalhaut's dust belt},} \nat,
  435, 1067, \dodoi{10.1038/nature03601}

% type= article
\bibitem[{B.~L. {Kamai} {et~al.}(2014){Kamai}, {Vrba}, {Stauffer}, \&
  {Stassun}}]{kamai2014}
{Kamai}, B.~L., {Vrba}, F.~J., {Stauffer}, J.~R., \& {Stassun}, K.~G. 2014,
  \bibinfo{title}{{New BVI $_{C}$ Photometry of Low-mass Pleiades Stars:
  Exploring the Effects of Rotation on Broadband Colors},} \aj, 148, 30,
  \dodoi{10.1088/0004-6256/148/2/30}

% type= article
\bibitem[{A. {Kaminski} {et~al.}(2025){Kaminski}, {Sabotta}, {Kemmer},
  {Chaturvedi}, {Burn}, {Morales}, {Caballero}, {Ribas}, {Reiners},
  {Quirrenbach}, {Amado}, {B{\'e}jar}, {Dreizler}, {Guenther}, {Hatzes},
  {Henning}, {K{\"u}rster}, {Montes}, {Nagel}, {Pall{\'e}}, {Pinter},
  {Reffert}, {Schlecker}, {Shan}, {Trifonov}, {Osorio}, \&
  {Zechmeister}}]{kaminski2025}
{Kaminski}, A., {Sabotta}, S., {Kemmer}, J., {et~al.} 2025,
  \bibinfo{title}{{The CARMENES search for exoplanets around M dwarfs:
  Occurrence rates of Earth-like planets around very low-mass stars},} \aap,
  696, A101, \dodoi{10.1051/0004-6361/202453381}

% type= article
\bibitem[{S.~C. {Keller} {et~al.}(2007){Keller}, {Schmidt}, {Bessell},
  {Conroy}, {Francis}, {Granlund}, {Kowald}, {Oates}, {Martin-Jones},
  {Preston}, {Tisserand}, {Vaccarella}, \& {Waterson}}]{keller2007}
{Keller}, S.~C., {Schmidt}, B.~P., {Bessell}, M.~S., {et~al.} 2007,
  \bibinfo{title}{{The SkyMapper Telescope and The Southern Sky Survey},}
  \pasa, 24, 1, \dodoi{10.1071/AS07001}

% type= article
\bibitem[{G.~M. {Kennedy} {et~al.}(2018){Kennedy}, {Bryden}, {Ardila}, {Eiroa},
  {Lestrade}, {Marshall}, {Matthews}, {Moro-Martin}, \& {Wyatt}}]{kennedy2018}
{Kennedy}, G.~M., {Bryden}, G., {Ardila}, D., {et~al.} 2018,
  \bibinfo{title}{{Kuiper belt analogues in nearby M-type planet-host
  systems},} \mnras, 476, 4584, \dodoi{10.1093/mnras/sty492}

% type= article
\bibitem[{S.~J. {Kenyon} \& B.~C. {Bromley}(2002){Kenyon} \&
  {Bromley}}]{kb2002signpost}
{Kenyon}, S.~J., \& {Bromley}, B.~C. 2002, \bibinfo{title}{{Dusty Rings:
  Signposts of Recent Planet Formation},} \apjl, 577, L35,
  \dodoi{10.1086/344084}

% type= article
\bibitem[{S.~J. {Kenyon} \& B.~C. {Bromley}(2004){Kenyon} \&
  {Bromley}}]{kb2004}
{Kenyon}, S.~J., \& {Bromley}, B.~C. 2004, \bibinfo{title}{{Detecting the Dusty
  Debris of Terrestrial Planet Formation},} \apjl, 602, L133,
  \dodoi{10.1086/382693}

% type= article
\bibitem[{S.~J. {Kenyon} \& B.~C. {Bromley}(2008){Kenyon} \&
  {Bromley}}]{kb2008}
{Kenyon}, S.~J., \& {Bromley}, B.~C. 2008, \bibinfo{title}{{Variations on
  Debris Disks: Icy Planet Formation at 30-150 AU for 1-3 M-sun Main-Sequence
  Stars},} \apjs, 179, 451, \dodoi{10.1086/591794}

% type= article
\bibitem[{S.~J. {Kenyon} \& L. {Hartmann}(1995){Kenyon} \& {Hartmann}}]{kh1995}
{Kenyon}, S.~J., \& {Hartmann}, L. 1995, \bibinfo{title}{{Pre-Main-Sequence
  Evolution in the Taurus-Auriga Molecular Cloud},} \apjs, 101, 117,
  \dodoi{10.1086/192235}

% type= article
\bibitem[{S.~J. {Kenyon} \& L.~W. {Hartmann}(1990){Kenyon} \&
  {Hartmann}}]{kh1990}
{Kenyon}, S.~J., \& {Hartmann}, L.~W. 1990, \bibinfo{title}{{On the Apparent
  Positions of T Tauri Stars in the H-R Diagram},} \apj, 349, 197,
  \dodoi{10.1086/168306}

% type= article
\bibitem[{S.~J. {Kenyon} {et~al.}(2016){Kenyon}, {Najita}, \&
  {Bromley}}]{knb2016}
{Kenyon}, S.~J., {Najita}, J.~R., \& {Bromley}, B.~C. 2016,
  \bibinfo{title}{{Rocky Planet Formation: Quick and Neat},} \apj, 831, 8,
  \dodoi{10.3847/0004-637X/831/1/8}

% type= article
\bibitem[{M.~F. {Kessler} {et~al.}(1996){Kessler}, {Steinz}, {Anderegg},
  {Clavel}, {Drechsel}, {Estaria}, {Faelker}, {Riedinger}, {Robson}, {Taylor},
  \& {Xim{\'e}nez de Ferr{\'a}n}}]{kessler1996}
{Kessler}, M.~F., {Steinz}, J.~A., {Anderegg}, M.~E., {et~al.} 1996,
  \bibinfo{title}{{The Infrared Space Observatory (ISO) mission.},} \aap, 315,
  L27

% type= article
\bibitem[{N.~V. {Kharchenko}(2001){Kharchenko}}]{kharchenko2001}
{Kharchenko}, N.~V. 2001, \bibinfo{title}{{All-sky compiled catalogue of 2.5
  million stars},} Kinematika i Fizika Nebesnykh Tel, 17, 409

% type= article
\bibitem[{N.~V. {Kharchenko} {et~al.}(2005){Kharchenko}, {Piskunov},
  {R{\"o}ser}, {Schilbach}, \& {Scholz}}]{kharchenko2005}
{Kharchenko}, N.~V., {Piskunov}, A.~E., {R{\"o}ser}, S., {Schilbach}, E., \&
  {Scholz}, R.~D. 2005, \bibinfo{title}{{Astrophysical parameters of Galactic
  open clusters},} \aap, 438, 1163, \dodoi{10.1051/0004-6361:20042523}

% type= article
\bibitem[{F. {Kiefer} {et~al.}(2018){Kiefer}, {Halbwachs}, {Lebreton},
  {Soubiran}, {Arenou}, {Pourbaix}, {Famaey}, {Guillout}, {Ibata}, \&
  {Mazeh}}]{kiefer2018}
{Kiefer}, F., {Halbwachs}, J.-L., {Lebreton}, Y., {et~al.} 2018,
  \bibinfo{title}{{Masses of the components of SB2 binaries observed with Gaia
  - IV. Accurate SB2 orbits for 14 binaries and masses of three
  binaries$^{*}$},} \mnras, 474, 731, \dodoi{10.1093/mnras/stx2794}

% type= article
\bibitem[{J.~S. {Kim} {et~al.}(2005){Kim}, {Hines}, {Backman}, {Hillenbrand},
  {Meyer}, {Rodmann}, {Moro-Mart{\'\i}n}, {Carpenter}, {Silverstone},
  {Bouwman}, {Mamajek}, {Wolf}, {Malhotra}, {Pascucci}, {Najita}, {Padgett},
  {Henning}, {Brooke}, {Cohen}, {Strom}, {Stobie}, {Engelbracht}, {Gordon},
  {Misselt}, {Morrison}, {Muzerolle}, \& {Su}}]{kim2005}
{Kim}, J.~S., {Hines}, D.~C., {Backman}, D.~E., {et~al.} 2005,
  \bibinfo{title}{{Formation and Evolution of Planetary Systems: Cold Outer
  Disks Associated with Sun-like Stars},} \apj, 632, 659,
  \dodoi{10.1086/432863}

% type= article
\bibitem[{S. {Kimeswenger} {et~al.}(2004){Kimeswenger}, {Lederle}, {Richichi},
  {Percheron}, {Paresce}, {Armsdorfer}, {Bacher}, {Cabrera-Lavers}, {Kausch},
  {Rassia}, {Schmeja}, {Tapken}, {Fouqu{\'e}}, {Maury}, \&
  {Epchtein}}]{kimeswenger2004}
{Kimeswenger}, S., {Lederle}, C., {Richichi}, A., {et~al.} 2004,
  \bibinfo{title}{{J - K DENIS photometry of a VLTI-selected sample of bright
  southern stars},} \aap, 413, 1037, \dodoi{10.1051/0004-6361:20031576}

% type= article
\bibitem[{J.~R. {King} {et~al.}(2003){King}, {Villarreal}, {Soderblom},
  {Gulliver}, \& {Adelman}}]{king2003}
{King}, J.~R., {Villarreal}, A.~R., {Soderblom}, D.~R., {Gulliver}, A.~F., \&
  {Adelman}, S.~J. 2003, \bibinfo{title}{{Stellar Kinematic Groups. II. A
  Reexamination of the Membership, Activity, and Age of the Ursa Major Group},}
  \aj, 125, 1980, \dodoi{10.1086/368241}

% type= article
\bibitem[{M. {Kiraga}(2012){Kiraga}}]{kiraga2012}
{Kiraga}, M. 2012, \bibinfo{title}{{ASAS Photometry of ROSAT Sources. I.
  Periodic Variable Stars Coincident with Bright Sources from the ROSAT All Sky
  Survey},} \actaa, 62, 67, \dodoi{10.48550/arXiv.1204.3825}

% type= article
\bibitem[{M. {Kiraga} \& K. {Stepien}(2013){Kiraga} \& {Stepien}}]{kiraga2013}
{Kiraga}, M., \& {Stepien}, K. 2013, \bibinfo{title}{{ASAS Photometry of ROSAT
  Sources. II. New Variables from the ASAS North Survey},} \actaa, 63, 53,
  \dodoi{10.48550/arXiv.1304.3236}

% type= article
\bibitem[{O. {Kochukhov} \& D. {Shulyak}(2019){Kochukhov} \&
  {Shulyak}}]{kochukhov2019}
{Kochukhov}, O., \& {Shulyak}, D. 2019, \bibinfo{title}{{Magnetic Field of the
  Eclipsing M-dwarf Binary YY Gem},} \apj, 873, 69,
  \dodoi{10.3847/1538-4357/ab06c5}

% type= article
\bibitem[{C. {Koen} {et~al.}(2010){Koen}, {Kilkenny}, {van Wyk}, \&
  {Marang}}]{koen2010}
{Koen}, C., {Kilkenny}, D., {van Wyk}, F., \& {Marang}, F. 2010,
  \bibinfo{title}{{UBV(RI)$_{C}$ JHK observations of Hipparcos-selected nearby
  stars},} \mnras, 403, 1949, \dodoi{10.1111/j.1365-2966.2009.16182.x}

% type= article
\bibitem[{D.~W. {Koerner} {et~al.}(2010){Koerner}, {Kim}, {Trilling}, {Larson},
  {Cotera}, {Stapelfeldt}, {Wahhaj}, {Fajardo-Acosta}, {Padgett}, \&
  {Backman}}]{koerner2010}
{Koerner}, D.~W., {Kim}, S., {Trilling}, D.~E., {et~al.} 2010,
  \bibinfo{title}{{New Debris Disk Candidates Around 49 Nearby Stars},} \apjl,
  710, L26, \dodoi{10.1088/2041-8205/710/1/L26}

% type= article
\bibitem[{J. {Koornneef}(1983){Koornneef}}]{koorneef1983}
{Koornneef}, J. 1983, \bibinfo{title}{{Near infrared photometry. I.
  Homogenization of near-infrared data from southern bright stars.},} \aaps,
  51, 489

% type= article
\bibitem[{M. {Kounkel} \& K. {Covey}(2019){Kounkel} \& {Covey}}]{kounkel2019}
{Kounkel}, M., \& {Covey}, K. 2019, \bibinfo{title}{{Untangling the Galaxy. I.
  Local Structure and Star Formation History of the Milky Way},} \aj, 158, 122,
  \dodoi{10.3847/1538-3881/ab339a}

% type= article
\bibitem[{M. {Kounkel} {et~al.}(2022){Kounkel}, {Mcbride}, {Stassun}, \&
  {Leigh}}]{kounkel2022}
{Kounkel}, M., {Mcbride}, A., {Stassun}, K.~G., \& {Leigh}, N. 2022,
  \bibinfo{title}{{Searching for young runaways across the sky},} \mnras, 517,
  1946, \dodoi{10.1093/mnras/stac2829}

% type= article
\bibitem[{G. {Kov{\'a}cs} {et~al.}(2014){Kov{\'a}cs}, {Hartman}, {Bakos},
  {Quinn}, {Penev}, {Latham}, {Bhatti}, {Csubry}, \& {de
  Val-Borro}}]{kovacs2014}
{Kov{\'a}cs}, G., {Hartman}, J.~D., {Bakos}, G.~{\'A}., {et~al.} 2014,
  \bibinfo{title}{{Stellar rotational periods in the planet hosting open
  cluster Praesepe},} \mnras, 442, 2081, \dodoi{10.1093/mnras/stu946}

% type= article
\bibitem[{A.~L. {Kraus} {et~al.}(2017){Kraus}, {Herczeg}, {Rizzuto}, {Mann},
  {Slesnick}, {Carpenter}, {Hillenbrand}, \& {Mamajek}}]{kraus2017}
{Kraus}, A.~L., {Herczeg}, G.~J., {Rizzuto}, A.~C., {et~al.} 2017,
  \bibinfo{title}{{The Greater Taurus-Auriga Ecosystem. I. There is a
  Distributed Older Population},} \apj, 838, 150,
  \dodoi{10.3847/1538-4357/aa62a0}

% type= article
\bibitem[{A.~L. {Kraus} \& L.~A. {Hillenbrand}(2007){Kraus} \&
  {Hillenbrand}}]{kraus2007}
{Kraus}, A.~L., \& {Hillenbrand}, L.~A. 2007, \bibinfo{title}{{The Stellar
  Populations of Praesepe and Coma Berenices},} \aj, 134, 2340,
  \dodoi{10.1086/522831}

% type= article
\bibitem[{A.~V. {Krivov} \& M.~C. {Wyatt}(2021){Krivov} \&
  {Wyatt}}]{krivov2021}
{Krivov}, A.~V., \& {Wyatt}, M.~C. 2021, \bibinfo{title}{{Solution to the
  debris disc mass problem: planetesimals are born small?},} \mnras, 500, 718,
  \dodoi{10.1093/mnras/staa2385}

% type= article
\bibitem[{M. {K{\"u}ker} {et~al.}(2019){K{\"u}ker}, {R{\"u}diger}, {Olah}, \&
  {Strassmeier}}]{kuker2019}
{K{\"u}ker}, M., {R{\"u}diger}, G., {Olah}, K., \& {Strassmeier}, K.~G. 2019,
  \bibinfo{title}{{Cycle period, differential rotation, and meridional flow for
  early M dwarf stars},} \aap, 622, A40, \dodoi{10.1051/0004-6361/201833173}

% type= article
\bibitem[{R. {Lallement} {et~al.}(2019){Lallement}, {Babusiaux}, {Vergely},
  {Katz}, {Arenou}, {Valette}, {Hottier}, \& {Capitanio}}]{lallement2019}
{Lallement}, R., {Babusiaux}, C., {Vergely}, J.~L., {et~al.} 2019,
  \bibinfo{title}{{Gaia-2MASS 3D maps of Galactic interstellar dust within 3
  kpc},} \aap, 625, A135, \dodoi{10.1051/0004-6361/201834695}

% type= article
\bibitem[{D.~L. {Lambert} \& B.~E. {Reddy}(2004){Lambert} \&
  {Reddy}}]{lambert2004}
{Lambert}, D.~L., \& {Reddy}, B.~E. 2004, \bibinfo{title}{{Lithium abundances
  of the local thin disc stars},} \mnras, 349, 757,
  \dodoi{10.1111/j.1365-2966.2004.07557.x}

% type= article
\bibitem[{A.~C. {Lanzafame} {et~al.}(2023){Lanzafame}, {Brugaletta},
  {Fr{\'e}mat}, {Sordo}, {Creevey}, {Andretta}, {Scandariato}, {Bus{\`a}},
  {Distefano}, {Korn}, {de Laverny}, {Recio-Blanco}, {Abreu Aramburu},
  {{\'A}lvarez}, {Andrae}, {Bailer-Jones}, {Bakker}, {Bellas-Velidis},
  {Bijaoui}, {Brouillet}, {Burlacu}, {Carballo}, {Casamiquela}, {Chaoul},
  {Chiavassa}, {Contursi}, {Cooper}, {Dafonte}, {Dapergolas}, {Delchambre},
  {Demouchy}, {Dharmawardena}, {Drimmel}, {Edvardsson}, {Fouesneau},
  {Garabato}, {Garc{\'\i}a-Lario}, {Garc{\'\i}a-Torres}, {Gavel}, {Gomez},
  {Gonz{\'a}lez-Santamar{\'\i}a}, {Hatzidimitriou}, {Heiter}, {Jean-Antoine
  Piccolo}, {Kontizas}, {Kordopatis}, {Lebreton}, {Licata}, {Lindstr{\o}m},
  {Livanou}, {Lobel}, {Lorca}, {Magdaleno Romeo}, {Manteiga}, {Marocco},
  {Marshall}, {Mary}, {Nicolas}, {Ordenovic}, {Pailler}, {Palicio},
  {Pallas-Quintela}, {Panem}, {Pichon}, {Poggio}, {Riclet}, {Robin}, {Rybizki},
  {Santove{\~n}a}, {Sarro}, {Schultheis}, {Segol}, {Silvelo}, {Slezak},
  {Smart}, {Soubiran}, {S{\"u}veges}, {Th{\'e}venin}, {Torralba Elipe}, {Ulla},
  {Utrilla}, {Vallenari}, {van Dillen}, {Zhao}, \& {Zorec}}]{lanzafame2023}
{Lanzafame}, A.~C., {Brugaletta}, E., {Fr{\'e}mat}, Y., {et~al.} 2023,
  \bibinfo{title}{{Gaia Data Release 3. Stellar chromospheric activity and mass
  accretion from Ca II IRT observed by the Radial Velocity Spectrometer},}
  \aap, 674, A30, \dodoi{10.1051/0004-6361/202244156}

% type= article
\bibitem[{W.~A. {Lawson} \& L.~A. {Crause}(2005){Lawson} \&
  {Crause}}]{lawson2005}
{Lawson}, W.~A., \& {Crause}, L.~A. 2005, \bibinfo{title}{{Rotation periods for
  stars of the TW Hydrae association: the evidence for two spatially and
  rotationally distinct pre-main-sequence populations},} \mnras, 357, 1399,
  \dodoi{10.1111/j.1365-2966.2005.08793.x}

% type= article
\bibitem[{R.~A. {Lee} {et~al.}(2024){Lee}, {Gaidos}, {van Saders}, {Feiden}, \&
  {Gagn{\'e}}}]{lee2024}
{Lee}, R.~A., {Gaidos}, E., {van Saders}, J., {Feiden}, G.~A., \& {Gagn{\'e}},
  J. 2024, \bibinfo{title}{{Revisiting the membership, multiplicity, and age of
  the Beta Pictoris Moving Group in the Gaia era},} \mnras, 528, 4760,
  \dodoi{10.1093/mnras/stae007}

% type= article
\bibitem[{P.~J.~T. {Leonard} \& M.~J. {Duncan}(1990){Leonard} \&
  {Duncan}}]{leonard1990}
{Leonard}, P. J.~T., \& {Duncan}, M.~J. 1990, \bibinfo{title}{{Runaway Stars
  from Young Star Clusters Containing Initial Binaries. II. A Mass Spectrum and
  a Binary Energy Spectrum},} \aj, 99, 608, \dodoi{10.1086/115354}

% type= article
\bibitem[{S. {L{\'e}pine} {et~al.}(2009){L{\'e}pine}, {Thorstensen}, {Shara},
  \& {Rich}}]{lepine2009}
{L{\'e}pine}, S., {Thorstensen}, J.~R., {Shara}, M.~M., \& {Rich}, R.~M. 2009,
  \bibinfo{title}{{New Neighbors: Parallaxes of 18 Nearby Stars Selected from
  the LSPM-North Catalog},} \aj, 137, 4109,
  \dodoi{10.1088/0004-6256/137/5/4109}

% type= article
\bibitem[{J.~F. {Lestrade} {et~al.}(2025){Lestrade}, {Matthews}, {Kennedy},
  {Sibthorpe}, {Wyatt}, {Booth}, {Greaves}, {Duch{\^e}ne}, {Moro-Mart{\'\i}n},
  \& {Jobic}}]{lestrade2025}
{Lestrade}, J.~F., {Matthews}, B.~C., {Kennedy}, G.~M., {et~al.} 2025,
  \bibinfo{title}{{Debris disks around M dwarfs: The Herschel DEBRIS survey},}
  \aap, 694, A123, \dodoi{10.1051/0004-6361/202451673}

% type= article
\bibitem[{L. {Li} {et~al.}(2023){Li}, {Wang}, {Chen}, \& {Jiang}}]{li2023}
{Li}, L., {Wang}, S., {Chen}, X., \& {Jiang}, Q. 2023, \bibinfo{title}{{The
  Ultraviolet to Mid-infrared Extinction Law of the Taurus Molecular Cloud
  Based on the Gaia DR3, GALEX, APASS, Pan-STARRS1, 2MASS, and WISE Surveys},}
  \apj, 956, 26, \dodoi{10.3847/1538-4357/aced8a}

% type= article
\bibitem[{J. {Lieman-Sifry} {et~al.}(2016){Lieman-Sifry}, {Hughes},
  {Carpenter}, {Gorti}, {Hales}, \& {Flaherty}}]{lieman2016}
{Lieman-Sifry}, J., {Hughes}, A.~M., {Carpenter}, J.~M., {et~al.} 2016,
  \bibinfo{title}{{Debris Disks in the Scorpius-Centaurus OB Association
  Resolved by ALMA},} \apj, 828, 25, \dodoi{10.3847/0004-637X/828/1/25}

% type= article
\bibitem[{F. {Ligni{\`e}res} {et~al.}(2009){Ligni{\`e}res}, {Petit},
  {B{\"o}hm}, \& {Auri{\`e}re}}]{lignieres2009}
{Ligni{\`e}res}, F., {Petit}, P., {B{\"o}hm}, T., \& {Auri{\`e}re}, M. 2009,
  \bibinfo{title}{{First evidence of a magnetic field on
  <ASTROBJ>Vega</ASTROBJ>. Towards a new class of magnetic A-type stars},}
  \aap, 500, L41, \dodoi{10.1051/0004-6361/200911996}

% type= article
\bibitem[{L. {Lindegren} {et~al.}(2021){Lindegren}, {Klioner}, {Hern{\'a}ndez},
  {Bombrun}, {Ramos-Lerate}, {Steidelm{\"u}ller}, {Bastian}, {Biermann}, {de
  Torres}, {Gerlach}, {Geyer}, {Hilger}, {Hobbs}, {Lammers}, {McMillan},
  {Stephenson}, {Casta{\~n}eda}, {Davidson}, {Fabricius}, {Gracia-Abril},
  {Portell}, {Rowell}, {Teyssier}, {Torra}, {Bartolom{\'e}}, {Clotet},
  {Garralda}, {Gonz{\'a}lez-Vidal}, {Torra}, {Abbas}, {Altmann}, {Anglada
  Varela}, {Balaguer-N{\'u}{\~n}ez}, {Balog}, {Barache}, {Becciani}, {Bernet},
  {Bertone}, {Bianchi}, {Bouquillon}, {Brown}, {Bucciarelli}, {Busonero},
  {Butkevich}, {Buzzi}, {Cancelliere}, {Carlucci}, {Charlot}, {Cioni},
  {Crosta}, {Crowley}, {del Peloso}, {del Pozo}, {Drimmel}, {Esquej}, {Fienga},
  {Fraile}, {Gai}, {Garcia-Reinaldos}, {Guerra}, {Hambly}, {Hauser},
  {Jan{\ss}en}, {Jordan}, {Kostrzewa-Rutkowska}, {Lattanzi}, {Liao}, {Licata},
  {Lister}, {L{\"o}ffler}, {Marchant}, {Masip}, {Mignard}, {Mints}, {Molina},
  {Mora}, {Morbidelli}, {Murphy}, {Pagani}, {Panuzzo}, {Pe{\~n}alosa Esteller},
  {Poggio}, {Re Fiorentin}, {Riva}, {Sagrist{\`a} Sell{\'e}s}, {Sanchez
  Gimenez}, {Sarasso}, {Sciacca}, {Siddiqui}, {Smart}, {Souami}, {Spagna},
  {Steele}, {Taris}, {Utrilla}, {van Reeven}, \& {Vecchiato}}]{lindegren2021}
{Lindegren}, L., {Klioner}, S.~A., {Hern{\'a}ndez}, J., {et~al.} 2021,
  \bibinfo{title}{{Gaia Early Data Release 3. The astrometric solution},} \aap,
  649, A2, \dodoi{10.1051/0004-6361/202039709}

% type= article
\bibitem[{J.~L. {Linsky}(2017){Linsky}}]{linsky2017}
{Linsky}, J.~L. 2017, \bibinfo{title}{{Stellar Model Chromospheres and
  Spectroscopic Diagnostics},} \araa, 55, 159,
  \dodoi{10.1146/annurev-astro-091916-055327}

% type= article
\bibitem[{J.~L. {Linsky} {et~al.}(1979){Linsky}, {Hunten}, {Sowell}, {Glackin},
  \& {Kelch}}]{linsky1979}
{Linsky}, J.~L., {Hunten}, D.~M., {Sowell}, R., {Glackin}, D.~L., \& {Kelch},
  W.~L. 1979, \bibinfo{title}{{Stellar model chromospheres. XI. A survey of CA
  II lam 8542 line profiles in late-type stars of differing chromospheric
  activity.},} \apjs, 41, 481, \dodoi{10.1086/190627}

% type= article
\bibitem[{J.~L. {Linsky} \& S. {Redfield}(2021){Linsky} \&
  {Redfield}}]{linsky2021}
{Linsky}, J.~L., \& {Redfield}, S. 2021, \bibinfo{title}{{Could the Local
  Cavity be an Irregularly Shaped Str{\"o}mgren Sphere?},} \apj, 920, 75,
  \dodoi{10.3847/1538-4357/ac1feb}

% type= article
\bibitem[{I. {Lipartito} {et~al.}(2021){Lipartito}, {Bailey}, {Brandt},
  {Mazin}, {Mateo}, {Spencer}, \& {Roederer}}]{lipartito2021}
{Lipartito}, I., {Bailey}, III, J.~I., {Brandt}, T.~D., {et~al.} 2021,
  \bibinfo{title}{{Orbital Parameters and Binary Properties of 37 FGK Stars in
  the Cores of Open Clusters NGC 2516 and NGC 2422},} \aj, 162, 285,
  \dodoi{10.3847/1538-3881/ac2ccd}

% type= article
\bibitem[{R. {Liseau} {et~al.}(2008){Liseau}, {Risacher}, {Brandeker}, {Eiroa},
  {Fridlund}, {Nilsson}, {Olofsson}, {Pilbratt}, \&
  {Th{\'e}bault}}]{liseau2008}
{Liseau}, R., {Risacher}, C., {Brandeker}, A., {et~al.} 2008,
  \bibinfo{title}{{q$^{1}$ Eridani: a solar-type star with a planet and a dust
  belt},} \aap, 480, L47, \dodoi{10.1051/0004-6361:20079276}

% type= article
\bibitem[{L. {Liu} \& X. {Pang}(2019){Liu} \& {Pang}}]{liu2019}
{Liu}, L., \& {Pang}, X. 2019, \bibinfo{title}{{A Catalog of Newly Identified
  Star Clusters in Gaia DR2},} \apjs, 245, 32, \dodoi{10.3847/1538-4365/ab530a}

% type= article
\bibitem[{M.~C. {Liu} {et~al.}(2004){Liu}, {Matthews}, {Williams}, \&
  {Kalas}}]{liu2004}
{Liu}, M.~C., {Matthews}, B.~C., {Williams}, J.~P., \& {Kalas}, P.~G. 2004,
  \bibinfo{title}{{A Submillimeter Search of Nearby Young Stars for Cold Dust:
  Discovery of Debris Disks around Two Low-Mass Stars},} \apj, 608, 526,
  \dodoi{10.1086/392531}

% type= article
\bibitem[{F. {Llorente de Andr{\'e}s} {et~al.}(2021){Llorente de Andr{\'e}s},
  {Chavero}, {de la Reza}, {Roca-F{\`a}brega}, \& {Cifuentes}}]{llorente2021}
{Llorente de Andr{\'e}s}, F., {Chavero}, C., {de la Reza}, R.,
  {Roca-F{\`a}brega}, S., \& {Cifuentes}, C. 2021, \bibinfo{title}{{The
  evolution of lithium in FGK dwarf stars. The lithium-rotation connection and
  the Li desert},} \aap, 654, A137, \dodoi{10.1051/0004-6361/202141339}

% type= article
\bibitem[{N. {Lodieu} {et~al.}(2019{\natexlab{a}}){Lodieu},
  {P{\'e}rez-Garrido}, {Smart}, \& {Silvotti}}]{lodieu2019b}
{Lodieu}, N., {P{\'e}rez-Garrido}, A., {Smart}, R.~L., \& {Silvotti}, R.
  2019{\natexlab{a}}, \bibinfo{title}{{A 5D view of the {\ensuremath{\alpha}}
  Per, Pleiades, and Praesepe clusters},} \aap, 628, A66,
  \dodoi{10.1051/0004-6361/201935533}

% type= article
\bibitem[{N. {Lodieu} {et~al.}(2019{\natexlab{b}}){Lodieu}, {Smart},
  {P{\'e}rez-Garrido}, \& {Silvotti}}]{lodieu2019a}
{Lodieu}, N., {Smart}, R.~L., {P{\'e}rez-Garrido}, A., \& {Silvotti}, R.
  2019{\natexlab{b}}, \bibinfo{title}{{A 3D view of the Hyades stellar and
  sub-stellar population},} \aap, 623, A35, \dodoi{10.1051/0004-6361/201834045}

% type= article
\bibitem[{R. {L{\'o}pez-Valdivia} {et~al.}(2015){L{\'o}pez-Valdivia},
  {Hern{\'a}ndez-{\'A}guila}, {Bertone}, {Ch{\'a}vez}, {Cruz-Saenz de Miera},
  \& {Amazo-G{\'o}mez}}]{lopezvaldivia2015}
{L{\'o}pez-Valdivia}, R., {Hern{\'a}ndez-{\'A}guila}, J.~B., {Bertone}, E.,
  {et~al.} 2015, \bibinfo{title}{{Lithium abundance in a sample of solar-like
  stars},} \mnras, 451, 4368, \dodoi{10.1093/mnras/stv1222}

% type= article
\bibitem[{D. {Lorenzo-Oliveira} {et~al.}(2016){Lorenzo-Oliveira}, {Porto de
  Mello}, {Dutra-Ferreira}, \& {Ribas}}]{lorenzooliveira2016}
{Lorenzo-Oliveira}, D., {Porto de Mello}, G.~F., {Dutra-Ferreira}, L., \&
  {Ribas}, I. 2016, \bibinfo{title}{{Fine structure of the age-chromospheric
  activity relation in solar-type stars. I. The Ca II infrared triplet:
  Absolute flux calibration},} \aap, 595, A11,
  \dodoi{10.1051/0004-6361/201628825}

% type= article
\bibitem[{D. {Lorenzo-Oliveira} {et~al.}(2018){Lorenzo-Oliveira}, {Freitas},
  {Mel{\'e}ndez}, {Bedell}, {Ram{\'\i}rez}, {Bean}, {Asplund}, {Spina},
  {Dreizler}, {Alves-Brito}, \& {Casagrande}}]{lorenzooliveira2018}
{Lorenzo-Oliveira}, D., {Freitas}, F.~C., {Mel{\'e}ndez}, J., {et~al.} 2018,
  \bibinfo{title}{{The Solar Twin Planet Search. The age-chromospheric activity
  relation},} \aap, 619, A73, \dodoi{10.1051/0004-6361/201629294}

% type= article
\bibitem[{J.~B. {Lovell} {et~al.}(2021){Lovell}, {Marino}, {Wyatt}, {Kennedy},
  {MacGregor}, {Stapelfeldt}, {Dent}, {Krist}, {Matr{\`a}}, {Kral},
  {Pani{\'c}}, {Pearce}, \& {Wilner}}]{lovell2021}
{Lovell}, J.~B., {Marino}, S., {Wyatt}, M.~C., {et~al.} 2021,
  \bibinfo{title}{{High-resolution ALMA and HST images of q$^{1}$ Eri: an
  asymmetric debris disc with an eccentric Jupiter},} \mnras, 506, 1978,
  \dodoi{10.1093/mnras/stab1678}

% type= article
\bibitem[{F.~J. {Low} {et~al.}(2005){Low}, {Smith}, {Werner}, {Chen}, {Krause},
  {Jura}, \& {Hines}}]{low2005}
{Low}, F.~J., {Smith}, P.~S., {Werner}, M., {et~al.} 2005,
  \bibinfo{title}{{Exploring Terrestrial Planet Formation in the TW Hydrae
  Association},} \apj, 631, 1170, \dodoi{10.1086/432640}

% type= article
\bibitem[{D. {Lubin} {et~al.}(2024){Lubin}, {Holden}, {Stock}, {Melis}, \&
  {Tytler}}]{lubin2024}
{Lubin}, D., {Holden}, B.~P., {Stock}, C., {Melis}, C., \& {Tytler}, D. 2024,
  \bibinfo{title}{{Hamilton Echelle Spectrograph Observations of Solar Analog
  Field Stars: Lithium Abundance and Activity},} \aj, 168, 240,
  \dodoi{10.3847/1538-3881/ad823d}

% type= article
\bibitem[{K.~L. {Luhman}(2022){Luhman}}]{luhman2022a}
{Luhman}, K.~L. 2022, \bibinfo{title}{{A Census of the Stellar Populations in
  the Sco-Cen Complex},} \aj, 163, 24, \dodoi{10.3847/1538-3881/ac35e2}

% type= article
\bibitem[{K.~L. {Luhman}(2023{\natexlab{a}}){Luhman}}]{luhman2023a}
{Luhman}, K.~L. 2023{\natexlab{a}}, \bibinfo{title}{{A Census of the Taurus
  Star-forming Region and Neighboring Associations with Gaia},} \aj, 165, 37,
  \dodoi{10.3847/1538-3881/ac9da3}

% type= article
\bibitem[{K.~L. {Luhman}(2023{\natexlab{b}}){Luhman}}]{luhman2023b}
{Luhman}, K.~L. 2023{\natexlab{b}}, \bibinfo{title}{{A Census of the TW Hya
  Association with Gaia},} \aj, 165, 269, \dodoi{10.3847/1538-3881/accf19}

% type= article
\bibitem[{K.~L. {Luhman} \& T.~L. {Esplin}(2022){Luhman} \&
  {Esplin}}]{luhman2022b}
{Luhman}, K.~L., \& {Esplin}, T.~L. 2022, \bibinfo{title}{{Spectroscopy of
  Candidate Members of the Sco-Cen Complex},} \aj, 163, 26,
  \dodoi{10.3847/1538-3881/ac35e4}

% type= article
\bibitem[{J.~C. {Lurie} {et~al.}(2014){Lurie}, {Henry}, {Jao}, {Quinn},
  {Winters}, {Ianna}, {Koerner}, {Riedel}, \& {Subasavage}}]{lurie2014}
{Lurie}, J.~C., {Henry}, T.~J., {Jao}, W.-C., {et~al.} 2014,
  \bibinfo{title}{{The Solar Neighborhood. XXXIV. a Search for Planets Orbiting
  Nearby M Dwarfs Using Astrometry},} \aj, 148, 91,
  \dodoi{10.1088/0004-6256/148/5/91}

% type= article
\bibitem[{W. {Lyra} {et~al.}(2006){Lyra}, {Moitinho}, {van der Bliek}, \&
  {Alves}}]{lyra2006}
{Lyra}, W., {Moitinho}, A., {van der Bliek}, N.~S., \& {Alves}, J. 2006,
  \bibinfo{title}{{On the difference between nuclear and contraction ages},}
  \aap, 453, 101, \dodoi{10.1051/0004-6361:20053894}

% type= article
\bibitem[{M.~A. {MacGregor} {et~al.}(2016){MacGregor}, {Wilner}, {Chandler},
  {Ricci}, {Maddison}, {Cranmer}, {Andrews}, {Hughes}, \&
  {Steele}}]{macgregor2016}
{MacGregor}, M.~A., {Wilner}, D.~J., {Chandler}, C., {et~al.} 2016,
  \bibinfo{title}{{Constraints on Planetesimal Collision Models in Debris
  Disks},} \apj, 823, 79, \dodoi{10.3847/0004-637X/823/2/79}

% type= article
\bibitem[{M.~A. {MacGregor} {et~al.}(2017){MacGregor}, {Matr{\`a}}, {Kalas},
  {Wilner}, {Pan}, {Kennedy}, {Wyatt}, {Duchene}, {Hughes}, {Rieke}, {Clampin},
  {Fitzgerald}, {Graham}, {Holland}, {Pani{\'c}}, {Shannon}, \&
  {Su}}]{macgregor2017}
{MacGregor}, M.~A., {Matr{\`a}}, L., {Kalas}, P., {et~al.} 2017,
  \bibinfo{title}{{A Complete ALMA Map of the Fomalhaut Debris Disk},} \apj,
  842, 8, \dodoi{10.3847/1538-4357/aa71ae}

% type= article
\bibitem[{M.~A. {MacGregor} {et~al.}(2022){MacGregor}, {Hurt}, {Stark},
  {Howard}, {Weinberger}, {Ren}, {Schneider}, {Choquet}, \&
  {Mawet}}]{macgregor2022}
{MacGregor}, M.~A., {Hurt}, S.~A., {Stark}, C.~C., {et~al.} 2022,
  \bibinfo{title}{{ALMA Images the Eccentric HD 53143 Debris Disk},} \apjl,
  933, L1, \dodoi{10.3847/2041-8213/ac7729}

% type= article
\bibitem[{E. {Magaudda} {et~al.}(2020){Magaudda}, {Stelzer}, {Covey}, {Raetz},
  {Matt}, \& {Scholz}}]{magaudda2020}
{Magaudda}, E., {Stelzer}, B., {Covey}, K.~R., {et~al.} 2020,
  \bibinfo{title}{{Relation of X-ray activity and rotation in M dwarfs and
  predicted time-evolution of the X-ray luminosity},} \aap, 638, A20,
  \dodoi{10.1051/0004-6361/201937408}

% type= article
\bibitem[{E. {Magaudda} {et~al.}(2022){Magaudda}, {Stelzer}, {Raetz},
  {Klutsch}, {Salvato}, \& {Wolf}}]{magaudda2022}
{Magaudda}, E., {Stelzer}, B., {Raetz}, S., {et~al.} 2022,
  \bibinfo{title}{{First eROSITA study of nearby M dwarfs and the
  rotation-activity relation in combination with TESS},} \aap, 661, A29,
  \dodoi{10.1051/0004-6361/202141617}

% type= article
\bibitem[{L. {Malo} {et~al.}(2014){Malo}, {Artigau}, {Doyon}, {Lafreni{\`e}re},
  {Albert}, \& {Gagn{\'e}}}]{malo2014}
{Malo}, L., {Artigau}, {\'E}., {Doyon}, R., {et~al.} 2014,
  \bibinfo{title}{{BANYAN. III. Radial Velocity, Rotation, and X-Ray Emission
  of Low-mass Star Candidates in Nearby Young Kinematic Groups},} \apj, 788,
  81, \dodoi{10.1088/0004-637X/788/1/81}

% type= article
\bibitem[{E.~E. {Mamajek} \& L.~A. {Hillenbrand}(2008){Mamajek} \&
  {Hillenbrand}}]{mamajek2008}
{Mamajek}, E.~E., \& {Hillenbrand}, L.~A. 2008, \bibinfo{title}{{Improved Age
  Estimation for Solar-Type Dwarfs Using Activity-Rotation Diagnostics},} \apj,
  687, 1264, \dodoi{10.1086/591785}

% type= article
\bibitem[{E.~E. {Mamajek} {et~al.}(1999){Mamajek}, {Lawson}, \&
  {Feigelson}}]{mamajek1999}
{Mamajek}, E.~E., {Lawson}, W.~A., \& {Feigelson}, E.~D. 1999,
  \bibinfo{title}{{The {\ensuremath{\eta}} Chamaeleontis Cluster: A Remarkable
  New Nearby Young Open Cluster},} \apjl, 516, L77, \dodoi{10.1086/312005}

% type= inproceedings
\bibitem[{C.~F. {Manara} {et~al.}(2023){Manara}, {Ansdell}, {Rosotti},
  {Hughes}, {Armitage}, {Lodato}, \& {Williams}}]{manara2023}
{Manara}, C.~F., {Ansdell}, M., {Rosotti}, G.~P., {et~al.} 2023,
  \bibinfo{title}{{Demographics of Young Stars and their Protoplanetary Disks:
  Lessons Learned on Disk Evolution and its Connection to Planet Formation},}
  in Astronomical Society of the Pacific Conference Series, Vol. 534,
  Protostars and Planets VII, ed. S.~{Inutsuka}, Y.~{Aikawa}, T.~{Muto},
  K.~{Tomida}, \& M.~{Tamura}, 539, \dodoi{10.48550/arXiv.2203.09930}

% type= article
\bibitem[{C.~F. {Manara} {et~al.}(2013){Manara}, {Testi}, {Rigliaco},
  {Alcal{\'a}}, {Natta}, {Stelzer}, {Biazzo}, {Covino}, {Covino}, {Cupani},
  {D'Elia}, \& {Randich}}]{manara2013}
{Manara}, C.~F., {Testi}, L., {Rigliaco}, E., {et~al.} 2013,
  \bibinfo{title}{{X-shooter spectroscopy of young stellar objects. II. Impact
  of chromospheric emission on accretion rate estimates},} \aap, 551, A107,
  \dodoi{10.1051/0004-6361/201220921}

% type= article
\bibitem[{S. {Marino} {et~al.}(2017){Marino}, {Wyatt}, {Kennedy}, {Holland},
  {Matr{\`a}}, {Shannon}, \& {Ivison}}]{marino2017}
{Marino}, S., {Wyatt}, M.~C., {Kennedy}, G.~M., {et~al.} 2017,
  \bibinfo{title}{{ALMA observations of the multiplanet system 61 Vir: what
  lies outside super-Earth systems?},} \mnras, 469, 3518,
  \dodoi{10.1093/mnras/stx1102}

% type= article
\bibitem[{S. {Marino} {et~al.}(2019){Marino}, {Yelverton}, {Booth}, {Faramaz},
  {Kennedy}, {Matr{\`a}}, \& {Wyatt}}]{marino2019}
{Marino}, S., {Yelverton}, B., {Booth}, M., {et~al.} 2019, \bibinfo{title}{{A
  gap in HD 92945's broad planetesimal disc revealed by ALMA},} \mnras, 484,
  1257, \dodoi{10.1093/mnras/stz049}

% type= article
\bibitem[{S. {Marino} {et~al.}(2018){Marino}, {Carpenter}, {Wyatt}, {Booth},
  {Casassus}, {Faramaz}, {Guzman}, {Hughes}, {Isella}, {Kennedy}, {Matr{\`a}},
  {Ricci}, \& {Corder}}]{marino2018}
{Marino}, S., {Carpenter}, J., {Wyatt}, M.~C., {et~al.} 2018,
  \bibinfo{title}{{A gap in the planetesimal disc around HD 107146 and
  asymmetric warm dust emission revealed by ALMA},} \mnras, 479, 5423,
  \dodoi{10.1093/mnras/sty1790}

% type= article
\bibitem[{S. {Marino} {et~al.}(2020){Marino}, {Zurlo}, {Faramaz}, {Milli},
  {Henning}, {Kennedy}, {Matr{\`a}}, {P{\'e}rez}, {Delorme}, {Cieza}, \&
  {Hughes}}]{marino2020}
{Marino}, S., {Zurlo}, A., {Faramaz}, V., {et~al.} 2020,
  \bibinfo{title}{{Insights into the planetary dynamics of HD 206893 with
  ALMA},} \mnras, 498, 1319, \dodoi{10.1093/mnras/staa2386}

% type= article
\bibitem[{S. {Marino} {et~al.}(2026){Marino}, {Matr{\`a}}, {Hughes},
  {Ehrhardt}, {Kennedy}, {del Burgo}, {Brennan}, {Han}, {Jankovic}, {Lovell},
  {Mac Manamon}, {Milli}, {Weber}, {Zawadzki}, {Bendahan-West}, {Fehr},
  {Mansell}, {Olofsson}, {Pearce}, {Bayo}, {Matthews}, {L{\"o}hne}, {Wyatt},
  {{\'A}brah{\'a}m}, {Bonduelle}, {Booth}, {Cataldi}, {Carpenter}, {Chiang},
  {Ertel}, {Hales}, {Henning}, {K{\'o}sp{\'a}l}, {Krivov}, {Luppe},
  {MacGregor}, {Marshall}, {Mo{\'o}r}, {P{\'e}rez}, {Sefilian}, {Sepulveda}, \&
  {Wilner}}]{marino2026}
{Marino}, S., {Matr{\`a}}, L., {Hughes}, A.~M., {et~al.} 2026,
  \bibinfo{title}{{The ALMA survey to Resolve exoKuiper belt Substructures
  (ARKS): I. Motivation, sample, data reduction, and results overview},} \aap,
  705, A195, \dodoi{10.1051/0004-6361/202556489}

% type= article
\bibitem[{S.~C. {Marsden} {et~al.}(2023){Marsden}, {Evensberget}, {Brown},
  {Neiner}, {Seach}, {Morin}, {Petit}, {Jeffers}, \& {Folsom}}]{marsden2023}
{Marsden}, S.~C., {Evensberget}, D., {Brown}, E.~L., {et~al.} 2023,
  \bibinfo{title}{{The magnetic field and stellar wind of the mature late-F
  star {\ensuremath{\chi}} Draconis A},} \mnras, 522, 792,
  \dodoi{10.1093/mnras/stad925}

% type= article
\bibitem[{J.~P. {Marshall} {et~al.}(2017){Marshall}, {Maddison}, {Thilliez},
  {Matthews}, {Wilner}, {Greaves}, \& {Holland}}]{marshall2017}
{Marshall}, J.~P., {Maddison}, S.~T., {Thilliez}, E., {et~al.} 2017,
  \bibinfo{title}{{New constraints on the millimetre emission of six debris
  discs},} \mnras, 468, 2719, \dodoi{10.1093/mnras/stx645}

% type= article
\bibitem[{J.~P. {Marshall} {et~al.}(2023){Marshall}, {Milli}, {Choquet}, {del
  Burgo}, {Kennedy}, {Kemper}, {Wyatt}, {Kral}, \& {Soummer}}]{marshall2023}
{Marshall}, J.~P., {Milli}, J., {Choquet}, E., {et~al.} 2023,
  \bibinfo{title}{{Stirred but not shaken: a multiwavelength view of HD 16743's
  debris disc},} \mnras, 521, 5940, \dodoi{10.1093/mnras/stad913}

% type= article
\bibitem[{J. {Martin} {et~al.}(2017){Martin}, {Fuhrmeister}, {Mittag},
  {Schmidt}, {Hempelmann}, {Gonz{\'a}lez-P{\'e}rez}, \& {Schmitt}}]{martin2017}
{Martin}, J., {Fuhrmeister}, B., {Mittag}, M., {et~al.} 2017,
  \bibinfo{title}{{The Ca II infrared triplet's performance as an activity
  indicator compared to Ca II H and K. Empirical relations to convert Ca II
  infrared triplet measurements to common activity indices},} \aap, 605, A113,
  \dodoi{10.1051/0004-6361/201630298}

% type= article
\bibitem[{R. {Mart{\'\i}nez-Arn{\'a}iz}
  {et~al.}(2010){Mart{\'\i}nez-Arn{\'a}iz}, {Maldonado}, {Montes}, {Eiroa}, \&
  {Montesinos}}]{martinez-arnaiz2010}
{Mart{\'\i}nez-Arn{\'a}iz}, R., {Maldonado}, J., {Montes}, D., {Eiroa}, C., \&
  {Montesinos}, B. 2010, \bibinfo{title}{{Chromospheric activity and rotation
  of FGK stars in the solar vicinity. An estimation of the radial velocity
  jitter},} \aap, 520, A79, \dodoi{10.1051/0004-6361/200913725}

% type= article
\bibitem[{C.~J. {Marvin} {et~al.}(2023){Marvin}, {Reiners},
  {Anglada-Escud{\'e}}, {Jeffers}, \& {Boro Saikia}}]{marvin2023}
{Marvin}, C.~J., {Reiners}, A., {Anglada-Escud{\'e}}, G., {Jeffers}, S.~V., \&
  {Boro Saikia}, S. 2023, \bibinfo{title}{{Absolute Ca II H \& K and H-alpha
  flux measurements of low-mass stars: Extending R'$_{HK}$ to M dwarfs},} \aap,
  671, A162, \dodoi{10.1051/0004-6361/201937306}

% type= article
\bibitem[{B.~D. {Mason} {et~al.}(2001){Mason}, {Wycoff}, {Hartkopf},
  {Douglass}, \& {Worley}}]{ORB62001b}
{Mason}, B.~D., {Wycoff}, G.~L., {Hartkopf}, W.~I., {Douglass}, G.~G., \&
  {Worley}, C.~E. 2001, \bibinfo{title}{{The 2001 US Naval Observatory Double
  Star CD-ROM. I. The Washington Double Star Catalog},} \aj, 122, 3466,
  \dodoi{10.1086/323920}

% type= article
\bibitem[{G.~S. {Mathews} {et~al.}(2013){Mathews}, {Pinte}, {Duch{\^e}ne},
  {Williams}, \& {M{\'e}nard}}]{mathews2013}
{Mathews}, G.~S., {Pinte}, C., {Duch{\^e}ne}, G., {Williams}, J.~P., \&
  {M{\'e}nard}, F. 2013, \bibinfo{title}{{A Herschel PACS survey of the dust
  and gas in Upper Scorpius disks},} \aap, 558, A66,
  \dodoi{10.1051/0004-6361/201321228}

% type= article
\bibitem[{G.~S. {Mathews} {et~al.}(2012){Mathews}, {Williams}, {M{\'e}nard},
  {Phillips}, {Duch{\^e}ne}, \& {Pinte}}]{mathews2012}
{Mathews}, G.~S., {Williams}, J.~P., {M{\'e}nard}, F., {et~al.} 2012,
  \bibinfo{title}{{The Late Stages of Protoplanetary Disk Evolution: A
  Millimeter Survey of Upper Scorpius},} \apj, 745, 23,
  \dodoi{10.1088/0004-637X/745/1/23}

% type= article
\bibitem[{S. {Mathur} {et~al.}(2025){Mathur}, {Santos}, {Claytor},
  {Garc{\'\i}a}, {Strugarek}, {Finley}, {Noraz}, {Amard}, {Beck}, {Bonanno},
  {Breton}, {Brun}, {Cao}, {Corsaro}, {Godoy-Rivera}, {Mathis},
  {Palakkatharappil}, {Pinsonneault}, \& {van Saders}}]{mathur2025}
{Mathur}, S., {Santos}, {\^A}. R.~G., {Claytor}, Z.~R., {et~al.} 2025,
  \bibinfo{title}{{Magnetic Activity Evolution of Solar-like Stars. II.
  S$_{ph}${\textendash}Ro Evolution of Kepler Main-sequence Targets},} \apj,
  982, 114, \dodoi{10.3847/1538-4357/adb8cc}

% type= article
\bibitem[{L. {Matr{\`a}} {et~al.}(2020){Matr{\`a}}, {Dent}, {Wilner}, {Marino},
  {Wyatt}, {Marshall}, {Su}, {Chavez}, {Hales}, {Hughes}, {Greaves}, \&
  {Corder}}]{matra2020}
{Matr{\`a}}, L., {Dent}, W. R.~F., {Wilner}, D.~J., {et~al.} 2020,
  \bibinfo{title}{{Dust Populations in the Iconic Vega Planetary System
  Resolved by ALMA},} \apj, 898, 146, \dodoi{10.3847/1538-4357/aba0a4}

% type= article
\bibitem[{L. {Matr{\`a}} {et~al.}(2025){Matr{\`a}}, {Marino}, {Wilner},
  {Kennedy}, {Booth}, {Krivov}, {Williams}, {Hughes}, {del Burgo}, {Carpenter},
  {Davies}, {Ertel}, {Kral}, {Lestrade}, {Marshall}, {Milli}, {{\"O}berg},
  {Pawellek}, {Sepulveda}, {Wyatt}, {Matthews}, \& {MacGregor}}]{matra2025}
{Matr{\`a}}, L., {Marino}, S., {Wilner}, D.~J., {et~al.} 2025,
  \bibinfo{title}{{REsolved ALMA and SMA Observations of Nearby Stars
  (REASONS): A population of 74 resolved planetesimal belts at millimetre
  wavelengths},} \aap, 693, A151, \dodoi{10.1051/0004-6361/202451397}

% type= inproceedings
\bibitem[{B.~C. {Matthews} {et~al.}(2014){Matthews}, {Krivov}, {Wyatt},
  {Bryden}, \& {Eiroa}}]{matthews2014}
{Matthews}, B.~C., {Krivov}, A.~V., {Wyatt}, M.~C., {Bryden}, G., \& {Eiroa},
  C. 2014, \bibinfo{title}{{Observations, Modeling, and Theory of Debris
  Disks},} in Protostars and Planet VI, ed. {Beuther, H., Klessen, R. S.,
  Dullemond, C.~P., \& Henning, T.} (The University of Arizona Press, Tucson,
  AZ), 521--544

% type= article
\bibitem[{B.~C. {Matthews} {et~al.}(2010){Matthews}, {Sibthorpe}, {Kennedy},
  {Phillips}, {Churcher}, {Duch{\^e}ne}, {Greaves}, {Lestrade}, {Moro-Martin},
  {Wyatt}, {Bastien}, {Biggs}, {Bouvier}, {Butner}, {Dent}, {di Francesco},
  {Eisl{\"o}ffel}, {Graham}, {Harvey}, {Hauschildt}, {Holland}, {Horner},
  {Ibar}, {Ivison}, {Johnstone}, {Kalas}, {Kavelaars}, {Rodriguez}, {Udry},
  {van der Werf}, {Wilner}, \& {Zuckerman}}]{matthews2010}
{Matthews}, B.~C., {Sibthorpe}, B., {Kennedy}, G., {et~al.} 2010,
  \bibinfo{title}{{Resolving debris discs in the far-infrared: Early highlights
  from the DEBRIS survey},} \aap, 518, L135,
  \dodoi{10.1051/0004-6361/201014667}

% type= article
\bibitem[{A. {McQuillan} {et~al.}(2014){McQuillan}, {Mazeh}, \&
  {Aigrain}}]{mcquillan2014}
{McQuillan}, A., {Mazeh}, T., \& {Aigrain}, S. 2014, \bibinfo{title}{{Rotation
  Periods of 34,030 Kepler Main-sequence Stars: The Full Autocorrelation
  Sample},} \apjs, 211, 24, \dodoi{10.1088/0067-0049/211/2/24}

% type= article
\bibitem[{S.~N. {Mellon} {et~al.}(2017){Mellon}, {Mamajek}, {Oberst}, \&
  {Pecaut}}]{mellon2017}
{Mellon}, S.~N., {Mamajek}, E.~E., {Oberst}, T.~E., \& {Pecaut}, M.~J. 2017,
  \bibinfo{title}{{Angular Momentum Evolution of Young Stars in the nearby
  Scorpius-Centaurus OB Association},} \apj, 844, 66,
  \dodoi{10.3847/1538-4357/aa77fb}

% type= article
\bibitem[{A. {Merloni} {et~al.}(2024){Merloni}, {Lamer}, {Liu}, {Ramos-Ceja},
  {Brunner}, {Bulbul}, {Dennerl}, {Doroshenko}, {Freyberg}, {Friedrich},
  {Gatuzz}, {Georgakakis}, {Haberl}, {Igo}, {Kreykenbohm}, {Liu}, {Maitra},
  {Malyali}, {Mayer}, {Nandra}, {Predehl}, {Robrade}, {Salvato}, {Sanders},
  {Stewart}, {Tub{\'\i}n-Arenas}, {Weber}, {Wilms}, {Arcodia}, {Artis},
  {Aschersleben}, {Avakyan}, {Aydar}, {Bahar}, {Balzer}, {Becker}, {Berger},
  {Boller}, {Bornemann}, {Br{\"u}ggen}, {Brusa}, {Buchner}, {Burwitz},
  {Camilloni}, {Clerc}, {Comparat}, {Coutinho}, {Czesla}, {Dannhauer},
  {Dauner}, {Dauser}, {Dietl}, {Dolag}, {Dwelly}, {Egg}, {Ehl}, {Freund},
  {Friedrich}, {Gaida}, {Garrel}, {Ghirardini}, {Gokus}, {Gr{\"u}nwald},
  {Grandis}, {Grotova}, {Gruen}, {Gueguen}, {H{\"a}mmerich}, {Hamaus},
  {Hasinger}, {Haubner}, {Homan}, {Ider Chitham}, {Joseph}, {Joyce},
  {K{\"o}nig}, {Kaltenbrunner}, {Khokhriakova}, {Kink}, {Kirsch}, {Kluge},
  {Knies}, {Krippendorf}, {Krumpe}, {Kurpas}, {Li}, {Liu}, {Locatelli},
  {Lorenz}, {M{\"u}ller}, {Magaudda}, {Mannes}, {McCall}, {Meidinger},
  {Michailidis}, {Migkas}, {Mu{\~n}oz-Giraldo}, {Musiimenta}, {Nguyen-Dang},
  {Ni}, {Olechowska}, {Ota}, {Pacaud}, {Pasini}, {Perinati}, {Pires},
  {Pommranz}, {Ponti}, {Poppenhaeger}, {P{\"u}hlhofer}, {Rau}, {Reh},
  {Reiprich}, {Roster}, {Saeedi}, {Santangelo}, {Sasaki}, {Schmitt},
  {Schneider}, {Schrabback}, {Schuster}, {Schwope}, {Seppi}, {Serim},
  {Shreeram}, {Sokolova-Lapa}, {Starck}, {Stelzer}, {Stierhof}, {Suleimanov},
  {Tenzer}, {Traulsen}, {Tr{\"u}mper}, {Tsuge}, {Urrutia}, {Veronica},
  {Waddell}, {Willer}, {Wolf}, {Yeung}, {Zainab}, {Zangrandi}, {Zhang},
  {Zhang}, \& {Zheng}}]{merloni2024}
{Merloni}, A., {Lamer}, G., {Liu}, T., {et~al.} 2024, \bibinfo{title}{{The
  SRG/eROSITA all-sky survey. First X-ray catalogues and data release of the
  western Galactic hemisphere},} \aap, 682, A34,
  \dodoi{10.1051/0004-6361/202347165}

% type= article
\bibitem[{J.~C. {Mermilliod}(1987{\natexlab{a}}){Mermilliod}}]{mermilliod1987a}
{Mermilliod}, J.~C. 1987{\natexlab{a}}, \bibinfo{title}{{UBV Photoelectric
  Photometry Catalogue (1986): I. The Original data},} \aaps, 71, 413

% type= article
\bibitem[{J.~C. {Mermilliod}(1987{\natexlab{b}}){Mermilliod}}]{mermilliod1987b}
{Mermilliod}, J.~C. 1987{\natexlab{b}}, \bibinfo{title}{{UBV photoelectric
  catalogue (1986). II. Analysis of the data.},} \aaps, 71, 119

% type= inproceedings
\bibitem[{J.-C. {Mermilliod}(1995){Mermilliod}}]{mermilliod1995}
{Mermilliod}, J.-C. 1995, \bibinfo{title}{{The Database for Galactic Open
  Clusters (BDA)},} in Information \& On-Line Data in Astronomy, ed. D.~{Egret}
  \& M.~A. {Albrecht}, Vol. 203 (Springer), 127,
  \dodoi{10.1007/978-94-011-0397-8_12}

% type= article
\bibitem[{J.~C. {Mermilliod} {et~al.}(2008){Mermilliod}, {Platais}, {James},
  {Grenon}, \& {Cargile}}]{mermilliod2008}
{Mermilliod}, J.~C., {Platais}, I., {James}, D.~J., {Grenon}, M., \& {Cargile},
  P.~A. 2008, \bibinfo{title}{{Membership, binarity, and rotation of F-G-K
  stars in the open cluster Blanco 1},} \aap, 485, 95,
  \dodoi{10.1051/0004-6361:20079072}

% type= article
\bibitem[{S. {Messina} {et~al.}(2017){Messina}, {Millward}, {Buccino}, {Zhang},
  {Medhi}, {Jofr{\'e}}, {Petrucci}, {Pi}, {Hambsch}, {Kehusmaa}, {Harlingten},
  {Artemenko}, {Curtis}, {Hentunen}, {Malo}, {Mauas}, {Monard}, {Muro Serrano},
  {Naves}, {Santallo}, {Savuskin}, \& {Tan}}]{messina2017}
{Messina}, S., {Millward}, M., {Buccino}, A., {et~al.} 2017,
  \bibinfo{title}{{The {\ensuremath{\beta}} Pictoris association: Catalog of
  photometric rotational periods of low-mass members and candidate members},}
  \aap, 600, A83, \dodoi{10.1051/0004-6361/201629152}

% type= article
\bibitem[{T.~S. {Metcalfe} {et~al.}(2010){Metcalfe}, {Basu}, {Henry},
  {Soderblom}, {Judge}, {Kn{\"o}lker}, {Mathur}, \& {Rempel}}]{metcalfe2010}
{Metcalfe}, T.~S., {Basu}, S., {Henry}, T.~J., {et~al.} 2010,
  \bibinfo{title}{{Discovery of a 1.6 Year Magnetic Activity Cycle in the
  Exoplanet Host Star {\ensuremath{\i}} Horologii},} \apjl, 723, L213,
  \dodoi{10.1088/2041-8205/723/2/L213}

% type= article
\bibitem[{N. {Meunier} {et~al.}(2022){Meunier}, {Kretzschmar}, {Gravet},
  {Mignon}, \& {Delfosse}}]{meunier2022a}
{Meunier}, N., {Kretzschmar}, M., {Gravet}, R., {Mignon}, L., \& {Delfosse}, X.
  2022, \bibinfo{title}{{Relationship between Ca and H{\ensuremath{\alpha}}
  chromospheric emission in F-G-K stars: Indication of stellar filaments?},}
  \aap, 658, A57, \dodoi{10.1051/0004-6361/202142120}

% type= article
\bibitem[{N. {Meunier} \& A.-M. {Lagrange}(2022){Meunier} \&
  {Lagrange}}]{meunier2022b}
{Meunier}, N., \& {Lagrange}, A.-M. 2022, \bibinfo{title}{{A new estimation of
  astrometric exoplanet detection limits in the habitable zone around nearby
  stars},} \aap, 659, A104, \dodoi{10.1051/0004-6361/202142702}

% type= article
\bibitem[{M.~R. {Meyer} {et~al.}(2006){Meyer}, {Hillenbrand}, {Backman},
  {Beckwith}, {Bouwman}, {Brooke}, {Carpenter}, {Cohen}, {Cortes}, {Crockett},
  {Gorti}, {Henning}, {Hines}, {Hollenbach}, {Kim}, {Lunine}, {Malhotra},
  {Mamajek}, {Metchev}, {Moro-Martin}, {Morris}, {Najita}, {Padgett},
  {Pascucci}, {Rodmann}, {Schlingman}, {Silverstone}, {Soderblom}, {Stauffer},
  {Stobie}, {Strom}, {Watson}, {Weidenschilling}, {Wolf}, \&
  {Young}}]{meyer2006}
{Meyer}, M.~R., {Hillenbrand}, L.~A., {Backman}, D., {et~al.} 2006,
  \bibinfo{title}{{The Formation and Evolution of Planetary Systems: Placing
  Our Solar System in Context with Spitzer},} \pasp, 118, 1690,
  \dodoi{10.1086/510099}

% type= article
\bibitem[{M.~R. {Meyer} {et~al.}(2008){Meyer}, {Carpenter}, {Mamajek},
  {Hillenbrand}, {Hollenbach}, {Moro-Martin}, {Kim}, {Silverstone}, {Najita},
  {Hines}, {Pascucci}, {Stauffer}, {Bouwman}, \& {Backman}}]{meyer2008}
{Meyer}, M.~R., {Carpenter}, J.~M., {Mamajek}, E.~E., {et~al.} 2008,
  \bibinfo{title}{{Evolution of Mid-Infrared Excess around Sun-like Stars:
  Constraints on Models of Terrestrial Planet Formation},} \apjl, 673, L181,
  \dodoi{10.1086/527470}

% type= article
\bibitem[{G. {Meynet} {et~al.}(1993){Meynet}, {Mermilliod}, \&
  {Maeder}}]{meynet1993}
{Meynet}, G., {Mermilliod}, J.~C., \& {Maeder}, A. 1993, \bibinfo{title}{{New
  dating of galactic open clusters.},} \aaps, 98, 477

% type= article
\bibitem[{G. {Micela} {et~al.}(1999){Micela}, {Sciortino}, {Harnden},
  {Kashyap}, {Rosner}, {Prosser}, {Damiani}, {Stauffer}, \&
  {Caillault}}]{micela1999}
{Micela}, G., {Sciortino}, S., {Harnden}, F.~R., J., {et~al.} 1999,
  \bibinfo{title}{{Deep ROSAT HRI observations of the Pleiades},} \aap, 341,
  751

% type= article
\bibitem[{F. {Middelkoop}(1982){Middelkoop}}]{middlekoop1982}
{Middelkoop}, F. 1982, \bibinfo{title}{{Magnetic structure in cool stars. IV -
  Rotation and CA II H and K emission of main-sequence stars},} \aap, 107, 31

% type= article
\bibitem[{L. {Mignon} {et~al.}(2023){Mignon}, {Meunier}, {Delfosse}, {Bonfils},
  {Santos}, {Forveille}, {Gaisn{\'e}}, {Astudillo-Defru}, {Lovis}, \&
  {Udry}}]{mignon2023}
{Mignon}, L., {Meunier}, N., {Delfosse}, X., {et~al.} 2023,
  \bibinfo{title}{{Characterisation of stellar activity of M dwarfs. I.
  Long-timescale variability in a large sample and detection of new cycles},}
  \aap, 675, A168, \dodoi{10.1051/0004-6361/202244249}

% type= article
\bibitem[{L. {Mignon} {et~al.}(2025){Mignon}, {Delfosse}, {Meunier},
  {Chaverot}, {Burn}, {Bonfils}, {Bouchy}, {Astudillo-Defru}, {Lo Curto},
  {Gaisne}, {Udry}, {Forveille}, {Segransan}, {Lovis}, {Santos}, \&
  {Mayor}}]{mignon2025}
{Mignon}, L., {Delfosse}, X., {Meunier}, N., {et~al.} 2025,
  \bibinfo{title}{{Radial velocity homogeneous analysis of M dwarfs observed
  with HARPS: II. Detection limits and planetary occurrence statistics},} \aap,
  700, A146, \dodoi{10.1051/0004-6361/202451142}

% type= article
\bibitem[{D. {Mikkola} {et~al.}(2023){Mikkola}, {McMillan}, \&
  {Hobbs}}]{mikkola2023}
{Mikkola}, D., {McMillan}, P.~J., \& {Hobbs}, D. 2023, \bibinfo{title}{{New
  stellar velocity substructures from Gaia DR3 proper motions},} \mnras, 519,
  1989, \dodoi{10.1093/mnras/stac3649}

% type= article
\bibitem[{N. {Miret-Roig} {et~al.}(2022){Miret-Roig}, {Galli}, {Olivares},
  {Bouy}, {Alves}, \& {Barrado}}]{miretroig2022}
{Miret-Roig}, N., {Galli}, P.~A.~B., {Olivares}, J., {et~al.} 2022,
  \bibinfo{title}{{The star formation history of Upper Scorpius and Ophiuchus.
  A 7D picture: positions, kinematics, and dynamical traceback ages},} \aap,
  667, A163, \dodoi{10.1051/0004-6361/202244709}

% type= article
\bibitem[{N. {Miret-Roig} {et~al.}(2020){Miret-Roig}, {Galli}, {Brandner},
  {Bouy}, {Barrado}, {Olivares}, {Antoja}, {Romero-G{\'o}mez}, {Figueras}, \&
  {Lillo-Box}}]{miretroig2020}
{Miret-Roig}, N., {Galli}, P.~A.~B., {Brandner}, W., {et~al.} 2020,
  \bibinfo{title}{{Dynamical traceback age of the {\ensuremath{\beta}} Pictoris
  moving group},} \aap, 642, A179, \dodoi{10.1051/0004-6361/202038765}

% type= article
\bibitem[{D.~G. {Monet} {et~al.}(2003){Monet}, {Levine}, {Canzian}, {Ables},
  {Bird}, {Dahn}, {Guetter}, {Harris}, {Henden}, {Leggett}, {Levison},
  {Luginbuhl}, {Martini}, {Monet}, {Munn}, {Pier}, {Rhodes}, {Riepe}, {Sell},
  {Stone}, {Vrba}, {Walker}, {Westerhout}, {Brucato}, {Reid}, {Schoening},
  {Hartley}, {Read}, \& {Tritton}}]{monet2003}
{Monet}, D.~G., {Levine}, S.~E., {Canzian}, B., {et~al.} 2003,
  \bibinfo{title}{{The USNO-B Catalog},} \aj, 125, 984, \dodoi{10.1086/345888}

% type= article
\bibitem[{T.~R. {Monroe} \& C.~A. {Pilachowski}(2010){Monroe} \&
  {Pilachowski}}]{monroe2010}
{Monroe}, T.~R., \& {Pilachowski}, C.~A. 2010, \bibinfo{title}{{Metallicities
  of Young Open Clusters. I. NGC 7160 and NGC 2232},} \aj, 140, 2109,
  \dodoi{10.1088/0004-6256/140/6/2109}

% type= article
\bibitem[{P. {Montegriffo} {et~al.}(2023){Montegriffo}, {De Angeli}, {Andrae},
  {Riello}, {Pancino}, {Sanna}, {Bellazzini}, {Evans}, {Carrasco}, {Sordo},
  {Busso}, {Cacciari}, {Jordi}, {van Leeuwen}, {Vallenari}, {Altavilla},
  {Barstow}, {Brown}, {Burgess}, {Castellani}, {Cowell}, {Davidson}, {De
  Luise}, {Delchambre}, {Diener}, {Fabricius}, {Fr{\'e}mat}, {Fouesneau},
  {Gilmore}, {Giuffrida}, {Hambly}, {Harrison}, {Hidalgo}, {Hodgkin},
  {Holland}, {Marinoni}, {Osborne}, {Pagani}, {Palaversa}, {Piersimoni},
  {Pulone}, {Ragaini}, {Rainer}, {Richards}, {Rowell}, {Ruz-Mieres}, {Sarro},
  {Walton}, \& {Yoldas}}]{montegriffo2023}
{Montegriffo}, P., {De Angeli}, F., {Andrae}, R., {et~al.} 2023,
  \bibinfo{title}{{Gaia Data Release 3. External calibration of BP/RP
  low-resolution spectroscopic data},} \aap, 674, A3,
  \dodoi{10.1051/0004-6361/202243880}

% type= article
\bibitem[{D. {Montes} {et~al.}(2001){Montes}, {L{\'o}pez-Santiago},
  {G{\'a}lvez}, {Fern{\'a}ndez-Figueroa}, {De Castro}, \&
  {Cornide}}]{montes2001}
{Montes}, D., {L{\'o}pez-Santiago}, J., {G{\'a}lvez}, M.~C., {et~al.} 2001,
  \bibinfo{title}{{Late-type members of young stellar kinematic groups - I.
  Single stars},} \mnras, 328, 45, \dodoi{10.1046/j.1365-8711.2001.04781.x}

% type= article
\bibitem[{B. {Montesinos} {et~al.}(2016){Montesinos}, {Eiroa}, {Krivov},
  {Marshall}, {Pilbratt}, {Liseau}, {Mora}, {Maldonado}, {Wolf}, {Ertel},
  {Bayo}, {Augereau}, {Heras}, {Fridlund}, {Danchi}, {Solano}, {Kirchschlager},
  {del Burgo}, \& {Montes}}]{montesinos2016}
{Montesinos}, B., {Eiroa}, C., {Krivov}, A.~V., {et~al.} 2016,
  \bibinfo{title}{{Incidence of debris discs around FGK stars in the solar
  neighbourhood},} \aap, 593, A51, \dodoi{10.1051/0004-6361/201628329}

% type= article
\bibitem[{A. {Mo{\'o}r} {et~al.}(2011){Mo{\'o}r}, {Pascucci}, {K{\'o}sp{\'a}l},
  {{\'A}brah{\'a}m}, {Csengeri}, {Kiss}, {Apai}, {Grady}, {Henning}, {Kiss},
  {Bayliss}, {Juh{\'a}sz}, {Kov{\'a}cs}, \& {Szalai}}]{moor2011a}
{Mo{\'o}r}, A., {Pascucci}, I., {K{\'o}sp{\'a}l}, {\'A}., {et~al.} 2011,
  \bibinfo{title}{{Structure and Evolution of Debris Disks Around F-type Stars.
  I. Observations, Database, and Basic Evolutionary Aspects},} \apjs, 193, 4,
  \dodoi{10.1088/0067-0049/193/1/4}

% type= article
\bibitem[{F.~Y. {Morales} {et~al.}(2016){Morales}, {Bryden}, {Werner}, \&
  {Stapelfeldt}}]{morales2016}
{Morales}, F.~Y., {Bryden}, G., {Werner}, M.~W., \& {Stapelfeldt}, K.~R. 2016,
  \bibinfo{title}{{Herschel-resolved Outer Belts of Two-belt Debris
  Disks{\textemdash}Evidence of Icy Grains},} \apj, 831, 97,
  \dodoi{10.3847/0004-637X/831/1/97}

% type= article
\bibitem[{F.~Y. {Morales} {et~al.}(2009){Morales}, {Werner}, {Bryden},
  {Plavchan}, {Stapelfeldt}, {Rieke}, {Su}, {Beichman}, {Chen}, {Grogan},
  {Kenyon}, {Moro-Martin}, \& {Wolf}}]{morales2009}
{Morales}, F.~Y., {Werner}, M.~W., {Bryden}, G., {et~al.} 2009,
  \bibinfo{title}{{Spitzer Mid-IR Spectra of Dust Debris Around A and Late B
  Type Stars: Asteroid Belt Analogs and Power-Law Dust Distributions},} \apj,
  699, 1067, \dodoi{10.1088/0004-637X/699/2/1067}

% type= article
\bibitem[{E. {Moraux} {et~al.}(2007){Moraux}, {Bouvier}, {Stauffer}, {Barrado y
  Navascu{\'e}s}, \& {Cuillandre}}]{moraux2007}
{Moraux}, E., {Bouvier}, J., {Stauffer}, J.~R., {Barrado y Navascu{\'e}s}, D.,
  \& {Cuillandre}, J.~C. 2007, \bibinfo{title}{{The lower mass function of the
  young open cluster Blanco 1: from 30 M$_{Jup}$ to 3
  M$_{{\ensuremath{\odot}}}$},} \aap, 471, 499,
  \dodoi{10.1051/0004-6361:20066308}

% type= book
\bibitem[{M. {Moshir} {et~al.}(1992){Moshir}, {Kopman}, \&
  {Conrow}}]{moshir1992}
{Moshir}, M., {Kopman}, G., \& {Conrow}, T.~A.~O. 1992, {IRAS Faint Source
  Survey, Explanatory supplement version 2} (Caltech Infrared Processing and
  Analysis Center, Pasadena, CA USA)

% type= article
\bibitem[{A. {Mucciarelli} {et~al.}(2021){Mucciarelli}, {Bellazzini}, \&
  {Massari}}]{mucciarelli2021}
{Mucciarelli}, A., {Bellazzini}, M., \& {Massari}, D. 2021,
  \bibinfo{title}{{Exploiting the Gaia EDR3 photometry to derive stellar
  temperatures},} \aap, 653, A90, \dodoi{10.1051/0004-6361/202140979}

% type= article
\bibitem[{P.~S. {Muirhead} {et~al.}(2022){Muirhead}, {Nordhaus}, \&
  {Drout}}]{muirhead2022}
{Muirhead}, P.~S., {Nordhaus}, J., \& {Drout}, M.~R. 2022,
  \bibinfo{title}{{Revised Stellar Parameters for V471 Tau, A Post-common
  Envelope Binary in the Hyades},} \aj, 163, 34,
  \dodoi{10.3847/1538-3881/ac390f}

% type= article
\bibitem[{H. {Murakami} {et~al.}(2007){Murakami}, {Baba}, {Barthel},
  {Clements}, {Cohen}, {Doi}, {Enya}, {Figueredo}, {Fujishiro}, {Fujiwara},
  {Fujiwara}, {Garcia-Lario}, {Goto}, {Hasegawa}, {Hibi}, {Hirao}, {Hiromoto},
  {Hong}, {Imai}, {Ishigaki}, {Ishiguro}, {Ishihara}, {Ita}, {Jeong}, {Jeong},
  {Kaneda}, {Kataza}, {Kawada}, {Kawai}, {Kawamura}, {Kessler}, {Kester},
  {Kii}, {Kim}, {Kim}, {Kobayashi}, {Koo}, {Kwon}, {Lee}, {Lorente}, {Makiuti},
  {Matsuhara}, {Matsumoto}, {Matsuo}, {Matsuura}, {M{\"U}ller}, {Murakami},
  {Nagata}, {Nakagawa}, {Naoi}, {Narita}, {Noda}, {Oh}, {Ohnishi}, {Ohyama},
  {Okada}, {Okuda}, {Oliver}, {Onaka}, {Ootsubo}, {Oyabu}, {Pak}, {Park},
  {Pearson}, {Rowan-Robinson}, {Saito}, {Sakon}, {Salama}, {Sato}, {Savage},
  {Serjeant}, {Shibai}, {Shirahata}, {Sohn}, {Suzuki}, {Takagi}, {Takahashi},
  {Tanab{\'E}}, {Takeuchi}, {Takita}, {Thomson}, {Uemizu}, {Ueno}, {Usui},
  {Verdugo}, {Wada}, {Wang}, {Watabe}, {Watarai}, {White}, {Yamamura},
  {Yamauchi}, \& {Yasuda}}]{murakami2007}
{Murakami}, H., {Baba}, H., {Barthel}, P., {et~al.} 2007, \bibinfo{title}{{The
  Infrared Astronomical Mission AKARI*},} \pasj, 59, S369,
  \dodoi{10.1093/pasj/59.sp2.S369}

% type= article
\bibitem[{S.~J. {Murphy} \& W.~A. {Lawson}(2015){Murphy} \&
  {Lawson}}]{murphy2015}
{Murphy}, S.~J., \& {Lawson}, W.~A. 2015, \bibinfo{title}{{New low-mass members
  of the Octans stellar association and an updated 30-40 Myr lithium age},}
  \mnras, 447, 1267, \dodoi{10.1093/mnras/stu2450}

% type= article
\bibitem[{J. {Najita} \& J.~P. {Williams}(2005){Najita} \&
  {Williams}}]{najita2005}
{Najita}, J., \& {Williams}, J.~P. 2005, \bibinfo{title}{{An 850
  {\ensuremath{\mu}}m Survey for Dust around Solar-Mass Stars},} \apj, 635,
  625, \dodoi{10.1086/497159}

% type= article
\bibitem[{J.~R. {Najita} {et~al.}(2022){Najita}, {Kenyon}, \&
  {Bromley}}]{nkb2022}
{Najita}, J.~R., {Kenyon}, S.~J., \& {Bromley}, B.~C. 2022,
  \bibinfo{title}{{From Pebbles and Planetesimals to Planets and Dust: The
  Protoplanetary Disk-Debris Disk Connection},} \apj, 925, 45,
  \dodoi{10.3847/1538-4357/ac37b6}

% type= article
\bibitem[{G. {Nandakumar} {et~al.}(2017){Nandakumar}, {Schultheis}, {Hayden},
  {Rojas-Arriagada}, {Kordopatis}, \& {Haywood}}]{nandakumar2017}
{Nandakumar}, G., {Schultheis}, M., {Hayden}, M., {et~al.} 2017,
  \bibinfo{title}{{Effects of the selection function on metallicity trends in
  spectroscopic surveys of the Milky Way},} \aap, 606, A97,
  \dodoi{10.1051/0004-6361/201731099}

% type= article
\bibitem[{T. {Naylor} {et~al.}(2002){Naylor}, {Totten}, {Jeffries}, {Pozzo},
  {Devey}, \& {Thompson}}]{naylor2002}
{Naylor}, T., {Totten}, E.~J., {Jeffries}, R.~D., {et~al.} 2002,
  \bibinfo{title}{{Optimal photometry for colour-magnitude diagrams and its
  application to NGC 2547},} \mnras, 335, 291,
  \dodoi{10.1046/j.1365-8711.2002.05592.x}

% type= article
\bibitem[{A. {Nederlander} {et~al.}(2021){Nederlander}, {Hughes}, {Fehr},
  {Flaherty}, {Su}, {Mo{\'o}r}, {Chiang}, {Andrews}, {Wilner}, \&
  {Marino}}]{nederlander2021}
{Nederlander}, A., {Hughes}, A.~M., {Fehr}, A.~J., {et~al.} 2021,
  \bibinfo{title}{{Resolving Structure in the Debris Disk around HD 206893 with
  ALMA},} \apj, 917, 5, \dodoi{10.3847/1538-4357/abdd32}

% type= article
\bibitem[{G. {Neugebauer} {et~al.}(1984){Neugebauer}, {Habing}, {van Duinen},
  {Aumann}, {Baud}, {Beichman}, {Beintema}, {Boggess}, {Clegg}, {de Jong},
  {Emerson}, {Gautier}, {Gillett}, {Harris}, {Hauser}, {Houck}, {Jennings},
  {Low}, {Marsden}, {Miley}, {Olnon}, {Pottasch}, {Raimond}, {Rowan-Robinson},
  {Soifer}, {Walker}, {Wesselius}, \& {Young}}]{neugebauer1984}
{Neugebauer}, G., {Habing}, H.~J., {van Duinen}, R., {et~al.} 1984,
  \bibinfo{title}{{The Infrared Astronomical Satellite (IRAS) mission.},}
  \apjl, 278, L1, \dodoi{10.1086/184209}

% type= article
\bibitem[{E.~R. {Newton} {et~al.}(2014){Newton}, {Charbonneau}, {Irwin},
  {Berta-Thompson}, {Rojas-Ayala}, {Covey}, \& {Lloyd}}]{newton2014}
{Newton}, E.~R., {Charbonneau}, D., {Irwin}, J., {et~al.} 2014,
  \bibinfo{title}{{Near-infrared Metallicities, Radial Velocities, and Spectral
  Types for 447 Nearby M Dwarfs},} \aj, 147, 20,
  \dodoi{10.1088/0004-6256/147/1/20}

% type= article
\bibitem[{E.~R. {Newton} {et~al.}(2017){Newton}, {Irwin}, {Charbonneau},
  {Berlind}, {Calkins}, \& {Mink}}]{newton2017}
{Newton}, E.~R., {Irwin}, J., {Charbonneau}, D., {et~al.} 2017,
  \bibinfo{title}{{The H{\ensuremath{\alpha}} Emission of Nearby M Dwarfs and
  its Relation to Stellar Rotation},} \apj, 834, 85,
  \dodoi{10.3847/1538-4357/834/1/85}

% type= article
\bibitem[{E.~R. {Newton} {et~al.}(2016){Newton}, {Irwin}, {Charbonneau},
  {Berta-Thompson}, {Dittmann}, \& {West}}]{newton2016}
{Newton}, E.~R., {Irwin}, J., {Charbonneau}, D., {et~al.} 2016,
  \bibinfo{title}{{The Rotation and Galactic Kinematics of Mid M Dwarfs in the
  Solar Neighborhood},} \apj, 821, 93, \dodoi{10.3847/0004-637X/821/2/93}

% type= article
\bibitem[{E.~R. {Newton} {et~al.}(2018){Newton}, {Mondrik}, {Irwin}, {Winters},
  \& {Charbonneau}}]{newton2018}
{Newton}, E.~R., {Mondrik}, N., {Irwin}, J., {Winters}, J.~G., \&
  {Charbonneau}, D. 2018, \bibinfo{title}{{New Rotation Period Measurements for
  M Dwarfs in the Southern Hemisphere: An Abundance of Slowly Rotating, Fully
  Convective Stars},} \aj, 156, 217, \dodoi{10.3847/1538-3881/aad73b}

% type= article
\bibitem[{R. {Nilsson} {et~al.}(2010){Nilsson}, {Liseau}, {Brandeker},
  {Olofsson}, {Pilbratt}, {Risacher}, {Rodmann}, {Augereau}, {Bergman},
  {Eiroa}, {Fridlund}, {Th{\'e}bault}, \& {White}}]{nilsson2010}
{Nilsson}, R., {Liseau}, R., {Brandeker}, A., {et~al.} 2010,
  \bibinfo{title}{{Kuiper belts around nearby stars},} \aap, 518, A40,
  \dodoi{10.1051/0004-6361/201014444}

% type= article
\bibitem[{A.~H. {Nisak} {et~al.}(2022){Nisak}, {White}, {Yep}, {Henry},
  {Paredes}, {James}, \& {Jao}}]{nisak2022}
{Nisak}, A.~H., {White}, R.~J., {Yep}, A., {et~al.} 2022,
  \bibinfo{title}{{Mapping out the Stellar Populations of IC 2602 and IC
  2391},} \aj, 163, 278, \dodoi{10.3847/1538-3881/ac63c3}

% type= article
\bibitem[{Z. {Niu} {et~al.}(2021){Niu}, {Yuan}, {Wang}, \& {Liu}}]{niu2021}
{Niu}, Z., {Yuan}, H., {Wang}, S., \& {Liu}, J. 2021, \bibinfo{title}{{Binary
  Fractions of G and K Dwarf Stars Based on Gaia EDR3 and LAMOST DR5: Impacts
  of the Chemical Abundances},} \apj, 922, 211,
  \dodoi{10.3847/1538-4357/ac2573}

% type= article
\bibitem[{B. {Nordstr{\"o}m} {et~al.}(2004){Nordstr{\"o}m}, {Mayor},
  {Andersen}, {Holmberg}, {Pont}, {J{\o}rgensen}, {Olsen}, {Udry}, \&
  {Mowlavi}}]{nordstrom2004}
{Nordstr{\"o}m}, B., {Mayor}, M., {Andersen}, J., {et~al.} 2004,
  \bibinfo{title}{{The Geneva-Copenhagen survey of the Solar neighbourhood.
  Ages, metallicities, and kinematic properties of {\ensuremath{\sim}}14 000 F
  and G dwarfs},} \aap, 418, 989, \dodoi{10.1051/0004-6361:20035959}

% type= article
\bibitem[{B.~J. {Norfolk} {et~al.}(2021){Norfolk}, {Maddison}, {Marshall},
  {Kennedy}, {Duch{\^e}ne}, {Wilner}, {Pinte}, {Mo{\'o}r}, {Matthews},
  {{\'A}brah{\'a}m}, {K{\'o}sp{\'a}l}, \& {van der Marel}}]{norfolk2021}
{Norfolk}, B.~J., {Maddison}, S.~T., {Marshall}, J.~P., {et~al.} 2021,
  \bibinfo{title}{{Four new planetesimals around typical and pre-main-sequence
  stars (PLATYPUS) debris discs at 8.8 mm},} \mnras, 507, 3139,
  \dodoi{10.1093/mnras/stab1901}

% type= article
\bibitem[{R.~W. {Noyes} {et~al.}(1984){Noyes}, {Hartmann}, {Baliunas},
  {Duncan}, \& {Vaughan}}]{noyes1984}
{Noyes}, R.~W., {Hartmann}, L.~W., {Baliunas}, S.~L., {Duncan}, D.~K., \&
  {Vaughan}, A.~H. 1984, \bibinfo{title}{{Rotation, convection, and magnetic
  activity in lower main-sequence stars.},} \apj, 279, 763,
  \dodoi{10.1086/161945}

% type= article
\bibitem[{A. {N{\'u}{\~n}ez} {et~al.}(2022){N{\'u}{\~n}ez}, {Ag{\"u}eros},
  {Covey}, {Douglas}, {Drake}, {Rampalli}, {Bowsher}, {Cargile}, {Kraus}, \&
  {Law}}]{nunez2022}
{N{\'u}{\~n}ez}, A., {Ag{\"u}eros}, M.~A., {Covey}, K.~R., {et~al.} 2022,
  \bibinfo{title}{{The Factory and the Beehive. IV. A Comprehensive Study of
  the Rotation X-Ray Activity Relation in Praesepe and the Hyades},} \apj, 931,
  45, \dodoi{10.3847/1538-4357/ac6517}

% type= article
\bibitem[{A. {N{\'u}{\~n}ez} {et~al.}(2024){N{\'u}{\~n}ez}, {Ag{\"u}eros},
  {Curtis}, {Covey}, {Douglas}, {Chu}, {DeLaurentiis}, {Wang}, \&
  {Drake}}]{nunez2024}
{N{\'u}{\~n}ez}, A., {Ag{\"u}eros}, M.~A., {Curtis}, J.~L., {et~al.} 2024,
  \bibinfo{title}{{The Factory and the Beehive. V. Chromospheric and Coronal
  Activity and Its Dependence on Rotation in Praesepe and the Hyades},} \apj,
  962, 12, \dodoi{10.3847/1538-4357/ad117e}

% type= article
\bibitem[{F. {Ochsenbein} {et~al.}(2000){Ochsenbein}, {Bauer}, \&
  {Marcout}}]{vizier2000}
{Ochsenbein}, F., {Bauer}, P., \& {Marcout}, J. 2000, \bibinfo{title}{{The
  VizieR database of astronomical catalogues},} \aaps, 143, 23,
  \dodoi{10.1051/aas:2000169}

% type= article
\bibitem[{R.~J. {Oelkers} {et~al.}(2016){Oelkers}, {Macri}, {Marshall},
  {DePoy}, {Lambas}, {Colazo}, \& {Stringer}}]{oelkers2016}
{Oelkers}, R.~J., {Macri}, L.~M., {Marshall}, J.~L., {et~al.} 2016,
  \bibinfo{title}{{A Wide-field Survey for Transiting Hot Jupiters and
  Eclipsing Pre-main-sequence Binaries in Young Stellar Associations},} \aj,
  152, 75, \dodoi{10.3847/0004-6256/152/3/75}

% type= article
\bibitem[{S. {Oh} \& N.~W. {Evans}(2020){Oh} \& {Evans}}]{oh2020}
{Oh}, S., \& {Evans}, N.~W. 2020, \bibinfo{title}{{Kinematic modelling of
  clusters with Gaia: the death throes of the Hyades},} \mnras, 498, 1920,
  \dodoi{10.1093/mnras/staa2381}

% type= article
\bibitem[{J. {Olivares} {et~al.}(2023){Olivares}, {Lodieu}, {Bejar}, {Martin},
  {Zerjal}, \& {Galli}}]{olivares2023}
{Olivares}, J., {Lodieu}, N., {Bejar}, V.~J.~S., {et~al.} 2023,
  \bibinfo{title}{{The cosmic waltz of Coma Berenices and Latyshev 2 (Group X).
  Membership, phase-space structure, mass, and energy distributions},} \aap,
  675, A28, \dodoi{10.1051/0004-6361/202244703}

% type= article
\bibitem[{J.~M. {Oliveira} {et~al.}(2003){Oliveira}, {Jeffries}, {Devey},
  {Barrado y Navascu{\'e}s}, {Naylor}, {Stauffer}, \& {Totten}}]{oliveira2003}
{Oliveira}, J.~M., {Jeffries}, R.~D., {Devey}, C.~R., {et~al.} 2003,
  \bibinfo{title}{{The lithium depletion boundary and the age of NGC 2547},}
  \mnras, 342, 651, \dodoi{10.1046/j.1365-8711.2003.06592.x}

% type= article
\bibitem[{J.~M. {Oliveira} {et~al.}(2002){Oliveira}, {Jeffries}, {Kenyon},
  {Thompson}, \& {Naylor}}]{oliveira2002}
{Oliveira}, J.~M., {Jeffries}, R.~D., {Kenyon}, M.~J., {Thompson}, S.~A., \&
  {Naylor}, T. 2002, \bibinfo{title}{{No disks around low-mass stars and brown
  dwarfs in the young sigma Orionis cluster?},} \aap, 382, L22,
  \dodoi{10.1051/0004-6361:20011778}

% type= article
\bibitem[{J.~M. {Oliveira} {et~al.}(2004){Oliveira}, {Jeffries}, \& {van
  Loon}}]{oliveira2004}
{Oliveira}, J.~M., {Jeffries}, R.~D., \& {van Loon}, J.~T. 2004,
  \bibinfo{title}{{An L'-band survey for circumstellar discs around low-mass
  stars in the young {\ensuremath{\sigma}} Orionis cluster},} \mnras, 347,
  1327, \dodoi{10.1111/j.1365-2966.2004.07315.x}

% type= article
\bibitem[{J. {Olofsson} {et~al.}(2018){Olofsson}, {van Holstein}, {Boccaletti},
  {Janson}, {Th{\'e}bault}, {Gratton}, {Lazzoni}, {Kral}, {Bayo}, {Canovas},
  {Caceres}, {Ginski}, {Pinte}, {Asensio-Torres}, {Chauvin}, {Desidera},
  {Henning}, {Langlois}, {Milli}, {Schlieder}, {Schreiber}, {Augereau},
  {Bonnefoy}, {Buenzli}, {Brandner}, {Durkan}, {Engler}, {Feldt}, {Godoy},
  {Grady}, {Hagelberg}, {Lagrange}, {Lannier}, {Ligi}, {Maire}, {Mawet},
  {M{\'e}nard}, {Mesa}, {Mouillet}, {Peretti}, {Perrot}, {Salter}, {Schmidt},
  {Sissa}, {Thalmann}, {Vigan}, {Abe}, {Feautrier}, {Le Mignant}, {Moulin},
  {Pavlov}, {Rabou}, {Rousset}, \& {Roux}}]{olofsson2018}
{Olofsson}, J., {van Holstein}, R.~G., {Boccaletti}, A., {et~al.} 2018,
  \bibinfo{title}{{Resolving faint structures in the debris disk around TWA 7.
  Tentative detections of an outer belt, a spiral arm, and a dusty cloud},}
  \aap, 617, A109, \dodoi{10.1051/0004-6361/201832583}

% type= article
\bibitem[{T.~J. {O'Neill} {et~al.}(2025){O'Neill}, {Goodman}, {Soler},
  {Zucker}, \& {Han}}]{oneill2025}
{O'Neill}, T.~J., {Goodman}, A.~A., {Soler}, J.~D., {Zucker}, C., \& {Han},
  J.~J. 2025, \bibinfo{title}{{A 3D Model of the Local Bubble's Magnetic Field:
  Insights from Dust and Starlight Polarization},} \apj, 988, 191,
  \dodoi{10.3847/1538-4357/ade306}

% type= article
\bibitem[{T.~J. {O'Neill} {et~al.}(2024){O'Neill}, {Zucker}, {Goodman}, \&
  {Edenhofer}}]{oneill2024}
{O'Neill}, T.~J., {Zucker}, C., {Goodman}, A.~A., \& {Edenhofer}, G. 2024,
  \bibinfo{title}{{The Local Bubble is a Local Chimney: A New Model from 3D
  Dust Mapping},} arXiv e-prints, arXiv:2403.04961,
  \dodoi{10.48550/arXiv.2403.04961}

% type= article
\bibitem[{C.~A. {Onken} {et~al.}(2024){Onken}, {Wolf}, {Bessell}, {Chang},
  {Luvaul}, {Tonry}, {White}, \& {Da Costa}}]{onken2024}
{Onken}, C.~A., {Wolf}, C., {Bessell}, M.~S., {et~al.} 2024,
  \bibinfo{title}{{SkyMapper Southern Survey: Data Release 4},} arXiv e-prints,
  arXiv:2402.02015, \dodoi{10.48550/arXiv.2402.02015}

% type= article
\bibitem[{G. {Pace} {et~al.}(2012){Pace}, {Castro}, {Mel{\'e}ndez},
  {Th{\'e}ado}, \& {do Nascimento}}]{pace2012}
{Pace}, G., {Castro}, M., {Mel{\'e}ndez}, J., {Th{\'e}ado}, S., \& {do
  Nascimento}, J.~D., J. 2012, \bibinfo{title}{{Lithium in M 67: From the main
  sequence to the red giant branch},} \aap, 541, A150,
  \dodoi{10.1051/0004-6361/201117704}

% type= article
\bibitem[{D.~L. {Padgett} {et~al.}(2008){Padgett}, {Rebull}, {Stapelfeldt},
  {Chapman}, {Lai}, {Mundy}, {Evans}, {Brooke}, {Cieza}, {Spiesman},
  {Noriega-Crespo}, {McCabe}, {Allen}, {Blake}, {Harvey}, {Huard},
  {J{\o}rgensen}, {Koerner}, {Myers}, {Sargent}, {Teuben}, {van Dishoeck},
  {Wahhaj}, \& {Young}}]{padgett2008}
{Padgett}, D.~L., {Rebull}, L.~M., {Stapelfeldt}, K.~R., {et~al.} 2008,
  \bibinfo{title}{{The Spitzer c2d Survey of Large, Nearby, Interstellar
  Clouds. VII. Ophiuchus Observed with MIPS},} \apj, 672, 1013,
  \dodoi{10.1086/523883}

% type= article
\bibitem[{X. {Pang} {et~al.}(2021){Pang}, {Li}, {Yu}, {Tang}, {Dinnbier},
  {Kroupa}, {Pasquato}, \& {Kouwenhoven}}]{pang2021}
{Pang}, X., {Li}, Y., {Yu}, Z., {et~al.} 2021, \bibinfo{title}{{3D Morphology
  of Open Clusters in the Solar Neighborhood with Gaia EDR 3: Its Relation to
  Cluster Dynamics},} \apj, 912, 162, \dodoi{10.3847/1538-4357/abeaac}

% type= article
\bibitem[{X. {Pang} {et~al.}(2022){Pang}, {Tang}, {Li}, {Yu}, {Wang}, {Li},
  {Li}, {Wang}, {Wang}, {Zhang}, {Pasquato}, \& {Kouwenhoven}}]{pang2022}
{Pang}, X., {Tang}, S.-Y., {Li}, Y., {et~al.} 2022, \bibinfo{title}{{3D
  Morphology of Open Clusters in the Solar Neighborhood with Gaia EDR 3. II.
  Hierarchical Star Formation Revealed by Spatial and Kinematic
  Substructures},} \apj, 931, 156, \dodoi{10.3847/1538-4357/ac674e}

% type= article
\bibitem[{N. {Pawellek} {et~al.}(2021){Pawellek}, {Wyatt}, {Matr{\`a}},
  {Kennedy}, \& {Yelverton}}]{pawellek2021}
{Pawellek}, N., {Wyatt}, M., {Matr{\`a}}, L., {Kennedy}, G., \& {Yelverton}, B.
  2021, \bibinfo{title}{{A {\ensuremath{\sim}}75 per cent occurrence rate of
  debris discs around F stars in the {\ensuremath{\beta}} Pic moving group},}
  \mnras, 502, 5390, \dodoi{10.1093/mnras/stab269}

% type= article
\bibitem[{T.~D. {Pearce} {et~al.}(2022){Pearce}, {Launhardt}, {Ostermann},
  {Kennedy}, {Gennaro}, {Booth}, {Krivov}, {Cugno}, {Henning}, {Quirrenbach},
  {Barcucci}, {Matthews}, {Ruh}, \& {Stone}}]{pearce2022}
{Pearce}, T.~D., {Launhardt}, R., {Ostermann}, R., {et~al.} 2022,
  \bibinfo{title}{{Planet populations inferred from debris discs. Insights from
  178 debris systems in the ISPY, LEECH, and LIStEN planet-hunting surveys},}
  \aap, 659, A135, \dodoi{10.1051/0004-6361/202142720}

% type= article
\bibitem[{M.~J. {Pecaut} \& E.~E. {Mamajek}(2013){Pecaut} \&
  {Mamajek}}]{pecaut2013}
{Pecaut}, M.~J., \& {Mamajek}, E.~E. 2013, \bibinfo{title}{{Intrinsic Colors,
  Temperatures, and Bolometric Corrections of Pre-main-sequence Stars},} \apjs,
  208, 9, \dodoi{10.1088/0067-0049/208/1/9}

% type= article
\bibitem[{M.~J. {Pecaut} \& E.~E. {Mamajek}(2016){Pecaut} \&
  {Mamajek}}]{pecaut2016}
{Pecaut}, M.~J., \& {Mamajek}, E.~E. 2016, \bibinfo{title}{{The star formation
  history and accretion-disc fraction among the K-type members of the
  Scorpius-Centaurus OB association},} \mnras, 461, 794,
  \dodoi{10.1093/mnras/stw1300}

% type= article
\bibitem[{V. {Perdelwitz} {et~al.}(2021){Perdelwitz}, {Mittag}, {Tal-Or},
  {Schmitt}, {Caballero}, {Jeffers}, {Reiners}, {Schweitzer}, {Trifonov},
  {Ribas}, {Quirrenbach}, {Amado}, {Seifert}, {Cifuentes},
  {Cort{\'e}s-Contreras}, {Montes}, {Revilla}, \&
  {Skrzypinski}}]{perdelwitz2021}
{Perdelwitz}, V., {Mittag}, M., {Tal-Or}, L., {et~al.} 2021,
  \bibinfo{title}{{CARMENES input catalog of M dwarfs. VI. A time-resolved Ca
  II H\&K catalog from archival data},} \aap, 652, A116,
  \dodoi{10.1051/0004-6361/202140889}

% type= article
\bibitem[{M.~A.~C. {Perryman} {et~al.}(1997){Perryman}, {Lindegren},
  {Kovalevsky}, {Hoeg}, {Bastian}, {Bernacca}, {Cr{\'e}z{\'e}}, {Donati},
  {Grenon}, {Grewing}, {van Leeuwen}, {van der Marel}, {Mignard}, {Murray}, {Le
  Poole}, {Schrijver}, {Turon}, {Arenou}, {Froeschl{\'e}}, \&
  {Petersen}}]{perryman1997}
{Perryman}, M.~A.~C., {Lindegren}, L., {Kovalevsky}, J., {et~al.} 1997,
  \bibinfo{title}{{The HIPPARCOS Catalogue},} \aap, 323, L49

% type= article
\bibitem[{E.~A. {Petigura} {et~al.}(2013){Petigura}, {Howard}, \&
  {Marcy}}]{petigura2013}
{Petigura}, E.~A., {Howard}, A.~W., \& {Marcy}, G.~W. 2013,
  \bibinfo{title}{{Prevalence of Earth-size planets orbiting Sun-like stars},}
  Proceedings of the National Academy of Science, 110, 19273,
  \dodoi{10.1073/pnas.1319909110}

% type= article
\bibitem[{P. {Petit} {et~al.}(2022){Petit}, {B{\"o}hm}, {Folsom},
  {Ligni{\`e}res}, \& {Cang}}]{petit2022}
{Petit}, P., {B{\"o}hm}, T., {Folsom}, C.~P., {Ligni{\`e}res}, F., \& {Cang},
  T. 2022, \bibinfo{title}{{A decade-long magnetic monitoring of Vega},} \aap,
  666, A20, \dodoi{10.1051/0004-6361/202143000}

% type= article
\bibitem[{P. {Petit} {et~al.}(2011){Petit}, {Ligni{\`e}res}, {Auri{\`e}re},
  {Wade}, {Alina}, {Ballot}, {B{\"o}hm}, {Jouve}, {Oza}, {Paletou}, \&
  {Th{\'e}ado}}]{petit2011}
{Petit}, P., {Ligni{\`e}res}, F., {Auri{\`e}re}, M., {et~al.} 2011,
  \bibinfo{title}{{Detection of a weak surface magnetic field on Sirius A: are
  all tepid stars magnetic?},} \aap, 532, L13,
  \dodoi{10.1051/0004-6361/201117573}

% type= article
\bibitem[{N.~M. {Phillips} {et~al.}(2010){Phillips}, {Greaves}, {Dent},
  {Matthews}, {Holland}, {Wyatt}, \& {Sibthorpe}}]{phillips2010}
{Phillips}, N.~M., {Greaves}, J.~S., {Dent}, W.~R.~F., {et~al.} 2010,
  \bibinfo{title}{{Target selection for the SUNS and DEBRIS surveys for debris
  discs in the solar neighbourhood},} \mnras, 403, 1089,
  \dodoi{10.1111/j.1365-2966.2009.15641.x}

% type= article
\bibitem[{L. {Piccotti} {et~al.}(2020){Piccotti}, {Docobo}, {Carini},
  {Tamazian}, {Brocato}, {Andrade}, \& {Campo}}]{piccotti2020}
{Piccotti}, L., {Docobo}, J.~{\'A}., {Carini}, R., {et~al.} 2020,
  \bibinfo{title}{{A study of the physical properties of SB2s with both the
  visual and spectroscopic orbits},} \mnras, 492, 2709,
  \dodoi{10.1093/mnras/stz3616}

% type= article
\bibitem[{I. {Pillitteri} {et~al.}(2006){Pillitteri}, {Micela}, {Damiani}, \&
  {Sciortino}}]{pillitteri2006}
{Pillitteri}, I., {Micela}, G., {Damiani}, F., \& {Sciortino}, S. 2006,
  \bibinfo{title}{{Deep X-ray survey of the young open cluster NGC 2516 with
  XMM-Newton},} \aap, 450, 993, \dodoi{10.1051/0004-6361:20054003}

% type= article
\bibitem[{I. {Pillitteri} {et~al.}(2004){Pillitteri}, {Micela}, {Sciortino},
  {Damiani}, \& {Harnden}}]{pillitteri2004}
{Pillitteri}, I., {Micela}, G., {Sciortino}, S., {Damiani}, F., \& {Harnden},
  F.~R., J. 2004, \bibinfo{title}{{XMM-Newton observations of the young open
  cluster Blanco 1. I. X-ray spectroscopy and photometry},} \aap, 421, 175,
  \dodoi{10.1051/0004-6361:20035869}

% type= article
\bibitem[{M. {Pinsonneault}(1997){Pinsonneault}}]{pinsonneault1997}
{Pinsonneault}, M. 1997, \bibinfo{title}{{Mixing in Stars},} \araa, 35, 557,
  \dodoi{10.1146/annurev.astro.35.1.557}

% type= article
\bibitem[{C.~V. {Pittman} {et~al.}(2025){Pittman}, {Espaillat}, {Robinson},
  {Thanathibodee}, {Lopez}, {Calvet}, {Zhu}, {Walter}, {Wendeborn}, {Manara},
  {Campbell-White}, {Claes}, {Fang}, {Frasca}, {Gameiro}, {Gangi},
  {Hern{\'a}ndez}, {K{\'o}sp{\'a}l}, {Mauc{\'o}}, {Muzerolle}, {Siwak},
  {Tychoniec}, \& {Venuti}}]{pittman2025}
{Pittman}, C.~V., {Espaillat}, C.~C., {Robinson}, C.~E., {et~al.} 2025,
  \bibinfo{title}{{The ODYSSEUS Survey. Characterizing Magnetospheric
  Geometries and Hotspot Structures in T Tauri Stars},} \apj, 992, 134,
  \dodoi{10.3847/1538-4357/adef35}

% type= article
\bibitem[{P. {Plavchan} {et~al.}(2009){Plavchan}, {Werner}, {Chen},
  {Stapelfeldt}, {Su}, {Stauffer}, \& {Song}}]{plavchan2009}
{Plavchan}, P., {Werner}, M.~W., {Chen}, C.~H., {et~al.} 2009,
  \bibinfo{title}{{New Debris Disks Around Young, Low-Mass Stars Discovered
  with the Spitzer Space Telescope},} \apj, 698, 1068,
  \dodoi{10.1088/0004-637X/698/2/1068}

% type= article
\bibitem[{A. {Poglitsch} {et~al.}(2010){Poglitsch}, {Waelkens}, {Geis},
  {Feuchtgruber}, {Vandenbussche}, {Rodriguez}, {Krause}, {Renotte}, {van
  Hoof}, {Saraceno}, {Cepa}, {Kerschbaum}, {Agn{\`e}se}, {Ali}, {Altieri},
  {Andreani}, {Augueres}, {Balog}, {Barl}, {Bauer}, {Belbachir}, {Benedettini},
  {Billot}, {Boulade}, {Bischof}, {Blommaert}, {Callut}, {Cara}, {Cerulli},
  {Cesarsky}, {Contursi}, {Creten}, {De Meester}, {Doublier}, {Doumayrou},
  {Duband}, {Exter}, {Genzel}, {Gillis}, {Gr{\"o}zinger}, {Henning},
  {Herreros}, {Huygen}, {Inguscio}, {Jakob}, {Jamar}, {Jean}, {de Jong},
  {Katterloher}, {Kiss}, {Klaas}, {Lemke}, {Lutz}, {Madden}, {Marquet},
  {Martignac}, {Mazy}, {Merken}, {Montfort}, {Morbidelli}, {M{\"u}ller},
  {Nielbock}, {Okumura}, {Orfei}, {Ottensamer}, {Pezzuto}, {Popesso},
  {Putzeys}, {Regibo}, {Reveret}, {Royer}, {Sauvage}, {Schreiber}, {Stegmaier},
  {Schmitt}, {Schubert}, {Sturm}, {Thiel}, {Tofani}, {Vavrek}, {Wetzstein},
  {Wieprecht}, \& {Wiezorrek}}]{poglitsch2010}
{Poglitsch}, A., {Waelkens}, C., {Geis}, N., {et~al.} 2010,
  \bibinfo{title}{{The Photodetector Array Camera and Spectrometer (PACS) on
  the Herschel Space Observatory},} \aap, 518, L2,
  \dodoi{10.1051/0004-6361/201014535}

% type= article
\bibitem[{M. {Popinchalk} {et~al.}(2023){Popinchalk}, {Faherty}, {Curtis},
  {Gagn{\'e}}, {Bardalez Gagliuffi}, {Vos}, {Ayala}, {Gonzales}, \&
  {Kiman}}]{popinchalk2023}
{Popinchalk}, M., {Faherty}, J.~K., {Curtis}, J.~L., {et~al.} 2023,
  \bibinfo{title}{{Examining the Rotation Period Distribution of the 40 Myr
  Tucana-Horologium Association with TESS},} \apj, 945, 114,
  \dodoi{10.3847/1538-4357/acb055}

% type= article
\bibitem[{D.~M. {Popper}(1990){Popper}}]{popper1990}
{Popper}, D.~M. 1990, \bibinfo{title}{{Orbits of Close Binaries with CA II H
  and K in Emission. III. Eleven More Systems},} \aj, 100, 247,
  \dodoi{10.1086/115511}

% type= article
\bibitem[{D. {Pourbaix} {et~al.}(2004){Pourbaix}, {Tokovinin}, {Batten},
  {Fekel}, {Hartkopf}, {Levato}, {Morrell}, {Torres}, \& {Udry}}]{SBC2004}
{Pourbaix}, D., {Tokovinin}, A.~A., {Batten}, A.~H., {et~al.} 2004,
  \bibinfo{title}{{S$_{B$^{9}$}$: The ninth catalogue of spectroscopic binary
  orbits},} \aap, 424, 727, \dodoi{10.1051/0004-6361:20041213}

% type= inproceedings
\bibitem[{P. {Predehl} {et~al.}(2006){Predehl}, {Hasinger}, {B{\"o}hringer},
  {Briel}, {Brunner}, {Churazov}, {Freyberg}, {Friedrich}, {Kendziorra},
  {Lutz}, {Meidinger}, {Pavlinsky}, {Pfeffermann}, {Santangelo}, {Schmitt},
  {Schuecker}, {Schwope}, {Steinmetz}, {Str{\"u}der}, {Sunyaev}, \&
  {Wilms}}]{predehl2006}
{Predehl}, P., {Hasinger}, G., {B{\"o}hringer}, H., {et~al.} 2006,
  \bibinfo{title}{{eROSITA},} in Society of Photo-Optical Instrumentation
  Engineers (SPIE) Conference Series, Vol. 6266, Space Telescopes and
  Instrumentation II: Ultraviolet to Gamma Ray, ed. M.~J.~L. {Turner} \&
  G.~{Hasinger}, 62660P, \dodoi{10.1117/12.670249}

% type= article
\bibitem[{T. {Preibisch} {et~al.}(2002){Preibisch}, {Brown}, {Bridges},
  {Guenther}, \& {Zinnecker}}]{preibisch2002}
{Preibisch}, T., {Brown}, A. G.~A., {Bridges}, T., {Guenther}, E., \&
  {Zinnecker}, H. 2002, \bibinfo{title}{{Exploring the Full Stellar Population
  of the Upper Scorpius OB Association},} \aj, 124, 404, \dodoi{10.1086/341174}

% type= article
\bibitem[{T. {Preibisch} {et~al.}(1998){Preibisch}, {Guenther}, {Zinnecker},
  {Sterzik}, {Frink}, \& {Roeser}}]{priebisch1998}
{Preibisch}, T., {Guenther}, E., {Zinnecker}, H., {et~al.} 1998,
  \bibinfo{title}{{A lithium-survey for pre-main sequence stars in the Upper
  Scorpius OB association},} \aap, 333, 619

% type= article
\bibitem[{L. {Prisinzano} {et~al.}(2003){Prisinzano}, {Micela}, {Sciortino}, \&
  {Favata}}]{prisinzano2003}
{Prisinzano}, L., {Micela}, G., {Sciortino}, S., \& {Favata}, F. 2003,
  \bibinfo{title}{{Luminosity and Mass Function of the Galactic open cluster
  NGC 2422},} \aap, 404, 927, \dodoi{10.1051/0004-6361:20030524}

% type= article
\bibitem[{A.~B.~A. {Queiroz} {et~al.}(2023){Queiroz}, {Anders}, {Chiappini},
  {Khalatyan}, {Santiago}, {Nepal}, {Steinmetz}, {Gallart}, {Valentini}, {Dal
  Ponte}, {Barbuy}, {P{\'e}rez-Villegas}, {Masseron}, {Fern{\'a}ndez-Trincado},
  {Khoperskov}, {Minchev}, {Fern{\'a}ndez-Alvar}, {Lane}, \&
  {Nitschelm}}]{queiroz2023}
{Queiroz}, A.~B.~A., {Anders}, F., {Chiappini}, C., {et~al.} 2023,
  \bibinfo{title}{{StarHorse results for spectroscopic surveys and Gaia DR3:
  Chrono-chemical populations in the solar vicinity, the genuine thick disk,
  and young alpha-rich stars},} \aap, 673, A155,
  \dodoi{10.1051/0004-6361/202245399}

% type= article
\bibitem[{D. {Raghavan} {et~al.}(2009){Raghavan}, {McAlister}, {Torres},
  {Latham}, {Mason}, {Boyajian}, {Baines}, {Williams}, {ten Brummelaar},
  {Farrington}, {Ridgway}, {Sturmann}, {Sturmann}, \& {Turner}}]{raghavan2009}
{Raghavan}, D., {McAlister}, H.~A., {Torres}, G., {et~al.} 2009,
  \bibinfo{title}{{The Visual Orbit of the 1.1 Day Spectroscopic Binary
  {\ensuremath{\sigma}}$^{2}$ Coronae Borealis from Interferometry at the Chara
  Array},} \apj, 690, 394, \dodoi{10.1088/0004-637X/690/1/394}

% type= article
\bibitem[{M. {Rainer} {et~al.}(2023){Rainer}, {Desidera}, {Borsa}, {Barbato},
  {Biazzo}, {Bonomo}, {Gratton}, {Messina}, {Scandariato}, {Affer}, {Benatti},
  {Carleo}, {Cabona}, {Covino}, {Lanza}, {Ligi}, {Maldonado}, {Mancini},
  {Nardiello}, {Sicilia}, {Sozzetti}, {Bignamini}, {Cosentino}, {Knapic},
  {Mart{\'\i}nez Fiorenzano}, {Molinari}, {Pedani}, \& {Poretti}}]{rainer2023}
{Rainer}, M., {Desidera}, S., {Borsa}, F., {et~al.} 2023, \bibinfo{title}{{The
  GAPS programme at TNG. XLIV. Projected rotational velocities of 273
  exoplanet-host stars observed with HARPS-N},} \aap, 676, A90,
  \dodoi{10.1051/0004-6361/202245564}

% type= article
\bibitem[{I. {Ram{\'\i}rez} {et~al.}(2012){Ram{\'\i}rez}, {Fish}, {Lambert}, \&
  {Allende Prieto}}]{ramirez2012}
{Ram{\'\i}rez}, I., {Fish}, J.~R., {Lambert}, D.~L., \& {Allende Prieto}, C.
  2012, \bibinfo{title}{{Lithium Abundances in nearby FGK Dwarf and Subgiant
  Stars: Internal Destruction, Galactic Chemical Evolution, and Exoplanets},}
  \apj, 756, 46, \dodoi{10.1088/0004-637X/756/1/46}

% type= article
\bibitem[{I. {Ram{\'\i}rez} {et~al.}(2014){Ram{\'\i}rez}, {Mel{\'e}ndez},
  {Bean}, {Asplund}, {Bedell}, {Monroe}, {Casagrande}, {Schirbel}, {Dreizler},
  {Teske}, {Tucci Maia}, {Alves-Brito}, \& {Baumann}}]{ramirez2014}
{Ram{\'\i}rez}, I., {Mel{\'e}ndez}, J., {Bean}, J., {et~al.} 2014,
  \bibinfo{title}{{The Solar Twin Planet Search. I. Fundamental parameters of
  the stellar sample},} \aap, 572, A48, \dodoi{10.1051/0004-6361/201424244}

% type= article
\bibitem[{R. {Rampalli} {et~al.}(2021){Rampalli}, {Ag{\"u}eros}, {Curtis},
  {Douglas}, {N{\'u}{\~n}ez}, {Cargile}, {Covey}, {Gosnell}, {Kraus}, {Law}, \&
  {Mann}}]{rampalli2021}
{Rampalli}, R., {Ag{\"u}eros}, M.~A., {Curtis}, J.~L., {et~al.} 2021,
  \bibinfo{title}{{Three K2 Campaigns Yield Rotation Periods for 1013 Stars in
  Praesepe},} \apj, 921, 167, \dodoi{10.3847/1538-4357/ac0c1e}

% type= article
\bibitem[{S. {Randich} {et~al.}(1996){Randich}, {Schmitt}, {Prosser}, \&
  {Stauffer}}]{randich1996}
{Randich}, S., {Schmitt}, J.~H.~M.~M., {Prosser}, C.~F., \& {Stauffer}, J.~R.
  1996, \bibinfo{title}{{The X-ray properties of the young open cluster around
  {\ensuremath{\alpha}} Persei.},} \aap, 305, 785

% type= article
\bibitem[{S. {Randich} {et~al.}(2022){Randich}, {Gilmore}, {Magrini}, {Sacco},
  {Jackson}, {Jeffries}, {Worley}, {Hourihane}, {Gonneau}, {Viscasillas
  Vazquez}, {Franciosini}, {Lewis}, {Alfaro}, {Allende Prieto}, {Bensby},
  {Blomme}, {Bragaglia}, {Flaccomio}, {Fran{\c{c}}ois}, {Irwin}, {Koposov},
  {Korn}, {Lanzafame}, {Pancino}, {Recio-Blanco}, {Smiljanic}, {Van Eck},
  {Zwitter}, {Asplund}, {Bonifacio}, {Feltzing}, {Binney}, {Drew}, {Ferguson},
  {Micela}, {Negueruela}, {Prusti}, {Rix}, {Vallenari}, {Bayo}, {Bergemann},
  {Biazzo}, {Carraro}, {Casey}, {Damiani}, {Frasca}, {Heiter}, {Hill},
  {Jofr{\'e}}, {de Laverny}, {Lind}, {Marconi}, {Martayan}, {Masseron},
  {Monaco}, {Morbidelli}, {Prisinzano}, {Sbordone}, {Sousa}, {Zaggia},
  {Adibekyan}, {Bonito}, {Caffau}, {Daflon}, {Feuillet}, {Gebran}, {Gonzalez
  Hernandez}, {Guiglion}, {Herrero}, {Lobel}, {Maiz Apellaniz}, {Merle},
  {Mikolaitis}, {Montes}, {Morel}, {Soubiran}, {Spina}, {Tabernero},
  {Tautvaisiene}, {Traven}, {Valentini}, {Van der Swaelmen}, {Villanova},
  {Wright}, {Abbas}, {Aguirre B{\o}rsen-Koch}, {Alves}, {Balaguer-Nunez},
  {Barklem}, {Barrado}, {Berlanas}, {Binks}, {Bressan}, {Capuzzo-Dolcetta},
  {Casagrande}, {Casamiquela}, {Collins}, {D'Orazi}, {Dantas}, {Debattista},
  {Delgado-Mena}, {Di Marcantonio}, {Drazdauskas}, {Evans}, {Famaey},
  {Franchini}, {Fr{\'e}mat}, {Friel}, {Fu}, {Geisler}, {Gerhard}, {Gonzalez
  Solares}, {Grebel}, {Gutierrez Albarran}, {Hatzidimitriou}, {Held},
  {Jim{\'e}nez-Esteban}, {J{\"o}nsson}, {Jordi}, {Khachaturyants},
  {Kordopatis}, {Kos}, {Lagarde}, {Mahy}, {Mapelli}, {Marfil}, {Martell},
  {Messina}, {Miglio}, {Minchev}, {Moitinho}, {Montalban}, {Monteiro},
  {Morossi}, {Mowlavi}, {Mucciarelli}, {Murphy}, {Nardetto}, {Ortolani},
  {Paletou}, {Palous}, {Paunzen}, {Pickering}, {Quirrenbach}, {Re Fiorentin},
  {Read}, {Romano}, {Ryde}, {Sanna}, {Santos}, {Seabroke}, {Spagna},
  {Steinmetz}, {Stonkut{\'e}}, {Sutorius}, {Th{\'e}venin}, {Tosi}, {Tsantaki},
  {Vink}, {Wright}, {Wyse}, {Zoccali}, {Zorec}, {Zucker}, \&
  {Walton}}]{randich2022}
{Randich}, S., {Gilmore}, G., {Magrini}, L., {et~al.} 2022,
  \bibinfo{title}{{The Gaia-ESO Public Spectroscopic Survey: Implementation,
  data products, open cluster survey, science, and legacy},} \aap, 666, A121,
  \dodoi{10.1051/0004-6361/202243141}

% type= article
\bibitem[{A. {Rathsam} {et~al.}(2023){Rathsam}, {Mel{\'e}ndez}, \& {Carvalho
  Silva}}]{rathsam2023}
{Rathsam}, A., {Mel{\'e}ndez}, J., \& {Carvalho Silva}, G. 2023,
  \bibinfo{title}{{Lithium depletion in solar analogs: age and mass effects},}
  \mnras, 525, 4642, \dodoi{10.1093/mnras/stad2589}

% type= article
\bibitem[{S. {Ratzenb{\"o}ck} {et~al.}(2023){Ratzenb{\"o}ck}, {Gro{\ss}schedl},
  {M{\"o}ller}, {Alves}, {Bomze}, \& {Meingast}}]{ratzenbock2023}
{Ratzenb{\"o}ck}, S., {Gro{\ss}schedl}, J.~E., {M{\"o}ller}, T., {et~al.} 2023,
  \bibinfo{title}{{Significance mode analysis (SigMA) for hierarchical
  structures. An application to the Sco-Cen OB association},} \aap, 677, A59,
  \dodoi{10.1051/0004-6361/202243690}

% type= article
\bibitem[{L.~M. {Rebull} {et~al.}(2020){Rebull}, {Stauffer}, {Cody},
  {Hillenbrand}, {Bouvier}, {Roggero}, \& {David}}]{rebull2020}
{Rebull}, L.~M., {Stauffer}, J.~R., {Cody}, A.~M., {et~al.} 2020,
  \bibinfo{title}{{Rotation of Low-mass Stars in Taurus with K2},} \aj, 159,
  273, \dodoi{10.3847/1538-3881/ab893c}

% type= article
\bibitem[{L.~M. {Rebull} {et~al.}(2018){Rebull}, {Stauffer}, {Cody},
  {Hillenbrand}, {David}, \& {Pinsonneault}}]{rebull2018}
{Rebull}, L.~M., {Stauffer}, J.~R., {Cody}, A.~M., {et~al.} 2018,
  \bibinfo{title}{{Rotation of Low-mass Stars in Upper Scorpius and
  {\ensuremath{\rho}} Ophiuchus with K2},} \aj, 155, 196,
  \dodoi{10.3847/1538-3881/aab605}

% type= article
\bibitem[{L.~M. {Rebull} {et~al.}(2017){Rebull}, {Stauffer}, {Hillenbrand},
  {Cody}, {Bouvier}, {Soderblom}, {Pinsonneault}, \& {Hebb}}]{rebull2017}
{Rebull}, L.~M., {Stauffer}, J.~R., {Hillenbrand}, L.~A., {et~al.} 2017,
  \bibinfo{title}{{Rotation of Late-type Stars in Praesepe with K2},} \apj,
  839, 92, \dodoi{10.3847/1538-4357/aa6aa4}

% type= article
\bibitem[{L.~M. {Rebull} {et~al.}(2022){Rebull}, {Stauffer}, {Hillenbrand},
  {Cody}, {Kruse}, \& {Powell}}]{rebull2022}
{Rebull}, L.~M., {Stauffer}, J.~R., {Hillenbrand}, L.~A., {et~al.} 2022,
  \bibinfo{title}{{Rotation of Low-mass Stars in Upper Centaurus-Lupus and
  Lower Centaurus-Crux with TESS},} \aj, 164, 80,
  \dodoi{10.3847/1538-3881/ac75f1}

% type= article
\bibitem[{L.~M. {Rebull} {et~al.}(2004){Rebull}, {Wolff}, \&
  {Strom}}]{rebull2004}
{Rebull}, L.~M., {Wolff}, S.~C., \& {Strom}, S.~E. 2004,
  \bibinfo{title}{{Stellar Rotation in Young Clusters: The First 4 Million
  Years},} \aj, 127, 1029, \dodoi{10.1086/380931}

% type= article
\bibitem[{L.~M. {Rebull} {et~al.}(2008){Rebull}, {Stapelfeldt}, {Werner},
  {Mannings}, {Chen}, {Stauffer}, {Smith}, {Song}, {Hines}, \&
  {Low}}]{rebull2008}
{Rebull}, L.~M., {Stapelfeldt}, K.~R., {Werner}, M.~W., {et~al.} 2008,
  \bibinfo{title}{{Spitzer MIPS Observations of Stars in the {$\beta$} Pictoris
  Moving Group},} \apj, 681, 1484, \dodoi{10.1086/588182}

% type= article
\bibitem[{L.~M. {Rebull} {et~al.}(2016){Rebull}, {Stauffer}, {Bouvier}, {Cody},
  {Hillenbrand}, {Soderblom}, {Valenti}, {Barrado}, {Bouy}, {Ciardi},
  {Pinsonneault}, {Stassun}, {Micela}, {Aigrain}, {Vrba}, {Somers},
  {Christiansen}, {Gillen}, \& {Collier Cameron}}]{rebull2016}
{Rebull}, L.~M., {Stauffer}, J.~R., {Bouvier}, J., {et~al.} 2016,
  \bibinfo{title}{{Rotation in the Pleiades with K2. I. Data and First
  Results},} \aj, 152, 113, \dodoi{10.3847/0004-6256/152/5/113}

% type= article
\bibitem[{I.~N. {Reid} {et~al.}(2002){Reid}, {Kilkenny}, \& {Cruz}}]{reid2002}
{Reid}, I.~N., {Kilkenny}, D., \& {Cruz}, K.~L. 2002, \bibinfo{title}{{Meeting
  the Cool Neighbors. II. Photometry of Southern NLTT Stars},} \aj, 123, 2822,
  \dodoi{10.1086/339700}

% type= article
\bibitem[{A. {Reiners} {et~al.}(2012){Reiners}, {Joshi}, \&
  {Goldman}}]{reiners2012}
{Reiners}, A., {Joshi}, N., \& {Goldman}, B. 2012, \bibinfo{title}{{A Catalog
  of Rotation and Activity in Early-M Stars},} \aj, 143, 93,
  \dodoi{10.1088/0004-6256/143/4/93}

% type= article
\bibitem[{A. {Reiners} {et~al.}(2018){Reiners}, {Zechmeister}, {Caballero},
  {Ribas}, {Morales}, {Jeffers}, {Sch{\"o}fer}, {Tal-Or}, {Quirrenbach},
  {Amado}, {Kaminski}, {Seifert}, {Abril}, {Aceituno}, {Alonso-Floriano},
  {Ammler-von Eiff}, {Antona}, {Anglada-Escud{\'e}}, {Anwand-Heerwart},
  {Arroyo-Torres}, {Azzaro}, {Baroch}, {Barrado}, {Bauer}, {Becerril},
  {B{\'e}jar}, {Ben{\'\i}tez}, {Berdinas}, {Bergond}, {Bl{\"u}mcke},
  {Brinkm{\"o}ller}, {del Burgo}, {Cano}, {C{\'a}rdenas V{\'a}zquez}, {Casal},
  {Cifuentes}, {Claret}, {Colom{\'e}}, {Cort{\'e}s-Contreras}, {Czesla},
  {D{\'\i}ez-Alonso}, {Dreizler}, {Feiz}, {Fern{\'a}ndez}, {Ferro},
  {Fuhrmeister}, {Galad{\'\i}-Enr{\'\i}quez}, {Garcia-Piquer}, {Garc{\'\i}a
  Vargas}, {Gesa}, {G{\'o}mez Galera}, {Gonz{\'a}lez Hern{\'a}ndez},
  {Gonz{\'a}lez-Peinado}, {Gr{\"o}zinger}, {Grohnert}, {Gu{\`a}rdia},
  {Guenther}, {Guijarro}, {de Guindos}, {Guti{\'e}rrez-Soto}, {Hagen},
  {Hatzes}, {Hauschildt}, {Hedrosa}, {Helmling}, {Henning}, {Hermelo},
  {Hern{\'a}ndez Arab{\'\i}}, {Hern{\'a}ndez Casta{\~n}o}, {Hern{\'a}ndez
  Hernando}, {Herrero}, {Huber}, {Huke}, {Johnson}, {de Juan}, {Kim}, {Klein},
  {Kl{\"u}ter}, {Klutsch}, {K{\"u}rster}, {Lafarga}, {Lamert}, {Lamp{\'o}n},
  {Lara}, {Laun}, {Lemke}, {Lenzen}, {Launhardt}, {L{\'o}pez del Fresno},
  {L{\'o}pez-Gonz{\'a}lez}, {L{\'o}pez-Puertas}, {L{\'o}pez Salas},
  {L{\'o}pez-Santiago}, {Luque}, {Mag{\'a}n Madinabeitia}, {Mall}, {Mancini},
  {Mandel}, {Marfil}, {Mar{\'\i}n Molina}, {Maroto Fern{\'a}ndez},
  {Mart{\'\i}n}, {Mart{\'\i}n-Ruiz}, {Marvin}, {Mathar}, {Mirabet}, {Montes},
  {Moreno-Raya}, {Moya}, {Mundt}, {Nagel}, {Naranjo}, {Nortmann}, {Nowak},
  {Ofir}, {Oreiro}, {Pall{\'e}}, {Panduro}, {Pascual}, {Passegger}, {Pavlov},
  {Pedraz}, {P{\'e}rez-Calpena}, {P{\'e}rez Medialdea}, {Perger}, {Perryman},
  {Pluto}, {Rabaza}, {Ram{\'o}n}, {Rebolo}, {Redondo}, {Reffert}, {Reinhart},
  {Rhode}, {Rix}, {Rodler}, {Rodr{\'\i}guez}, {Rodr{\'\i}guez-L{\'o}pez},
  {Rodr{\'\i}guez Trinidad}, {Rohloff}, {Rosich}, {Sadegi},
  {S{\'a}nchez-Blanco}, {S{\'a}nchez Carrasco}, {S{\'a}nchez-L{\'o}pez},
  {Sanz-Forcada}, {Sarkis}, {Sarmiento}, {Sch{\"a}fer}, {Schmitt}, {Schiller},
  {Schweitzer}, {Solano}, {Stahl}, {Strachan}, {St{\"u}rmer}, {Su{\'a}rez},
  {Tabernero}, {Tala}, {Trifonov}, {Tulloch}, {Ulbrich}, {Veredas}, {Vico
  Linares}, {Vilardell}, {Wagner}, {Winkler}, {Wolthoff}, {Xu}, {Yan}, \&
  {Zapatero Osorio}}]{reiners2018}
{Reiners}, A., {Zechmeister}, M., {Caballero}, J.~A., {et~al.} 2018,
  \bibinfo{title}{{The CARMENES search for exoplanets around M dwarfs.
  High-resolution optical and near-infrared spectroscopy of 324 survey stars},}
  \aap, 612, A49, \dodoi{10.1051/0004-6361/201732054}

% type= article
\bibitem[{A. {Reiners} {et~al.}(2022){Reiners}, {Shulyak}, {K{\"a}pyl{\"a}},
  {Ribas}, {Nagel}, {Zechmeister}, {Caballero}, {Shan}, {Fuhrmeister},
  {Quirrenbach}, {Amado}, {Montes}, {Jeffers}, {Azzaro}, {B{\'e}jar},
  {Chaturvedi}, {Henning}, {K{\"u}rster}, \& {Pall{\'e}}}]{reiners2022}
{Reiners}, A., {Shulyak}, D., {K{\"a}pyl{\"a}}, P.~J., {et~al.} 2022,
  \bibinfo{title}{{Magnetism, rotation, and nonthermal emission in cool stars.
  Average magnetic field measurements in 292 M dwarfs},} \aap, 662, A41,
  \dodoi{10.1051/0004-6361/202243251}

% type= article
\bibitem[{J.~H. {Rhee} {et~al.}(2007){Rhee}, {Song}, \& {Zuckerman}}]{rhee2007}
{Rhee}, J.~H., {Song}, I., \& {Zuckerman}, B. 2007, \bibinfo{title}{{EF
  Chamaeleontis: Warm Dust Orbiting a Nearby 10 Myr Old Star},} \apj, 671, 616,
  \dodoi{10.1086/520760}

% type= article
\bibitem[{B. {Riaz} {et~al.}(2006){Riaz}, {Mullan}, \& {Gizis}}]{riaz2006}
{Riaz}, B., {Mullan}, D.~J., \& {Gizis}, J.~E. 2006, \bibinfo{title}{{Spitzer
  Observations of Nearby M Dwarfs},} \apj, 650, 1133, \dodoi{10.1086/507446}

% type= article
\bibitem[{A.~R. {Riedel} {et~al.}(2014){Riedel}, {Finch}, {Henry},
  {Subasavage}, {Jao}, {Malo}, {Rodriguez}, {White}, {Gies}, {Dieterich},
  {Winters}, {Davison}, {Nelan}, {Blunt}, {Cruz}, {Rice}, \&
  {Ianna}}]{riedel2014}
{Riedel}, A.~R., {Finch}, C.~T., {Henry}, T.~J., {et~al.} 2014,
  \bibinfo{title}{{The Solar Neighborhood. XXXIII. Parallax Results from the
  CTIOPI 0.9 m Program: Trigonometric Parallaxes of Nearby Low-mass Active and
  Young Systems},} \aj, 147, 85, \dodoi{10.1088/0004-6256/147/4/85}

% type= article
\bibitem[{G.~H. {Rieke} {et~al.}(2004){Rieke}, {Young}, {Engelbracht}, {Kelly},
  {Low}, {Haller}, {Beeman}, {Gordon}, {Stansberry}, {Misselt}, {Cadien},
  {Morrison}, {Rivlis}, {Latter}, {Noriega-Crespo}, {Padgett}, {Stapelfeldt},
  {Hines}, {Egami}, {Muzerolle}, {Alonso-Herrero}, {Blaylock}, {Dole}, {Hinz},
  {Le Floc'h}, {Papovich}, {P{\'e}rez-Gonz{\'a}lez}, {Smith}, {Su}, {Bennett},
  {Frayer}, {Henderson}, {Lu}, {Masci}, {Pesenson}, {Rebull}, {Rho}, {Keene},
  {Stolovy}, {Wachter}, {Wheaton}, {Werner}, \& {Richards}}]{rieke2004}
{Rieke}, G.~H., {Young}, E.~T., {Engelbracht}, C.~W., {et~al.} 2004,
  \bibinfo{title}{{The Multiband Imaging Photometer for Spitzer (MIPS)},}
  \apjs, 154, 25, \dodoi{10.1086/422717}

% type= article
\bibitem[{G.~H. {Rieke} {et~al.}(2005){Rieke}, {Su}, {Stansberry}, {Trilling},
  {Bryden}, {Muzerolle}, {White}, {Gorlova}, {Young}, {Beichman},
  {Stapelfeldt}, \& {Hines}}]{rieke2005}
{Rieke}, G.~H., {Su}, K.~Y.~L., {Stansberry}, J.~A., {et~al.} 2005,
  \bibinfo{title}{{Decay of Planetary Debris Disks},} \apj, 620, 1010,
  \dodoi{10.1086/426937}

% type= article
\bibitem[{M. {Riello} {et~al.}(2021){Riello}, {De Angeli}, {Evans},
  {Montegriffo}, {Carrasco}, {Busso}, {Palaversa}, {Burgess}, {Diener},
  {Davidson}, {Rowell}, {Fabricius}, {Jordi}, {Bellazzini}, {Pancino},
  {Harrison}, {Cacciari}, {van Leeuwen}, {Hambly}, {Hodgkin}, {Osborne},
  {Altavilla}, {Barstow}, {Brown}, {Castellani}, {Cowell}, {De Luise},
  {Gilmore}, {Giuffrida}, {Hidalgo}, {Holland}, {Marinoni}, {Pagani},
  {Piersimoni}, {Pulone}, {Ragaini}, {Rainer}, {Richards}, {Sanna}, {Walton},
  {Weiler}, \& {Yoldas}}]{riello2021}
{Riello}, M., {De Angeli}, F., {Evans}, D.~W., {et~al.} 2021,
  \bibinfo{title}{{Gaia Early Data Release 3. Photometric content and
  validation},} \aap, 649, A3, \dodoi{10.1051/0004-6361/202039587}

% type= article
\bibitem[{P. {Riviere-Marichalar} {et~al.}(2013){Riviere-Marichalar}, {Pinte},
  {Barrado}, {Thi}, {Eiroa}, {Kamp}, {Montesinos}, {Donaldson}, {Augereau},
  {Hu{\'e}lamo}, {Roberge}, {Ardila}, {Sandell}, {Williams}, {Dent}, {Menard},
  {Lillo-Box}, \& {Duch{\^e}ne}}]{riviere2013}
{Riviere-Marichalar}, P., {Pinte}, C., {Barrado}, D., {et~al.} 2013,
  \bibinfo{title}{{Gas and dust in the TW Hydrae association as seen by the
  Herschel Space Observatory},} \aap, 555, A67,
  \dodoi{10.1051/0004-6361/201321506}

% type= article
\bibitem[{P. {Riviere-Marichalar} {et~al.}(2014){Riviere-Marichalar},
  {Barrado}, {Montesinos}, {Duch{\^e}ne}, {Bouy}, {Pinte}, {Menard},
  {Donaldson}, {Eiroa}, {Krivov}, {Kamp}, {Mendigut{\'\i}a}, {Dent}, \&
  {Lillo-Box}}]{riviere2014}
{Riviere-Marichalar}, P., {Barrado}, D., {Montesinos}, B., {et~al.} 2014,
  \bibinfo{title}{{Gas and dust in the beta Pictoris moving group as seen by
  the Herschel Space Observatory},} \aap, 565, A68,
  \dodoi{10.1051/0004-6361/201322901}

% type= article
\bibitem[{P. {Riviere-Marichalar} {et~al.}(2015){Riviere-Marichalar},
  {Elliott}, {Rebollido}, {Bayo}, {Ribas}, {Mer{\'\i}n}, {Kamp}, {Dent}, \&
  {Montesinos}}]{riviere2015}
{Riviere-Marichalar}, P., {Elliott}, P., {Rebollido}, I., {et~al.} 2015,
  \bibinfo{title}{{Herschel-PACS observations of discs in the
  {\ensuremath{\eta}} Chamaeleontis association},} \aap, 584, A22,
  \dodoi{10.1051/0004-6361/201526584}

% type= article
\bibitem[{J. {Robrade} {et~al.}(2022){Robrade}, {Czesla}, {Freund}, {Schmitt},
  \& {Schneider}}]{robrade2022}
{Robrade}, J., {Czesla}, S., {Freund}, S., {Schmitt}, J.~H.~M.~M., \&
  {Schneider}, P.~C. 2022, \bibinfo{title}{{eROSITA X-ray scan of the
  {\ensuremath{\eta}} Chamaeleontis cluster. Member study and search for
  dispersed low-mass stars},} \aap, 661, A34,
  \dodoi{10.1051/0004-6361/202141124}

% type= article
\bibitem[{V. {Roccatagliata} {et~al.}(2009){Roccatagliata}, {Henning}, {Wolf},
  {Rodmann}, {Corder}, {Carpenter}, {Meyer}, \& {Dowell}}]{roccatagliata2009}
{Roccatagliata}, V., {Henning}, T., {Wolf}, S., {et~al.} 2009,
  \bibinfo{title}{{Long-wavelength observations of debris discs around sun-like
  stars},} \aap, 497, 409, \dodoi{10.1051/0004-6361/200811018}

% type= article
\bibitem[{V. {Roccatagliata} {et~al.}(2024){Roccatagliata}, {Sicilia-Aguilar},
  {Kim}, {Campbell-White}, {Fang}, {Murphy}, {Wolf}, {Lawson}, {Henning}, \&
  {Bouwman}}]{roccatagliata2024}
{Roccatagliata}, V., {Sicilia-Aguilar}, A., {Kim}, M., {et~al.} 2024,
  \bibinfo{title}{{Protoplanetary and debris disks in the {\ensuremath{\eta}}
  Chamaeleontis Association. A submillimeter survey obtained with APEX/LABOCA
  observations},} \aap, 682, A63, \dodoi{10.1051/0004-6361/202346655}

% type= article
\bibitem[{B. {Rojas-Ayala} {et~al.}(2012){Rojas-Ayala}, {Covey}, {Muirhead}, \&
  {Lloyd}}]{rojas-ayala2012}
{Rojas-Ayala}, B., {Covey}, K.~R., {Muirhead}, P.~S., \& {Lloyd}, J.~P. 2012,
  \bibinfo{title}{{Metallicity and Temperature Indicators in M Dwarf K-band
  Spectra: Testing New and Updated Calibrations with Observations of 133 Solar
  Neighborhood M Dwarfs},} \apj, 748, 93, \dodoi{10.1088/0004-637X/748/2/93}

% type= article
\bibitem[{N.~G. {Roman}(1949){Roman}}]{roman1949}
{Roman}, N.~G. 1949, \bibinfo{title}{{The Ursa Major Group.},} \apj, 110, 205,
  \dodoi{10.1086/145199}

% type= article
\bibitem[{D. {Romano} {et~al.}(2021){Romano}, {Magrini}, {Randich}, {Casali},
  {Bonifacio}, {Jeffries}, {Matteucci}, {Franciosini}, {Spina}, {Guiglion},
  {Chiappini}, {Mucciarelli}, {Ventura}, {Grisoni}, {Bellazzini}, {Bensby},
  {Bragaglia}, {de Laverny}, {Korn}, {Martell}, {Tautvaisiene}, {Carraro},
  {Gonneau}, {Jofr{\'e}}, {Pancino}, {Smiljanic}, {Vallenari}, {Fu},
  {Guti{\'e}rrez Albarr{\'a}n}, {Jim{\'e}nez-Esteban}, {Montes}, {Damiani},
  {Bergemann}, \& {Worley}}]{romano2021}
{Romano}, D., {Magrini}, L., {Randich}, S., {et~al.} 2021, \bibinfo{title}{{The
  Gaia-ESO Survey: Galactic evolution of lithium from iDR6},} \aap, 653, A72,
  \dodoi{10.1051/0004-6361/202141340}

% type= article
\bibitem[{F. {Royer} {et~al.}(2002{\natexlab{a}}){Royer}, {Gerbaldi},
  {Faraggiana}, \& {G{\'o}mez}}]{royer2002a}
{Royer}, F., {Gerbaldi}, M., {Faraggiana}, R., \& {G{\'o}mez}, A.~E.
  2002{\natexlab{a}}, \bibinfo{title}{{Rotational velocities of A-type stars.
  I. Measurement of v sin i in the southern hemisphere},} \aap, 381, 105,
  \dodoi{10.1051/0004-6361:20011422}

% type= article
\bibitem[{F. {Royer} {et~al.}(2002{\natexlab{b}}){Royer}, {Grenier}, {Baylac},
  {G{\'o}mez}, \& {Zorec}}]{royer2002b}
{Royer}, F., {Grenier}, S., {Baylac}, M.~O., {G{\'o}mez}, A.~E., \& {Zorec}, J.
  2002{\natexlab{b}}, \bibinfo{title}{{Rotational velocities of A-type stars in
  the northern hemisphere. II. Measurement of v sin i},} \aap, 393, 897,
  \dodoi{10.1051/0004-6361:20020943}

% type= article
\bibitem[{F. {Royer} {et~al.}(2007){Royer}, {Zorec}, \&
  {G{\'o}mez}}]{royer2007}
{Royer}, F., {Zorec}, J., \& {G{\'o}mez}, A.~E. 2007,
  \bibinfo{title}{{Rotational velocities of A-type stars. III. Velocity
  distributions},} \aap, 463, 671, \dodoi{10.1051/0004-6361:20065224}

% type= article
\bibitem[{M. {Rugel} {et~al.}(2018){Rugel}, {Fedele}, \& {Herczeg}}]{rugel2018}
{Rugel}, M., {Fedele}, D., \& {Herczeg}, G. 2018, \bibinfo{title}{{X-shooter
  observations of low-mass stars in the {\ensuremath{\eta}} Chamaeleontis
  association},} \aap, 609, A70, \dodoi{10.1051/0004-6361/201630111}

% type= article
\bibitem[{T. {Ryabchikova} {et~al.}(2022){Ryabchikova}, {Zvyagintsev},
  {Tkachenko}, {Tsymbal}, {Pakhomov}, \& {Semenko}}]{ryabchikova2022}
{Ryabchikova}, T., {Zvyagintsev}, S., {Tkachenko}, A., {et~al.} 2022,
  \bibinfo{title}{{Fundamental parameters and abundance analysis of the
  components in the SB2 system HD 60803},} \mnras, 509, 202,
  \dodoi{10.1093/mnras/stab2891}

% type= article
\bibitem[{N.~N. {Samus'} {et~al.}(2017){Samus'}, {Kazarovets}, {Durlevich},
  {Kireeva}, \& {Pastukhova}}]{samus2017}
{Samus'}, N.~N., {Kazarovets}, E.~V., {Durlevich}, O.~V., {Kireeva}, N.~N., \&
  {Pastukhova}, E.~N. 2017, \bibinfo{title}{{General catalogue of variable
  stars: Version GCVS 5.1},} Astronomy Reports, 61, 80,
  \dodoi{10.1134/S1063772917010085}

% type= article
\bibitem[{A.~R.~G. {Santos} {et~al.}(2025){Santos}, {Godoy-Rivera}, {Mathur},
  {Breton}, {Garc{\'\i}a}, \& {Cunha}}]{santos2025}
{Santos}, A.~R.~G., {Godoy-Rivera}, D., {Mathur}, S., {et~al.} 2025,
  \bibinfo{title}{{Signature of spin-down stalling in stellar magnetic
  activity: The case of the open cluster NGC 6811},} \aap, 697, A177,
  \dodoi{10.1051/0004-6361/202554030}

% type= article
\bibitem[{D.~R.~G. {Schleicher} {et~al.}(2023){Schleicher}, {Hidalgo}, \&
  {Galli}}]{schleicher2023}
{Schleicher}, D. R.~G., {Hidalgo}, J.~P., \& {Galli}, D. 2023,
  \bibinfo{title}{{Survival of fossil fields during the pre-main sequence
  evolution of intermediate-mass stars},} \aap, 678, A204,
  \dodoi{10.1051/0004-6361/202346809}

% type= article
\bibitem[{J.~H.~M.~M. {Schmitt} {et~al.}(2022){Schmitt}, {Czesla}, {Freund},
  {Robrade}, \& {Schneider}}]{schmitt2022}
{Schmitt}, J.~H.~M.~M., {Czesla}, S., {Freund}, S., {Robrade}, J., \&
  {Schneider}, P.~C. 2022, \bibinfo{title}{{X-raying the Sco-Cen OB
  association: The low-mass stellar population revealed by eROSITA},} \aap,
  661, A40, \dodoi{10.1051/0004-6361/202141132}

% type= article
\bibitem[{P. {Sch{\"o}fer} {et~al.}(2019){Sch{\"o}fer}, {Jeffers}, {Reiners},
  {Shulyak}, {Fuhrmeister}, {Johnson}, {Zechmeister}, {Ribas}, {Quirrenbach},
  {Amado}, {Caballero}, {Anglada-Escud{\'e}}, {Bauer}, {B{\'e}jar},
  {Cort{\'e}s-Contreras}, {Dreizler}, {Guenther}, {Kaminski}, {K{\"u}rster},
  {Lafarga}, {Montes}, {Morales}, {Pedraz}, \& {Tal-Or}}]{schofer2019}
{Sch{\"o}fer}, P., {Jeffers}, S.~V., {Reiners}, A., {et~al.} 2019,
  \bibinfo{title}{{The CARMENES search for exoplanets around M dwarfs. Activity
  indicators at visible and near-infrared wavelengths},} \aap, 623, A44,
  \dodoi{10.1051/0004-6361/201834114}

% type= article
\bibitem[{C. {Schr{\"o}der} {et~al.}(2009{\natexlab{a}}){Schr{\"o}der},
  {Reiners}, \& {Schmitt}}]{schroeder2009}
{Schr{\"o}der}, C., {Reiners}, A., \& {Schmitt}, J.~H.~M.~M.
  2009{\natexlab{a}}, \bibinfo{title}{{Ca II HK emission in rapidly rotating
  stars. Evidence for an onset of the solar-type dynamo},} \aap, 493, 1099,
  \dodoi{10.1051/0004-6361:200810377}

% type= article
\bibitem[{C. {Schr{\"o}der} {et~al.}(2009{\natexlab{b}}){Schr{\"o}der},
  {Reiners}, \& {Schmitt}}]{schroder2009}
{Schr{\"o}der}, C., {Reiners}, A., \& {Schmitt}, J.~H.~M.~M.
  2009{\natexlab{b}}, \bibinfo{title}{{Ca II HK emission in rapidly rotating
  stars. Evidence for an onset of the solar-type dynamo},} \aap, 493, 1099,
  \dodoi{10.1051/0004-6361:200810377}

% type= article
\bibitem[{S. {Sciortino} {et~al.}(1998){Sciortino}, {Damiani}, {Favata}, \&
  {Micela}}]{sciortino1998}
{Sciortino}, S., {Damiani}, F., {Favata}, F., \& {Micela}, G. 1998,
  \bibinfo{title}{{An X-ray study of the PMS population of the Upper Sco-Cen
  association},} \aap, 332, 825

% type= article
\bibitem[{V. {See} {et~al.}(2016){See}, {Jardine}, {Vidotto}, {Donati}, {Boro
  Saikia}, {Bouvier}, {Fares}, {Folsom}, {Gregory}, {Hussain}, {Jeffers},
  {Marsden}, {Morin}, {Moutou}, {do Nascimento}, {Petit}, \& {Waite}}]{see2016}
{See}, V., {Jardine}, M., {Vidotto}, A.~A., {et~al.} 2016, \bibinfo{title}{{The
  connection between stellar activity cycles and magnetic field topology},}
  \mnras, 462, 4442, \dodoi{10.1093/mnras/stw2010}

% type= article
\bibitem[{E. {Semenova} {et~al.}(2020){Semenova}, {Bergemann}, {Deal},
  {Serenelli}, {Hansen}, {Gallagher}, {Bayo}, {Bensby}, {Bragaglia}, {Carraro},
  {Morbidelli}, {Pancino}, \& {Smiljanic}}]{semenova2020}
{Semenova}, E., {Bergemann}, M., {Deal}, M., {et~al.} 2020,
  \bibinfo{title}{{The Gaia-ESO survey: 3D NLTE abundances in the open cluster
  NGC 2420 suggest atomic diffusion and turbulent mixing are at the origin of
  chemical abundance variations},} \aap, 643, A164,
  \dodoi{10.1051/0004-6361/202038833}

% type= article
\bibitem[{A.~G. {Sepulveda} {et~al.}(2019){Sepulveda}, {Matr{\`a}}, {Kennedy},
  {del Burgo}, {{\"O}berg}, {Wilner}, {Marino}, {Booth}, {Carpenter}, {Davies},
  {Dent}, {Ertel}, {Lestrade}, {Marshall}, {Milli}, {Wyatt}, {MacGregor}, \&
  {Matthews}}]{sepulveda2019}
{Sepulveda}, A.~G., {Matr{\`a}}, L., {Kennedy}, G.~M., {et~al.} 2019,
  \bibinfo{title}{{The REASONS Survey: Resolved Millimeter Observations of a
  Large Debris Disk around the Nearby F Star HD 170773},} \apj, 881, 84,
  \dodoi{10.3847/1538-4357/ab2b98}

% type= article
\bibitem[{P. {Sestito} \& S. {Randich}(2005){Sestito} \&
  {Randich}}]{sestito2005}
{Sestito}, P., \& {Randich}, S. 2005, \bibinfo{title}{{Time scales of Li
  evolution: a homogeneous analysis of open clusters from ZAMS to late-MS},}
  \aap, 442, 615, \dodoi{10.1051/0004-6361:20053482}

% type= article
\bibitem[{D.~M. {Sfeir} {et~al.}(1999){Sfeir}, {Lallement}, {Crifo}, \&
  {Welsh}}]{sfeir1999}
{Sfeir}, D.~M., {Lallement}, R., {Crifo}, F., \& {Welsh}, B.~Y. 1999,
  \bibinfo{title}{{Mapping the contours of the Local bubble: preliminary
  results},} \aap, 346, 785

% type= article
\bibitem[{Y. {Shan} {et~al.}(2024){Shan}, {Revilla}, {Skrzypinski}, {Dreizler},
  {B{\'e}jar}, {Caballero}, {Cardona Guill{\'e}n}, {Cifuentes}, {Fuhrmeister},
  {Reiners}, {Vanaverbeke}, {Ribas}, {Quirrenbach}, {Amado}, {Aceituno},
  {Casanova}, {Cort{\'e}s-Contreras}, {Dubois}, {Gorrini}, {Henning},
  {Herrero}, {Jeffers}, {Kemmer}, {Lalitha}, {Lodieu}, {Logie}, {L{\'o}pez
  Gonz{\'a}lez}, {Mart{\'\i}n-Ruiz}, {Montes}, {Morales}, {Nagel}, {Pall{\'e}},
  {Perdelwitz}, {P{\'e}rez-Torres}, {Pollacco}, {Rau},
  {Rodr{\'\i}guez-L{\'o}pez}, {Rodr{\'\i}guez}, {Sch{\"o}fer}, {Seifert},
  {Sota}, {Zapatero Osorio}, \& {Zechmeister}}]{shan2024}
{Shan}, Y., {Revilla}, D., {Skrzypinski}, S.~L., {et~al.} 2024,
  \bibinfo{title}{{CARMENES input catalog of M dwarfs. VII. New rotation
  periods for the survey stars and their correlations with stellar activity},}
  \aap, 684, A9, \dodoi{10.1051/0004-6361/202346794}

% type= article
\bibitem[{I. {Sheret} {et~al.}(2004){Sheret}, {Dent}, \& {Wyatt}}]{sheret2004}
{Sheret}, I., {Dent}, W.~R.~F., \& {Wyatt}, M.~C. 2004,
  \bibinfo{title}{{Submillimetre observations and modelling of Vega-type
  stars},} \mnras, 348, 1282, \dodoi{10.1111/j.1365-2966.2004.07448.x}

% type= article
\bibitem[{W.~H. {Sherry} {et~al.}(2008){Sherry}, {Walter}, {Wolk}, \&
  {Adams}}]{sherry2008}
{Sherry}, W.~H., {Walter}, F.~M., {Wolk}, S.~J., \& {Adams}, N.~R. 2008,
  \bibinfo{title}{{Main-Sequence Fitting Distance to the {\ensuremath{\sigma}}
  Ori Cluster},} \aj, 135, 1616, \dodoi{10.1088/0004-6256/135/4/1616}

% type= article
\bibitem[{B. {Sibthorpe} {et~al.}(2018){Sibthorpe}, {Kennedy}, {Wyatt},
  {Lestrade}, {Greaves}, {Matthews}, \& {Duch{\^e}ne}}]{sibthorpe2018}
{Sibthorpe}, B., {Kennedy}, G.~M., {Wyatt}, M.~C., {et~al.} 2018,
  \bibinfo{title}{{Analysis of the Herschel DEBRIS Sun-like star sample},}
  \mnras, 475, 3046, \dodoi{10.1093/mnras/stx3188}

% type= article
\bibitem[{N. {Siegler} {et~al.}(2007){Siegler}, {Muzerolle}, {Young}, {Rieke},
  {Mamajek}, {Trilling}, {Gorlova}, \& {Su}}]{siegler2007}
{Siegler}, N., {Muzerolle}, J., {Young}, E.~T., {et~al.} 2007,
  \bibinfo{title}{{Spitzer 24 {$\mu$}m Observations of Open Cluster IC 2391 and
  Debris Disk Evolution of FGK Stars},} \apj, 654, 580, \dodoi{10.1086/509042}

% type= article
\bibitem[{J.~M. {Sierchio} {et~al.}(2010){Sierchio}, {Rieke}, {Su}, {Plavchan},
  {Stauffer}, \& {Gorlova}}]{sierchio2010}
{Sierchio}, J.~M., {Rieke}, G.~H., {Su}, K.~Y.~L., {et~al.} 2010,
  \bibinfo{title}{{Debris Disks around Solar-type Stars: Observations of the
  Pleiades with the Spitzer Space Telescope},} \apj, 712, 1421,
  \dodoi{10.1088/0004-637X/712/2/1421}

% type= article
\bibitem[{J. {Silaj} \& J.~D. {Landstreet}(2014){Silaj} \&
  {Landstreet}}]{silaj2014}
{Silaj}, J., \& {Landstreet}, J.~D. 2014, \bibinfo{title}{{Accurate age
  determinations of several nearby open clusters containing magnetic Ap
  stars},} \aap, 566, A132, \dodoi{10.1051/0004-6361/201321468}

% type= article
\bibitem[{M.~D. {Silverstone} {et~al.}(2006){Silverstone}, {Meyer}, {Mamajek},
  {Hines}, {Hillenbrand}, {Najita}, {Pascucci}, {Bouwman}, {Kim}, {Carpenter},
  {Stauffer}, {Backman}, {Moro-Martin}, {Henning}, {Wolf}, {Brooke}, \&
  {Padgett}}]{silverstone2006}
{Silverstone}, M.~D., {Meyer}, M.~R., {Mamajek}, E.~E., {et~al.} 2006,
  \bibinfo{title}{{Formation and Evolution of Planetary Systems (FEPS):
  Primordial Warm Dust Evolution from 3 to 30 Myr around Sun-like Stars},}
  \apj, 639, 1138, \dodoi{10.1086/499418}

% type= article
\bibitem[{M.~F. {Skrutskie} {et~al.}(2006){Skrutskie}, {Cutri}, {Stiening},
  {Weinberg}, {Schneider}, {Carpenter}, {Beichman}, {Capps}, {Chester},
  {Elias}, {Huchra}, {Liebert}, {Lonsdale}, {Monet}, {Price}, {Seitzer},
  {Jarrett}, {Kirkpatrick}, {Gizis}, {Howard}, {Evans}, {Fowler}, {Fullmer},
  {Hurt}, {Light}, {Kopan}, {Marsh}, {McCallon}, {Tam}, {Van Dyk}, \&
  {Wheelock}}]{skrutskie2006}
{Skrutskie}, M.~F., {Cutri}, R.~M., {Stiening}, R., {et~al.} 2006,
  \bibinfo{title}{{The Two Micron All Sky Survey (2MASS)},} \aj, 131, 1163,
  \dodoi{10.1086/498708}

% type= article
\bibitem[{B.~J. {Smith} {et~al.}(2004){Smith}, {Price}, \& {Baker}}]{smith2004}
{Smith}, B.~J., {Price}, S.~D., \& {Baker}, R.~I. 2004, \bibinfo{title}{{The
  COBE DIRBE Point Source Catalog},} \apjs, 154, 673, \dodoi{10.1086/423248}

% type= article
\bibitem[{D.~R. {Soderblom} {et~al.}(1993{\natexlab{a}}){Soderblom}, {Jones},
  {Balachandran}, {Stauffer}, {Duncan}, {Fedele}, \& {Hudon}}]{soderblom1993c}
{Soderblom}, D.~R., {Jones}, B.~F., {Balachandran}, S., {et~al.}
  1993{\natexlab{a}}, \bibinfo{title}{{The Evolution of the Lithium Abundances
  of Solar-Type Stars. III. The Pleiades},} \aj, 106, 1059,
  \dodoi{10.1086/116704}

% type= article
\bibitem[{D.~R. {Soderblom} \& M. {Mayor}(1993){Soderblom} \&
  {Mayor}}]{soderblom1993a}
{Soderblom}, D.~R., \& {Mayor}, M. 1993, \bibinfo{title}{{Stellar Kinematic
  Groups. I. The URSA Major Group},} \aj, 105, 226, \dodoi{10.1086/116422}

% type= article
\bibitem[{D.~R. {Soderblom} {et~al.}(1993{\natexlab{b}}){Soderblom},
  {Stauffer}, {Hudon}, \& {Jones}}]{soderblom1993b}
{Soderblom}, D.~R., {Stauffer}, J.~R., {Hudon}, J.~D., \& {Jones}, B.~F.
  1993{\natexlab{b}}, \bibinfo{title}{{Rotation and Chromospheric Emission
  among F, G, and K Dwarfs of the Pleiades},} \apjs, 85, 315,
  \dodoi{10.1086/191767}

% type= article
\bibitem[{A.~P. {Sousa} {et~al.}(2023){Sousa}, {Bouvier}, {Alencar}, {Donati},
  {Dougados}, {Alecian}, {Carmona}, {Rebull}, {Cook}, {Artigau}, {Fouqu{\'e}},
  \& {Doyon}}]{sousa2023}
{Sousa}, A.~P., {Bouvier}, J., {Alencar}, S.~H.~P., {et~al.} 2023,
  \bibinfo{title}{{New insights on the near-infrared veiling of young stars
  using CFHT/SPIRou data},} \aap, 670, A142,
  \dodoi{10.1051/0004-6361/202244720}

% type= article
\bibitem[{D. {Souto} {et~al.}(2021){Souto}, {Cunha}, \& {Smith}}]{souto2021}
{Souto}, D., {Cunha}, K., \& {Smith}, V.~V. 2021, \bibinfo{title}{{A
  Metallicity Study of F, G, K, and M Dwarfs in the Coma Berenices Open Cluster
  from the APOGEE Survey},} \apj, 917, 11, \dodoi{10.3847/1538-4357/abfdb5}

% type= article
\bibitem[{P.~V. {Souza dos Santos} {et~al.}(2024){Souza dos Santos}, {Porto de
  Mello}, {Costa-Bhering}, {Lorenzo-Oliveira}, {Almeida-Fernandes},
  {Dutra-Ferreira}, \& {Ribas}}]{souzadossantos2024}
{Souza dos Santos}, P.~V., {Porto de Mello}, G.~F., {Costa-Bhering}, E.,
  {et~al.} 2024, \bibinfo{title}{{Fine structure of the age-chromospheric
  activity relation in solar-type stars: II. H{\ensuremath{\alpha}} line},}
  \mnras, 532, 563, \dodoi{10.1093/mnras/stae1532}

% type= article
\bibitem[{K.~G. {Stassun} \& G. {Torres}(2016){Stassun} \&
  {Torres}}]{stassun2016}
{Stassun}, K.~G., \& {Torres}, G. 2016, \bibinfo{title}{{Evidence for a
  Systematic Offset of -0.25 mas in the Gaia DR1 Parallaxes},} \apjl, 831, L6,
  \dodoi{10.3847/2041-8205/831/1/L6}

% type= article
\bibitem[{K.~G. {Stassun} \& G. {Torres}(2018){Stassun} \&
  {Torres}}]{stassun2018}
{Stassun}, K.~G., \& {Torres}, G. 2018, \bibinfo{title}{{Evidence for a
  Systematic Offset of -80 {\ensuremath{\mu}}as in the Gaia DR2 Parallaxes},}
  \apj, 862, 61, \dodoi{10.3847/1538-4357/aacafc}

% type= article
\bibitem[{K.~G. {Stassun} {et~al.}(2019){Stassun}, {Oelkers}, {Paegert},
  {Torres}, {Pepper}, {De Lee}, {Collins}, {Latham}, {Muirhead}, {Chittidi},
  {Rojas-Ayala}, {Fleming}, {Rose}, {Tenenbaum}, {Ting}, {Kane}, {Barclay},
  {Bean}, {Brassuer}, {Charbonneau}, {Ge}, {Lissauer}, {Mann}, {McLean},
  {Mullally}, {Narita}, {Plavchan}, {Ricker}, {Sasselov}, {Seager}, {Sharma},
  {Shiao}, {Sozzetti}, {Stello}, {Vanderspek}, {Wallace}, \&
  {Winn}}]{stassun2019}
{Stassun}, K.~G., {Oelkers}, R.~J., {Paegert}, M., {et~al.} 2019,
  \bibinfo{title}{{The Revised TESS Input Catalog and Candidate Target List},}
  \aj, 158, 138, \dodoi{10.3847/1538-3881/ab3467}

% type= article
\bibitem[{J.~R. {Stauffer} {et~al.}(1994){Stauffer}, {Caillault}, {Gagne},
  {Prosser}, \& {Hartmann}}]{stauffer1994}
{Stauffer}, J.~R., {Caillault}, J.~P., {Gagne}, M., {Prosser}, C.~F., \&
  {Hartmann}, L.~W. 1994, \bibinfo{title}{{A Deep Imaging Survey of the
  Pleiades with ROSAT},} \apjs, 91, 625, \dodoi{10.1086/191951}

% type= article
\bibitem[{J.~R. {Stauffer} \& L.~W. {Hartmann}(1987){Stauffer} \&
  {Hartmann}}]{stauffer1987}
{Stauffer}, J.~R., \& {Hartmann}, L.~W. 1987, \bibinfo{title}{{The Distribution
  of Rotational Velocities for Low-Mass Stars in the Pleiades},} \apj, 318,
  337, \dodoi{10.1086/165371}

% type= article
\bibitem[{J.~R. {Stauffer} {et~al.}(1997){Stauffer}, {Hartmann}, {Prosser},
  {Randich}, {Balachandran}, {Patten}, {Simon}, \& {Giampapa}}]{stauffer1997}
{Stauffer}, J.~R., {Hartmann}, L.~W., {Prosser}, C.~F., {et~al.} 1997,
  \bibinfo{title}{{Rotational Velocities and Chromospheric/Coronal Activity of
  Low-Mass Stars in the Young Open Clusters IC 2391 and IC 2602},} \apj, 479,
  776, \dodoi{10.1086/303930}

% type= article
\bibitem[{J.~R. {Stauffer} {et~al.}(1998){Stauffer}, {Schultz}, \&
  {Kirkpatrick}}]{stauffer1998}
{Stauffer}, J.~R., {Schultz}, G., \& {Kirkpatrick}, J.~D. 1998,
  \bibinfo{title}{{Keck Spectra of Pleiades Brown Dwarf Candidates and a
  Precise Determination of the Lithium Depletion Edge in the Pleiades},} \apjl,
  499, L199, \dodoi{10.1086/311379}

% type= article
\bibitem[{J.~R. {Stauffer} {et~al.}(2005){Stauffer}, {Rebull}, {Carpenter},
  {Hillenbrand}, {Backman}, {Meyer}, {Kim}, {Silverstone}, {Young}, {Hines},
  {Soderblom}, {Mamajek}, {Morris}, {Bouwman}, \& {Strom}}]{stauffer2005}
{Stauffer}, J.~R., {Rebull}, L.~M., {Carpenter}, J., {et~al.} 2005,
  \bibinfo{title}{{Spitzer Space Telescope Observations of G Dwarfs in the
  Pleiades: Circumstellar Debris Disks at 100 Myr Age},} \aj, 130, 1834,
  \dodoi{10.1086/444420}

% type= article
\bibitem[{J.~R. {Stauffer} {et~al.}(2007){Stauffer}, {Hartmann}, {Fazio},
  {Allen}, {Patten}, {Lowrance}, {Hurt}, {Rebull}, {Cutri}, {Ramirez}, {Young},
  {Rieke}, {Gorlova}, {Muzerolle}, {Slesnick}, \& {Skrutskie}}]{stauffer2007}
{Stauffer}, J.~R., {Hartmann}, L.~W., {Fazio}, G.~G., {et~al.} 2007,
  \bibinfo{title}{{Near- and Mid-Infrared Photometry of the Pleiades and a New
  List of Substellar Candidate Members},} \apjs, 172, 663,
  \dodoi{10.1086/518961}

% type= article
\bibitem[{J.~R. {Stauffer} {et~al.}(2010){Stauffer}, {Rebull}, {James},
  {Noriega-Crespo}, {Strom}, {Wolk}, {Carpenter}, {Barrado y Navascues},
  {Micela}, {Backman}, \& {Cargile}}]{stauffer2010}
{Stauffer}, J.~R., {Rebull}, L.~M., {James}, D., {et~al.} 2010,
  \bibinfo{title}{{Debris Disks of Members of the Blanco 1 Open Cluster},}
  \apj, 719, 1859, \dodoi{10.1088/0004-637X/719/2/1859}

% type= article
\bibitem[{A. {Steele} {et~al.}(2016){Steele}, {Hughes}, {Carpenter}, {Ricarte},
  {Andrews}, {Wilner}, \& {Chiang}}]{steele2016}
{Steele}, A., {Hughes}, A.~M., {Carpenter}, J., {et~al.} 2016,
  \bibinfo{title}{{Resolved Millimeter-wavelength Observations of Debris Disks
  around Solar-type Stars},} \apj, 816, 27, \dodoi{10.3847/0004-637X/816/1/27}

% type= article
\bibitem[{M. {Steinmetz} {et~al.}(2020){Steinmetz}, {Matijevic}, {Enke},
  {Zwitter}, {Guiglion}, {McMillan}, {Kordopatis}, {Valentini}, {Chiappini},
  {Casagrande}, {Wojno}, {Anguiano}, {Bienaym{\'e}}, {Bijaoui}, {Binney},
  {Burton}, {Cass}, {de Laverny}, {Fiegert}, {Freeman}, {Fulbright}, {Gibson},
  {Gilmore}, {Grebel}, {Helmi}, {Kunder}, {Munari}, {Navarro}, {Parker},
  {Ruchti}, {Recio-Blanco}, {Reid}, {Seabroke}, {Siviero}, {Siebert}, {Stupar},
  {Watson}, {Williams}, {Wyse}, {Anders}, {Antoja}, {Birko}, {Bland-Hawthorn},
  {Bossini}, {Garc{\'\i}a}, {Carrillo}, {Chaplin}, {Elsworth}, {Famaey},
  {Gerhard}, {Jofre}, {Just}, {Mathur}, {Miglio}, {Minchev}, {Monari},
  {Mosser}, {Ritter}, {Rodrigues}, {Scholz}, {Sharma}, {Sysoliatina}, \& {RAVE
  Collaboration}}]{steinmetz2020}
{Steinmetz}, M., {Matijevic}, G., {Enke}, H., {et~al.} 2020,
  \bibinfo{title}{{The Sixth Data Release of the Radial Velocity Experiment
  (RAVE). I. Survey Description, Spectra, and Radial Velocities},} \aj, 160,
  82, \dodoi{10.3847/1538-3881/ab9ab9}

% type= article
\bibitem[{K. {Strassmeier} {et~al.}(2000){Strassmeier}, {Washuettl}, {Granzer},
  {Scheck}, \& {Weber}}]{strass2000}
{Strassmeier}, K., {Washuettl}, A., {Granzer}, T., {Scheck}, M., \& {Weber}, M.
  2000, \bibinfo{title}{{The Vienna-KPNO search for Doppler-imaging candidate
  stars. I. A catalog of stellar-activity indicators for 1058 late-type
  Hipparcos stars},} \aaps, 142, 275, \dodoi{10.1051/aas:2000328}

% type= article
\bibitem[{K.~Y.~L. {Su} {et~al.}(2006){Su}, {Rieke}, {Stansberry}, {Bryden},
  {Stapelfeldt}, {Trilling}, {Muzerolle}, {Beichman}, {Moro-Martin}, {Hines},
  \& {Werner}}]{su2006}
{Su}, K.~Y.~L., {Rieke}, G.~H., {Stansberry}, J.~A., {et~al.} 2006,
  \bibinfo{title}{{Debris Disk Evolution around A Stars},} \apj, 653, 675,
  \dodoi{10.1086/508649}

% type= article
\bibitem[{D. {Sullivan} {et~al.}(2022){Sullivan}, {Wilner}, {Matr{\`a}},
  {Wyatt}, {Andrews}, {MacGregor}, \& {Matthews}}]{sullivan2022}
{Sullivan}, D., {Wilner}, D.~J., {Matr{\`a}}, L., {et~al.} 2022,
  \bibinfo{title}{{An ALMA 1.3 millimeter Search for Debris Disks around
  Solar-type Stars in the Pleiades},} \aj, 164, 100,
  \dodoi{10.3847/1538-3881/ac80c5}

% type= article
\bibitem[{H. {Sung} {et~al.}(2002){Sung}, {Bessell}, {Lee}, \&
  {Lee}}]{sung2002}
{Sung}, H., {Bessell}, M.~S., {Lee}, B.-W., \& {Lee}, S.-G. 2002,
  \bibinfo{title}{{The Open Cluster NGC 2516. I. Optical Photometry},} \aj,
  123, 290, \dodoi{10.1086/324729}

% type= article
\bibitem[{N. {Susemiehl} \& M.~R. {Meyer}(2022){Susemiehl} \&
  {Meyer}}]{susemiehl2022}
{Susemiehl}, N., \& {Meyer}, M.~R. 2022, \bibinfo{title}{{Constraints on the
  orbital separation distribution and binary fraction of M dwarfs},} \aap, 657,
  A48, \dodoi{10.1051/0004-6361/202038582}

% type= article
\bibitem[{S. {Talon} \& C. {Charbonnel}(2005){Talon} \&
  {Charbonnel}}]{talon2005}
{Talon}, S., \& {Charbonnel}, C. 2005, \bibinfo{title}{{Hydrodynamical stellar
  models including rotation, internal gravity waves, and atomic diffusion. I.
  Formalism and tests on Pop I dwarfs},} \aap, 440, 981,
  \dodoi{10.1051/0004-6361:20053020}

% type= article
\bibitem[{J. {Tang} {et~al.}(2014){Tang}, {Bressan}, {Rosenfield}, {Slemer},
  {Marigo}, {Girardi}, \& {Bianchi}}]{jtang2014}
{Tang}, J., {Bressan}, A., {Rosenfield}, P., {et~al.} 2014,
  \bibinfo{title}{{New PARSEC evolutionary tracks of massive stars at low
  metallicity: testing canonical stellar evolution in nearby star-forming dwarf
  galaxies},} \mnras, 445, 4287, \dodoi{10.1093/mnras/stu2029}

% type= article
\bibitem[{S.-Y. {Tang} {et~al.}(2018){Tang}, {Chen}, {Chiang}, {Jose},
  {Herczeg}, \& {Goldman}}]{tang2018}
{Tang}, S.-Y., {Chen}, W.~P., {Chiang}, P.~S., {et~al.} 2018,
  \bibinfo{title}{{Characterization of Stellar and Substellar Members in the
  Coma Berenices Star Cluster},} \apj, 862, 106,
  \dodoi{10.3847/1538-4357/aacb7a}

% type= article
\bibitem[{S.-Y. {Tang} {et~al.}(2019){Tang}, {Pang}, {Yuan}, {Chen}, {Hong},
  {Goldman}, {Just}, {Shukirgaliyev}, \& {Lin}}]{tang2019}
{Tang}, S.-Y., {Pang}, X., {Yuan}, Z., {et~al.} 2019,
  \bibinfo{title}{{Discovery of Tidal Tails in Disrupting Open Clusters: Coma
  Berenices and a Neighbor Stellar Group},} \apj, 877, 12,
  \dodoi{10.3847/1538-4357/ab13b0}

% type= article
\bibitem[{A. {Tanner} {et~al.}(2020){Tanner}, {Plavchan}, {Bryden}, {Kennedy},
  {Matr{\'a}}, {Cronin-Coltsmann}, {Lowrance}, {Henry}, {Riaz}, {Gizis},
  {Riedel}, \& {Choquet}}]{tanner2020}
{Tanner}, A., {Plavchan}, P., {Bryden}, G., {et~al.} 2020,
  \bibinfo{title}{{Herschel Observations of Disks around Late-type Stars},}
  \pasp, 132, 084401, \dodoi{10.1088/1538-3873/ab895f}

% type= article
\bibitem[{B.~J. {Taylor}(2006){Taylor}}]{taylor2006}
{Taylor}, B.~J. 2006, \bibinfo{title}{{The Benchmark Cluster Reddening Project.
  I. Reddening Values for the Hyades, Coma, and Praesepe},} \aj, 132, 2453,
  \dodoi{10.1086/508610}

% type= article
\bibitem[{P. {Testa} {et~al.}(2015){Testa}, {Saar}, \& {Drake}}]{testa2015}
{Testa}, P., {Saar}, S.~H., \& {Drake}, J.~J. 2015, \bibinfo{title}{{Stellar
  activity and coronal heating: an overview of recent results},} Philosophical
  Transactions of the Royal Society of London Series A, 373, 20140259,
  \dodoi{10.1098/rsta.2014.0259}

% type= article
\bibitem[{N.~D. {Thureau} {et~al.}(2014){Thureau}, {Greaves}, {Matthews},
  {Kennedy}, {Phillips}, {Booth}, {Duch{\^e}ne}, {Horner}, {Rodriguez},
  {Sibthorpe}, \& {Wyatt}}]{thureau2014}
{Thureau}, N.~D., {Greaves}, J.~S., {Matthews}, B.~C., {et~al.} 2014,
  \bibinfo{title}{{An unbiased study of debris discs around A-type stars with
  Herschel},} \mnras, 445, 2558, \dodoi{10.1093/mnras/stu1864}

% type= article
\bibitem[{T.~L. {Tobin} {et~al.}(2024){Tobin}, {Currie}, {Li}, {Chilcote},
  {Brandt}, {Lacy}, {Kuzuhara}, {Vincent}, {El Morsy}, {Deo}, {Williams},
  {Guyon}, {Lozi}, {Vievard}, {Skaf}, {Ahn}, {Groff}, {Kasdin}, {Uyama},
  {Tamura}, {Gibbs}, {Lewis}, {Bowens-Rubin}, {Salama}, {An}, \&
  {Chen}}]{tobin2024}
{Tobin}, T.~L., {Currie}, T., {Li}, Y., {et~al.} 2024,
  \bibinfo{title}{{Direct-imaging Discovery of a Substellar Companion Orbiting
  the Accelerating Variable Star HIP 39017},} \aj, 167, 205,
  \dodoi{10.3847/1538-3881/ad3077}

% type= article
\bibitem[{A. {Tokovinin}(2018){Tokovinin}}]{tokovinin2018}
{Tokovinin}, A. 2018, \bibinfo{title}{{The Updated Multiple Star Catalog},}
  \apjs, 235, 6, \dodoi{10.3847/1538-4365/aaa1a5}

% type= article
\bibitem[{A. {Tokovinin} {et~al.}(2006){Tokovinin}, {Thomas}, {Sterzik}, \&
  {Udry}}]{tokovinin2006}
{Tokovinin}, A., {Thomas}, S., {Sterzik}, M., \& {Udry}, S. 2006,
  \bibinfo{title}{{Tertiary companions to close spectroscopic binaries},} \aap,
  450, 681, \dodoi{10.1051/0004-6361:20054427}

% type= incollection
\bibitem[{C.~A.~O. {Torres} {et~al.}(2008){Torres}, {Quast}, {Melo}, \&
  {Sterzik}}]{torres2008}
{Torres}, C.~A.~O., {Quast}, G.~R., {Melo}, C.~H.~F., \& {Sterzik}, M.~F. 2008,
  \bibinfo{title}{{Young Nearby Loose Associations},} in Handbook of Star
  Forming Regions, Volume II, ed. B.~{Reipurth}, Vol.~5 (ASP Press, San
  Francisco, CA USA), 757, \dodoi{10.48550/arXiv.0808.3362}

% type= article
\bibitem[{G. {Torres} {et~al.}(2021){Torres}, {Latham}, \&
  {Quinn}}]{torres2021}
{Torres}, G., {Latham}, D.~W., \& {Quinn}, S.~N. 2021,
  \bibinfo{title}{{Long-term Spectroscopic Survey of the Pleiades Cluster: The
  Binary Population},} \apj, 921, 117, \dodoi{10.3847/1538-4357/ac1585}

% type= article
\bibitem[{G. {Torres} \& I. {Ribas}(2002){Torres} \& {Ribas}}]{torres2002}
{Torres}, G., \& {Ribas}, I. 2002, \bibinfo{title}{{Absolute Dimensions of the
  M-Type Eclipsing Binary YY Geminorum (Castor C): A Challenge to Evolutionary
  Models in the Lower Main Sequence},} \apj, 567, 1140, \dodoi{10.1086/338587}

% type= article
\bibitem[{G. {Torres} {et~al.}(2022){Torres}, {Schaefer}, {Monnier}, {Anugu},
  {Davies}, {Ennis}, {Farrington}, {Gardner}, {Klement}, {Kraus}, {Labdon},
  {Lanthermann}, {Le Bouquin}, {Setterholm}, \& {ten Brummelaar}}]{torres2022}
{Torres}, G., {Schaefer}, G.~H., {Monnier}, J.~D., {et~al.} 2022,
  \bibinfo{title}{{The Orbits and Dynamical Masses of the Castor System},}
  \apj, 941, 8, \dodoi{10.3847/1538-4357/ac9d8d}

% type= article
\bibitem[{G. {Torres} {et~al.}(2024){Torres}, {Schaefer}, {Stefanik}, {Latham},
  {Boden}, {Anugu}, {Jones}, {Klement}, {Kraus}, {Lanthermann}, \&
  {Monnier}}]{torres2024}
{Torres}, G., {Schaefer}, G.~H., {Stefanik}, R.~P., {et~al.} 2024,
  \bibinfo{title}{{Orbits and Dynamical Masses for Six Binary Systems in the
  Hyades Cluster},} \apj, 971, 31, \dodoi{10.3847/1538-4357/ad54b2}

% type= article
\bibitem[{D.~E. {Trilling} {et~al.}(2007){Trilling}, {Stansberry},
  {Stapelfeldt}, {Rieke}, {Su}, {Gray}, {Corbally}, {Bryden}, {Chen}, {Boden},
  \& {Beichman}}]{trill2007}
{Trilling}, D.~E., {Stansberry}, J.~A., {Stapelfeldt}, K.~R., {et~al.} 2007,
  \bibinfo{title}{{Debris disks in main-sequence binary systems.},} \apj, 658,
  1289, \dodoi{10.1086/511668}

% type= article
\bibitem[{D.~E. {Trilling} {et~al.}(2008){Trilling}, {Bryden}, {Beichman},
  {Rieke}, {Su}, {Stansberry}, {Blaylock}, {Stapelfeldt}, {Beeman}, \&
  {Haller}}]{trill2008}
{Trilling}, D.~E., {Bryden}, G., {Beichman}, C.~A., {et~al.} 2008,
  \bibinfo{title}{{Debris Disks around Sun-like Stars},} \apj, 674, 1086,
  \dodoi{10.1086/525514}

% type= article
\bibitem[{L. {Tu} {et~al.}(2015){Tu}, {Johnstone}, {G{\"u}del}, \&
  {Lammer}}]{tu2015}
{Tu}, L., {Johnstone}, C.~P., {G{\"u}del}, M., \& {Lammer}, H. 2015,
  \bibinfo{title}{{The extreme ultraviolet and X-ray Sun in Time: High-energy
  evolutionary tracks of a solar-like star},} \aap, 577, L3,
  \dodoi{10.1051/0004-6361/201526146}

% type= article
\bibitem[{C. {Turon} {et~al.}(1993){Turon}, {Creze}, {Egret}, {Gomez},
  {Grenon}, {Jahrei{\ss}}, {Requieme}, {Argue}, {Bec-Borsenberger},
  {Dommanget}, {Mennessier}, {Arenou}, {Chareton}, {Crifo}, {Mermilliod},
  {Morin}, {Nicolet}, {Nys}, {Prevot}, {Rousseau}, {Perryman}, {Arlot},
  {Baglin}, {Barthes}, {Baylac}, {Brosche}, {Burnet}, {Delhaye}, {Dettbarn},
  {Erbach}, {Figueras}, {Fricke}, {Helmer}, {Hemenway}, {Jordi}, {Lampens},
  {Lederle}, {Lub}, {Manfroid}, {Mattci}, {Mazurier}, {Mermilliod}, {Morrison},
  {Murray}, {Oblak}, {Perie}, {Pernier}, {Le Poole}, {Quijano}, {Rapaport},
  {Sellier}, {Torra}, {Tucholke}, {de Vegt}, {Argyle}, {Bacchus}, {Baron},
  {Calaf}, {Cordoni}, {Fabricius}, {Feaugas}, {Fehlberg}, {Florkowski}, {de
  Geus}, {Gibbs}, {Hartmann}, {Jauncey}, {Johnston}, {Marouard}, {Mekkas},
  {Muinos}, {Nunez}, {Ochsenbein}, {de Orus}, {Paredes}, {Penston}, {Petersen},
  {Peyrin}, {Robin}, {Roman}, {Rossello}, {Schwan}, {Sinachopoulos}, {White},
  {Zacharias}, {Hog}, {Kovalevsky}, {van Leeuwen}, {Lindegren}, {Schutz}, \&
  {Schrijver}}]{turon1993}
{Turon}, C., {Creze}, M., {Egret}, D., {et~al.} 1993, \bibinfo{title}{{Version
  2 of the HIPPARCOS Input Catalogue},} Bulletin d'Information du Centre de
  Donnees Stellaires, 43, 5

% type= article
\bibitem[{K. {Ujjwal} {et~al.}(2020){Ujjwal}, {Kartha}, {Mathew}, {Manoj}, \&
  {Narang}}]{ujjwal2020}
{Ujjwal}, K., {Kartha}, S.~S., {Mathew}, B., {Manoj}, P., \& {Narang}, M. 2020,
  \bibinfo{title}{{Analysis of Membership Probability in Nearby Young Moving
  Groups with Gaia DR2},} \aj, 159, 166, \dodoi{10.3847/1538-3881/ab76d6}

% type= article
\bibitem[{L.~E. {Urban} {et~al.}(2012){Urban}, {Rieke}, {Su}, \&
  {Trilling}}]{urban2012}
{Urban}, L.~E., {Rieke}, G., {Su}, K., \& {Trilling}, D.~E. 2012,
  \bibinfo{title}{{The Incidence of Debris Disks at 24 {\ensuremath{\mu}}m and
  670 Myr},} \apj, 750, 98, \dodoi{10.1088/0004-637X/750/2/98}

% type= article
\bibitem[{S.~E. {Urban} {et~al.}(1998){Urban}, {Corbin}, \&
  {Wycoff}}]{urban1998}
{Urban}, S.~E., {Corbin}, T.~E., \& {Wycoff}, G.~L. 1998, \bibinfo{title}{{The
  ACT Reference Catalog},} \aj, 115, 2161, \dodoi{10.1086/300344}

% type= book
\bibitem[{W.~F. {van Altena} {et~al.}(1995){van Altena}, {Lee}, \&
  {Hoffleit}}]{vanaltena1995}
{van Altena}, W.~F., {Lee}, J.~T., \& {Hoffleit}, E.~D. 1995, {The general
  catalogue of trigonometric [stellar] parallaxes} (Yale University
  Observatory, New Haven, CT USA)

% type= article
\bibitem[{G.~T. {van Belle} {et~al.}(2021){van Belle}, {von Braun}, {Ciardi},
  {Pilyavsky}, {Buckingham}, {Boden}, {Clark}, {Hartman}, {van Belle},
  {Bucknew}, \& {Cole}}]{vanbelle2021}
{van Belle}, G.~T., {von Braun}, K., {Ciardi}, D.~R., {et~al.} 2021,
  \bibinfo{title}{{Direct Measurements of Giant Star Effective Temperatures and
  Linear Radii: Calibration against Spectral Types and V - K Color},} \apj,
  922, 163, \dodoi{10.3847/1538-4357/ac1687}

% type= article
\bibitem[{F. {van Leeuwen}(2007){van Leeuwen}}]{vanleeuwen2007}
{van Leeuwen}, F. 2007, \bibinfo{title}{{Validation of the new Hipparcos
  reduction},} \aap, 474, 653, \dodoi{10.1051/0004-6361:20078357}

% type= article
\bibitem[{F. {van Leeuwen} {et~al.}(1997){van Leeuwen}, {Evans}, {Grenon},
  {Grossmann}, {Mignard}, \& {Perryman}}]{vanleeuwen1997}
{van Leeuwen}, F., {Evans}, D.~W., {Grenon}, M., {et~al.} 1997,
  \bibinfo{title}{{The HIPPARCOS mission: photometric data.},} \aap, 323, L61

% type= article
\bibitem[{A.~H. {Vaughan} \& G.~W. {Preston}(1980){Vaughan} \&
  {Preston}}]{vaughan1980}
{Vaughan}, A.~H., \& {Preston}, G.~W. 1980, \bibinfo{title}{{A survey of
  chromospheric CA II H and K emission in field stars of the solar
  neighborhood.},} \pasp, 92, 385, \dodoi{10.1086/130683}

% type= incollection
\bibitem[{D. {Veras}(2021){Veras}}]{veras2021}
{Veras}, D. 2021, \bibinfo{title}{{Planetary Systems Around White Dwarfs},} in
  Oxford Research Encyclopedia of Planetary Science (IOP Publishing Ltd,
  Bristol, UK), 1, \dodoi{10.1093/acrefore/9780190647926.013.238}

% type= article
\bibitem[{L. {Vican} {et~al.}(2016){Vican}, {Schneider}, {Bryden}, {Melis},
  {Zuckerman}, {Rhee}, \& {Song}}]{vican2016}
{Vican}, L., {Schneider}, A., {Bryden}, G., {et~al.} 2016,
  \bibinfo{title}{{Herschel Observations of Dusty Debris Disks},} \apj, 833,
  263, \dodoi{10.3847/1538-4357/833/2/263}

% type= inproceedings
\bibitem[{W. {Voges} {et~al.}(2001){Voges}, {Boller}, {Englhauser}, {Freyberg},
  \& {Supper}}]{voges2001}
{Voges}, W., {Boller}, T., {Englhauser}, J., {Freyberg}, M., \& {Supper}, R.
  2001, \bibinfo{title}{{The ROSAT X-ray Database from All-Sky Survey and
  Pointed Observations},} in Astronomical Society of the Pacific Conference
  Series, Vol. 225, Virtual Observatories of the Future, ed. R.~J. {Brunner},
  S.~G. {Djorgovski}, \& A.~S. {Szalay}, 234

% type= article
\bibitem[{W. {Voges} {et~al.}(1999){Voges}, {Aschenbach}, {Boller},
  {Br{\"a}uninger}, {Briel}, {Burkert}, {Dennerl}, {Englhauser}, {Gruber},
  {Haberl}, {Hartner}, {Hasinger}, {K{\"u}rster}, {Pfeffermann}, {Pietsch},
  {Predehl}, {Rosso}, {Schmitt}, {Tr{\"u}mper}, \& {Zimmermann}}]{voges1999}
{Voges}, W., {Aschenbach}, B., {Boller}, T., {et~al.} 1999,
  \bibinfo{title}{{The ROSAT all-sky survey bright source catalogue},} \aap,
  349, 389, \dodoi{10.48550/arXiv.astro-ph/9909315}

% type= article
\bibitem[{W. {Voges} {et~al.}(2000){Voges}, {Aschenbach}, {Boller},
  {Brauninger}, {Briel}, {Burkert}, {Dennerl}, {Englhauser}, {Gruber},
  {Haberl}, {Hartner}, {Hasinger}, {Pfeffermann}, {Pietsch}, {Predehl},
  {Schmitt}, {Trumper}, \& {Zimmermann}}]{voges2000}
{Voges}, W., {Aschenbach}, B., {Boller}, T., {et~al.} 2000,
  \bibinfo{title}{{Rosat All-Sky Survey Faint Source Catalogue},} \iaucirc,
  7432, 3

% type= article
\bibitem[{F.~M. {Walter} {et~al.}(1994){Walter}, {Vrba}, {Mathieu}, {Brown}, \&
  {Myers}}]{walter1994}
{Walter}, F.~M., {Vrba}, F.~J., {Mathieu}, R.~D., {Brown}, A., \& {Myers},
  P.~C. 1994, \bibinfo{title}{{X-Ray Sources in Regions of Star Formation. V.
  The Low Mass Stars of the Upper Scorpius Associations},} \aj, 107, 692,
  \dodoi{10.1086/116889}

% type= article
\bibitem[{J.~J. {Wang} {et~al.}(1995){Wang}, {Chen}, {Zhao}, \&
  {Jiang}}]{wang1995}
{Wang}, J.~J., {Chen}, L., {Zhao}, J.~H., \& {Jiang}, P.~F. 1995,
  \bibinfo{title}{{High-precision study of proper motions and membership of 924
  stars in the central region of Praesepe.},} \aaps, 113, 419

% type= article
\bibitem[{R. {Wang} {et~al.}(2020){Wang}, {Luo}, {Chen}, {Hou}, {Zhang},
  {Zhao}, {Li}, {Hou}, \& {LAMOST MRS Collaboration}}]{wang2020}
{Wang}, R., {Luo}, A.~L., {Chen}, J.-J., {et~al.} 2020,
  \bibinfo{title}{{SPCANet: Stellar Parameters and Chemical Abundances Network
  for LAMOST-II Medium Resolution Survey},} \apj, 891, 23,
  \dodoi{10.3847/1538-4357/ab6dea}

% type= article
\bibitem[{S. {Wang} \& X. {Chen}(2019){Wang} \& {Chen}}]{wang2019}
{Wang}, S., \& {Chen}, X. 2019, \bibinfo{title}{{The Optical to Mid-infrared
  Extinction Law Based on the APOGEE, Gaia DR2, Pan-STARRS1, SDSS, APASS,
  2MASS, and WISE Surveys},} \apj, 877, 116, \dodoi{10.3847/1538-4357/ab1c61}

% type= article
\bibitem[{S. {Wang} {et~al.}(2016){Wang}, {Liu}, {Qiu}, {Bai}, {Yang}, {Guo},
  \& {Zhang}}]{wang2016}
{Wang}, S., {Liu}, J., {Qiu}, Y., {et~al.} 2016, \bibinfo{title}{{CHANDRA ACIS
  Survey of X-Ray Point Sources: The Source Catalog},} \apjs, 224, 40,
  \dodoi{10.3847/0067-0049/224/2/40}

% type= article
\bibitem[{M.~A. {Weber} {et~al.}(2023){Weber}, {Schunker}, {Jouve}, \&
  {I{\c{s}}{\i}k}}]{weber2023}
{Weber}, M.~A., {Schunker}, H., {Jouve}, L., \& {I{\c{s}}{\i}k}, E. 2023,
  \bibinfo{title}{{Understanding Active Region Origins and Emergence on the Sun
  and Other Cool Stars},} \ssr, 219, 63, \dodoi{10.1007/s11214-023-01006-5}

% type= article
\bibitem[{M. {Weiler} {et~al.}(2023){Weiler}, {Carrasco}, {Fabricius}, \&
  {Jordi}}]{weiler2023}
{Weiler}, M., {Carrasco}, J.~M., {Fabricius}, C., \& {Jordi}, C. 2023,
  \bibinfo{title}{{Analysing spectral lines in Gaia low-resolution spectra},}
  \aap, 671, A52, \dodoi{10.1051/0004-6361/202244764}

% type= article
\bibitem[{E.~W. {Weis}(1993){Weis}}]{weis1993}
{Weis}, E.~W. 1993, \bibinfo{title}{{Photometry of Dwarf K and M Stars},} \aj,
  105, 1962, \dodoi{10.1086/116571}

% type= article
\bibitem[{M.~C. {Weisskopf} {et~al.}(2002){Weisskopf}, {Brinkman}, {Canizares},
  {Garmire}, {Murray}, \& {Van Speybroeck}}]{weisskopf2002}
{Weisskopf}, M.~C., {Brinkman}, B., {Canizares}, C., {et~al.} 2002,
  \bibinfo{title}{{An Overview of the Performance and Scientific Results from
  the Chandra X-Ray Observatory},} \pasp, 114, 1, \dodoi{10.1086/338108}

% type= article
\bibitem[{B.~Y. {Welsh} \& R.~L. {Shelton}(2009){Welsh} \&
  {Shelton}}]{welsh2009}
{Welsh}, B.~Y., \& {Shelton}, R.~L. 2009, \bibinfo{title}{{The trouble with the
  Local Bubble},} \apss, 323, 1, \dodoi{10.1007/s10509-009-0053-3}

% type= article
\bibitem[{J.~A. {White} {et~al.}(2018){White}, {Boley}, {MacGregor}, {Hughes},
  \& {Wilner}}]{white2018}
{White}, J.~A., {Boley}, A.~C., {MacGregor}, M.~A., {Hughes}, A.~M., \&
  {Wilner}, D.~J. 2018, \bibinfo{title}{{ALMA and VLA observations of the HD
  141569 system},} \mnras, 474, 4500, \dodoi{10.1093/mnras/stx3098}

% type= article
\bibitem[{R.~J. {White} {et~al.}(2007){White}, {Gabor}, \&
  {Hillenbrand}}]{white2007}
{White}, R.~J., {Gabor}, J.~M., \& {Hillenbrand}, L.~A. 2007,
  \bibinfo{title}{{High-Dispersion Optical Spectra of Nearby Stars Younger Than
  the Sun},} \aj, 133, 2524, \dodoi{10.1086/514336}

% type= article
\bibitem[{J.~P. {Williams} \& S.~M. {Andrews}(2006){Williams} \&
  {Andrews}}]{williams2006}
{Williams}, J.~P., \& {Andrews}, S.~M. 2006, \bibinfo{title}{{The Dust
  Properties of Eight Debris Disk Candidates as Determined by Submillimeter
  Photometry},} \apj, 653, 1480, \dodoi{10.1086/508919}

% type= article
\bibitem[{J.~P. {Williams} {et~al.}(2004){Williams}, {Najita}, {Liu},
  {Bottinelli}, {Carpenter}, {Hillenbrand}, {Meyer}, \&
  {Soderblom}}]{williams2004}
{Williams}, J.~P., {Najita}, J., {Liu}, M.~C., {et~al.} 2004,
  \bibinfo{title}{{Detection of Cool Dust around the G2 V Star HD 107146},}
  \apj, 604, 414, \dodoi{10.1086/381721}

% type= article
\bibitem[{D.~J. {Wilner} {et~al.}(2018){Wilner}, {MacGregor}, {Andrews},
  {Hughes}, {Matthews}, \& {Su}}]{wilner2018}
{Wilner}, D.~J., {MacGregor}, M.~A., {Andrews}, S.~M., {et~al.} 2018,
  \bibinfo{title}{{Resolved Millimeter Observations of the HR 8799 Debris
  Disk},} \apj, 855, 56, \dodoi{10.3847/1538-4357/aaacd7}

% type= article
\bibitem[{O.~C. {Wilson}(1963){Wilson}}]{wilson1963}
{Wilson}, O.~C. 1963, \bibinfo{title}{{A Probable Correlation Between
  Chromospheric Activity and Age in Main-Sequence Stars.},} \apj, 138, 832,
  \dodoi{10.1086/147689}

% type= article
\bibitem[{I. {Winnick} {et~al.}(2025){Winnick}, {Yana Galarza}, {Reggiani},
  {Ferreira}, {Baraffe}, {Lorenzo-Oliveira}, {Oyague}, {Valle}, {Trujillo
  Diaz}, {Leigh}, {Flores Trivigno}, {Lopez-Valdivia}, {Carvalho Silva},
  {Martioli}, \& {Perottoni}}]{winnick2025}
{Winnick}, I., {Yana Galarza}, J., {Reggiani}, H., {et~al.} 2025,
  \bibinfo{title}{{Extreme Lithium Depletion in Solar Twins: Challenging
  Non-Standard Mixing Models},} arXiv e-prints, arXiv:2508.16513,
  \dodoi{10.48550/arXiv.2508.16513}

% type= article
\bibitem[{M.~L. {Wood} {et~al.}(2023){Wood}, {Mann}, {Barber}, {Bush}, {Kraus},
  {Tofflemire}, {Vanderburg}, {Newton}, {Feiden}, {Zhou}, {Bouma}, {Quinn},
  {Armstrong}, {Osborn}, {Adibekyan}, {Mena}, {Sousa}, {Gagn{\'e}}, {Fields},
  {Milburn}, {Thao}, {Schmidt}, {Gnilka}, {Howell}, {Law}, {Ziegler},
  {Brice{\~n}o}, {Ricker}, {Vanderspek}, {Latham}, {Seager}, {Winn}, {Jenkins},
  {Schlieder}, {Osborn}, {Twicken}, {Ciardi}, \& {Huang}}]{wood2023}
{Wood}, M.~L., {Mann}, A.~W., {Barber}, M.~G., {et~al.} 2023,
  \bibinfo{title}{{TESS Hunt for Young and Maturing Exoplanets (THYME). IX. A
  27 Myr Extended Population of Lower Centaurus Crux with a Transiting
  Two-planet System},} \aj, 165, 85, \dodoi{10.3847/1538-3881/aca8fc}

% type= article
\bibitem[{C.~E. {Worley} \& G.~G. {Douglass}(1997){Worley} \&
  {Douglass}}]{WDS1997}
{Worley}, C.~E., \& {Douglass}, G.~G. 1997, \bibinfo{title}{{The Washington
  Double Star Catalog (WDS, 1996.0)},} \aaps, 125, 523,
  \dodoi{10.1051/aas:1997239}

% type= article
\bibitem[{G. {Worthey} \& H.-c. {Lee}(2011){Worthey} \& {Lee}}]{worthey2011}
{Worthey}, G., \& {Lee}, H.-c. 2011, \bibinfo{title}{{An Empirical UBV RI JHK
  Color-Temperature Calibration for Stars},} \apjs, 193, 1,
  \dodoi{10.1088/0067-0049/193/1/1}

% type= article
\bibitem[{C.~O. {Wright} {et~al.}(2003){Wright}, {Egan}, {Kraemer}, \&
  {Price}}]{wright2003}
{Wright}, C.~O., {Egan}, M.~P., {Kraemer}, K.~E., \& {Price}, S.~D. 2003,
  \bibinfo{title}{{The Tycho-2 Spectral Type Catalog},} \aj, 125, 359,
  \dodoi{10.1086/345511}

% type= article
\bibitem[{E.~L. {Wright} {et~al.}(2010){Wright}, {Eisenhardt}, {Mainzer},
  {Ressler}, {Cutri}, {Jarrett}, {Kirkpatrick}, {Padgett}, {McMillan},
  {Skrutskie}, {Stanford}, {Cohen}, {Walker}, {Mather}, {Leisawitz}, {Gautier},
  {McLean}, {Benford}, {Lonsdale}, {Blain}, {Mendez}, {Irace}, {Duval}, {Liu},
  {Royer}, {Heinrichsen}, {Howard}, {Shannon}, {Kendall}, {Walsh}, {Larsen},
  {Cardon}, {Schick}, {Schwalm}, {Abid}, {Fabinsky}, {Naes}, \&
  {Tsai}}]{wright2010}
{Wright}, E.~L., {Eisenhardt}, P. R.~M., {Mainzer}, A.~K., {et~al.} 2010,
  \bibinfo{title}{{The Wide-field Infrared Survey Explorer (WISE): Mission
  Description and Initial On-orbit Performance},} \aj, 140, 1868,
  \dodoi{10.1088/0004-6256/140/6/1868}

% type= article
\bibitem[{J.~T. {Wright} {et~al.}(2004){Wright}, {Marcy}, {Butler}, \&
  {Vogt}}]{wright2004}
{Wright}, J.~T., {Marcy}, G.~W., {Butler}, R.~P., \& {Vogt}, S.~S. 2004,
  \bibinfo{title}{{Chromospheric Ca II Emission in Nearby F, G, K, and M
  Stars},} \apjs, 152, 261, \dodoi{10.1086/386283}

% type= incollection
\bibitem[{M.~C. {Wyatt}(2021){Wyatt}}]{wyatt2021}
{Wyatt}, M.~C. 2021, \bibinfo{title}{{Debris Disks},} in ExoFrontiers; Big
  Questions in Exoplanetary Science, ed. N.~{Madhusudhan} (IOP Publishing Ltd,
  Bristol, UK), 15--1, \dodoi{10.1088/2514-3433/abfa8fch15}

% type= article
\bibitem[{M.~C. {Wyatt} {et~al.}(2007){Wyatt}, {Smith}, {Su}, {Rieke},
  {Greaves}, {Beichman}, \& {Bryden}}]{wyatt2007}
{Wyatt}, M.~C., {Smith}, R., {Su}, K.~Y.~L., {et~al.} 2007,
  \bibinfo{title}{{Steady State Evolution of Debris Disks around A Stars},}
  \apj, 663, 365, \dodoi{10.1086/518404}

% type= article
\bibitem[{M. {Xiang} {et~al.}(2019){Xiang}, {Ting}, {Rix}, {Sandford}, {Buder},
  {Lind}, {Liu}, {Shi}, \& {Zhang}}]{xiang2019}
{Xiang}, M., {Ting}, Y.-S., {Rix}, H.-W., {et~al.} 2019,
  \bibinfo{title}{{Abundance Estimates for 16 Elements in 6 Million Stars from
  LAMOST DR5 Low-Resolution Spectra},} \apjs, 245, 34,
  \dodoi{10.3847/1538-4365/ab5364}

% type= article
\bibitem[{B. {Yanny} {et~al.}(2009){Yanny}, {Rockosi}, {Newberg}, {Knapp},
  {Adelman-McCarthy}, {Alcorn}, {Allam}, {Allende Prieto}, {An}, {Anderson},
  {Anderson}, {Bailer-Jones}, {Bastian}, {Beers}, {Bell}, {Belokurov},
  {Bizyaev}, {Blythe}, {Bochanski}, {Boroski}, {Brinchmann}, {Brinkmann},
  {Brewington}, {Carey}, {Cudworth}, {Evans}, {Evans}, {Gates}, {G{\"a}nsicke},
  {Gillespie}, {Gilmore}, {Nebot Gomez-Moran}, {Grebel}, {Greenwell}, {Gunn},
  {Jordan}, {Jordan}, {Harding}, {Harris}, {Hendry}, {Holder}, {Ivans},
  {Ivezic}, {Jester}, {Johnson}, {Kent}, {Kleinman}, {Kniazev}, {Krzesinski},
  {Kron}, {Kuropatkin}, {Lebedeva}, {Lee}, {French Leger}, {L{\'e}pine},
  {Levine}, {Lin}, {Long}, {Loomis}, {Lupton}, {Malanushenko}, {Malanushenko},
  {Margon}, {Martinez-Delgado}, {McGehee}, {Monet}, {Morrison}, {Munn},
  {Neilsen}, {Nitta}, {Norris}, {Oravetz}, {Owen}, {Padmanabhan}, {Pan},
  {Peterson}, {Pier}, {Platson}, {Re Fiorentin}, {Richards}, {Rix}, {Schlegel},
  {Schneider}, {Schreiber}, {Schwope}, {Sibley}, {Simmons}, {Snedden}, {Allyn
  Smith}, {Stark}, {Stauffer}, {Steinmetz}, {Stoughton}, {SubbaRao}, {Szalay},
  {Szkody}, {Thakar}, {Sivarani}, {Tucker}, {Uomoto}, {Vanden Berk}, {Vidrih},
  {Wadadekar}, {Watters}, {Wilhelm}, {Wyse}, {Yarger}, \& {Zucker}}]{yanny2009}
{Yanny}, B., {Rockosi}, C., {Newberg}, H.~J., {et~al.} 2009,
  \bibinfo{title}{{SEGUE: A Spectroscopic Survey of 240,000 Stars with g =
  14-20},} \aj, 137, 4377, \dodoi{10.1088/0004-6256/137/5/4377}

% type= article
\bibitem[{L. {Ye} {et~al.}(2024){Ye}, {Bi}, {Zhang}, {Sun}, {Long}, {Ge}, {Li},
  {Zhang}, {Chen}, {Li}, {Zhou}, \& {Xiang}}]{ye2024}
{Ye}, L., {Bi}, S., {Zhang}, J., {et~al.} 2024, \bibinfo{title}{{Relations of
  Rotation and Chromospheric Activity to Stellar Age for FGK Dwarfs from Kepler
  and LAMOST},} \apjs, 271, 19, \dodoi{10.3847/1538-4365/ad1eee}

% type= article
\bibitem[{B. {Yelverton} {et~al.}(2020){Yelverton}, {Kennedy}, \&
  {Su}}]{yelverton2020}
{Yelverton}, B., {Kennedy}, G.~M., \& {Su}, K. Y.~L. 2020, \bibinfo{title}{{No
  significant correlation between radial velocity planet presence and debris
  disc properties},} \mnras, 495, 1943, \dodoi{10.1093/mnras/staa1316}

% type= article
\bibitem[{B. {Yelverton} {et~al.}(2019){Yelverton}, {Kennedy}, {Su}, \&
  {Wyatt}}]{yelverton2019}
{Yelverton}, B., {Kennedy}, G.~M., {Su}, K. Y.~L., \& {Wyatt}, M.~C. 2019,
  \bibinfo{title}{{A statistically significant lack of debris discs in medium
  separation binary systems},} \mnras, 488, 3588, \dodoi{10.1093/mnras/stz1927}

% type= article
\bibitem[{E.~T. {Young} {et~al.}(2004){Young}, {Lada}, {Teixeira}, {Muzerolle},
  {Muench}, {Stauffer}, {Beichman}, {Rieke}, {Hines}, {Su}, {Engelbracht},
  {Gordon}, {Misselt}, {Morrison}, {Stansberry}, \& {Kelly}}]{young2004}
{Young}, E.~T., {Lada}, C.~J., {Teixeira}, P., {et~al.} 2004,
  \bibinfo{title}{{Spitzer Observations of NGC 2547: The Disk Population at 25
  Million Years},} \apjs, 154, 428, \dodoi{10.1086/422688}

% type= article
\bibitem[{H. {Yu} {et~al.}(2020){Yu}, {Shao}, {Diaferio}, \& {Li}}]{yu2020}
{Yu}, H., {Shao}, Z., {Diaferio}, A., \& {Li}, L. 2020,
  \bibinfo{title}{{Unveiling the Hierarchical Structure of Open Star Clusters:
  The Perseus Double Cluster},} \apj, 899, 144,
  \dodoi{10.3847/1538-4357/aba8f3}

% type= article
\bibitem[{N. {Zacharias} {et~al.}(2013){Zacharias}, {Finch}, {Girard},
  {Henden}, {Bartlett}, {Monet}, \& {Zacharias}}]{zacharias2013}
{Zacharias}, N., {Finch}, C.~T., {Girard}, T.~M., {et~al.} 2013,
  \bibinfo{title}{{The Fourth US Naval Observatory CCD Astrograph Catalog
  (UCAC4)},} \aj, 145, 44, \dodoi{10.1088/0004-6256/145/2/44}

% type= article
\bibitem[{N. {Zacharias} {et~al.}(2004){Zacharias}, {Urban}, {Zacharias},
  {Wycoff}, {Hall}, {Monet}, \& {Rafferty}}]{zacharias2004}
{Zacharias}, N., {Urban}, S.~E., {Zacharias}, M.~I., {et~al.} 2004,
  \bibinfo{title}{{The Second US Naval Observatory CCD Astrograph Catalog
  (UCAC2)},} \aj, 127, 3043, \dodoi{10.1086/386353}

% type= article
\bibitem[{M. {Zerjal} {et~al.}(2023){Zerjal}, {Ireland}, {Crundall},
  {Krumholz}, \& {Rains}}]{zerjal2023}
{Zerjal}, M., {Ireland}, M.~J., {Crundall}, T.~D., {Krumholz}, M.~R., \&
  {Rains}, A.~D. 2023, \bibinfo{title}{{CHRONOSTAR - II. Kinematic age and
  substructure of the Scorpius-Centaurus OB2 association},} \mnras, 519, 3992,
  \dodoi{10.1093/mnras/stac3693}

% type= article
\bibitem[{M. {Zerjal} {et~al.}(2019){Zerjal}, {Ireland}, {Nordlander}, {Lin},
  {Buder}, {Casagrande}, {Cotar}, {de Silva}, {Horner}, {Martell}, {Traven},
  {Zwitter}, \& {Galah Collaboration}}]{zerjal2019}
{Zerjal}, M., {Ireland}, M.~J., {Nordlander}, T., {et~al.} 2019,
  \bibinfo{title}{{The GALAH Survey: lithium-strong KM dwarfs},} \mnras, 484,
  4591, \dodoi{10.1093/mnras/stz296}

% type= article
\bibitem[{M. {Zerjal} {et~al.}(2021){Zerjal}, {Rains}, {Ireland}, {Zhou},
  {Kammerer}, {Wallace}, {Orenstein}, {Nordlander}, {Abbot}, \&
  {Chang}}]{zerjal2021}
{Zerjal}, M., {Rains}, A.~D., {Ireland}, M.~J., {et~al.} 2021,
  \bibinfo{title}{{A spectroscopically confirmed Gaia-selected sample of 318
  new young stars within {\ensuremath{\sim}}200 pc},} \mnras, 503, 938,
  \dodoi{10.1093/mnras/stab513}

% type= article
\bibitem[{R. {Zhang} \& H. {Yuan}(2023){Zhang} \& {Yuan}}]{rzhang2023}
{Zhang}, R., \& {Yuan}, H. 2023, \bibinfo{title}{{Empirical Temperature- and
  Extinction-dependent Extinction Coefficients for the GALEX, Pan-STARRS 1,
  Gaia, SDSS, 2MASS, and WISE Passbands},} \apjs, 264, 14,
  \dodoi{10.3847/1538-4365/ac9dfa}

% type= article
\bibitem[{W. {Zhang} {et~al.}(2022){Zhang}, {Zhang}, {He}, {Song}, {Luo}, \&
  {Zhang}}]{wzhang2022}
{Zhang}, W., {Zhang}, J., {He}, H., {et~al.} 2022, \bibinfo{title}{{Stellar
  Chromospheric Activity Database of Solar-like Stars Based on the LAMOST
  Low-Resolution Spectroscopic Survey},} \apjs, 263, 12,
  \dodoi{10.3847/1538-4365/ac9406}

% type= article
\bibitem[{X. {Zhang} {et~al.}(2023){Zhang}, {Green}, \& {Rix}}]{xzhang2023}
{Zhang}, X., {Green}, G.~M., \& {Rix}, H.-W. 2023, \bibinfo{title}{{Parameters
  of 220 million stars from Gaia BP/RP spectra},} \mnras, 524, 1855,
  \dodoi{10.1093/mnras/stad1941}

% type= article
\bibitem[{Y. {Zhang} {et~al.}(2020){Zhang}, {Tang}, {Chen}, {Pang}, \&
  {Liu}}]{zhang2020}
{Zhang}, Y., {Tang}, S.-Y., {Chen}, W.~P., {Pang}, X., \& {Liu}, J.~Z. 2020,
  \bibinfo{title}{{Diagnosing the Stellar Population and Tidal Structure of the
  Blanco 1 Star Cluster},} \apj, 889, 99, \dodoi{10.3847/1538-4357/ab63d4}

% type= article
\bibitem[{W. {Zhu} \& S. {Dong}(2021){Zhu} \& {Dong}}]{zhu2021}
{Zhu}, W., \& {Dong}, S. 2021, \bibinfo{title}{{Exoplanet Statistics and
  Theoretical Implications},} \araa, 59, 291,
  \dodoi{10.1146/annurev-astro-112420-020055}

% type= article
\bibitem[{G. {Zills} {et~al.}(2024){Zills}, {Criscuoli}, {Bertello}, \&
  {Pevtsov}}]{zills2024}
{Zills}, G., {Criscuoli}, S., {Bertello}, L., \& {Pevtsov}, A. 2024,
  \bibinfo{title}{{Sun-as-a-star variability of H{\ensuremath{\alpha}} and Ca
  II 854.2 nm lines},} Frontiers in Astronomy and Space Sciences, 10, 1328364,
  \dodoi{10.3389/fspas.2023.1328364}

% type= article
\bibitem[{J. {Zorec} \& F. {Royer}(2012){Zorec} \& {Royer}}]{zorec2012}
{Zorec}, J., \& {Royer}, F. 2012, \bibinfo{title}{{Rotational velocities of
  A-type stars. IV. Evolution of rotational velocities},} \aap, 537, A120,
  \dodoi{10.1051/0004-6361/201117691}

% type= article
\bibitem[{C. {Zucker} {et~al.}(2025){Zucker}, {Saydjari}, {Speagle},
  {Schlafly}, {Green}, {Benjamin}, {Peek}, {Edenhofer}, {Goodman}, {Kuhn}, \&
  {Finkbeiner}}]{zucker2025}
{Zucker}, C., {Saydjari}, A.~K., {Speagle}, J.~S., {et~al.} 2025,
  \bibinfo{title}{{A Deep, High-angular-resolution 3D Dust Map of the Southern
  Galactic Plane},} \apj, 992, 39, \dodoi{10.3847/1538-4357/adfbe6}

% type= article
\bibitem[{B. {Zuckerman} {et~al.}(2006){Zuckerman}, {Bessell}, {Song}, \&
  {Kim}}]{zuckerman2006}
{Zuckerman}, B., {Bessell}, M.~S., {Song}, I., \& {Kim}, S. 2006,
  \bibinfo{title}{{The Carina-Near Moving Group},} \apjl, 649, L115,
  \dodoi{10.1086/508060}

% type= article
\bibitem[{B. {Zuckerman} {et~al.}(2011){Zuckerman}, {Rhee}, {Song}, \&
  {Bessell}}]{zuckerman2011}
{Zuckerman}, B., {Rhee}, J.~H., {Song}, I., \& {Bessell}, M.~S. 2011,
  \bibinfo{title}{{The Tucana/Horologium, Columba, AB Doradus, and Argus
  Associations: New Members and Dusty Debris Disks},} \apj, 732, 61,
  \dodoi{10.1088/0004-637X/732/2/61}

% type= article
\bibitem[{B. {Zuckerman} \& I. {Song}(2004){Zuckerman} \&
  {Song}}]{zucksong2004}
{Zuckerman}, B., \& {Song}, I. 2004, \bibinfo{title}{{Young Stars Near the
  Sun},} \araa, 42, 685, \dodoi{10.1146/annurev.astro.42.053102.134111}

% type= article
\bibitem[{B. {Zuckerman} {et~al.}(2001{\natexlab{a}}){Zuckerman}, {Song},
  {Bessell}, \& {Webb}}]{zuckerman2001a}
{Zuckerman}, B., {Song}, I., {Bessell}, M.~S., \& {Webb}, R.~A.
  2001{\natexlab{a}}, \bibinfo{title}{{ZZZ The {\ensuremath{\beta}} Pictoris
  Moving Group},} \apjl, 562, L87, \dodoi{10.1086/337968}

% type= article
\bibitem[{B. {Zuckerman} {et~al.}(2001{\natexlab{b}}){Zuckerman}, {Song}, \&
  {Webb}}]{zuckerman2001b}
{Zuckerman}, B., {Song}, I., \& {Webb}, R.~A. 2001{\natexlab{b}},
  \bibinfo{title}{{Tucana Association},} \apj, 559, 388, \dodoi{10.1086/322305}

% type= article
\bibitem[{B. {Zuckerman} {et~al.}(2013){Zuckerman}, {Vican}, {Song}, \&
  {Schneider}}]{zuckerman2013}
{Zuckerman}, B., {Vican}, L., {Song}, I., \& {Schneider}, A. 2013,
  \bibinfo{title}{{Young Stars near Earth: The Octans-Near Association and
  Castor Moving Group},} \apj, 778, 5, \dodoi{10.1088/0004-637X/778/1/5}

% type= article
\bibitem[{B. {Zuckerman} {et~al.}(2001{\natexlab{c}}){Zuckerman}, {Webb},
  {Schwartz}, \& {Becklin}}]{zuckerman2001c}
{Zuckerman}, B., {Webb}, R.~A., {Schwartz}, M., \& {Becklin}, E.~E.
  2001{\natexlab{c}}, \bibinfo{title}{{The TW Hydrae Association: Discovery of
  T Tauri Star Members Near HR 4796},} \apjl, 549, L233, \dodoi{10.1086/319155}

% type= article
\bibitem[{F. {Zuo} {et~al.}(2024){Zuo}, {Luo}, {Du}, {Li}, {Jones}, {Song},
  {Kong}, \& {Guo}}]{zuo2024}
{Zuo}, F., {Luo}, A.-L., {Du}, B., {et~al.} 2024, \bibinfo{title}{{Projected
  Rotational Velocities for LAMOST Stars with Effective Temperatures Lower than
  9000 K},} \apjs, 271, 4, \dodoi{10.3847/1538-4365/ad1eeb}

\end{thebibliography}

\end{document}